\documentclass{egPubl}
\usepackage{egsgp2021}
 
\SpecialIssuePaper         % uncomment for final version of Computer Graphics Forum, special issue
\usepackage[T1]{fontenc}
\usepackage[utf8]{inputenc}
\DeclareUnicodeCharacter{0301}{\'}
\usepackage{dfadobe}  

\biberVersion
\BibtexOrBiblatex
\usepackage[backend=biber,bibstyle=EG,citestyle=alphabetic,backref=true]{biblatex} 
\electronicVersion
\PrintedOrElectronic

\ifpdf \usepackage[pdftex]{graphicx} \pdfcompresslevel=9
\else \usepackage[dvips]{graphicx} \fi

\usepackage{egweblnk}

\usepackage{graphicx}
\usepackage{amsmath, amssymb, mathtools}
\usepackage[normalem]{ulem}

\usepackage{newtxtext}
\usepackage{newtxmath}
\usepackage{bm}

\usepackage{color}
\usepackage{cleveref}
\usepackage{pgfplots}
\pgfplotsset{compat=newest}
\usepackage{pgfplotstable}
\usepgfplotslibrary{statistics}

\usepackage{microtype}
\usepackage{tabularx}
\usepackage{siunitx,wrapfig}

\newcommand{\R}{\mathbb{R}}
\newcommand{\Sym}{\operatorname{Sym}}

\DeclareMathOperator*{\argmin}{arg\,min}
\DeclareMathOperator{\tr}{tr}

\newtheorem{definition}{Definition}
\newtheorem{notation}{Notation}

\newtheorem{lemma}{Lemma}
\newtheorem{remark}{Remark}

\DeclareGraphicsExtensions{.pdf,.png}

\title{Frame Field Operators}
\author[D.\ Palmer \& O.\ Stein \& J.\ Solomon]
    {\parbox{\textwidth}{\centering D.\ Palmer$^{1}$\orcid{0000-0002-1931-5673}
            and O.\ Stein$^{1}$%\orcid{} 
            and J.\ Solomon$^1$\orcid{0000-0002-7701-7586}
            }
            \\
    {\parbox{\textwidth}{\centering $^1$Massachusetts Institute of Technology, United States}
    }
}
\teaser{
\newcommand{\imageheight}{0.2\linewidth}
\newcommand{\miniheight}{0.1\linewidth}
  \centering
  \begin{tabular}{ccccc}
  \includegraphics[height=\imageheight]{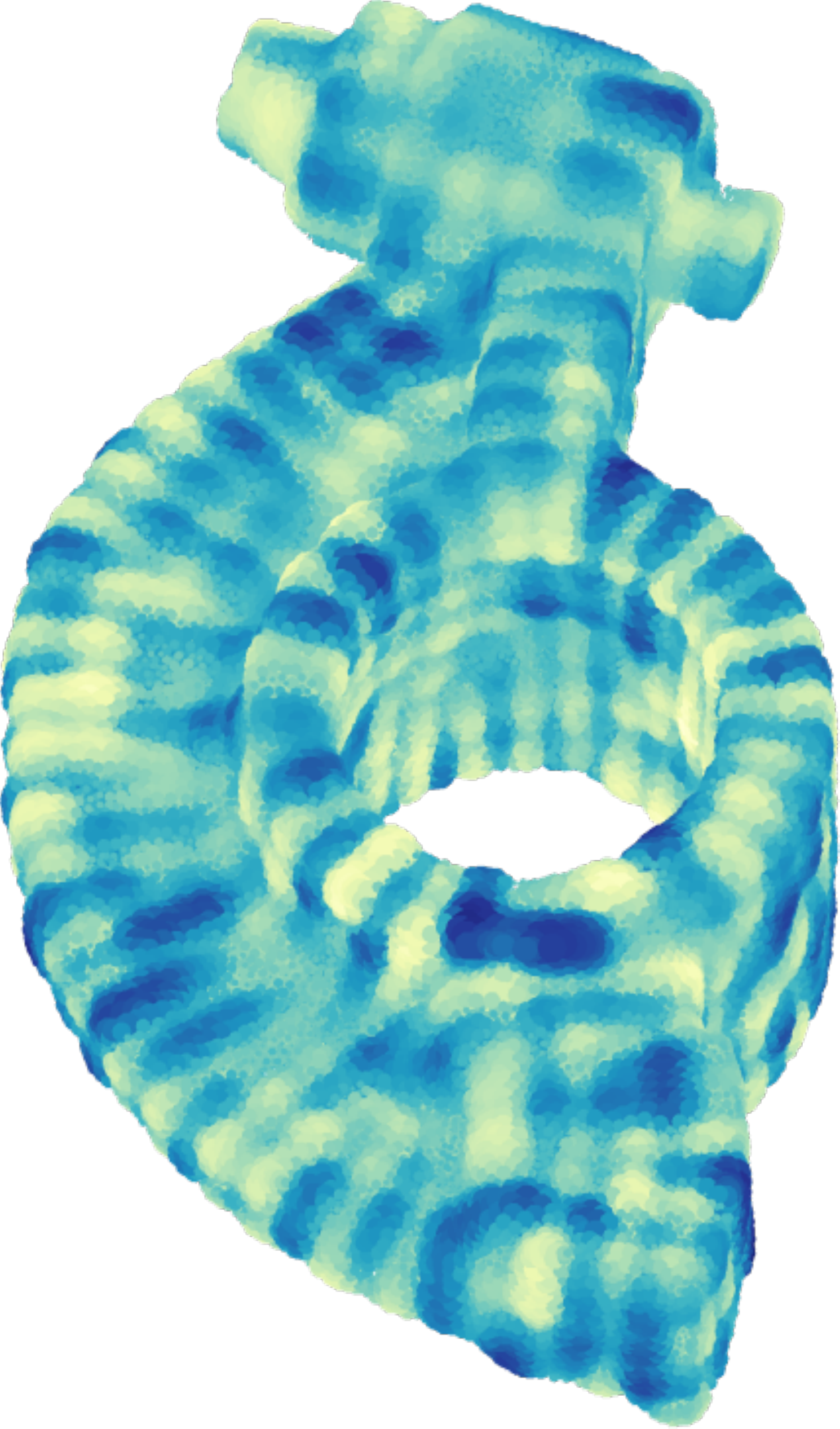} &
  \includegraphics[height=\imageheight]{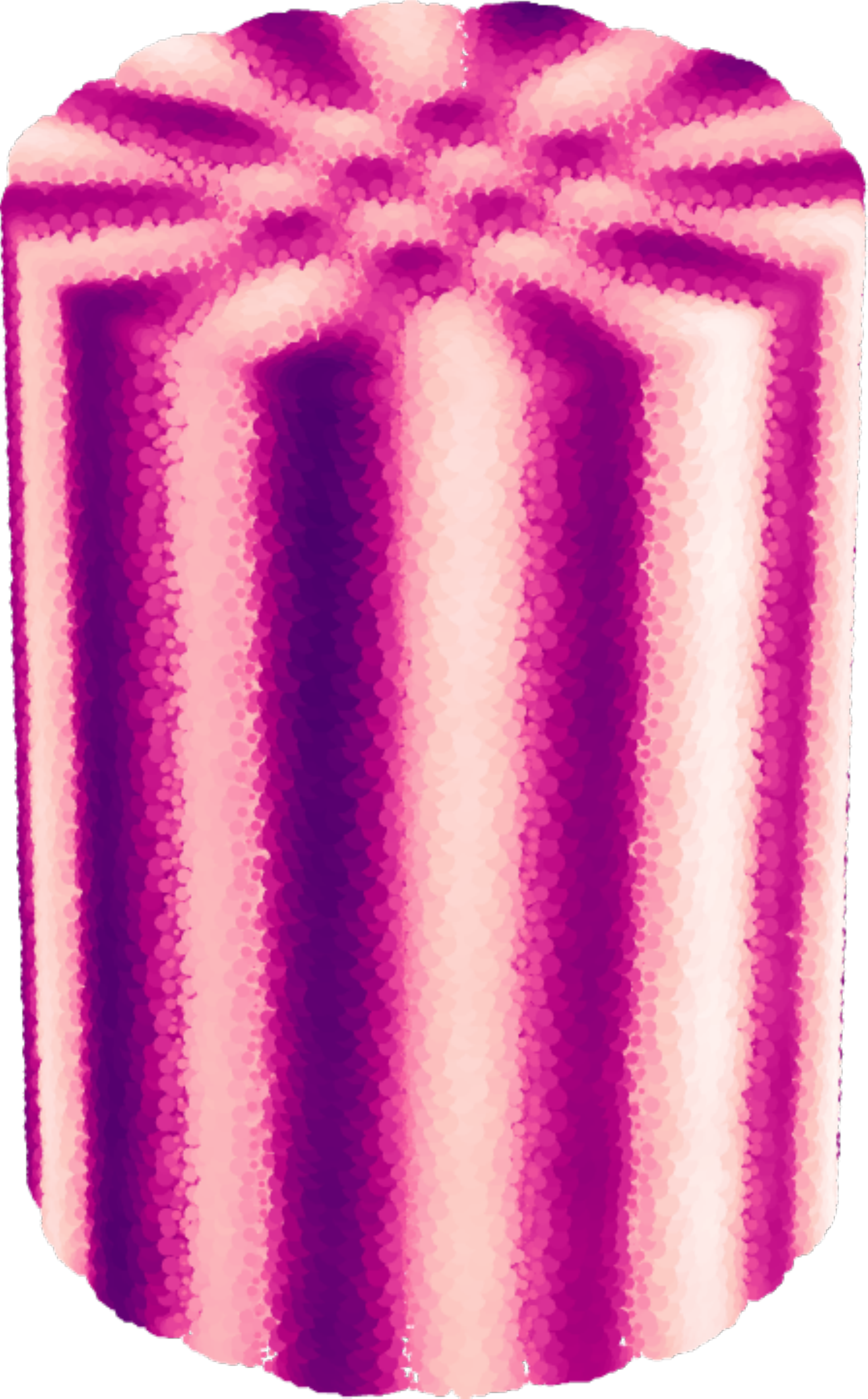} &
  \includegraphics[height=\imageheight]{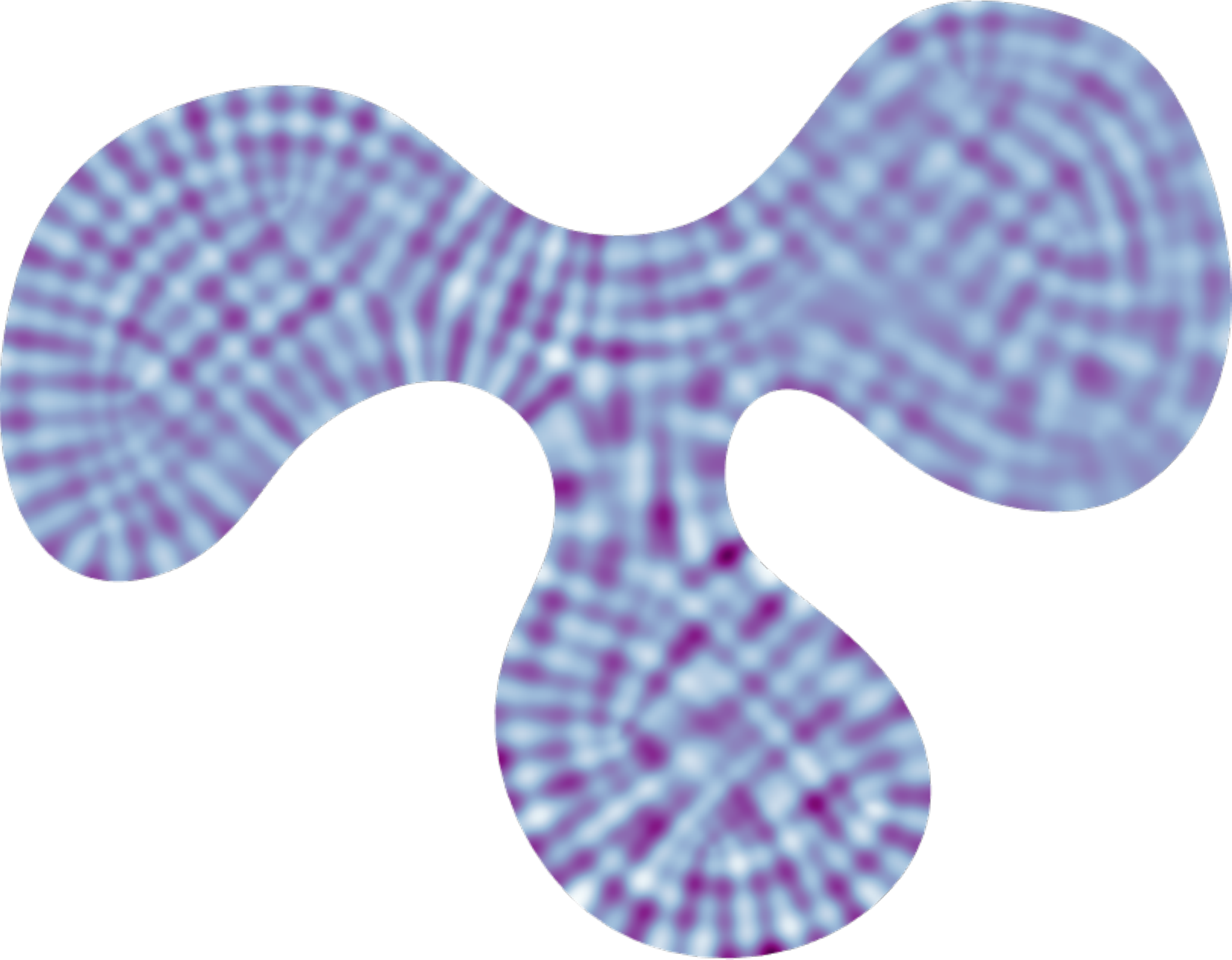} &
  \includegraphics[height=\imageheight]{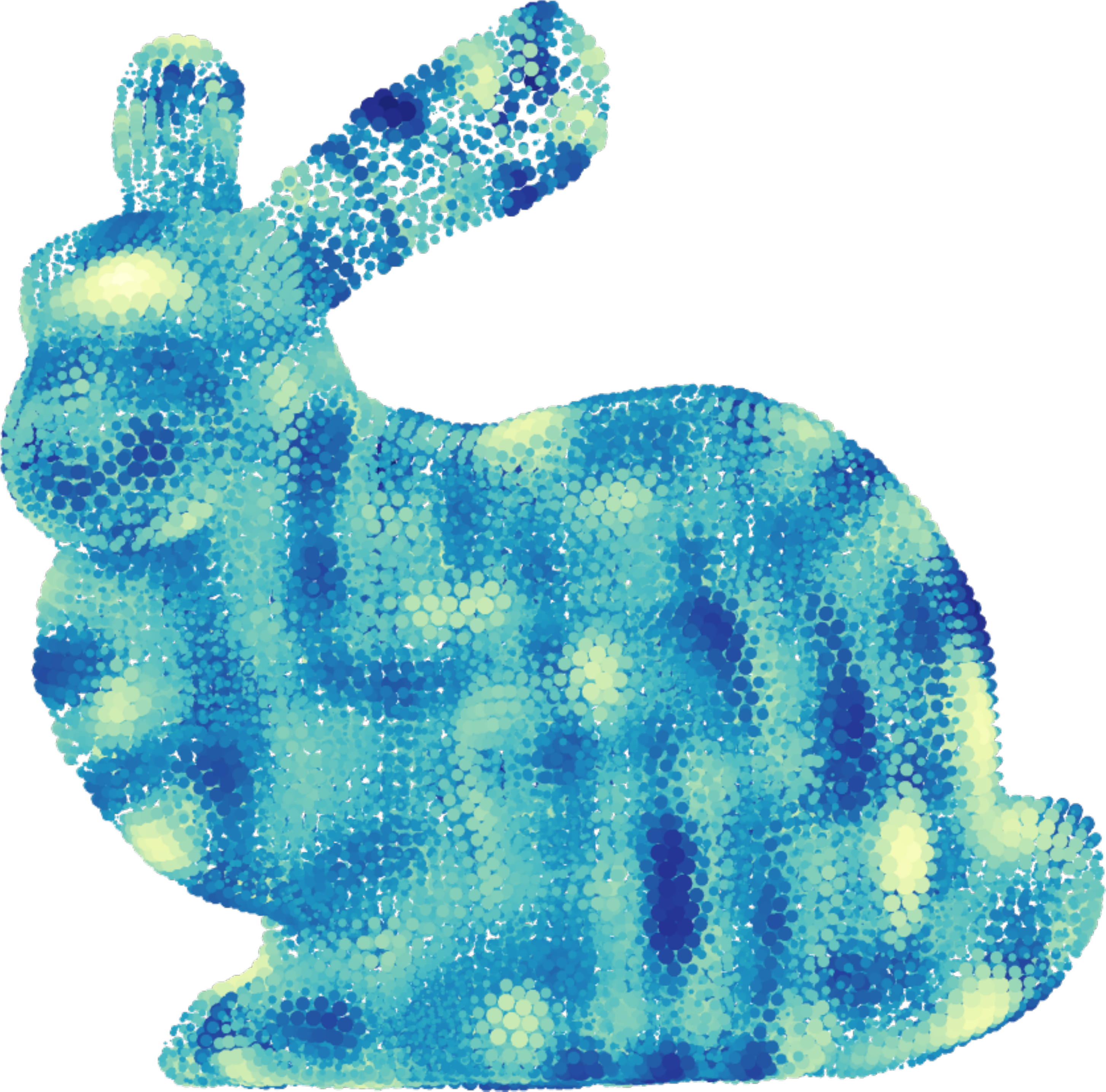} &
  \includegraphics[height=\imageheight]{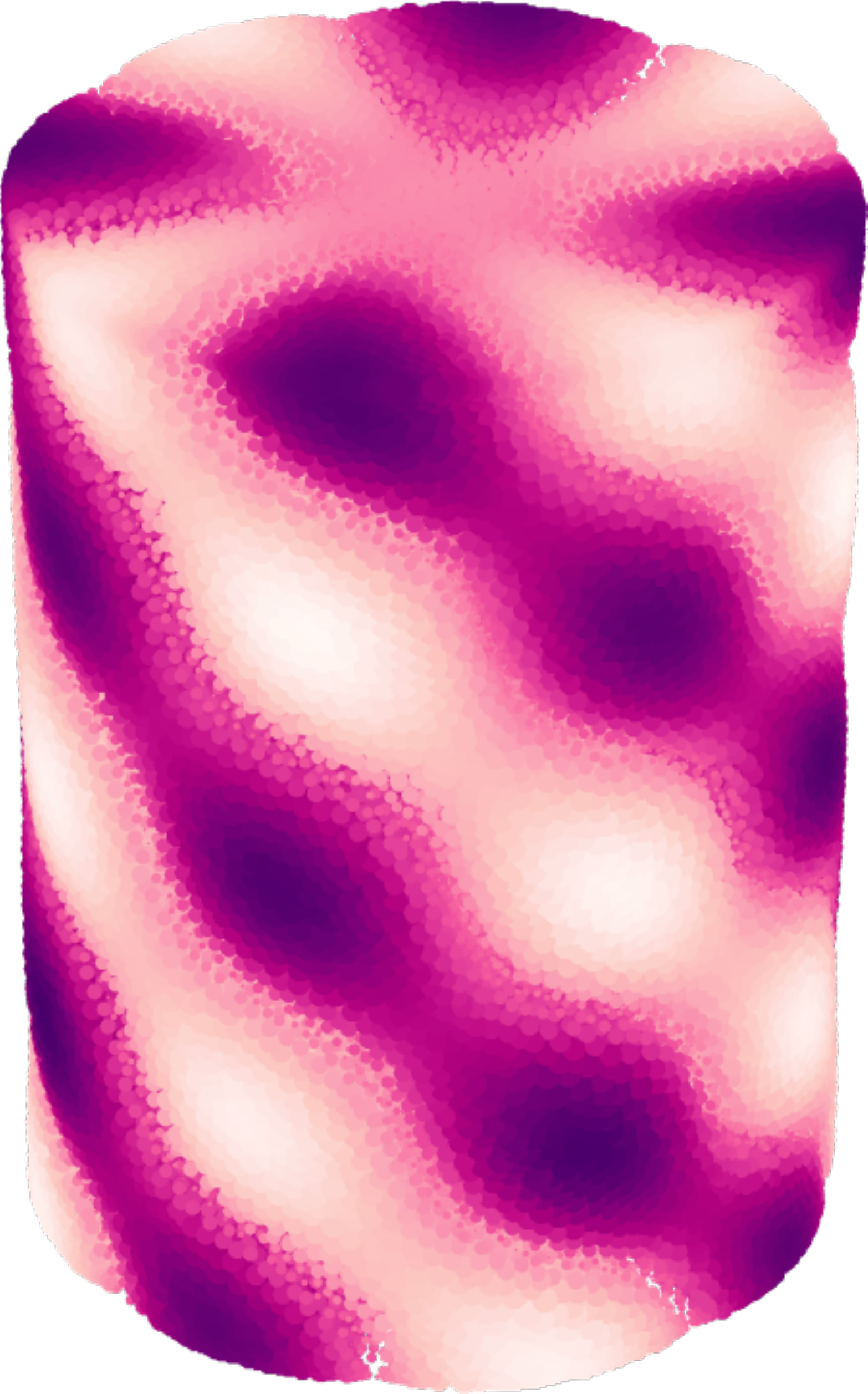} \\
  \includegraphics[height=\miniheight]{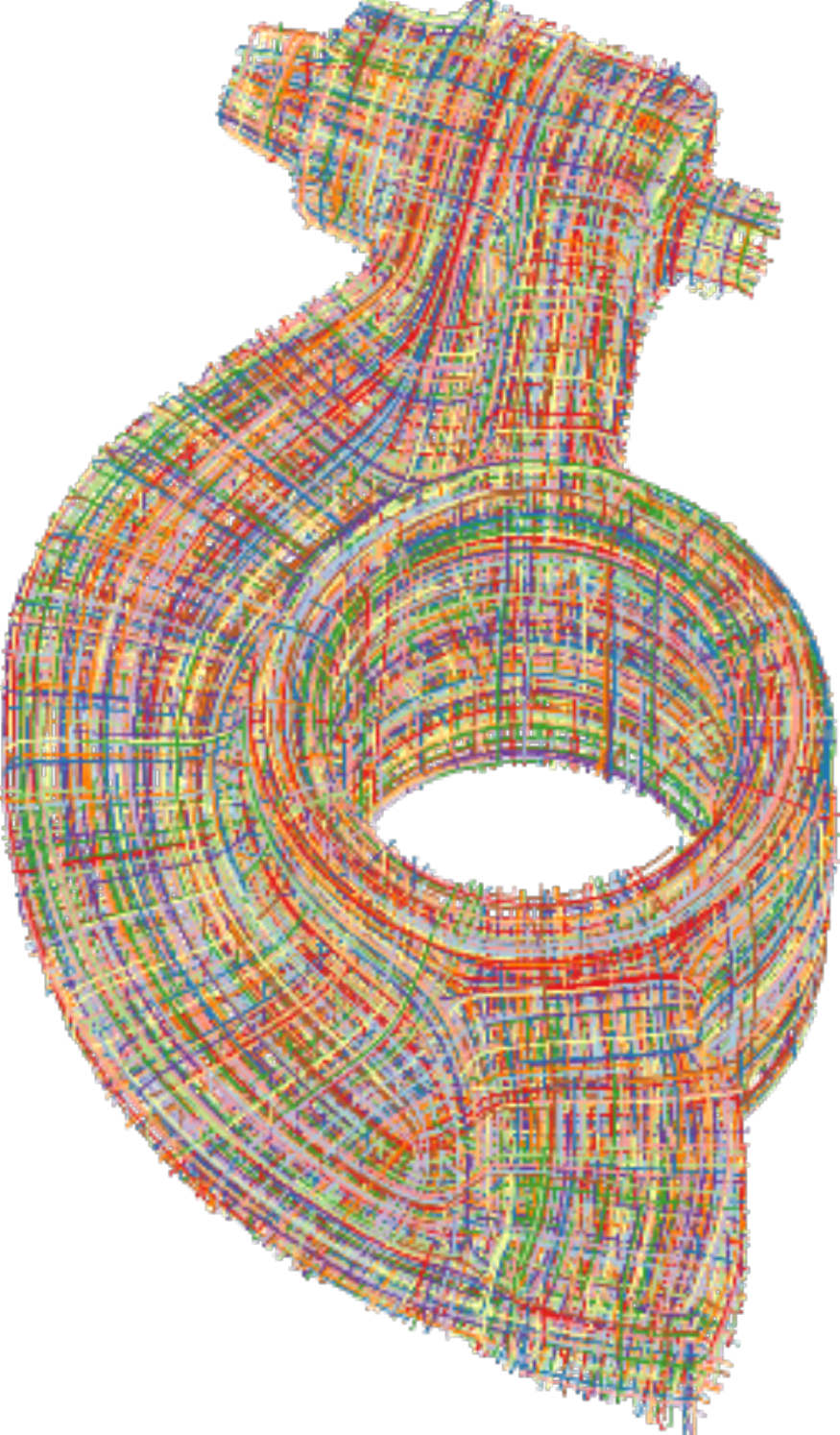} &
  \includegraphics[height=\miniheight]{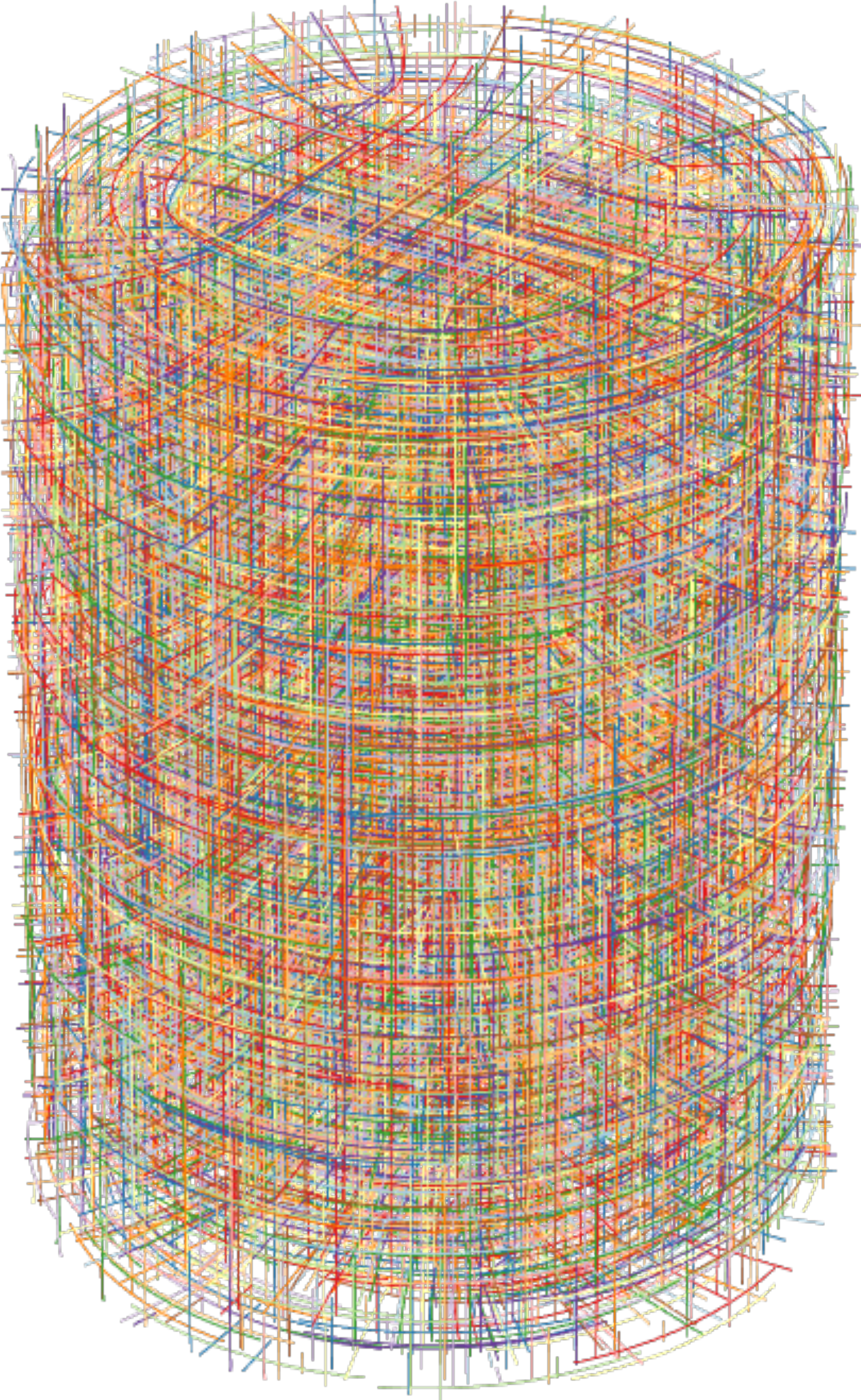} &
  \includegraphics[height=\miniheight]{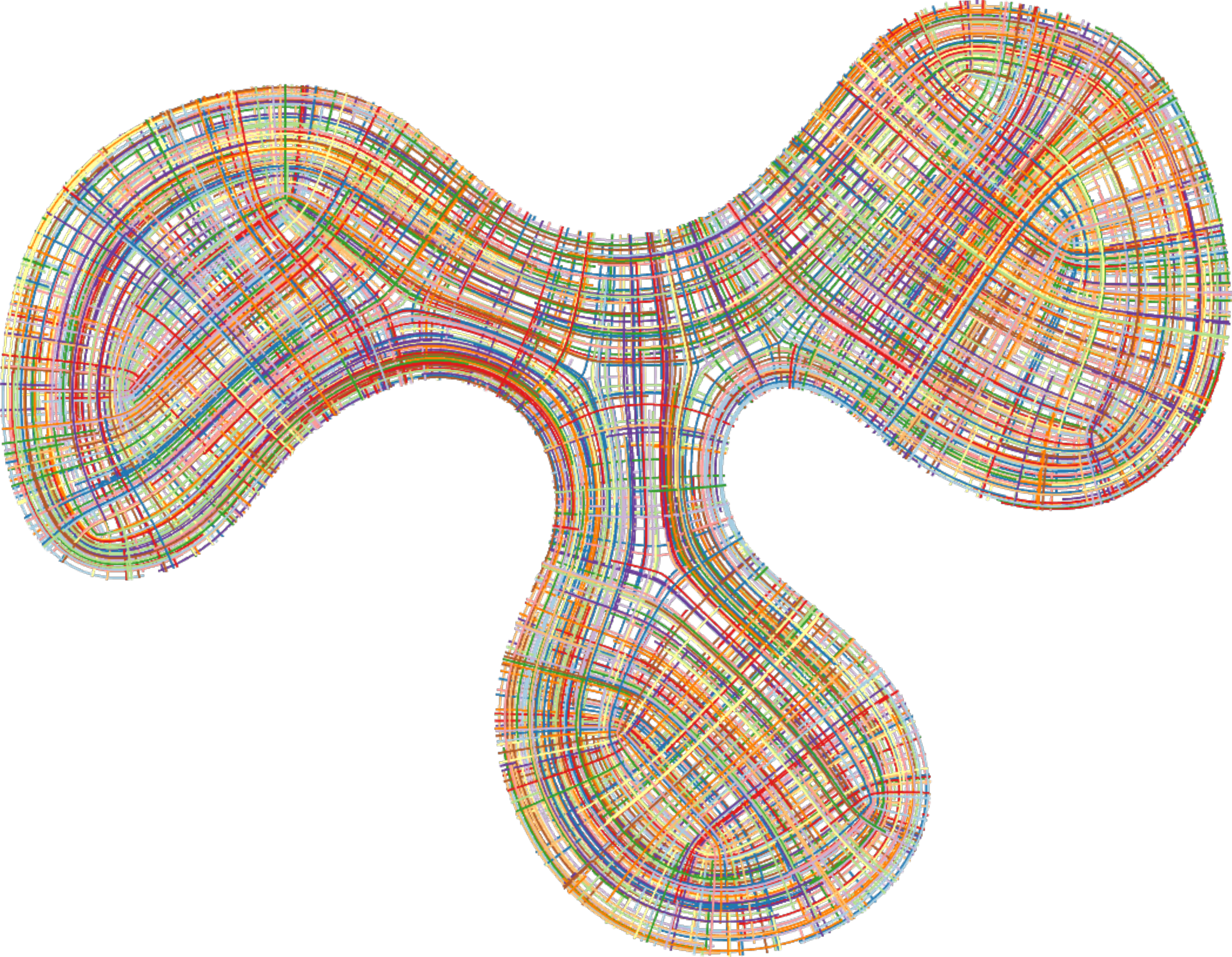} &
  \includegraphics[height=\miniheight]{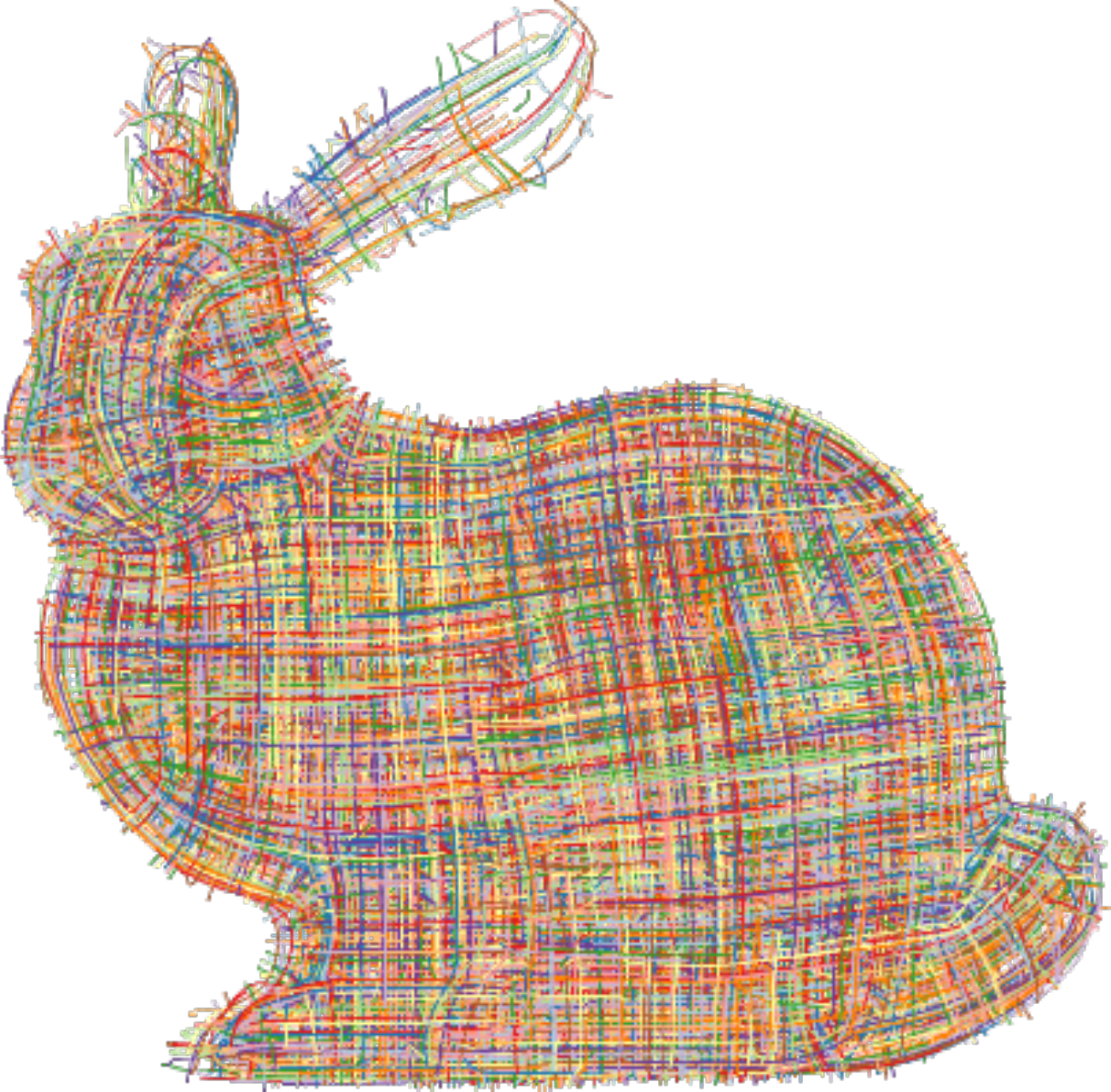} &
  \includegraphics[height=\miniheight]{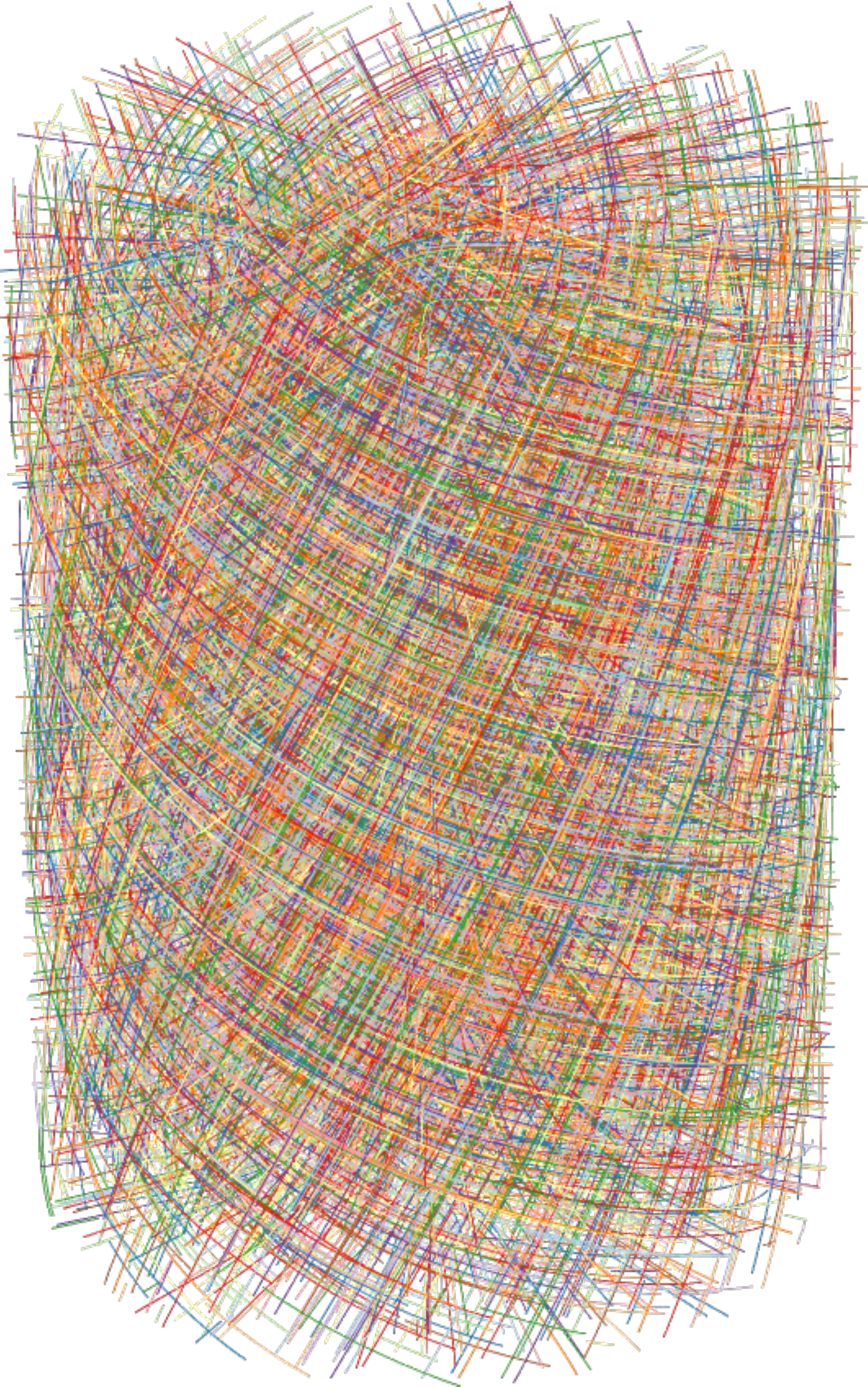}
  \end{tabular}
  \rule{0pt}{150pt}
  \caption{We introduce a \emph{frame field operator} parametrized by a planar or volumetric frame field, which measures function variation in the frame directions. Eigenfunctions of the operator (top) oscillate strongly along frame directions (bottom).}
\label{fig:teaser}
}

\begin{document}
\maketitle

\begin{abstract}
    Differential operators are widely used in geometry processing for problem domains like spectral shape analysis, data interpolation, parametrization and mapping, and meshing. In addition to the ubiquitous cotangent Laplacian, anisotropic second-order operators, as well as higher-order operators such as the Bilaplacian, have been discretized for specialized applications. In this paper, we study a class of operators that generalizes the fourth-order Bilaplacian to support anisotropic behavior. The anisotropy is parametrized by a \emph{symmetric frame field}, first studied in connection with quadrilateral and hexahedral meshing, which allows for fine-grained control of local directions of variation. We discretize these operators using a mixed finite element scheme, verify convergence of the discretization, study the behavior of the operator under pullback, and present potential applications.
\end{abstract}

\section{Introduction}

Differential operators and their discrete counterparts are widely used in geometry processing algorithms for data smoothing, interpolation, simulation, mapping, and numerous other applications. Starting from the classical Laplace-Beltrami operator \(\Delta\) on surfaces, specialized applications have demanded a wider class of operators. For example, higher-order operators such as the Bilaplacian \(\Delta^2\) may be used where the second-order smoothness of Laplacian solutions is insufficient. While the classical Laplacian is isotropic, operators incorporating anisotropy can be used for pattern generation, computing specialized distances, fluid simulation, and other applications.

Although anisotropic operators studied in geometry processing have generally been second-order, higher-order operators admit a richer variety of anisotropic behaviors. In particular, the coefficients of a fourth-order operator form a fourth-order symmetric tensor field, which can have symmetries that are not representable in second order. This allows a fourth-order operator to express, for example, multiple equally-preferred directions of variation at each point.

In this work, we construct anisotropic operators from the fourth-order \emph{symmetric frame fields} employed in quadrilateral and hexahedral meshing, which encode multiple directions at each point in a domain. Symmetric frame fields are tensor fields that have local quadrilateral or octahedral symmetry. While the only second-order tensors with this symmetry are scaled identity tensors---which are actually \emph{isotropic}---fourth-order \emph{octahedral} and more general \emph{odeco} tensor fields are nondegenerate and have been applied widely in quadrilateral and hexahedral meshing. Because of their relationship to meshing, many algorithms exist for designing and manipulating these fields.

Through a variational framework, we construct a family of \emph{frame field operators}---fourth-order elliptic differential operators acting on scalar functions, which measure variation in the directions of their associated frame fields. We define these operators in both planar domains---where they are related to the \emph{orthotropic thin plate} operators of elastic physics---and volumetric domains. Our frame field operators provide a link between the realms of frame field design and anisotropic elliptic operators.

We discretize frame field operators using a mixed finite element approach. While the linear finite elements commonly used in surface and volumetric geometry processing are well-suited to the discretization of second-order partial differential equations---for which the weak and variational forms only involve first derivatives---higher-order operators do not fit as neatly into this framework. Recently, the method of \emph{mixed finite elements} has shown promise in discretizing the Bilaplacian on planar domains and triangulated surfaces. Here, we apply an analogous method to our much larger class of fourth-order frame field operators on both planar and volumetric domains. Our discretization naturally generalizes recent discretizations of the Bilaplacian: a single parameter controls the degree of anisotropy, and when it is set to one, we recover the Bilaplacian. Moreover, we expand the palette of boundary conditions for discrete fourth-order operators by showing how to impose Neumann boundary conditions variationally, by restricting the space of Lagrange multipliers. We evaluate our discrete frame field operators numerically, study their properties, and outline a range of potential applications.

\paragraph*{Contributions}

In this work, we
\begin{itemize}
    \item introduce a new class of \emph{frame field operators} parametrized by planar and volumetric frame fields, which measure function variation along frame directions;
    \item design a mixed finite element discretization for frame field operators, including a natural way to impose desired boundary conditions;
    \item empirically validate the expected convergence and behavior of discrete frame field operators;
    \item show examples of PDE solutions, eigenfunctions, and assorted potential applications; and
    \item provide a \textsc{matlab} implementation in supplemental material.
\end{itemize}

\begin{figure*}
\centering
\newcommand{\imagewidth}{0.23\textwidth}
\begin{tabularx}{\textwidth}{c|ccc}
    \includegraphics[width=\imagewidth]{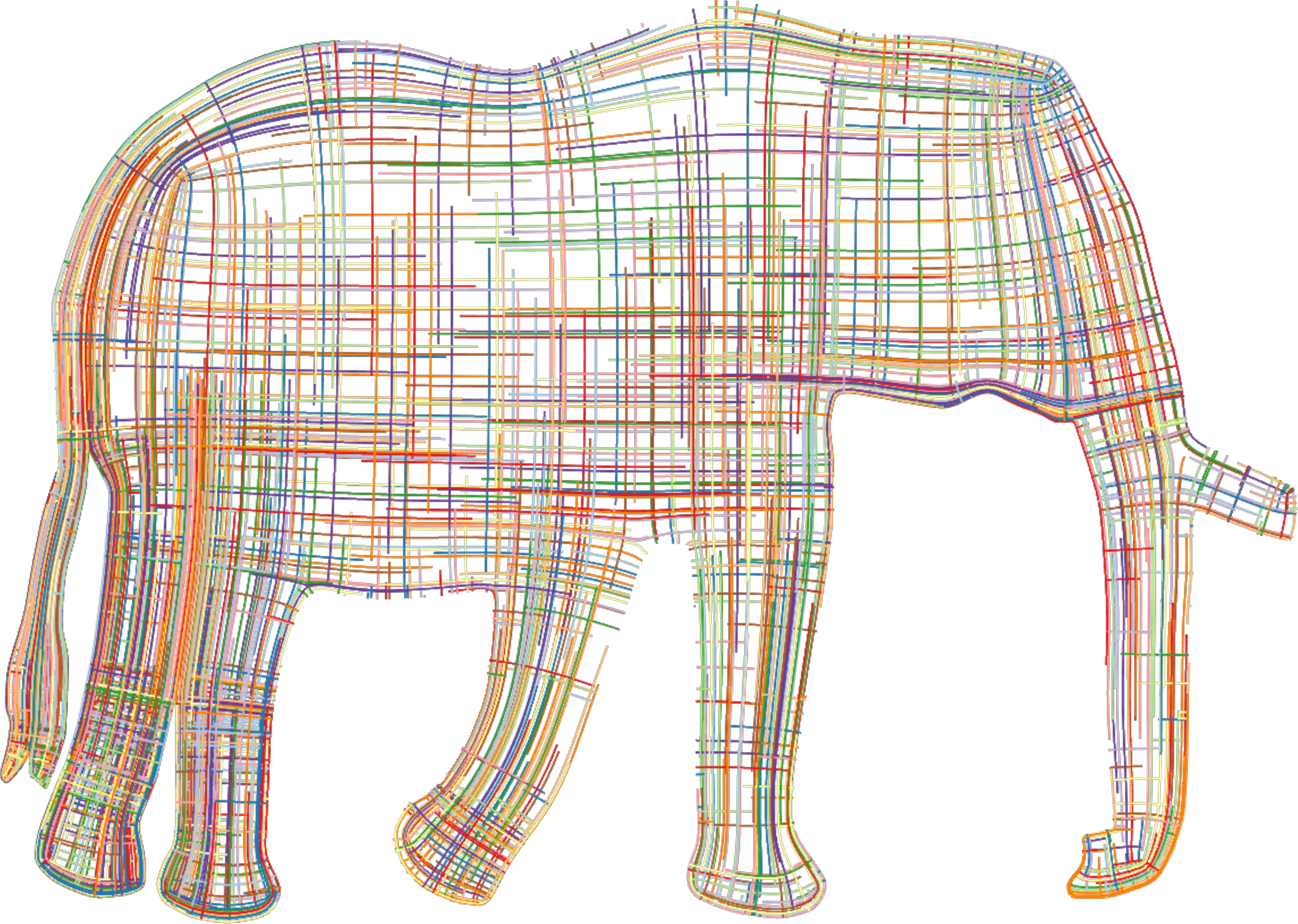} &
    \includegraphics[width=\imagewidth]{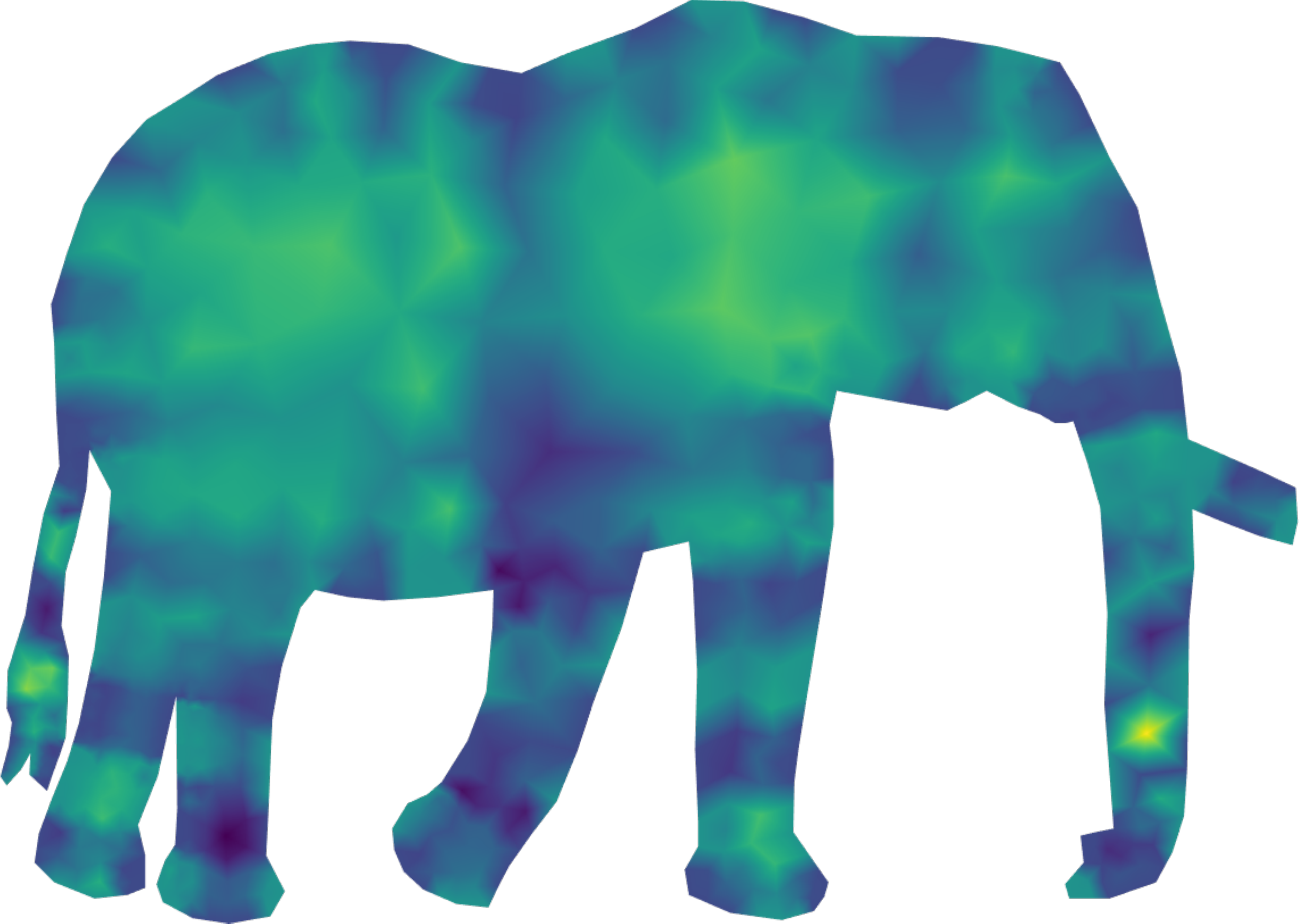} &
    \includegraphics[width=\imagewidth]{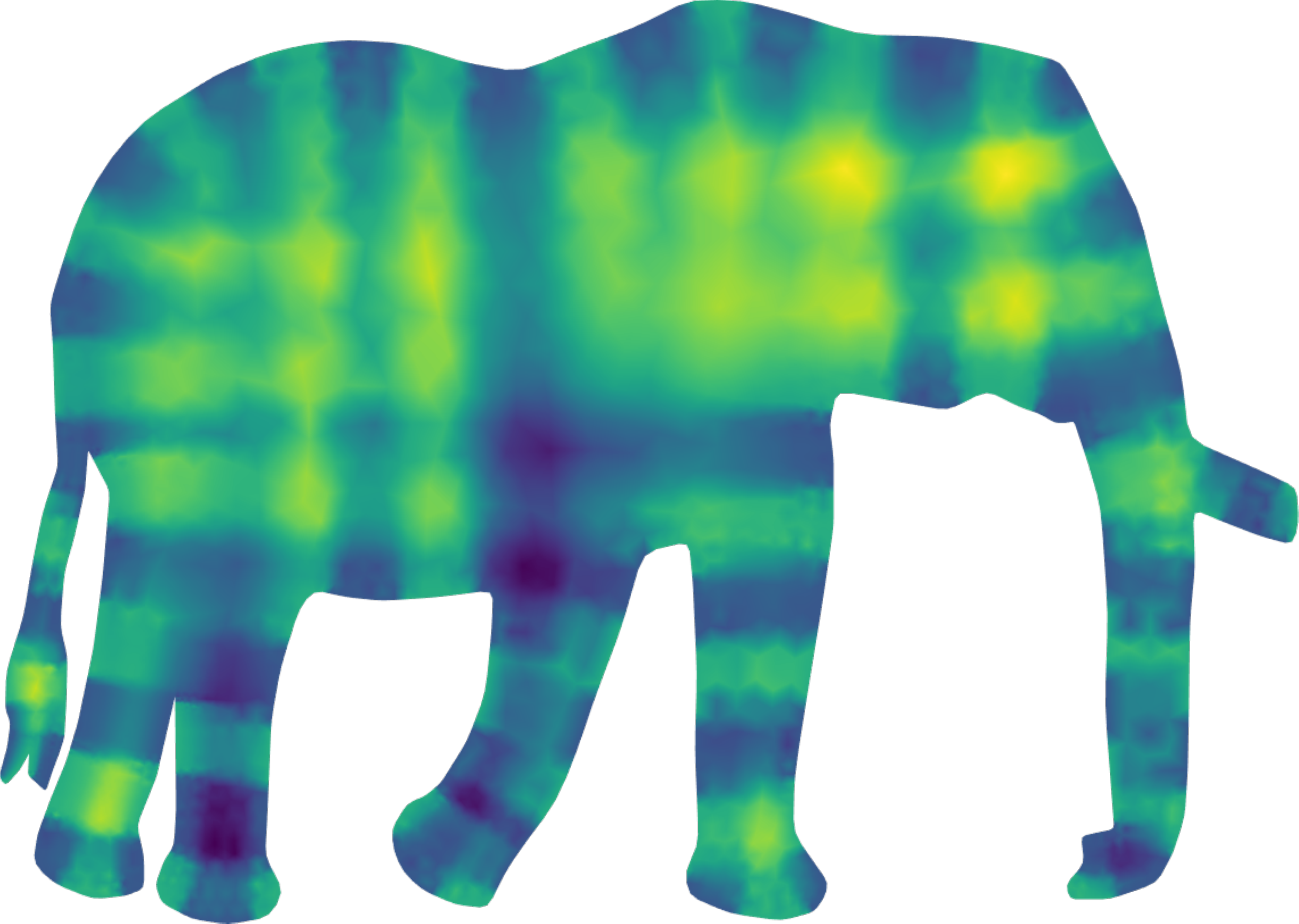} &
    \includegraphics[width=\imagewidth]{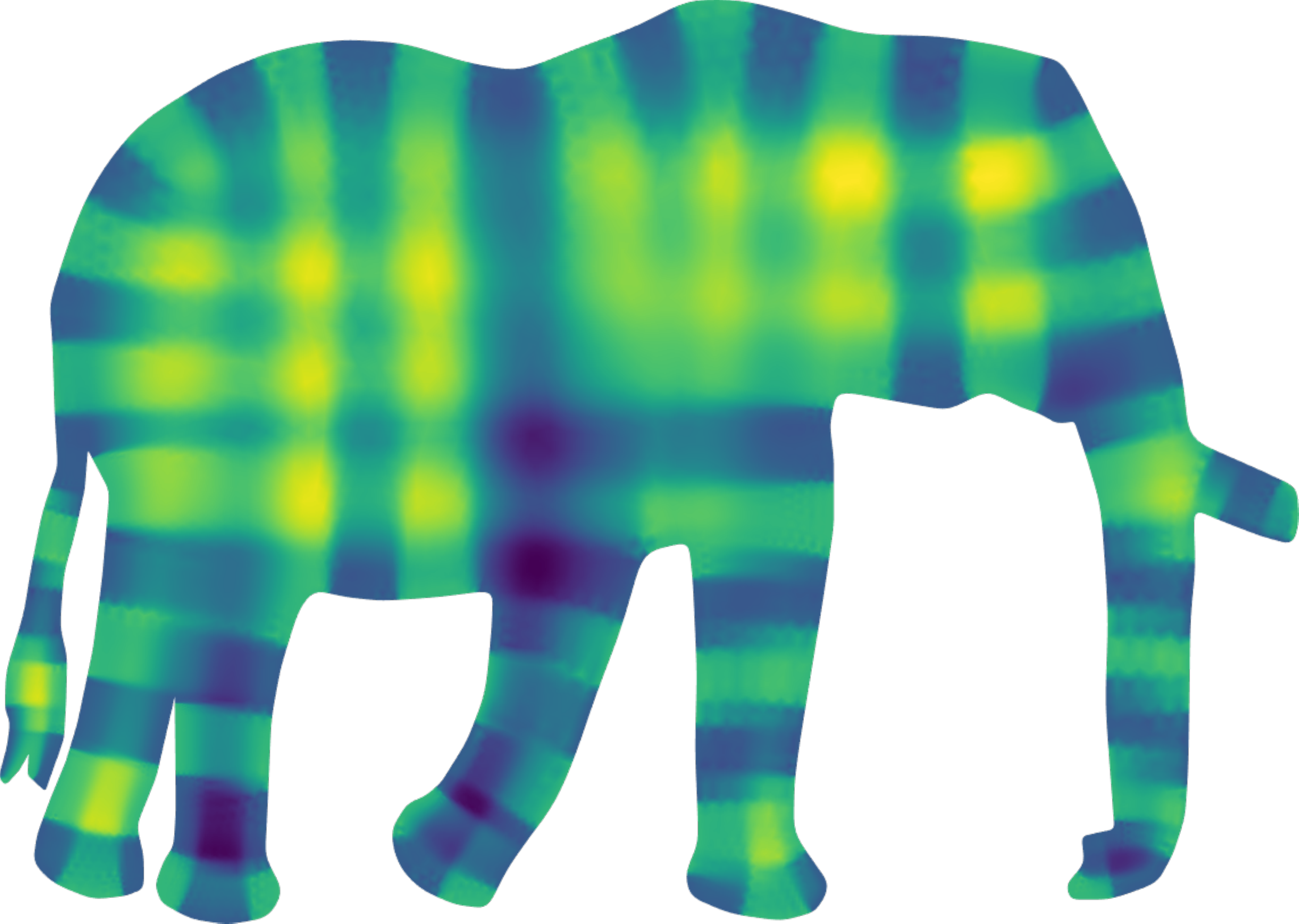} \\
  & $L_{\text{mean}} = \num{1.851e-02}$ & $L_{\text{mean}} = \num{9.255e-03}$ & $L_{\text{mean}} = \num{4.636e-03}$ \\
  & \includegraphics[width=\imagewidth]{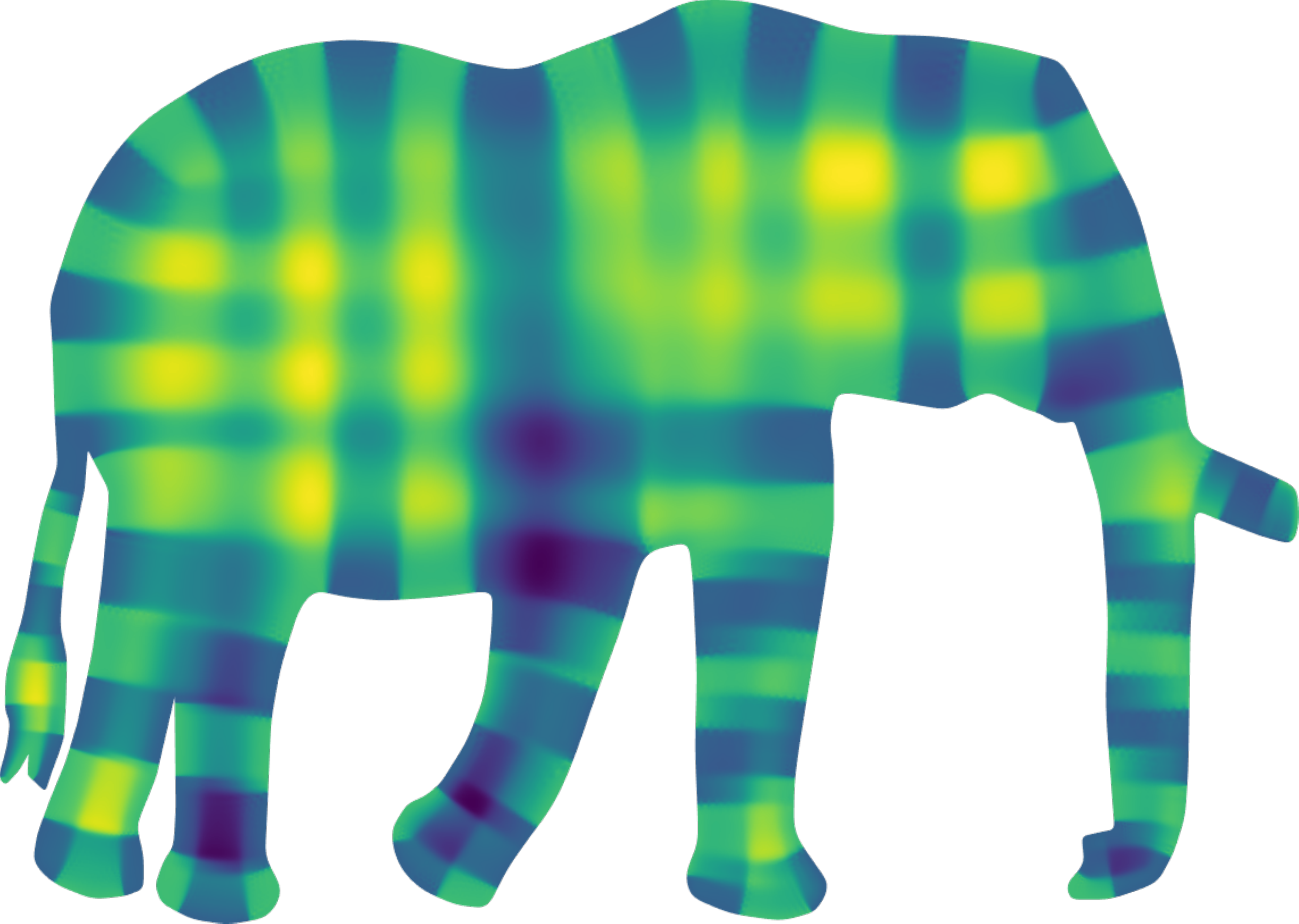} &
    \includegraphics[width=\imagewidth]{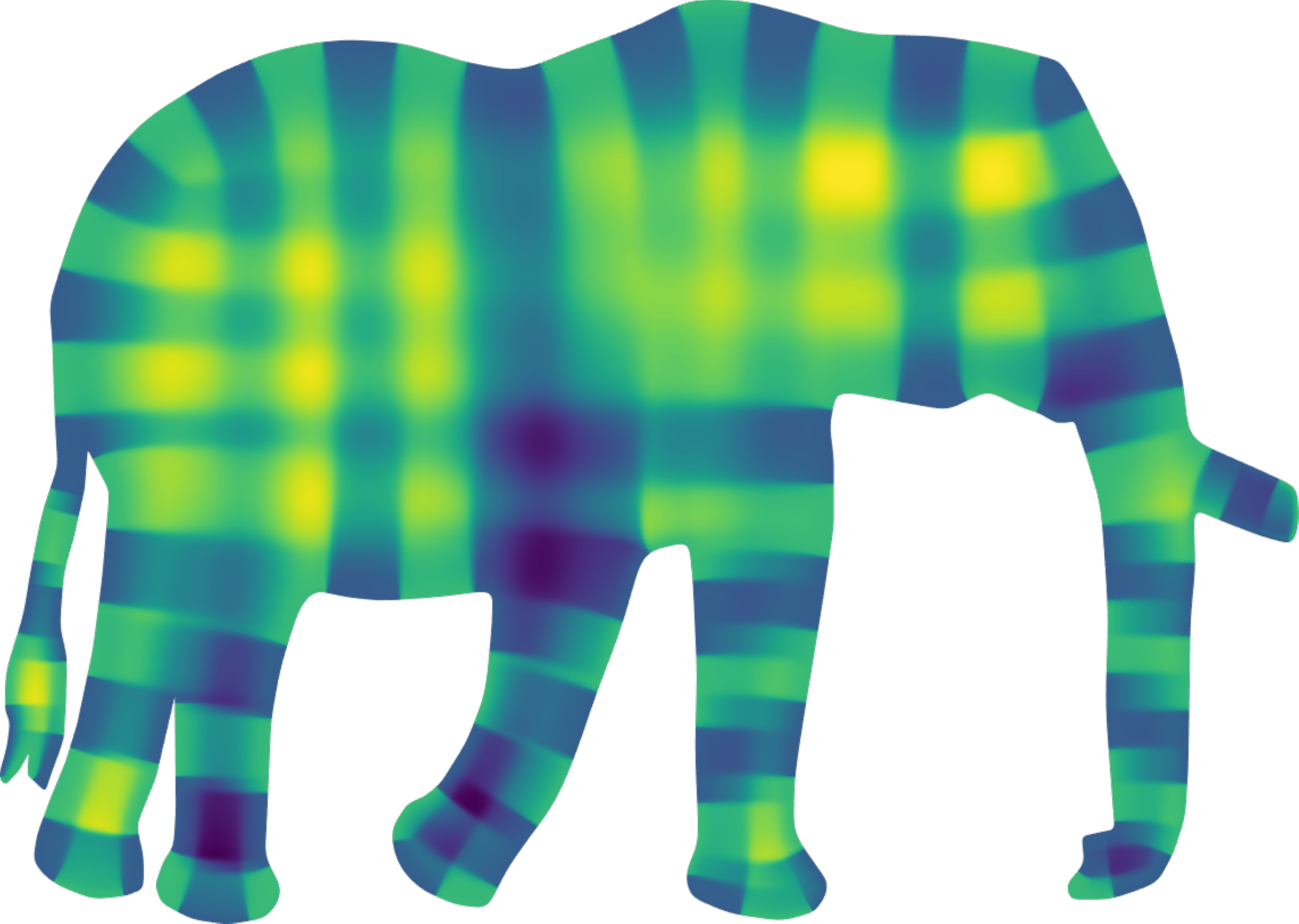} &
    \includegraphics[width=\imagewidth]{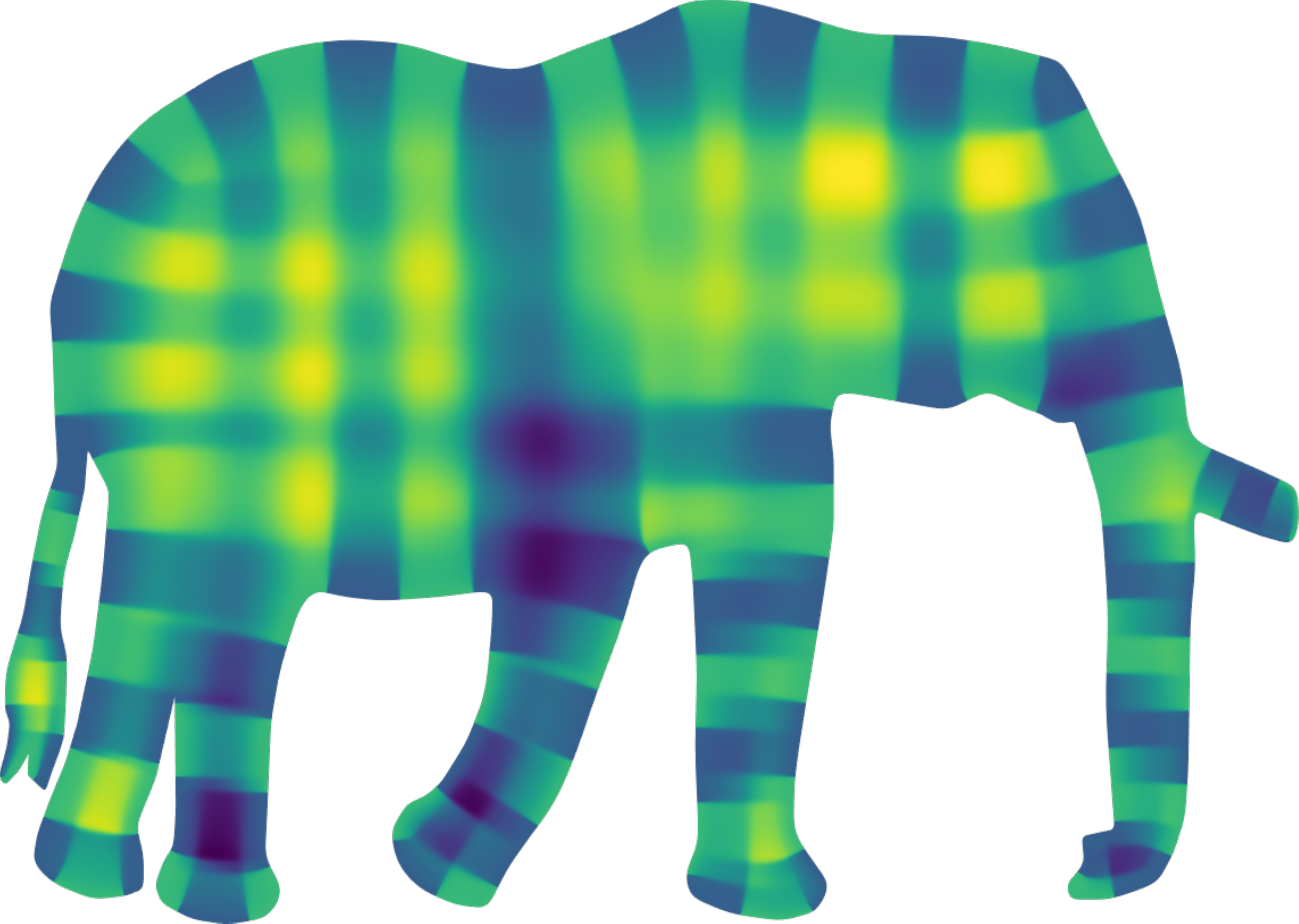} \\
  & $L_{\text{mean}} = \num{2.322e-03}$ & $L_{\text{mean}} = \num{1.162e-03}$ & $L_{\text{mean}} = \num{5.814e-04}$ \\
    \includegraphics[width=\imagewidth]{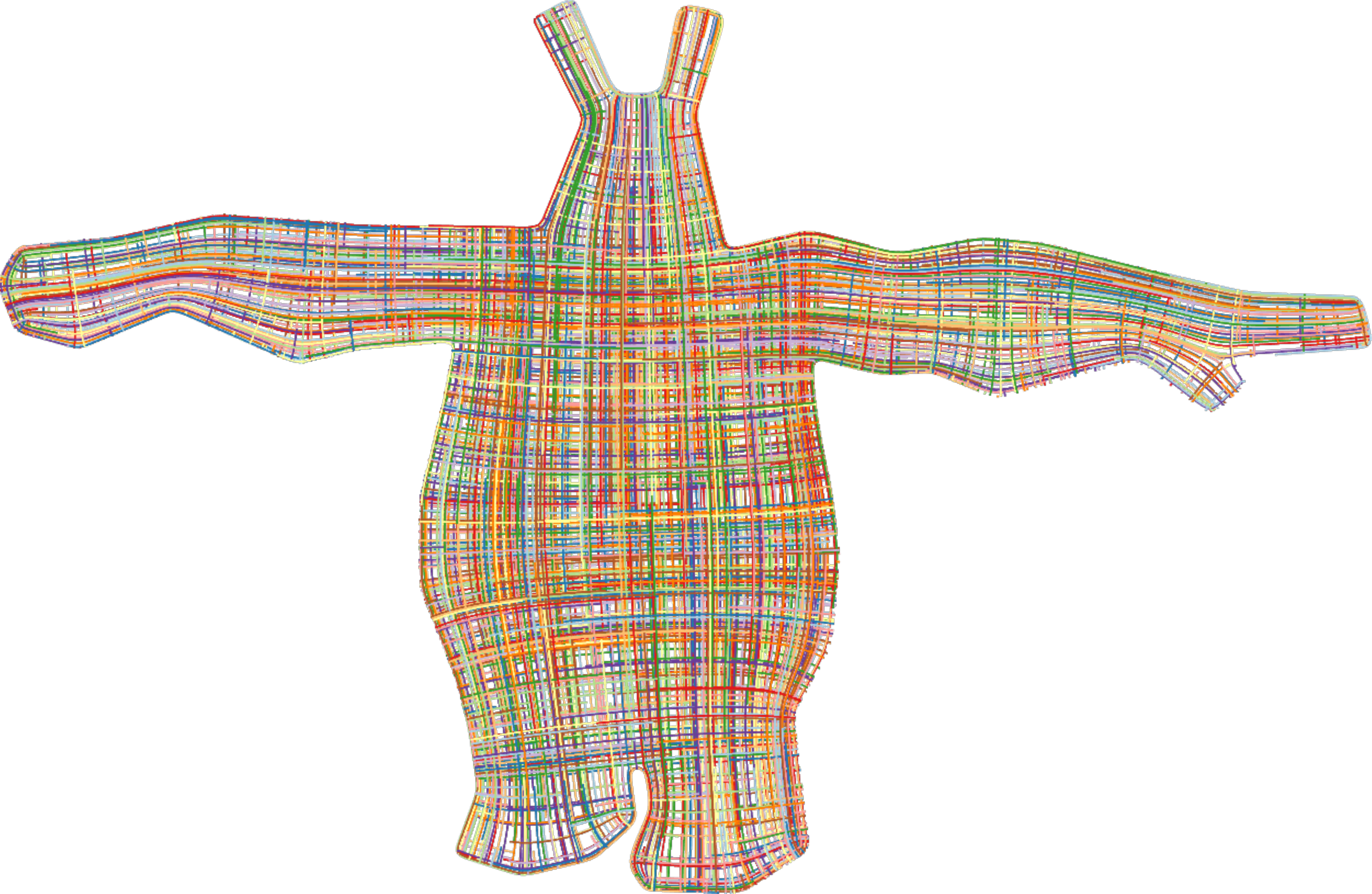} &
    \includegraphics[width=\imagewidth]{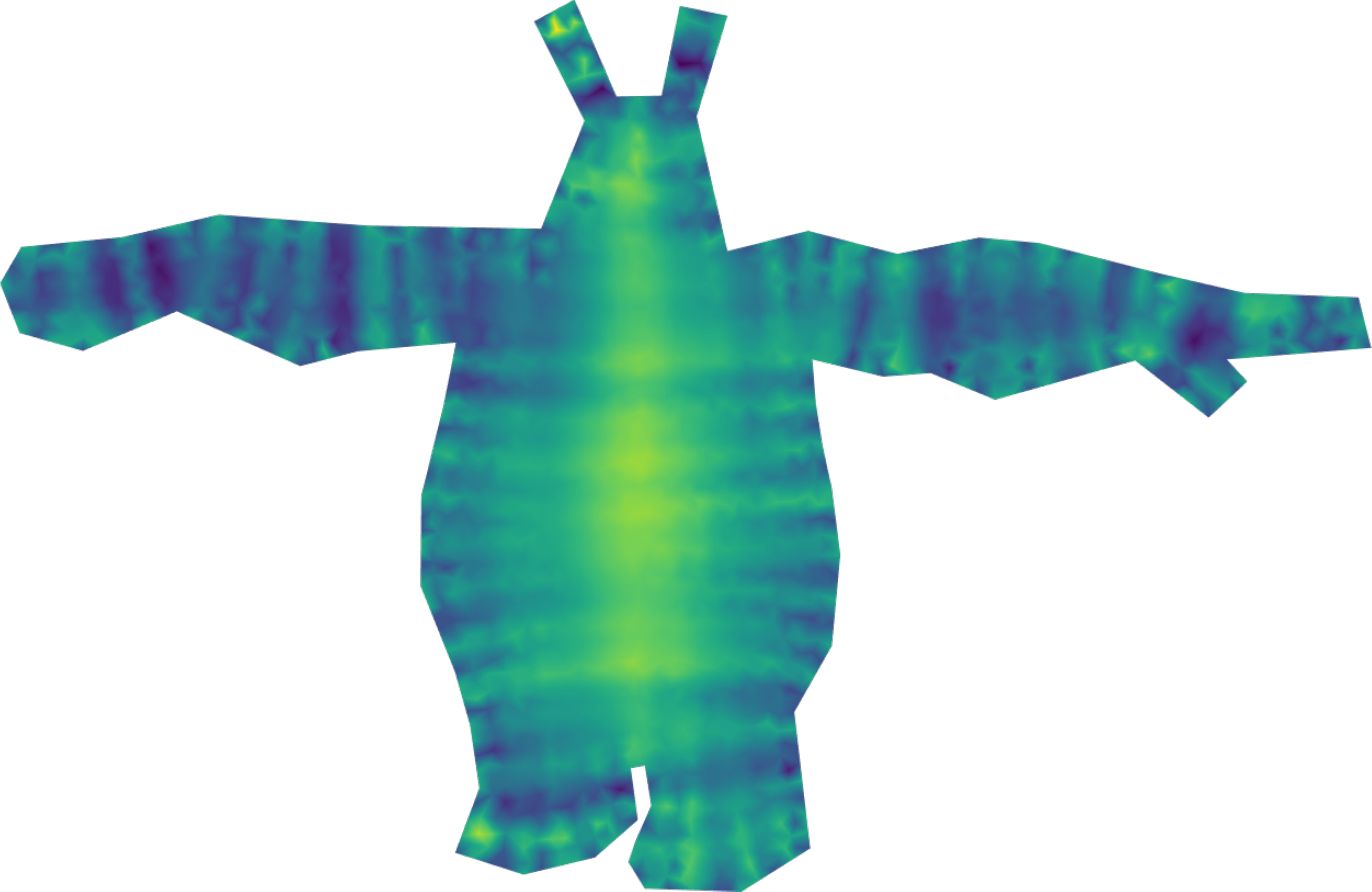} &
    \includegraphics[width=\imagewidth]{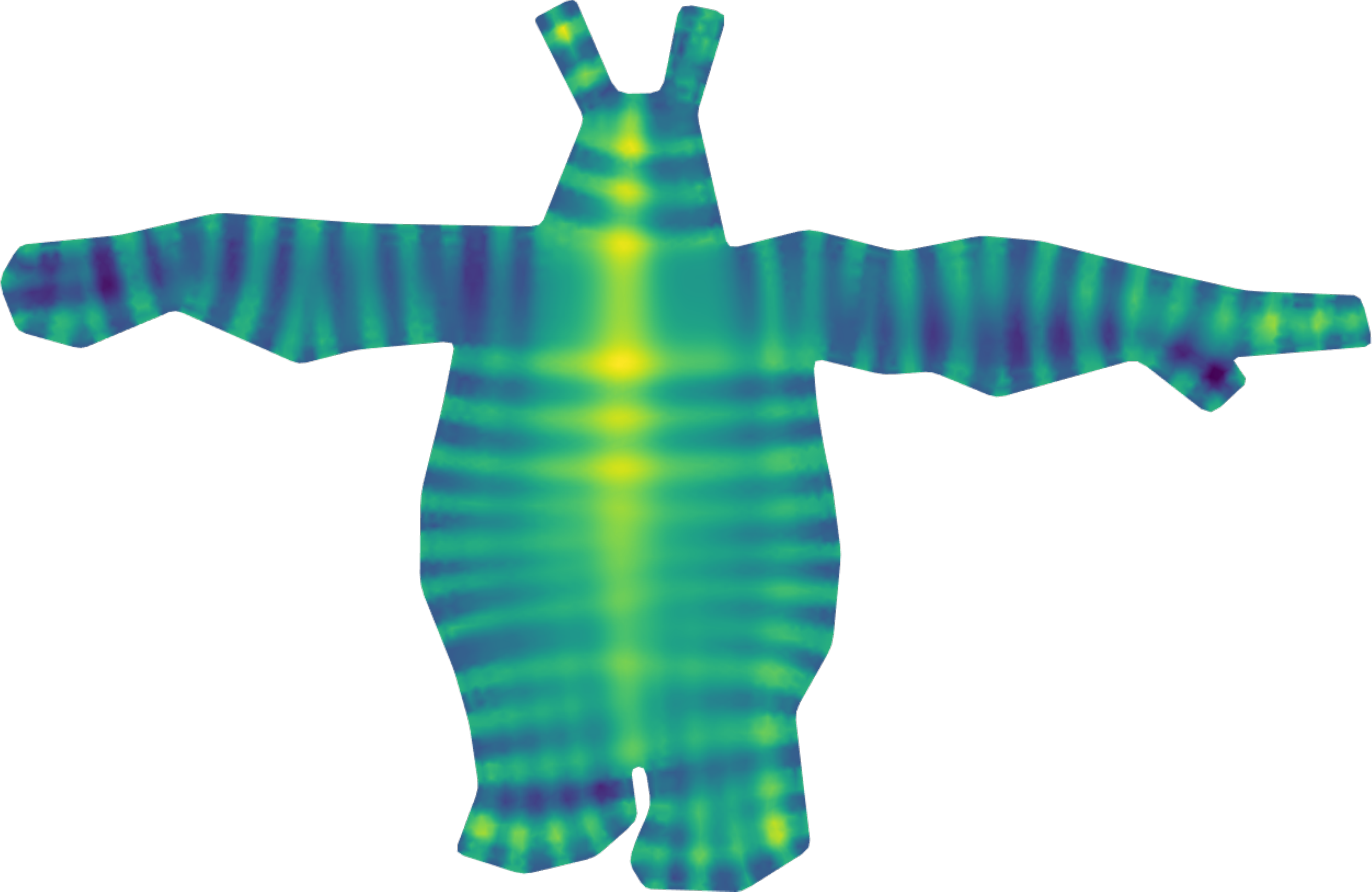} &
    \includegraphics[width=\imagewidth]{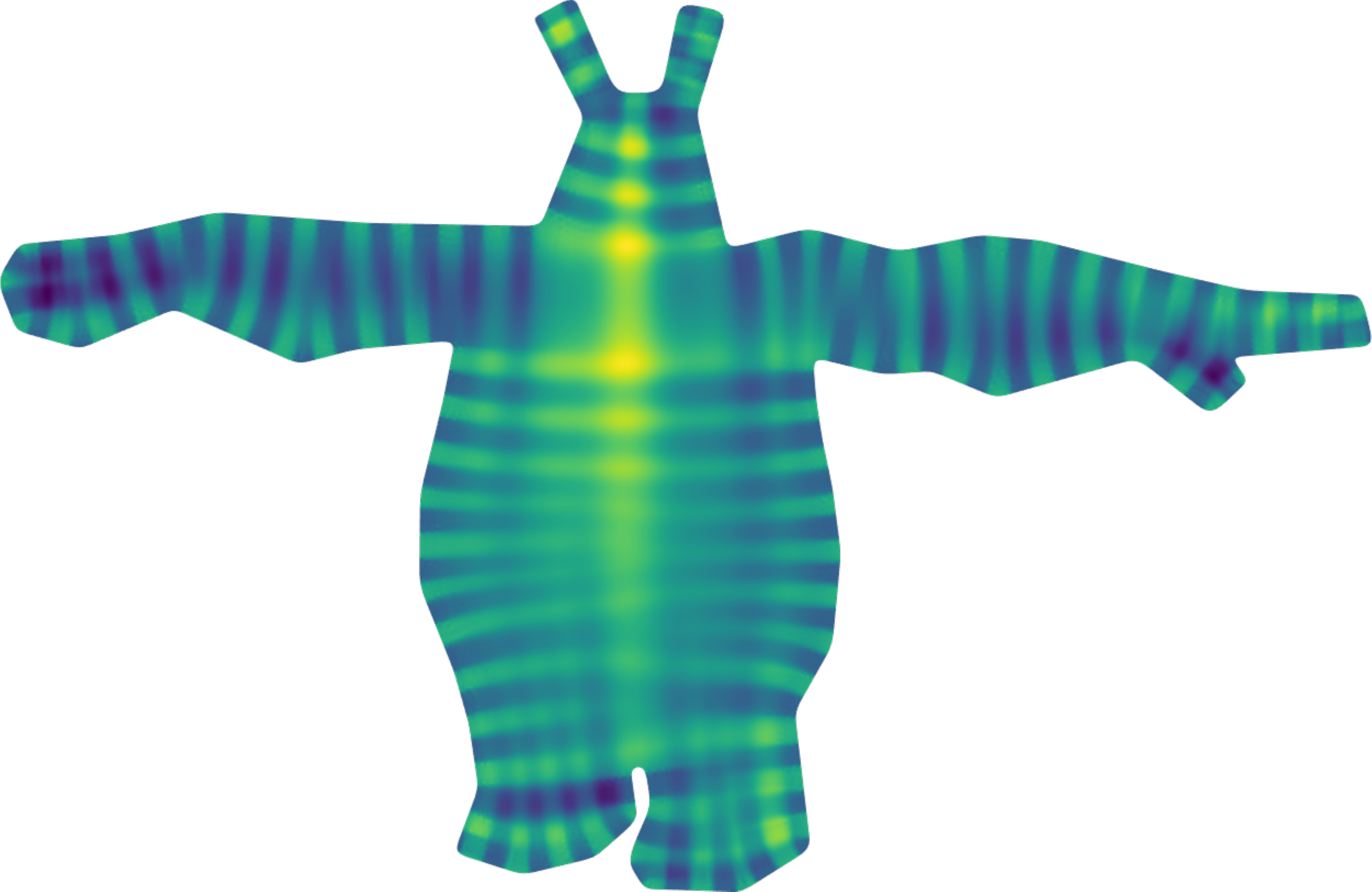} \\
  & $L_{\text{mean}} = \num{1.199e-01}$ & $L_{\text{mean}} = \num{5.890e-02}$ & $L_{\text{mean}} = \num{2.932e-02}$ \\
  & \includegraphics[width=\imagewidth]{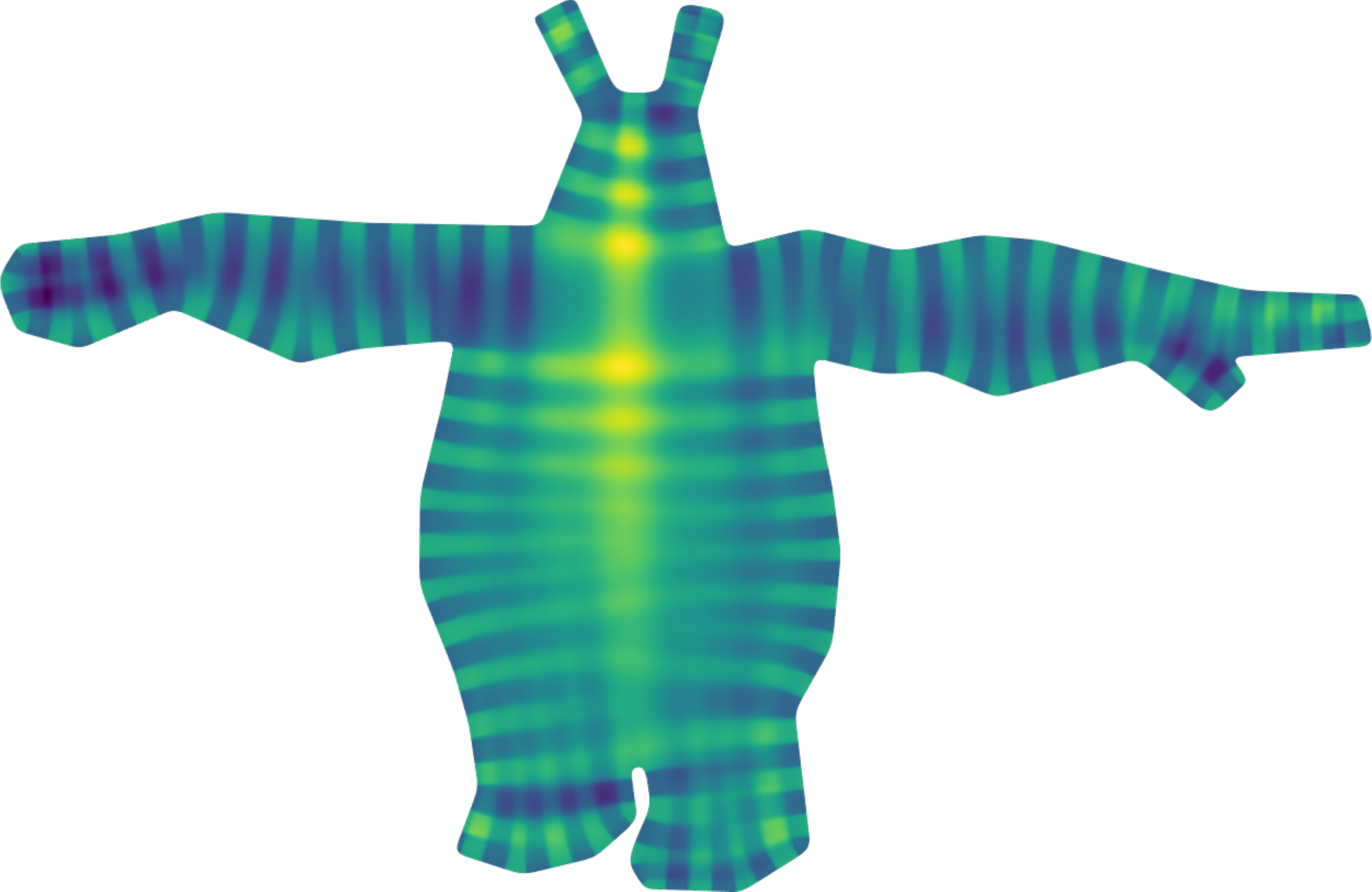} &
    \includegraphics[width=\imagewidth]{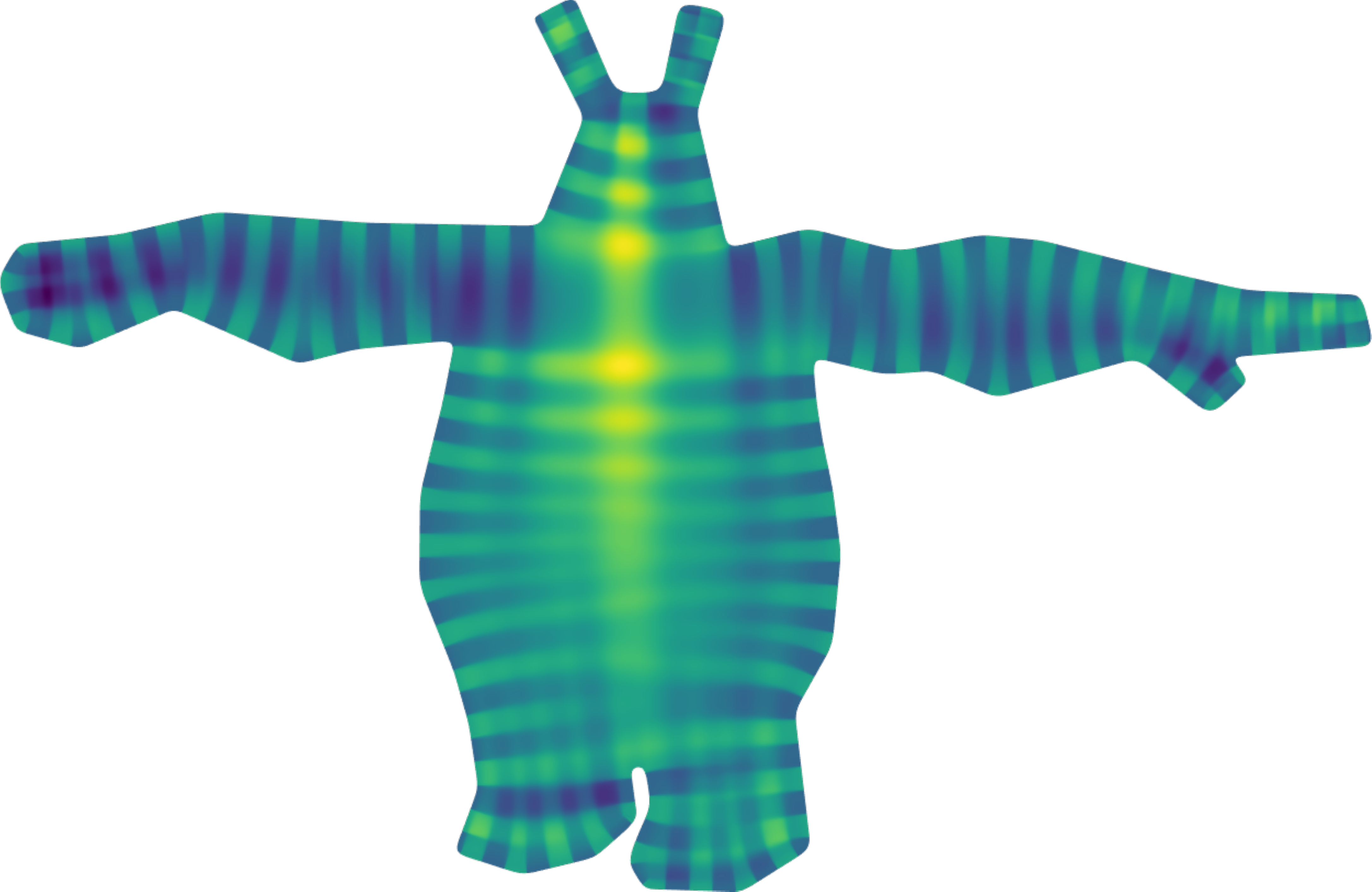} &
    \includegraphics[width=\imagewidth]{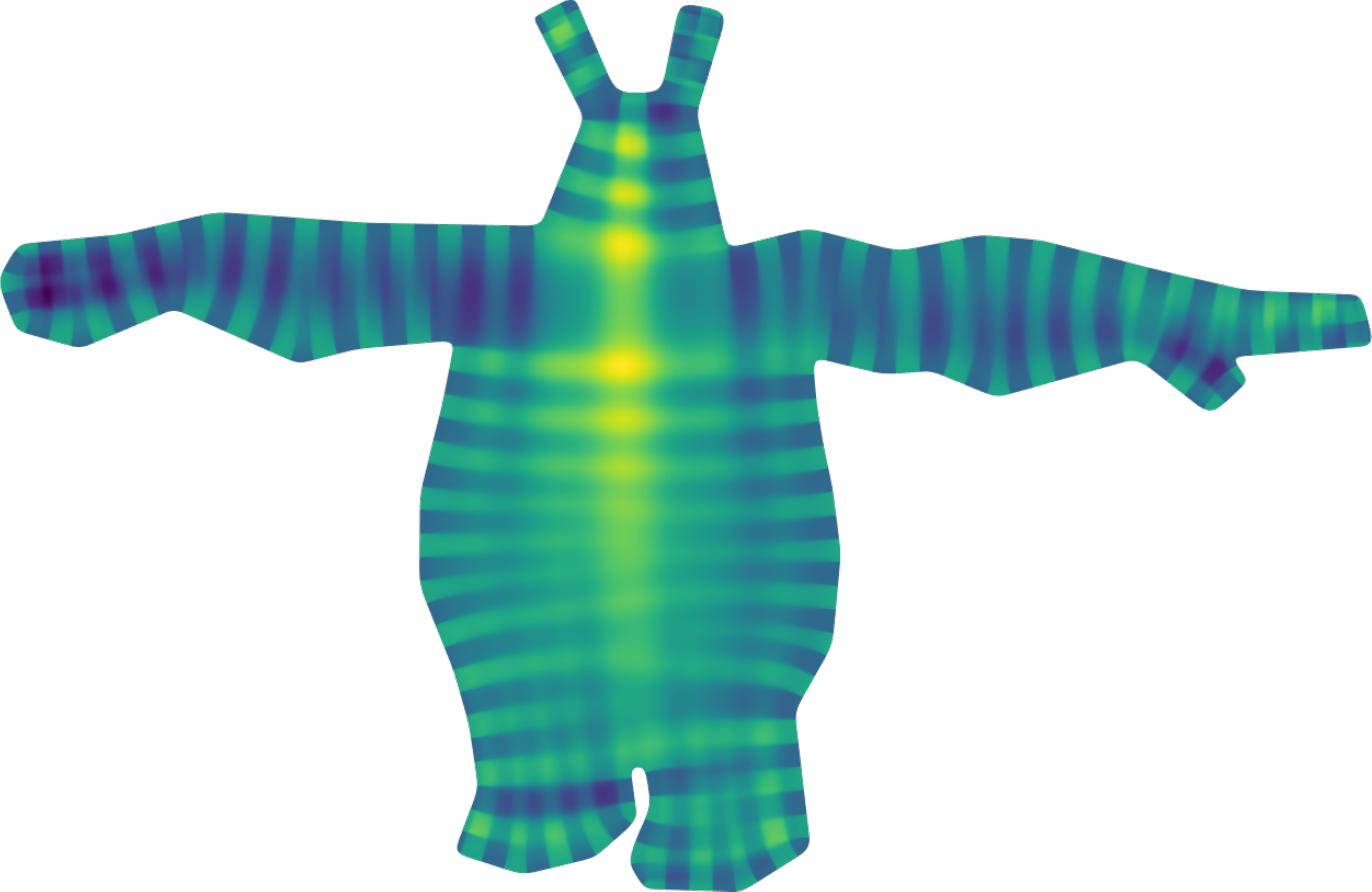} \\
  & $L_{\text{mean}} = \num{1.464e-02}$ & $L_{\text{mean}} = \num{7.315e-03}$ & $L_{\text{mean}} = \num{3.657e-03}$ \\
\end{tabularx}
\caption{Solutions to Dirichlet boundary-value problems converge under mesh refinement (Loop subdivision) as the mean edge length $L_{\text{mean}}$ decreases. Here the boundary conditions consist of a square wave in boundary arc length.}
\label{fig:cvg-2d}    
\end{figure*}

\section{Related Work}
\subsection{Phase Field Models} 

%\justin{this section is missing references}

Our work is loosely inspired by \emph{phase field models} from physics. These models often appear as ``mesoscale'' or ``effective theory'' abstractions in the physics of materials and pattern formation, wherein the physical state of a system is encoded in a \emph{phase field}, and the average behavior of microscale components is encoded in a PDE. Powerful tools from analysis can then help illuminate large-scale structural properties of the system. The freedom to specify the PDE affords enough flexibility to model a vast array of phenomena. A general reference on phase field modeling can be found in \cite{provatas2011phase}.
%\justin{Can you connect this language to language that geometry processing people would know?  E.g., maybe a phase field is analogous to some construction in frame field design papers?}

%Phase field models of \emph{partitions} arise in, for example, solidification problems, where a continuous fluid phase transitions to a solid phase consisting of discrete crystal grains. The basic idea of the phase field model is that a partition can be encoded in the isosurfaces of the phase field in a dynamically meaningful way. 

\subsection{Scalar Fields in Meshing}\label{sec:meshingrelatedwork}
Phase field-like models have appeared in geometry processing in \emph{Morse-based quadrangulation} methods, which employ a scalar oscillatory field to encode the combinatorial structure of a mesh; these controllable oscillatory fields have partly inspired our work.
We do not attempt to provide a complete overview of quadrilateral or hexahedral meshing; surveys can be found in \cite{Bommes2013,Campen2017,Amenta2019}.

In Morse-based meshing, a quadrilateral mesh is extracted from the \emph{Morse complex} or \emph{Morse-Smale complex} (MSC) of a scalar function (Morse function). The Morse complex is a topological skeleton whose study appears in Morse theory, where it is used to connect the topology of a manifold to the critical points and gradient flows of functions on it. A good reference on Morse theory and the Morse complex appears in \cite[Chapter 8]{Jost2017}. Prior to the advent of Morse-based meshing, the Morse complex of triangle meshes was studied in \cite{banchoff1970critical}, and the Morse-Smale complex on simplicial complexes was introduced and refined in \cite{edelsbrunner2000topological,edelsbrunner2001hierarchical,bremer2003multi,edelsbrunner2003morse}. \cite{Ni2004} proposed a method of computing ``fair'' Morse functions for use in mesh cutting and clustering.

In the work of Dong et al.\ \cite{Dong2006}, Laplacian eigenfunctions are used as Morse functions because of their ubiquity and smoothness. An argument from basic properties of the MSC in two dimensions shows that the mesh elements will be quadrilateral. Huang et al.\ \cite{Huang2008} extend this approach to allow orientation and alignment control by relaxing the Laplacian eigenproblem to a so-called \emph{quasi-eigenproblem} and introducing an objective term measuring alignment to a vector field. Alignment is measured against an ordinary vector field as symmetrized cross-field representations were not well-developed at the time.

\cite{ling2011spectral} studies the boundary conditions for Morse quadrangulation. The first- and second-order boundary conditions mean the Laplacian eigenproblem must still be relaxed to a least-squares quasi-eigenproblem. In our work, the use of a fourth-order operator means greater flexibility in choosing boundary conditions when solving a simple linear eigenproblem. \cite{Ling2014} refines the spectral approach by extending the alignment objective to allow separate control of local oscillation frequency in two orthogonal directions. This is a proxy for control of quad element size along each direction.
%In our work, such control is encoded in the basic operator itself through the use of \emph{odeco} tensor fields.\justin{careful with previous sentence unless you have meshing examples}

Other works solve more complicated nonlinear optimization problems to compute higher-quality Morse functions at the cost of complexity and performance. \cite{Zhang2010} extends the scalar Morse function to a section of a four-dimensional vector bundle to prevent degeneracy of the independent oscillation directions. Fang et al.\ \cite{Fang2018} combine this approach with a piecewise mesh construction approach that seeks the ``best of both worlds'' of parametrization-based and Morse-based meshing algorithms.

%\textcolor{red}{ODED: You should mention that there are a variety of other ways to mesh using differential operators, and then reference a survey or course \cite{Bommes2013,Campen2017,Amenta2019}.}
%\drp{Added the survey references to the beginning of this subsection. Apart from field-based and Morse-based meshing, are there other operator-based approaches I should be aware of?}

\subsection{Frame Fields and Tensor Fields}
Another key ingredient of our work is a tensor representation for symmetric frame fields, which have also grown up in the meshing literature.

Cross fields on surfaces have been applied extensively to the quad meshing problem. References can be found in the aforementioned meshing surveys as well as \cite{vaxman2016}. Cross field representations on planar domains and surfaces typically make use of the special structure available in two dimensions to represent crosses by complex numbers. More general non-orthogonal frame fields on surfaces have also been considered in \cite{panozzo2014,diamanti2014,diamanti2015}.

More recently, volumetric frame fields have become popular in the hex meshing literature. The challenge of representing volumetric frames with their complicated non-Abelian symmetries has been tackled in various ways in the hex-meshing literature. Early papers propose using special homogeneous polynomials or their coefficients in a basis of spherical harmonics to represent fields with local octahedral symmetry \cite{huang,RaySokolov2,Solomon}. More recently, \cite{chemin} suggests thinking of octahedral fields as special symmetric fourth-order tensor fields. \cite{palmer2020} proposes viewing octahedral fields as a subclass of more general tensor fields called \emph{odeco fields}, which can represent frame fields with independent scale along each frame axis. Other tensor field design problems have been explored in \cite{shen2016,palacios2017}. The homogeneous polynomial and symmetric tensor field representations naturally suggest thinking of a frame field as the principal symbol of a partial differential operator, which we explore in this work.

In field-based meshing, the recovery of a mesh from a field is generally mediated by a parametrization---i.e., a map from the domain to be meshed into Euclidean space, through which a lattice is then pulled back to yield a hex mesh. The frame field is viewed as encoding the derivatives of the parametrization up to symmetry, and the map is optimized to agree with the field in least-squares. Parametrization approaches have proven successful in 2D \cite{bommes2009mixed,bommes2013integer} and made inroads in the volumetric setting as well \cite{nieser2011cubecover,Lyon:2016:HRH}. In \Cref{sec:paramconnection}, we examine the frame field operator in the special case of frame fields associated to a parametrization.

\subsection{Discrete Bilaplacian}

The Bilaplacian, defined as the square of the scalar Laplace-Beltrami operator $\Delta$, is a popular fourth-order differential operator in geometry processing.
It is used for applications in surface fairing \cite{Desbrun1999}, surface deformation \cite{Sorkine2004},
data interpolation \cite{Jacobson2012}, data smoothing \cite{Weinkauf2011},
the computation of smooth distances \cite{Lipman2010}, skinning and character animation
\cite{Jacobson2011}, physical simulation \cite{Bergou2006}, and more \cite{Sykora2014,Andrews2011}.
The Bilaplacian with zero Neumann boundary is often discretized using mixed finite elements for the Laplacian \cite{Jacobson2010}, although other approaches are also popular for different boundary conditions \cite{Bergou2006,Stein2018,Stein2020}.

We consider a more general class of fourth order operators beyond the Bilaplacian, whose principal symbols are constructed from \emph{orthogonally decomposable} tensor fields, of which the Bilaplacian is a special case.
Our discretization generalizes the mixed finite element discretization for the Hessian energy on flat domains of Stein et al.\ \cite{Stein2018}, which is based on the classical mixed finite element method for the biharmonic equation \cite{Scholz1978}. We show how to engineer desired boundary conditions using a suitable choice of Lagrange multipliers. When our ellipticity parameter $\epsilon$ is set to $1$ and appropriate boundary conditions are chosen, we recover the Hessian energy discretization.
% \textcolor{red}{ODED: remove this text if we don't end up discussing Stein2020}
% The Crouzeix-Raviart discretization of Stein et al.\ \cite{Stein2020} cannot be adapted to our operators as easily {\color{red}(see Figure XXX)}.

\subsection{Discrete Anisotropic Operators}
Several works in geometry processing have studied elliptic operators with built-in anisotropy and their applications.
The survey of Wang and Solomon \cite[Section 5.7]{Wang2019} provides a comprehensive overview of anisotropic Laplacians, their discretization, and their applications.
Anisotropic operators have seen use in anisotropic meshing \cite{Fu2014}, coloring vector graphics \cite{Finch2011}, elasticity simulation \cite{Kim2019}, and surface reconstruction \cite{Yu2013}.

Azencot et al. \cite{azencot2013operator} show how to represent discrete tangent vector fields by their first-order directional derivative operators on scalar functions. \cite{azencot2017consistent} extend this representation to two-dimensional cross fields by transforming them into vector fields using the complex power approach. While these operators discretize a directional (first) derivative, our operator is an anisotropic linear elliptic operator that measures function variation in all frame directions at once.  As a result, our construction yields operators whose eigenfunctions oscillate in alignment with field directions, providing a fundamentally different means of encoding a frame field in a linear operator that generalizes to the volumetric setting.

\begin{figure}
\centering
\newcommand{\imgwidth}{0.1\textwidth}
\pgfplotstableread[col sep=comma, skip first n=1]{figures/horse-ev.csv}{\evs}
\begin{tikzpicture}
    \begin{loglogaxis}[
        mark size = 0.5pt,
        width = \columnwidth,
        height = 0.7\columnwidth,
        grid = none,
        xlabel = {Eigenvalue at Level 6},
        ylabel = {Eigenvalue Error},
        ylabel near ticks,
        legend pos = south east,
        every tick label/.append style = {font=\tiny}]
    \addplot [only marks, red, mark = *] table [x index = 5, y expr = {abs(\thisrowno{0}-\thisrowno{5})}] {\evs};
    \addlegendentry{Level 1};
    \addplot [only marks, green!70!black, mark = *] table [x index = 5, y expr = {abs(\thisrowno{1}-\thisrowno{5})}] {\evs};
    \addlegendentry{Level 2};
    \addplot [only marks, blue, mark = *] table [x index = 5, y expr = {abs(\thisrowno{2}-\thisrowno{5})}] {\evs};
    \addlegendentry{Level 3};
    \addplot [only marks, orange, mark = *] table [x index = 5, y expr = {abs(\thisrowno{3}-\thisrowno{5})}] {\evs};
    \addlegendentry{Level 4};
    \addplot [only marks, violet, mark = *] table [x index = 5, y expr = {abs(\thisrowno{4}-\thisrowno{5})}] {\evs};
    \addlegendentry{Level 5};
\end{loglogaxis}
\end{tikzpicture}
\caption{As the mesh is refined via Loop subdivision, eigenvalues converge. Here, the spectrum for a frame field operator on the \textbf{horse} is shown over five levels of Loop subdivision. Eigenvalue error is computed against the result on the sixth level of subdivision. The underlying frame field is the same as that depicted in \Cref{fig:dist}.}
\label{fig:eigs-2d}    
\end{figure}
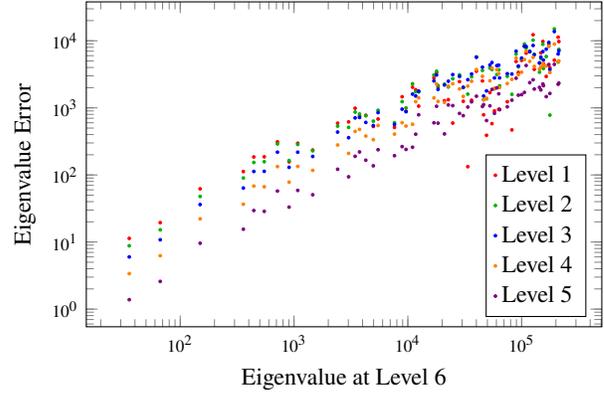

\section{Theory}

In this section, we will detail the construction of an elliptic operator measuring alignment to a symmetric frame field and examine its properties. We will also examine the behavior of the frame field operator for a frame field arising from a parametrization map.

%\oded{what pullback? You might want to add a bit of detail here (leaving it out is ok for abstract/intro, but at this point I'm not sure).}

\subsection{Preliminaries} 

We aim to construct an operator that measures alignment to a given frame field. First, we recall the definition of symmetric tensors and tensor fields, which will be used to build the coefficients of our operator.

\begin{definition}[tensors, tensor fields]
The space of \textbf{symmetric $k$th-order tensors} on $\R^n$ is the symmetric tensor product of $k$ copies of $\R^n$, notated as $\Sym^k \R^n$. Elements $T \in \Sym^k \R^n$ are given by sets of coefficients $T_{i_1\dots i_k}$ invariant under all permutations of the indices $i_1,\dots,i_k$.
A \textbf{symmetric fourth-order tensor field} on a domain $\Omega \subset \R^n$ is a continuous map $T : \Omega \to \Sym^4 \R^n$.
\end{definition}

\begin{notation}
In what follows, all tensors will be symmetric unless otherwise specified. We will make liberal use of the Einstein summation convention in formulas with Latin indices. We will also use the tensor contraction notation $A : T$ defined by
\begin{equation} (A : T)_{kl} \coloneqq A_{ij} T_{ijkl} \end{equation}
when $A$ and $T$ are (symmetric) second- and fourth-order tensors, respectively.
\end{notation}
We now focus on a particular class of tensor fields, popularized in hexahedral meshing for encoding collections of orthogonal directions \cite{chemin, palmer2020}:

\begin{definition}[odeco tensors, odeco fields]
A fourth-order symmetric tensor $T$ is \textbf{orthogonally decomposable (odeco)} if it can be written
\begin{equation}
    \sum_\alpha w_\alpha (\xi^\alpha)^{\otimes 4}
\end{equation}
for some orthonormal set of vectors $\xi^\alpha \in \R^n$, where $\otimes$ denotes the tensor product \cite{Robeva}. The $\xi^\alpha$ are known as the \textbf{orthogonal components} or \textbf{generalized eigenvectors} of $T$, the latter being in reference to the property that
\[ T_{ijkl} \xi^{\alpha}_i = w_\alpha \xi^{\alpha}_j \xi^{\alpha}_k \xi^{\alpha}_l. \]
Similarly, the $w_\alpha$ are known as the \textbf{weights} or \textbf{generalized eigenvalues} of $T$.
The odeco tensors form an algebraic variety known as an \emph{odeco variety} \cite{Robeva,Boralevi}. An \textbf{odeco field} is a field of odeco tensors \cite{palmer2020}.
\end{definition}

A particularly important subset of odeco tensors are those that are symmetric under permutation of their components. We refer to these as (conformal) octahedral due to their octahedral symmetry when the base dimension is three. We will focus our efforts on such fields, though our constructions also generalize to odeco fields.

\begin{definition}[octahedral, conformal octahedral tensors]
A tensor $T$ is \textbf{conformal octahedral} if it is odeco with equal weights $w_\alpha = w \ge 0$. $T$ is \textbf{octahedral} if all $w_\alpha = 1$. The octahedral tensors form a smooth variety known as the \emph{octahedral variety} \cite{palmer2020}. The conformal octahedral tensors occupy a cone over the octahedral variety.
We define \textbf{octahedral fields} and \textbf{conformal octahedral fields} as fields valued in the octahedral and conformal octahedral tensors, respectively.
\end{definition}
\begin{remark}
A generic odeco tensor $T = \sum_\alpha w_\alpha (\xi^{\alpha})^{\otimes 4}$ encodes its components $\{\xi^\alpha\}$ up to sign, as changing the sign of $\xi^\alpha$ has no effect on the coefficients of $T$. If $T$ is conformal octahedral with positive weight, it encodes its components up to permutation and sign. Thus odeco and (conformal) octahedral fields are generally known as \textbf{symmetric frame fields}. For more discussion and algorithms for computing such fields, see \cite{palmer2020}.
\end{remark}

Finally, we recall a useful norm on fourth-order tensors:
\begin{definition}[tensor spectral norm]
    By analogy to the operator norm for second order tensors, one can define a spectral norm on symmetric tensors \cite{friedland2020spectral} as
    \begin{equation}
        \|T\| \coloneqq \max_{\|v\| = 1} T_{ijkl} v_i v_j v_k v_l.
    \end{equation}
\end{definition}
%\justin{is this definition yours or something standard? if it's the latter maybe mention this and add a citation to a text, and move this definition to the previous subsection (my fault!).  if it's even marginally new, then leave it here.}
%\drp{I don't know if this is standard. I just checked and it doesn't seem to appear in Elina's odeco papers. However, I still think it might belong in the preliminaries section.}
When $T$ is odeco with weights $w_\alpha$, $\|T\| = \max_\alpha |w_\alpha|$. In particular, when $T$ is conformal octahedral with weight $w$, $\|T\| = w$, and $T/\|T\|$ is octahedral.

\subsection{Variational Problem}

Our task is to define a variational problem, whose optimality conditions will yield the desired frame field operator. The variational problem will also lead directly to our mixed finite element discretization in \Cref{sec.discretization}. The functional we define will measure alignment to a frame field. To do this, we first show that the tensor field itself encodes this alignment.
%\justin{I started a new section to signal that new things are happening here.  Add a paragraph outlining your approach here.  In particular, the text below refers to ``our variational problem'' but doesn't answer ``for what?'' early on.}
%
%\justin{Our first task is to define XYZ, which will be used in our formulation to do ABC:}

\begin{definition}[alignment]
Let $T$ be an odeco tensor with orthogonal components $\xi^\alpha$. We say a second-order symmetric tensor $S$ is \textbf{aligned} with $T$ if the $\xi^\alpha$ are eigenvectors of $S$.
\end{definition}
If $u$ is a scalar function and $S = \nabla^2u$ is its Hessian, then alignment between $S$ and $T$ expresses the intuitive idea that the primary directions of curvature of $u$ occur along the components of $T$.

In what follows, we will focus on the case of conformal octahedral tensors and fields. The case of general odeco tensors is similar but messier. To motivate our variational problem, we will view a fourth order tensor as inducing a quadratic form on second-order tensors. The following property of conformal octahedral tensors says that this quadratic form measures alignment:
\begin{lemma}
    Let $T$ be a conformal octahedral tensor with components $\xi^\alpha$. Then for any given set of distinct eigenvalues $\lambda_i(S)$, the quadratic form $S : T : S$ is maximized over $S$ when the eigenvectors of $S$ agree with the $\xi^\alpha$.
    \begin{proof}
        Using the fact that the $\xi^\alpha$ form an orthonormal basis, we expand the expression for $T$ to get
        \begin{align}
            \frac{1}{\|T\|} \, S : T : S &= \sum_\alpha (\xi^\alpha)^\top S (\xi^\alpha)(\xi^\alpha)^\top S (\xi^\alpha) \\
            &\le \sum_{\alpha,\beta} (\xi^\alpha)^\top S (\xi^\beta)(\xi^\beta)^\top S (\xi^\alpha) \\
            \intertext{as all terms are nonnegative}
            &= \sum_{\alpha} (\xi^\alpha)^\top S \left( \sum_\beta (\xi^\beta)(\xi^\beta)^\top \right) S (\xi^\alpha) \\
            &= \sum_{\alpha} (\xi^\alpha)^\top S^2 (\xi^\alpha) \\
            \intertext{as the parenthesized tensor is the identity by orthonormality of $\xi^\beta$}
            &= \tr\left(S^2 \sum_\alpha \xi^\alpha (\xi^\alpha)^\transp \right) \\
            &= \tr S^2 \\
            &= \sum_{i} \lambda_i(S)^2,
        \end{align}
        with equality if the $\xi^\alpha$ are eigenvectors of $S$.
    \end{proof}
\end{lemma}
Let $\Omega \subset \R^n$ be a compact domain and $T$ a conformal octahedral field on $\Omega$. Intuitively, we want to define a functional $\mathcal{E}_T(u)$ that will be \emph{minimized} when the scalar function $u : \Omega \to \R$ oscillates in alignment with the frame field. Since $S : T : S$ is \emph{maximized} when $S$ is aligned to $T$, we might be tempted to define $\mathcal{E}_T(u) \coloneqq -(\nabla^2 u) : T : (\nabla^2 u)$. However, this quadratic form is negative and degenerate. In particular,
\begin{equation}
    (\xi^\alpha \otimes \xi^\beta + \xi^\beta \otimes \xi^\alpha) : T : (\xi^\alpha \otimes \xi^\beta + \xi^\beta \otimes \xi^\alpha) = 0
\end{equation}
when $\alpha \ne \beta$.
To get an elliptic differential operator, we want to define a positive, nondegenerate functional. To this end, we first choose some $\epsilon \in (0, 1]$ and define a modified tensor field
\begin{equation}T^\epsilon \coloneqq \|T\|\mathbb{I} - (1 - \epsilon) T, \end{equation}
where $\mathbb{I}$ denotes the fourth-order identity tensor whose characteristic property is that $\mathbb{I} : S = S$ for any symmetric second-order tensor $S$. Now we can define a functional as follows:
\begin{definition}[frame field functional]
The \textbf{frame field functional} associated to $T$ with ellipticity $\epsilon > 0$ is given by
\begin{equation} 
\begin{aligned} \mathcal{E}_{T,\epsilon}(u) &= \frac{1}{2} \int_\Omega (\nabla^2 u) : T^\epsilon : (\nabla^2 u) \; d\Omega \\
&= \frac{1}{2} \int_\Omega \|T\| \|\nabla^2 u\|_F^2 - (1-\epsilon)(\nabla^2 u) : T : (\nabla^2 u) \; d\Omega \end{aligned} \end{equation}
for $u \in H^2(\Omega)$, where $\nabla^2 u$ denotes the Hessian of $u$, and $\|\cdot \|_F$ is the pointwise Frobenius norm.
\end{definition}
In summary, for any conformal octahedral field, we have constructed a nondegenerate functional $\mathcal{E}_{T,\epsilon}$ that preserves the alignment-measuring properties of the quadratic form $(\nabla^2 u) : T : (\nabla^2 u)$. Intuitively, $\mathcal{E}_{T, \epsilon}$ wants the Hessian of $u$ to align to the frame directions, and this effect becomes stronger as $\epsilon \to 0$. When $\epsilon = 1$, $\mathcal{E}_{T,1}$ is the \emph{Hessian energy} (see \cite{Stein2018}), for which the Euler-Lagrange equation is the biharmonic equation.

\begin{figure}
\newcommand{\imagewidth}{0.3\columnwidth}
\centering
\pgfplotstableread[col sep=comma]{figures/horse-mean-length.csv}{\data}
\begin{tikzpicture}
    \begin{loglogaxis}[
        width = \columnwidth,
        height = 0.7\columnwidth,
        enlarge x limits = 0.1,
        enlarge y limits = 0.2,
        grid = none,
        xlabel = {Mean Edge Length},
        ylabel = {Absolute Eigenvalue Error},
        ylabel near ticks,
        x dir = reverse,
        every tick label/.append style = {font=\tiny},
        legend pos = south west,
        legend style = {nodes = {scale = 0.5, transform shape}}]
    \addplot+ table [x index = 0, y index = 10] {\data};
    \addlegendentry{$\lambda_{10}$}
    \addplot+ table [x index = 0, y index = 20] {\data};
    \addlegendentry{$\lambda_{20}$}
    \addplot+ table [x index = 0, y index = 30] {\data};
    \addlegendentry{$\lambda_{30}$}
    \addplot+ table [x index = 0, y index = 40] {\data};
    \addlegendentry{$\lambda_{40}$}
    \addplot+ table [x index = 0, y index = 50] {\data};
    \addlegendentry{$\lambda_{50}$}
    \addplot+ table [x index = 0, y index = 60] {\data};
    \addlegendentry{$\lambda_{60}$}
%    node [pos = 0, inner sep=0pt] {\includegraphics[width = \imagewidth]{figures/clover-e64-lvl1}}
%    node [pos = 0.5] {\includegraphics[width = \imagewidth]{figures/clover-e64-lvl2}}
%    node [pos = 1] {\includegraphics[width = \imagewidth]{figures/clover-e64-lvl3}};
%    \node[inner sep=0pt] at (axis cs:0.0189, 0.0520) {\includegraphics[width = \imagewidth]{figures/clover-e64-lvl1}};
%    \node[inner sep=0pt] at (axis cs:0.0094, 0.0149) {\includegraphics[width = \imagewidth]{figures/clover-e64-lvl2}};
%    \node[inner sep=0pt] at (axis cs:0.0047, 0.0035) {\includegraphics[width = \imagewidth]{figures/clover-e64-lvl3}};
    \end{loglogaxis}
\end{tikzpicture}%
\\
\includegraphics[width=\imagewidth]{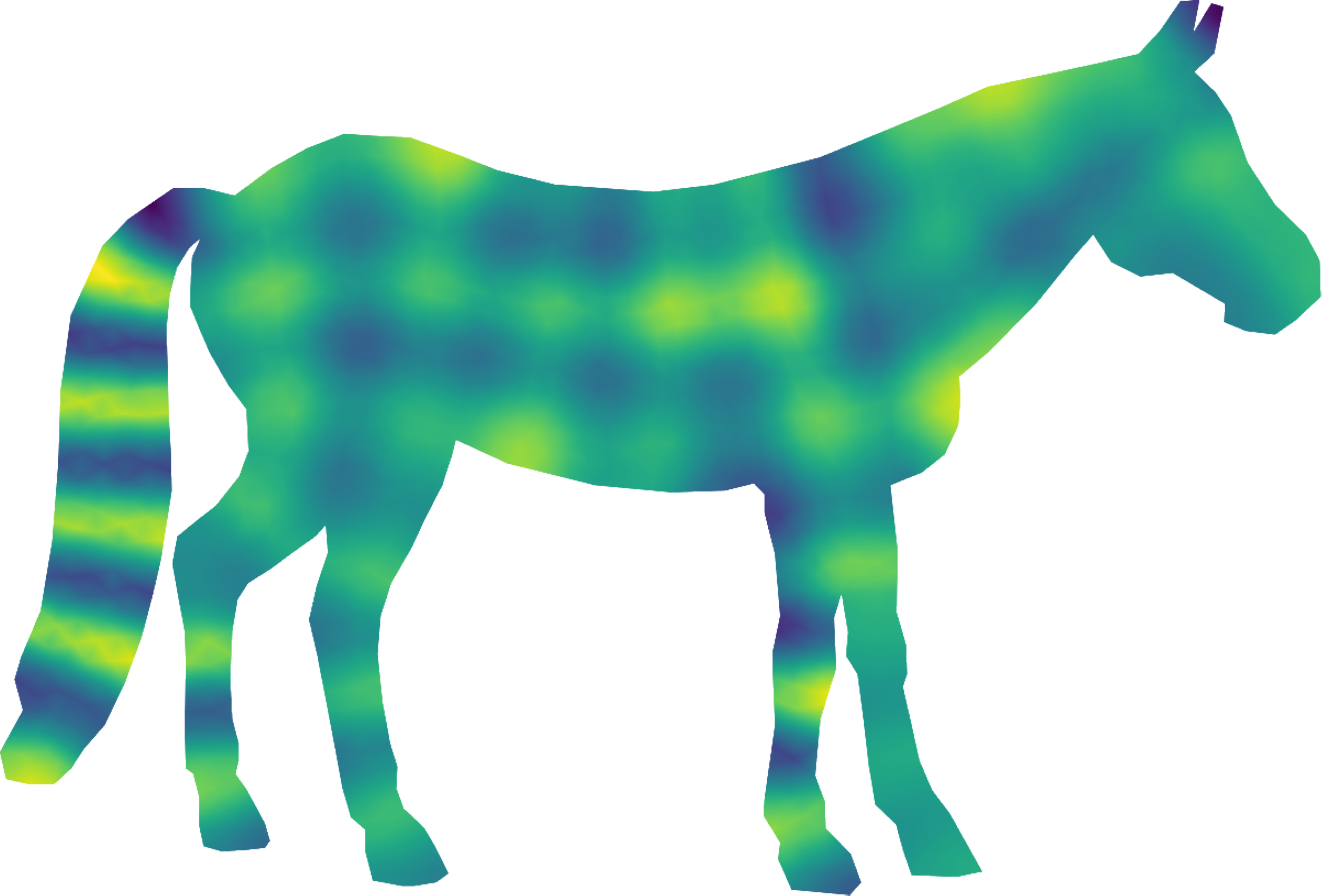}
\includegraphics[width=\imagewidth]{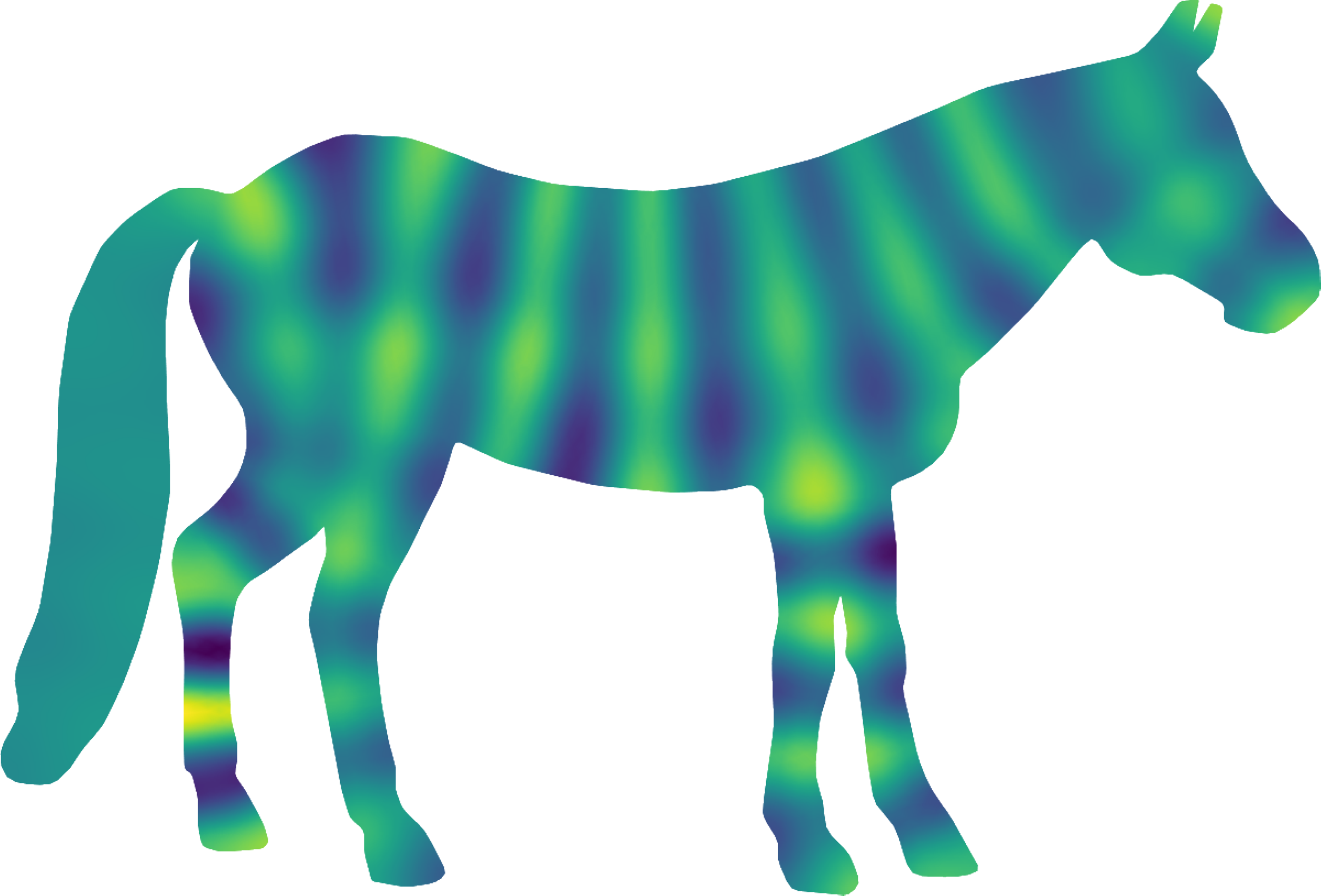}
\includegraphics[width=\imagewidth]{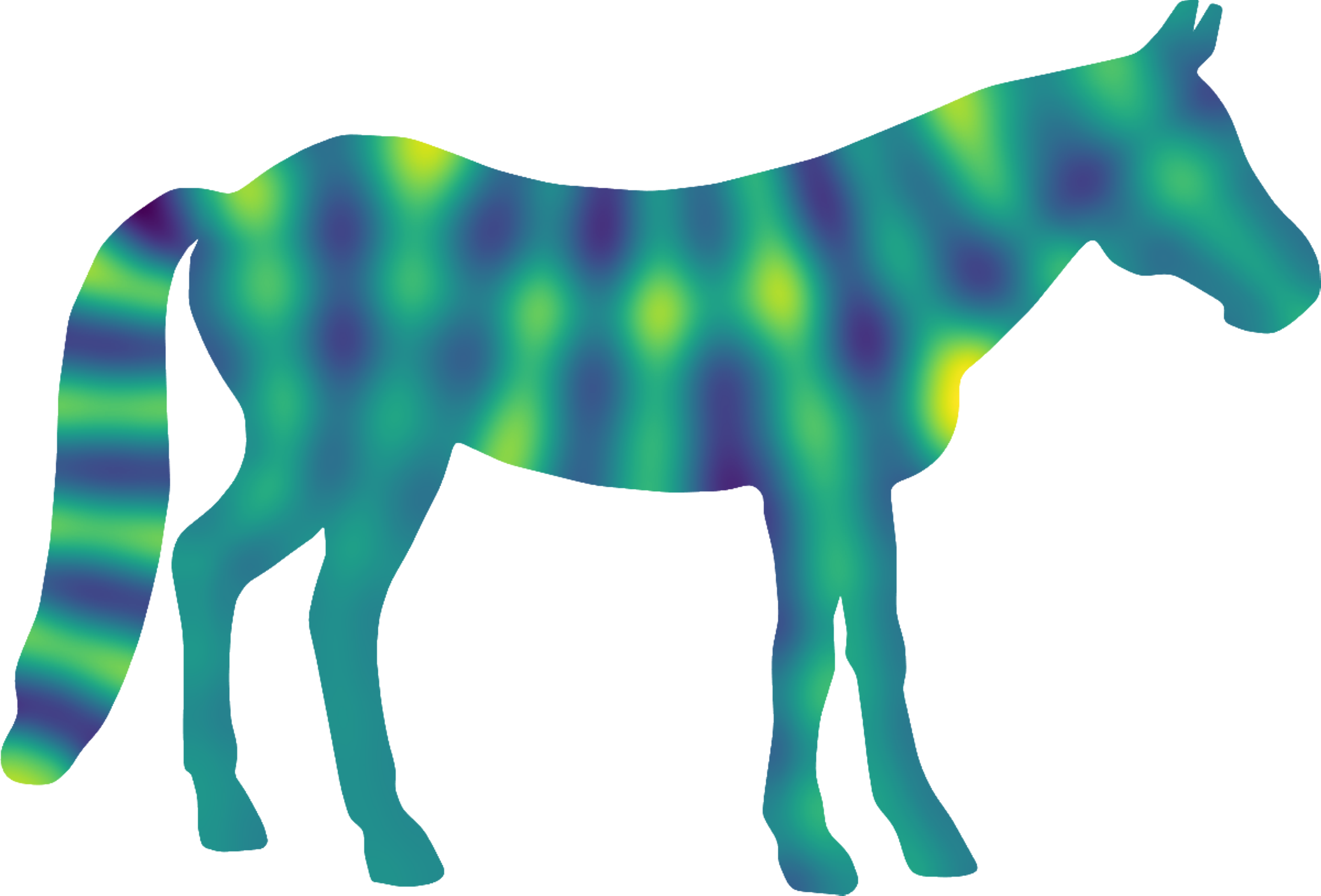} \\
\includegraphics[width=\imagewidth]{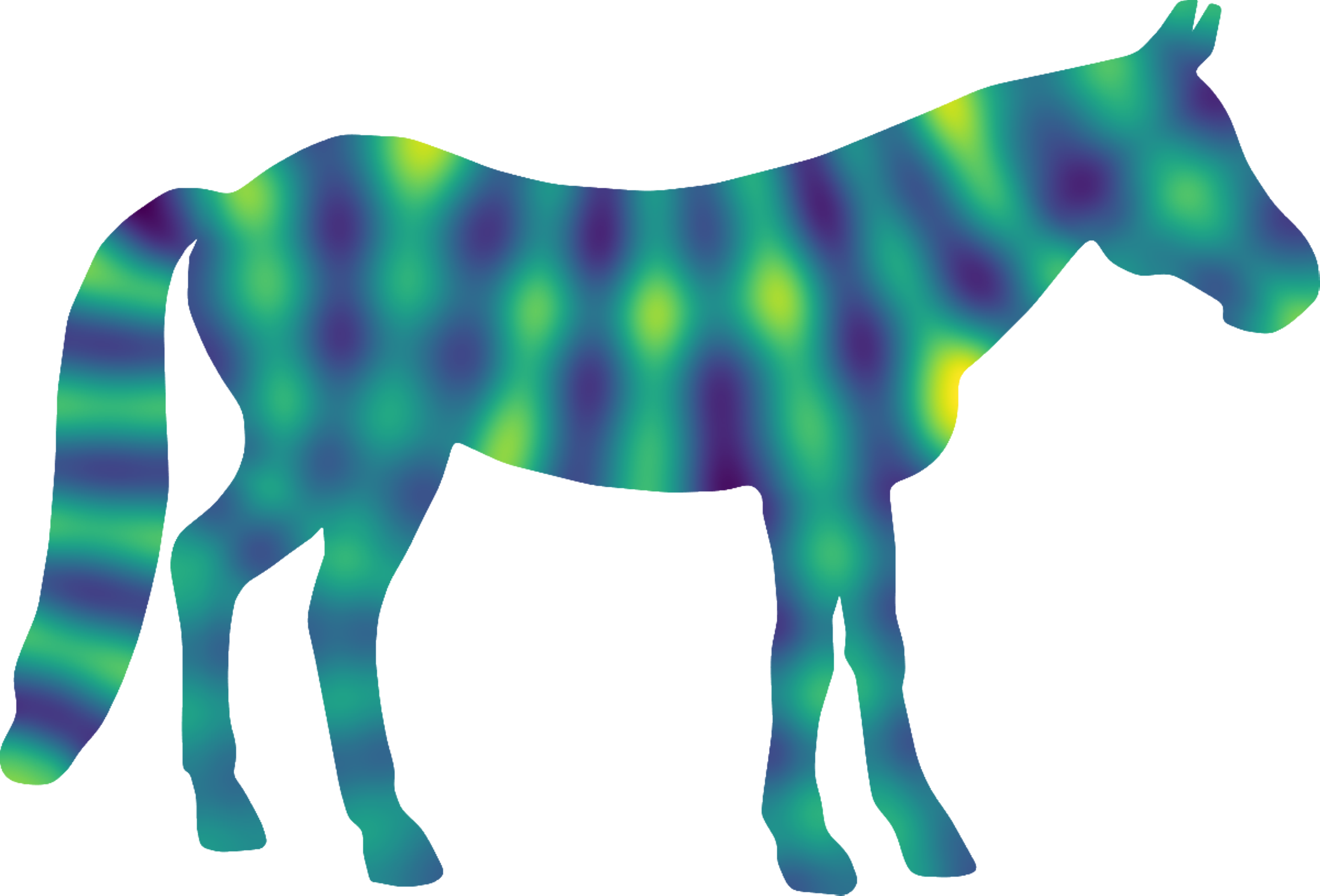}
\includegraphics[width=\imagewidth]{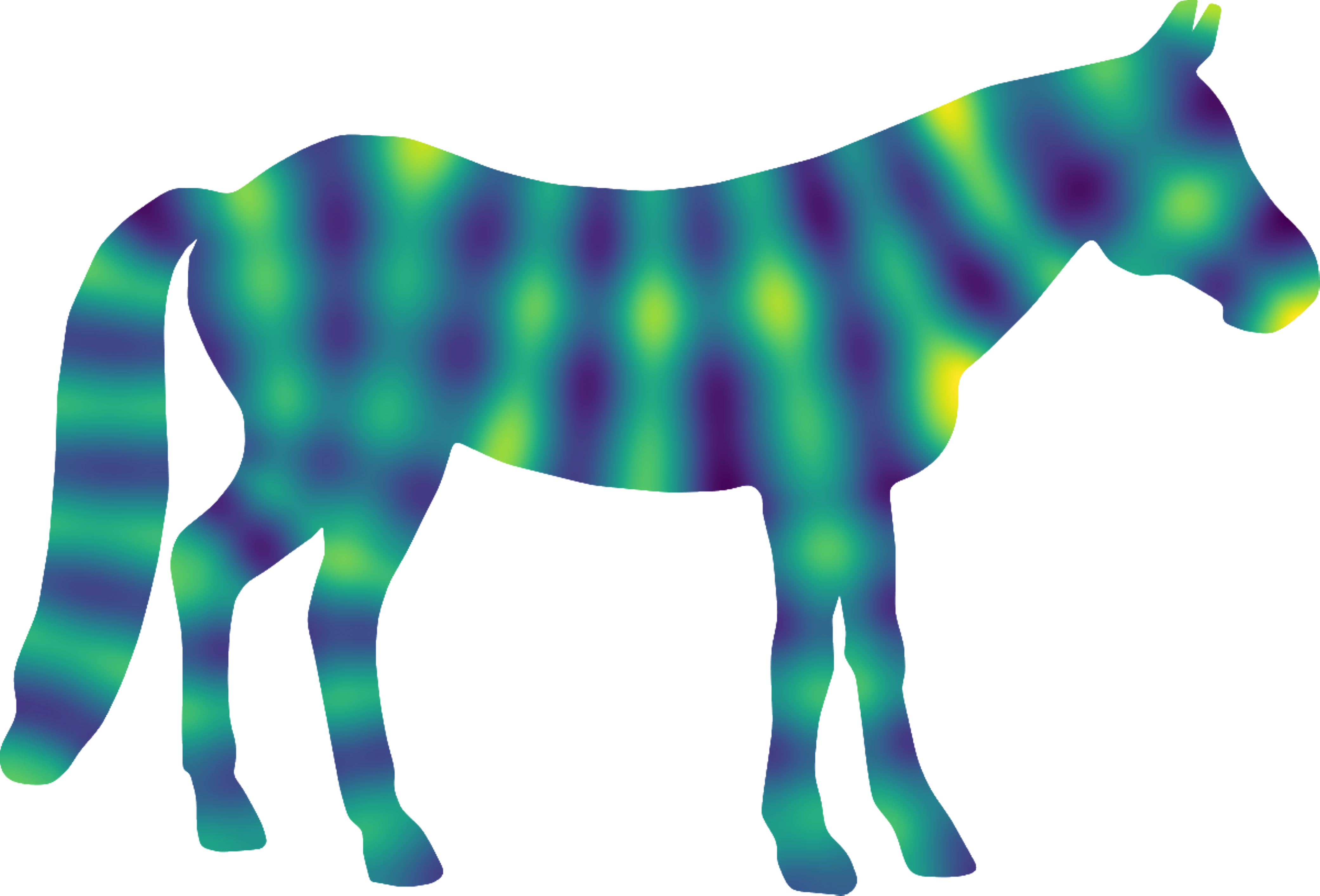}
\includegraphics[width=\imagewidth]{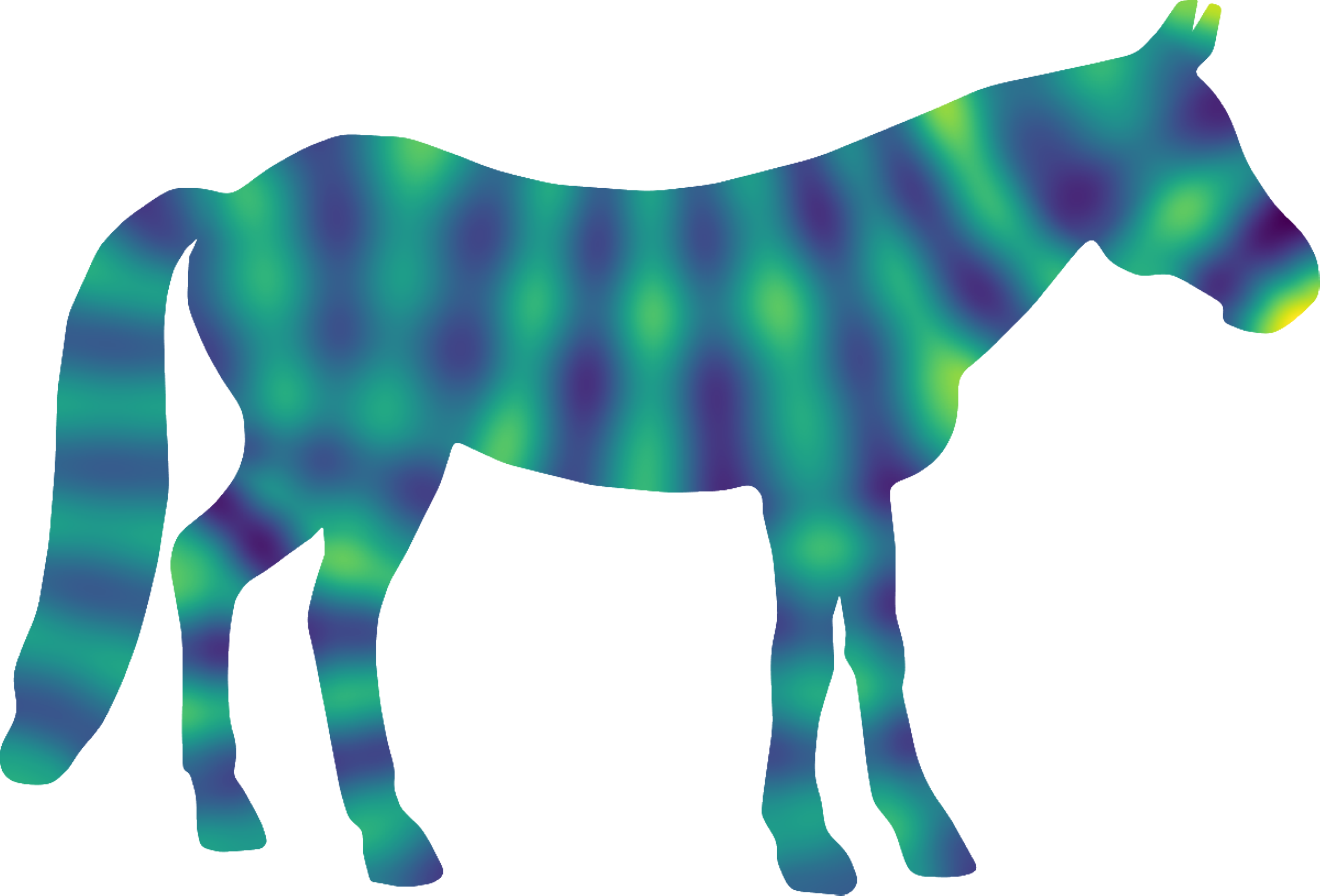}
\caption{As the mesh is refined, frame field eigenvalues and eigenfunctions converge. Eigenvalue error (top) is measured against the value at the lowest resolution. The $64$th eigenfunction is shown over six mesh resolutions (left to right, top to bottom). The underlying frame field on the \textbf{horse} is the same as that depicted in \Cref{fig:dist}.}
\label{fig:eigf-2d}
\end{figure}

\subsection{Euler-Lagrange Equations}
In the previous section, we proposed a functional $\mathcal{E}_{T,\epsilon}$ measuring the alignment of the variation directions of a scalar function $u$ to a given frame field $T$. Now we follow a standard procedure to extract a differential operator from this variational formulation, and we show that the resulting \emph{frame field operator} is elliptic.
%\justin{Short paragraph reminding us that your goal in this section is to get an operator out of the energy in the previous paragraph}
 
Taking the first variation of $\mathcal{E}_{T,\epsilon}$ with respect to $u$ yields the following Euler-Lagrange equations:
\begin{equation} 0 = \int_\Omega \sum_{i, j, k, l} T^{\epsilon}_{ijkl} (\partial_i \partial_j u)(\partial_k \partial_l v) \; d\Omega, \end{equation}
for any smooth test function $v \in C^\infty(\Omega)$. Integrating by parts yields
\begin{equation}
\begin{aligned}
0 &= \int_{\partial\Omega} T^{\epsilon}_{ijkl} (\partial_i \partial_j u)n_k \partial_l v \; dA - \int_{\Omega} \partial_k (T^{\epsilon}_{ijkl} (\partial_i \partial_j u)) \partial_l v \; d\Omega \\
&= \int_{\partial\Omega} [T^{\epsilon}_{ijkl} (\partial_i \partial_j u)n_k \partial_l v - \partial_k(T^{\epsilon}_{ijkl} (\partial_i \partial_j u)) n_l v] \; dA \\
&\quad + \int_{\Omega} \partial_k \partial_l (T^{\epsilon}_{ijkl} \partial_i \partial_j u) v\; d\Omega.
\end{aligned}
\end{equation}
Eliminating the test function $v$, we obtain the following PDE with natural boundary conditions:
\begin{align}
    \partial_k \partial_l(T^{\epsilon}_{ijkl}\partial_i\partial_j u) &= 0 \label{eq.pde1}\\
    n_i T^{\epsilon}_{ijkl} \partial^2_{jk} u &= 0 \quad \text{on }\partial \Omega \label{eq.bdry2} \\
    n_i \partial_j (T^{\epsilon}_{ijkl} \partial^2_{kl} u) &= 0 \quad \text{on }\partial\Omega. \label{eq.bdry3}
\end{align}
Accordingly, we define our differential operator as follows:
\begin{definition}[frame field operator] The \textbf{frame field operator} associated to a conformal octahedral frame field with ellipticity $\epsilon$ is given by
\begin{equation} \mathcal{A}_{T,\epsilon} u = \partial_k \partial_l (T^{\epsilon}_{ijkl} \partial_i \partial_j u). \end{equation}
\end{definition}
The fourth-order term will have coefficients $T^{\epsilon}_{ijkl}$, i.e., the \emph{principal symbol} of $\mathcal{A}_{T,\epsilon}$ is given by the polynomial
\begin{equation}\begin{aligned}
\sigma_P(\mathcal{A}_{T,\epsilon})(x, \zeta) &= T^{\epsilon}_{ijkl}(x)\zeta_i\zeta_j\zeta_k\zeta_l \\
&= \|T(x)\|\left(\|\zeta\|^4 - (1-\epsilon)\sum_{\alpha} (\xi^\alpha(x) \cdot \zeta)^4 \right) \\
&\ge \epsilon \|T(x)\| \|\zeta\|^4,
\end{aligned}\end{equation}

\vspace{-.1in} % need to start a new paragraph here or wrapfigure freaks out, so i added some negative vspace to correct for it a little
\begin{wrapfigure}[4]{r}{0.2\linewidth}
    \vspace{-.35in}\centering
    \includegraphics[width=.95\linewidth]{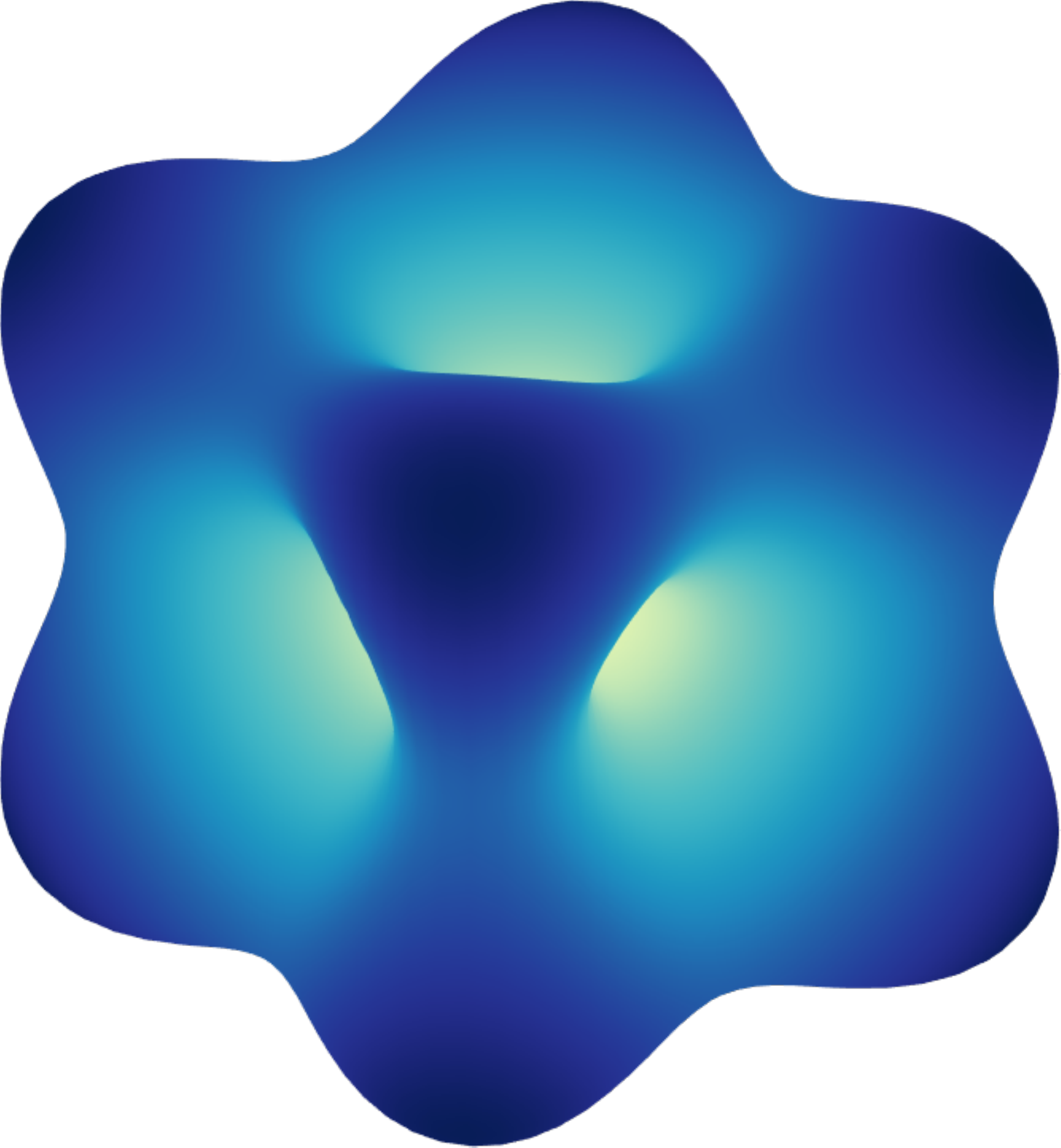}
\end{wrapfigure}
\noindent where $\zeta \in T_x^*\R^n = \R^n$,
confirming that $\mathcal{A}_{T,\epsilon}$ is \emph{elliptic}. Moreover, if $\|T(x)\| \ge C > 0$ for all $x \in \Omega$, then $\mathcal{A}_{T,\epsilon}$ is \emph{uniformly elliptic}. An example of $\sigma_P(\mathcal{A}_{T,\epsilon})$ is shown at right (inset) as a plot over the unit sphere $\|\zeta\| = 1$.

To sum up, we have shown that solutions to the variational problem $\min_u \mathcal{E}_{T,\epsilon}(u)$ satisfy a fourth-order elliptic PDE $\mathcal{A}_{T,\epsilon} u = 0$ and corresponding natural boundary conditions. Intuitively, the solutions to this PDE are functions ``most aligned'' to the frame field $T$.

\subsection{Eigenproblem}
A wider variety of field-aligned functions can be obtained by solving an eigenproblem for $\mathcal{A}_T$. Imposing the nondegeneracy constraint $\|u\|_{L^2(\Omega)} = 1$, we obtain the Lagrangian
\begin{equation}
    \mathcal{E}_{T,\epsilon}(u) - \lambda (\|u\|^2 - 1),
\end{equation}
whose Euler-Lagrange equations consist of the eigenvalue problem
\begin{equation} \mathcal{A}_{T,\epsilon} u = \partial_i \partial_j (T^\epsilon_{ijkl}  \partial_k \partial_l u) = \lambda u, \label{eq.eig} \end{equation}
together with the natural boundary conditions \eqref{eq.bdry2}--\eqref{eq.bdry3}.

\subsection{Boundary-Aligned Frame Fields}

%\justin{intro paragraph here.  what's about to happen and why?}
Frame fields encountered in hex meshing satisfy an alignment condition at the boundary of a domain. Given a boundary-aligned frame field, the natural boundary condition \eqref{eq.bdry2} simplifies considerably.

\begin{definition}[boundary-aligned frame field] A frame field $T$ on a domain $\Omega$ is \textbf{boundary-aligned} if the boundary normal $n(x)$ is a generalized eigenvector of $T(x)$ for all $x \in \partial \Omega$:
\begin{equation}
T_{ijkl}(x) n_l(x) = w(x) n_i(x) n_j(x) n_k(x) \text{ for all } x \in \partial \Omega. \label{eq.bdry_align}
\end{equation}
\end{definition}
Suppose that $T$ is boundary-aligned. Then from \eqref{eq.bdry2} and \eqref{eq.bdry_align}, the second-order natural boundary condition reduces to
\begin{equation}
\begin{aligned}
0 &= n_i T^{\epsilon}_{ijkl} \partial^2_{jk} u \\
  &= \|T\| n_i \partial^2_{ij} - (1-\epsilon)n_i T_{ijkl} \partial^2_{jk} u \\
  &= \|T\| n_i \partial^2_{ij} - (1-\epsilon)\|T\| n_j n_k n_l \partial^2_{jk} u \\
  &= \|T\| (I - (1 - \epsilon) n n^\top) (\nabla^2 u) n.
\end{aligned}
\end{equation}
Observing that $(I - n n^\top)$ is positive definite, we obtain the reduced second-order boundary conditions
\begin{equation}
(\nabla^2 u) n = 0 \quad \text{on } \partial \Omega.
\label{eq.bdry2reduced}    
\end{equation}
Intuitively, when $T$ is boundary-aligned, the natural boundary condition \eqref{eq.bdry2reduced} says that $u$ is \emph{linear} along the normal direction at the boundary, and moreover that its normal derivative is constant over $\partial \Omega$. Notice the similarity to the natural boundary conditions studied in \cite{Stein2018}.

\subsection{Relationship to Parametrization}\label{sec:paramconnection}

In parametrization-based quad and hex meshing, frame fields enter as a way to encode derivatives of a parametrization up to some symmetry, either quadrilateral or octahedral (see e.g., \cite{bommes2009mixed,bommes2013integer,nieser2011cubecover,liu2018}). In this section, we explore the properties of the frame field operator associated to a frame field arising from a parametrization. We motivate why we might expect high-frequency eigenfunctions of such an operator to have local lattice-like structure.
%\justin{Link the definition below to past work here}

\begin{figure}
\centering
%\begin{tikzpicture}
%\begin{axis}[%
%view={0}{90},
%width=\columnwidth,
%height=\columnwidth,
%%scale only axis,
%colormap name = viridis,
%%xmin=-31.4159265358979,
%%xmax=31.4159265358979,
%%ymin=-31.4159265358979,
%%ymax=31.4159265358979,
%%axis background/.style={fill=white},
%axis x line*=bottom,
%axis y line*=left
%]
%\addplot3[scatter, only marks, mark=*, mark size=1.7pt, scatter src = {z}, domain=-10*pi:10*pi, y domain=-10*pi:10*pi, samples=41] {max(-3, ln(0.01 * (x^4 + y^4) + 2 * x^2 * y^2))}; %{(x, y, )}
%\end{axis}
%\end{tikzpicture} \\
\pgfplotstableread[col sep=comma, skip first n=1]{figures/square-ev.csv}{\evs}
\begin{tikzpicture}
\begin{loglogaxis}[
        mark size = 0.5pt,
        width = \columnwidth,
        height = 0.7\columnwidth,
        grid = none,
        xlabel = {Analytic Eigenvalue},
        ylabel = {Eigenvalue Error},
        ylabel near ticks,
        legend pos = south east,
        every tick label/.append style = {font=\tiny}]
    \addplot [only marks, red, mark = *] table [x index = 0, y expr = {abs(\thisrowno{1}-\thisrowno{0})}] {\evs};
    \addlegendentry{$L_{\text{mean}} = 0.03$};
    \addplot [only marks, green!70!black, mark = *] table [x index = 0, y expr = {abs(\thisrowno{2}-\thisrowno{0})}] {\evs};
    \addlegendentry{$L_{\text{mean}} = 0.025$};
    \addplot [only marks, blue, mark = *] table [x index = 0, y expr = {abs(\thisrowno{3}-\thisrowno{0})}] {\evs};
    \addlegendentry{$L_{\text{mean}} = 0.02$};
    \addplot [only marks, orange, mark = *] table [x index = 0, y expr = {abs(\thisrowno{4}-\thisrowno{0})}] {\evs};
    \addlegendentry{$L_{\text{mean}} = 0.015$};
\end{loglogaxis}
\end{tikzpicture}
\caption{For a constant frame field on the square $[-1, 1]^2$, eigenvalues can be computed analytically using the Fourier transform. The spectrum of our discrete operator converges to the analytic spectrum as mesh resolution increases. $L_{\text{mean}}$ indicates mean mesh edge length.}
\label{fig:square}    
\end{figure}
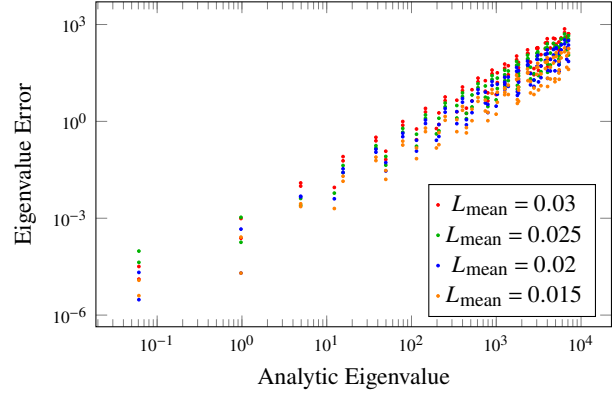

\begin{definition}[map frame fields]
Suppose $f : \Omega \to f(\Omega) \subset \R^n$ is a diffeomorphism. Let $df$ be the differential of $f$. The \textbf{map frame field} associated to the map $f$ is defined as follows:
\begin{equation}
\begin{aligned}
(\mathrm{T}f)_{ijkl} &\coloneqq \sum_\alpha df^\alpha_i df^\alpha_j df^\alpha_k df^\alpha_l \\
&= \sum_\alpha (\partial_i f^\alpha) (\partial_j f^\alpha) (\partial_k f^\alpha) (\partial_l f^\alpha),
\end{aligned}
\end{equation}
where $\partial_i = \partial/\partial x_i$ is differentiation with respect to the $i$th coordinate in $\Omega$. The \textbf{inverse map frame field} or \textbf{map coframe field} is given by
\begin{equation} (\hat{\mathrm{T}}f)^{ijkl} \coloneqq \sum_\alpha (df^{-1})^i_\alpha (df^{-1})^i_\alpha (df^{-1})^\alpha_k (df^{-1})^l_\alpha, \end{equation}
where $df^{-1}$ denotes the matrix inverse of $df$.
\end{definition}
\begin{remark}
    Observe that $\hat{\mathrm{T}}f$ is the image of the constant coframe field $\sum_\alpha (e_\alpha)^{\otimes 4}$ under the natural pullback map.
\end{remark}
\begin{remark}
    For $\mathrm{T}f$ and $\hat{\mathrm{T}}f$ to be conformal octahedral frame fields, the map $f$ must be conformal. Moreover, for $\mathrm{T}f$ and $\hat{\mathrm{T}}f$ to be octahedral, $f$ must be a rigid motion, and $\mathrm{T}f$ and $\hat{\mathrm{T}}f$ will then be constant.
\end{remark}

Now suppose $f$ is conformal, let $v : \R^n \to \R$, and let $u = v \circ f$ be its pullback to $\Omega$. Then, modulo terms of lower differential order,
\begin{equation}
\partial^4_{ijkl} u \equiv df_i^{a} df_j^{b} df_k^{c}df_l^{d} (\partial^4_{abcd} v) \circ f.    
\end{equation}
Let $J = df$ and $\hat{J} = df^{-1}$ so that $\hat{J}_a^b J_b^c = \delta_a^c$. Then
\begin{equation} \begin{aligned}
(\hat{\mathrm{T}}f)^{ijkl} \partial^4_{ijkl} u &\equiv \sum_\alpha \hat{J}^i_\alpha \hat{J}^j_\alpha \hat{J}^k_\alpha \hat{J}^l_\alpha J_i^{a} J_j^{b} J_k^{c}J_l^{d} (\partial^4_{abcd} v) \circ f \\
&= \sum_{\alpha} \delta_\alpha^a\delta_\alpha^b\delta_\alpha^c\delta_\alpha^d (\partial^4_{abcd} v) \circ f \\
&= \sum_\alpha \partial^4_{\alpha \alpha \alpha \alpha} v \circ f. \end{aligned} \end{equation}
Also,
\begin{equation} \begin{aligned}
\partial^4_{ijij} u &\equiv (\partial_i f^{a}) J_i^{a} J_j^{b} J_k^{c}J_l^{d} (\partial^4_{abcd} v) \circ f \\
&= \| J^a \|^2 \| J^b \|^2 \delta_{ac}\delta_{bd} [(\partial^4_{abcd} v) \circ f] \\
&= \|\mathrm{T} f\| \partial^4_{abab} v \circ f, \end{aligned} \end{equation}
where we have used that $f$ is conformal.
Hence, evaluating the full frame field operator on $u$ is equivalent at highest order to evaluating a constant frame field operator on $v$:
\begin{equation}
    \mathcal{A}_{\hat{\mathrm{T}}f, \epsilon} u \equiv (\partial^4_{abab} - (1 - \epsilon) \partial^4_{aaaa}) v \circ f + \text{lower order},
\end{equation}
where we have used that $\|\hat{\mathrm{T}}f\|\|\mathrm{T}f\| = 1$. This says that---up to terms of lower differential order---the frame field operator associated to a map coframe field pushes forward through the map to the constant frame field operator.

From a hex meshing perspective, if $f$ is a parametrization carrying our frame field to a constant frame field, we might hope eigenfunctions whose critical points lie on a lattice to be pulled back to eigenfunctions whose critical points form a hex mesh. The above analysis tells us that this is true in the high-frequency limit---indeed,
at high frequencies, the highest-order derivatives will dominate, and $\mathcal{A}_{\hat{\mathrm{T}}f,\epsilon}$ will approach the pullback of the constant frame field operator on $f(\Omega)$. We should therefore expect that eigenfunctions of $\mathcal{A}_{\hat{\mathrm{T}}f,\epsilon}$ will increasingly look like warped copies of constant frame field eigenfunctions as their frequency increases. Even at relatively low frequencies, this appears to be borne out empirically (see \Cref{fig:warp,fig:warp-evs}).

%\justin{paragraph here summarizing what just happened and giving intution.  why is this calculation so exciting?  what did we learn?  is it verified in the experiments somewhere (and if so give a forward reference)}

\begin{figure}
\centering
\pgfplotstableread[col sep=comma]{figures/sphere-mean-length.csv}{\data}
\begin{tikzpicture}
\newcommand{\imagewidth}{0.15\columnwidth}
    \begin{semilogyaxis}[
        width = \columnwidth,
        height = 0.7\columnwidth,
        enlarge x limits = 0.1,
        enlarge y limits = 0.2,
        grid = none,
        xlabel = {Mean Edge Length},
        ylabel = {Absolute Eigenvalue Error},
        ylabel near ticks,
        x dir = reverse,
        every tick label/.append style = {font=\tiny},
        legend pos = south west,
        legend style = {nodes = {scale = 0.5, transform shape}}]
    \addplot+ table [x index = 0, y index = 10] {\data};
    \addlegendentry{$\lambda_{10}$}
    \addplot+ table [x index = 0, y index = 20] {\data};
    \addlegendentry{$\lambda_{20}$}
    \addplot+ table [x index = 0, y index = 30] {\data};
    \addlegendentry{$\lambda_{30}$}
    \addplot+ table [x index = 0, y index = 40] {\data};
    \addlegendentry{$\lambda_{40}$}
    \addplot+ table [x index = 0, y index = 50] {\data};
    \addlegendentry{$\lambda_{50}$}
    \addplot+ table [x index = 0, y index = 60] {\data};
    \addlegendentry{$\lambda_{60}$}
    
    \node[inner sep=0pt] at (axis cs:0.1, 5e3) {\includegraphics[width = \imagewidth]{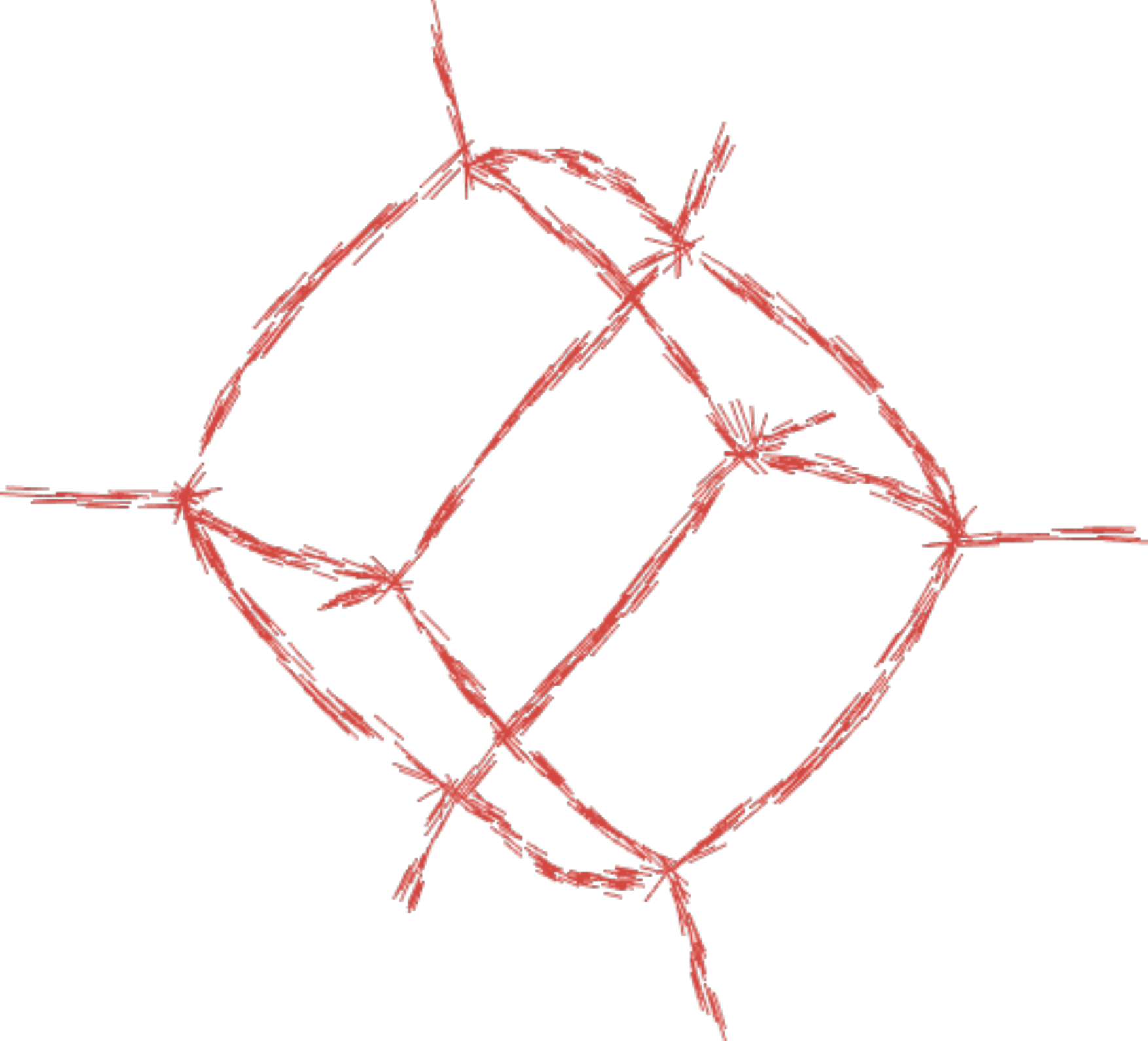}};
    \node[inner sep=0pt] at (axis cs:10^-1.15, 5e3) {\includegraphics[width = \imagewidth]{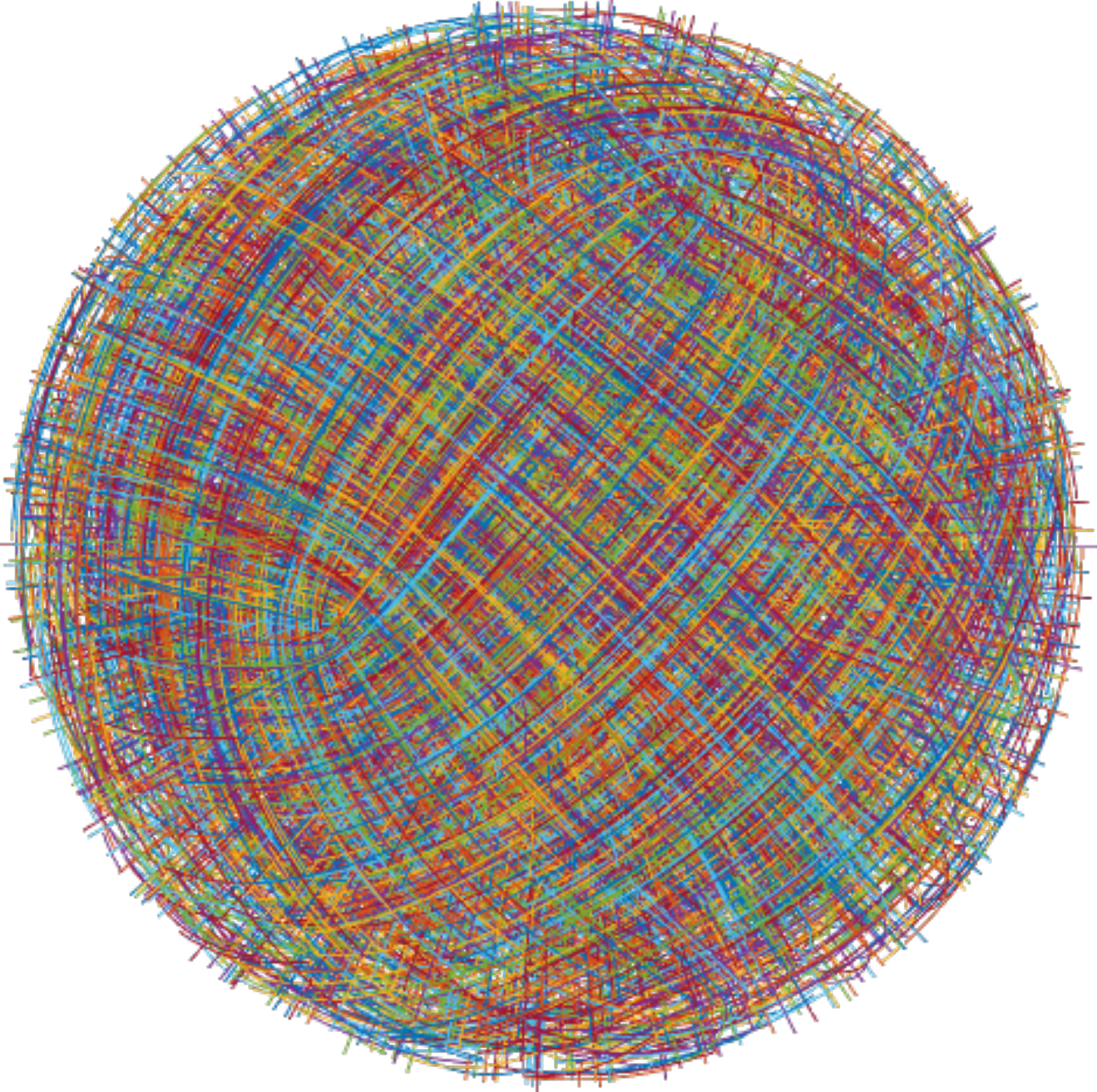}};
    \end{semilogyaxis}
\end{tikzpicture}
\caption{Convergence of the frame field operator spectrum on a ball in 3D, for the frame field shown (inset). We compare to eigenvalues on a finer mesh with mean edge length $\num{0.0680}$.}
\label{fig:cvg-3d}    
\end{figure}

\begin{figure}
\centering
\newcommand{\imagewidth}{0.24\columnwidth}
\setlength\tabcolsep{1.5pt}
\begin{tabular}{@{}c|ccc@{}}
\includegraphics[width=\imagewidth]{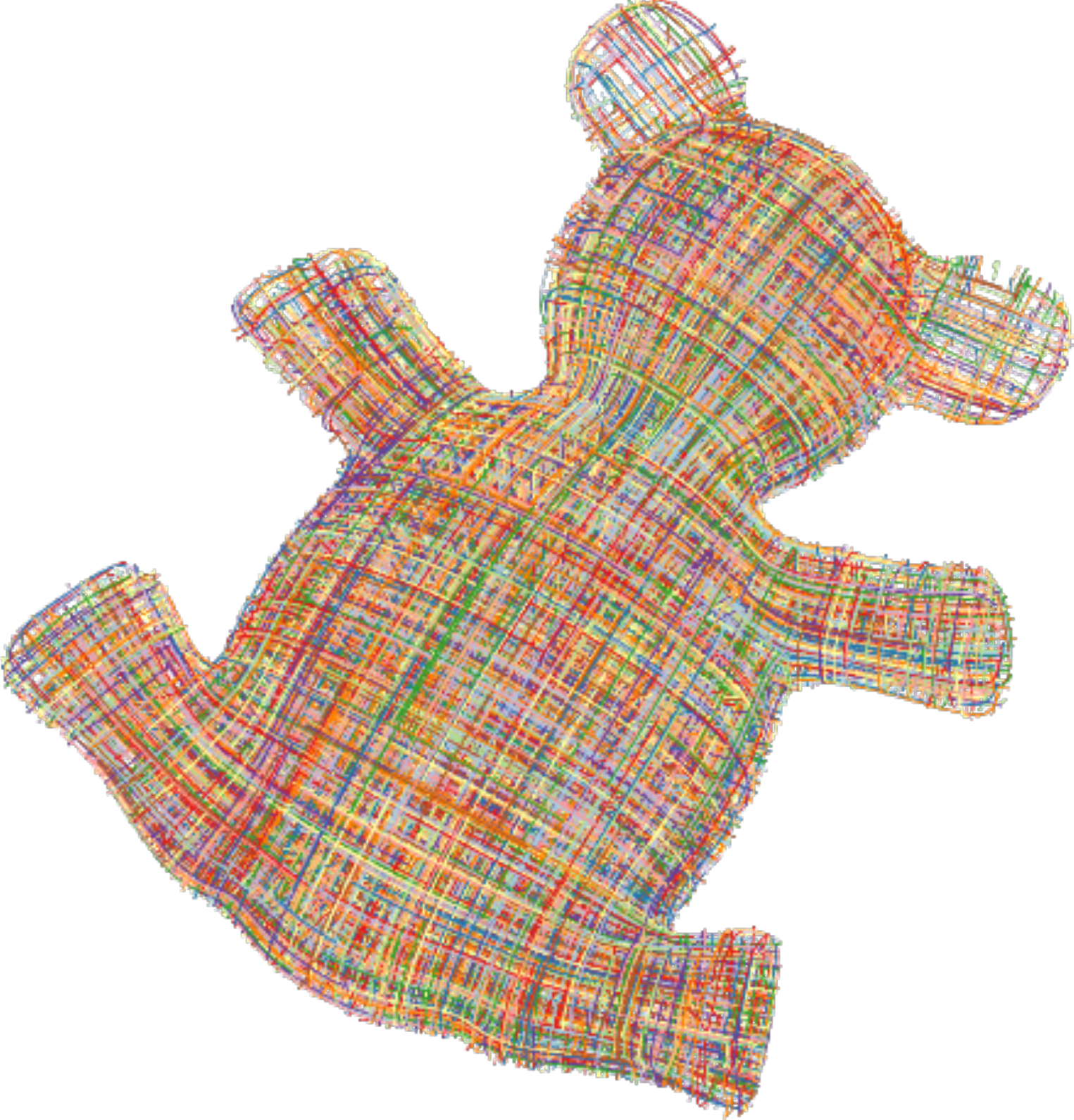} &
\includegraphics[width=\imagewidth]{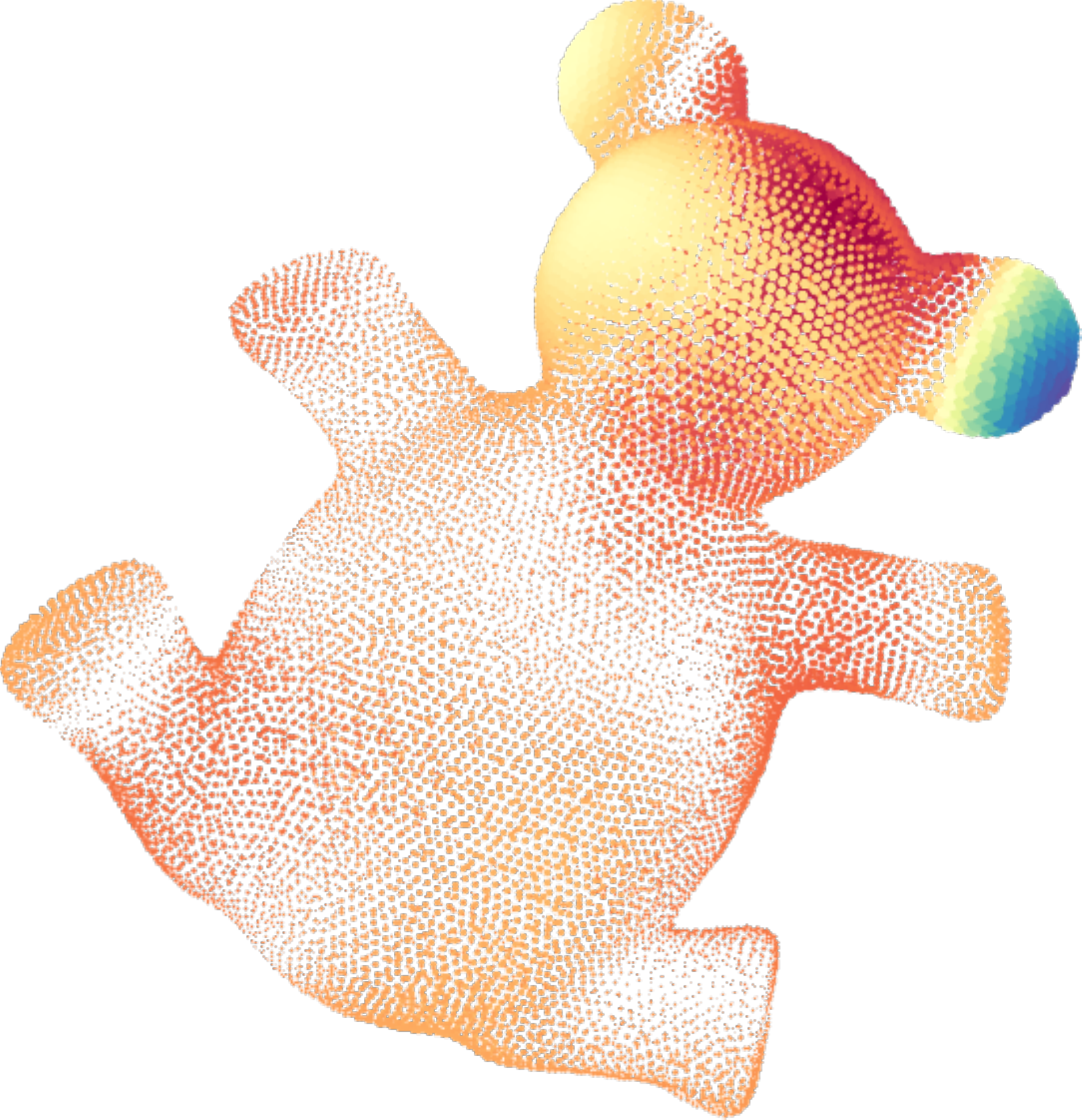} &
\includegraphics[width=\imagewidth]{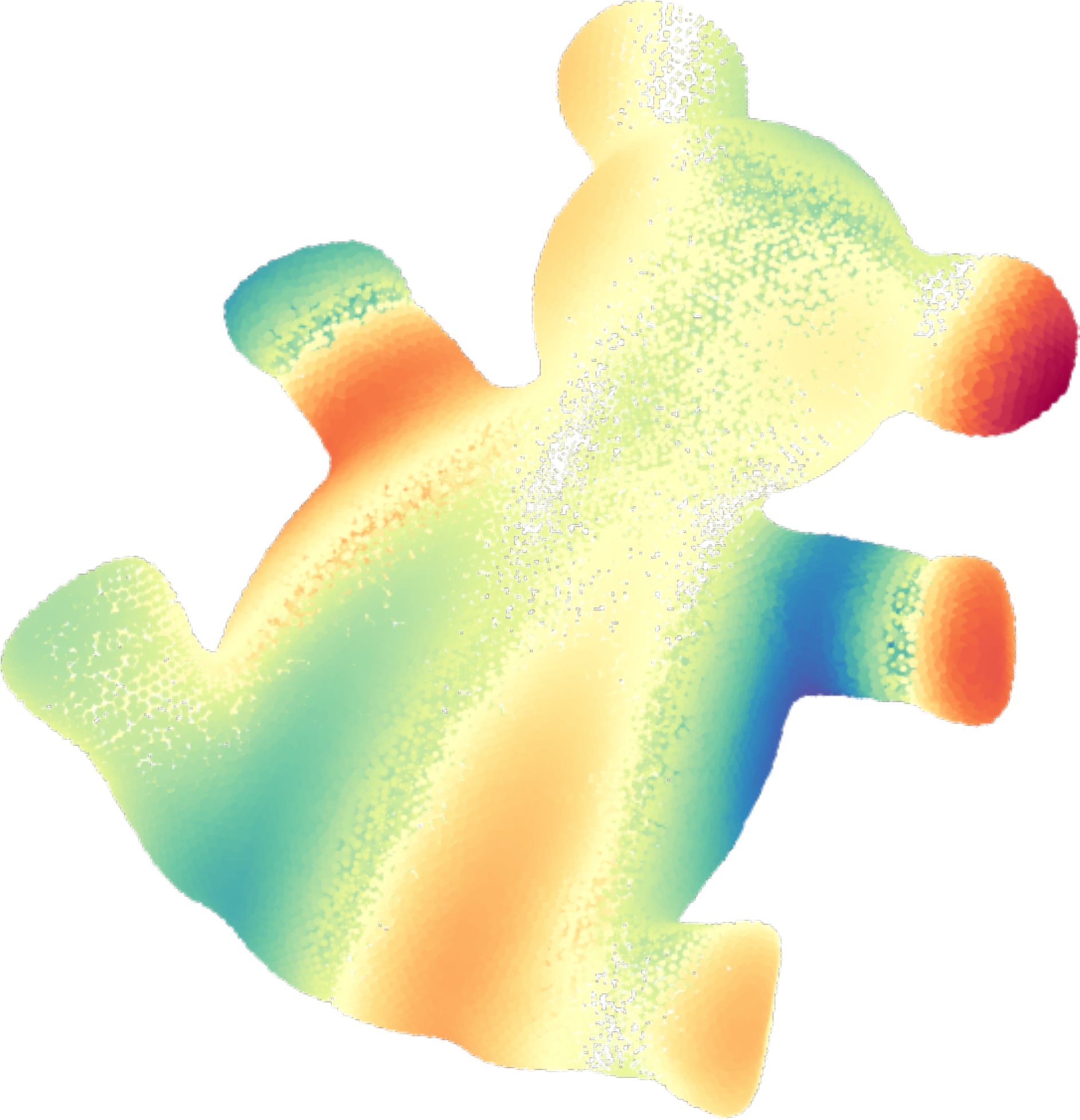} &
\includegraphics[width=\imagewidth]{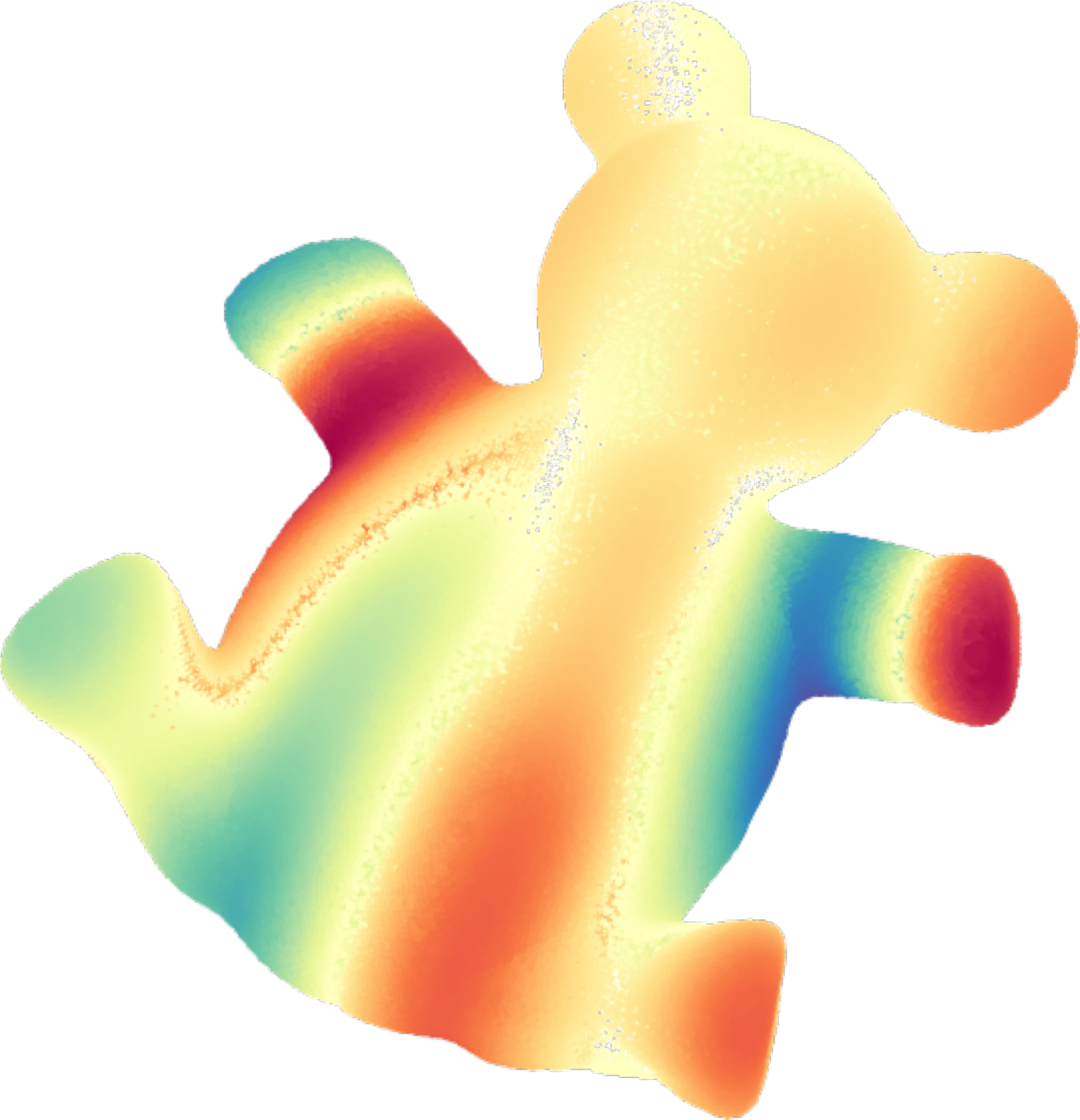} \\
\includegraphics[width=\imagewidth]{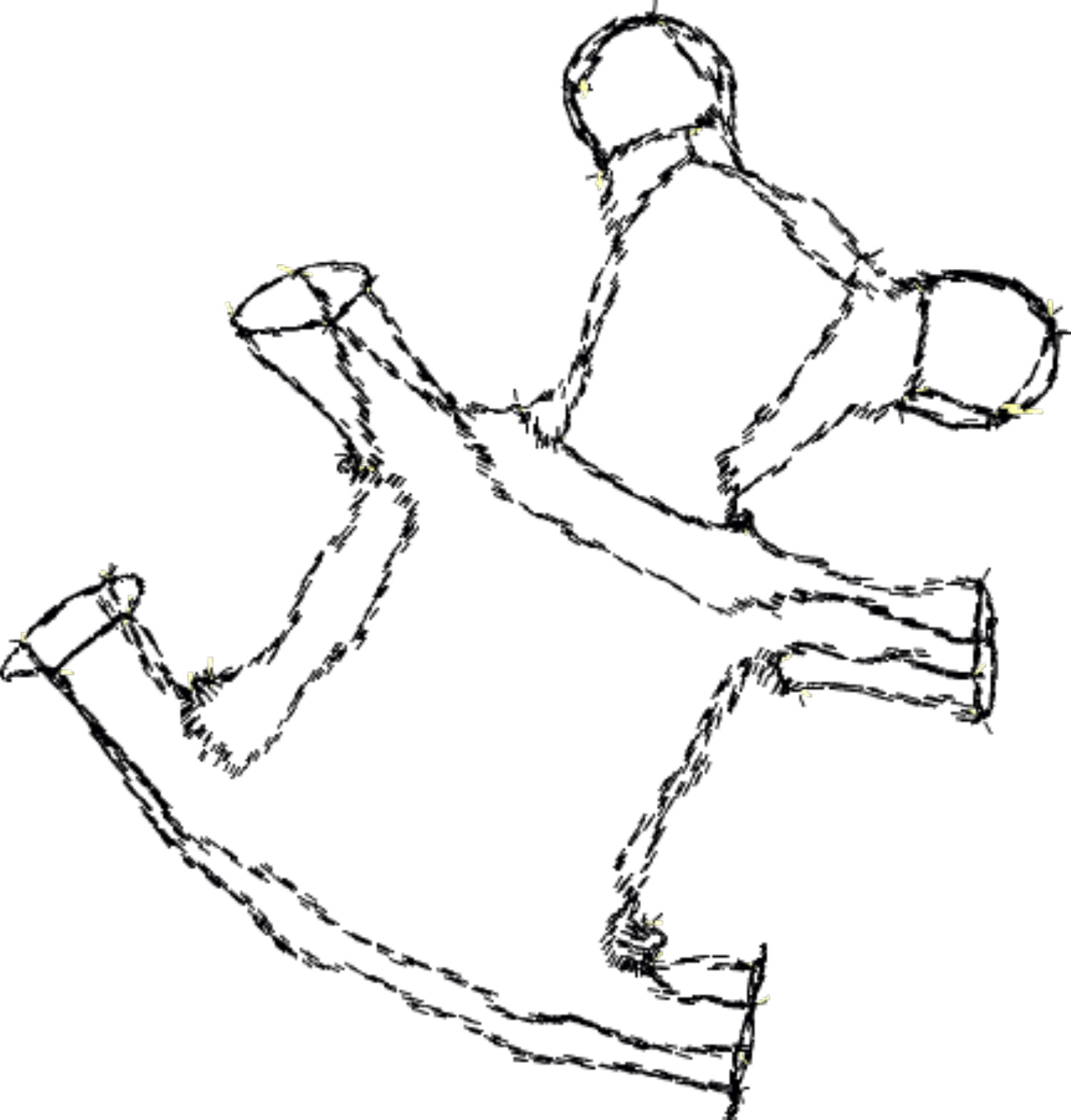} &
\includegraphics[width=\imagewidth]{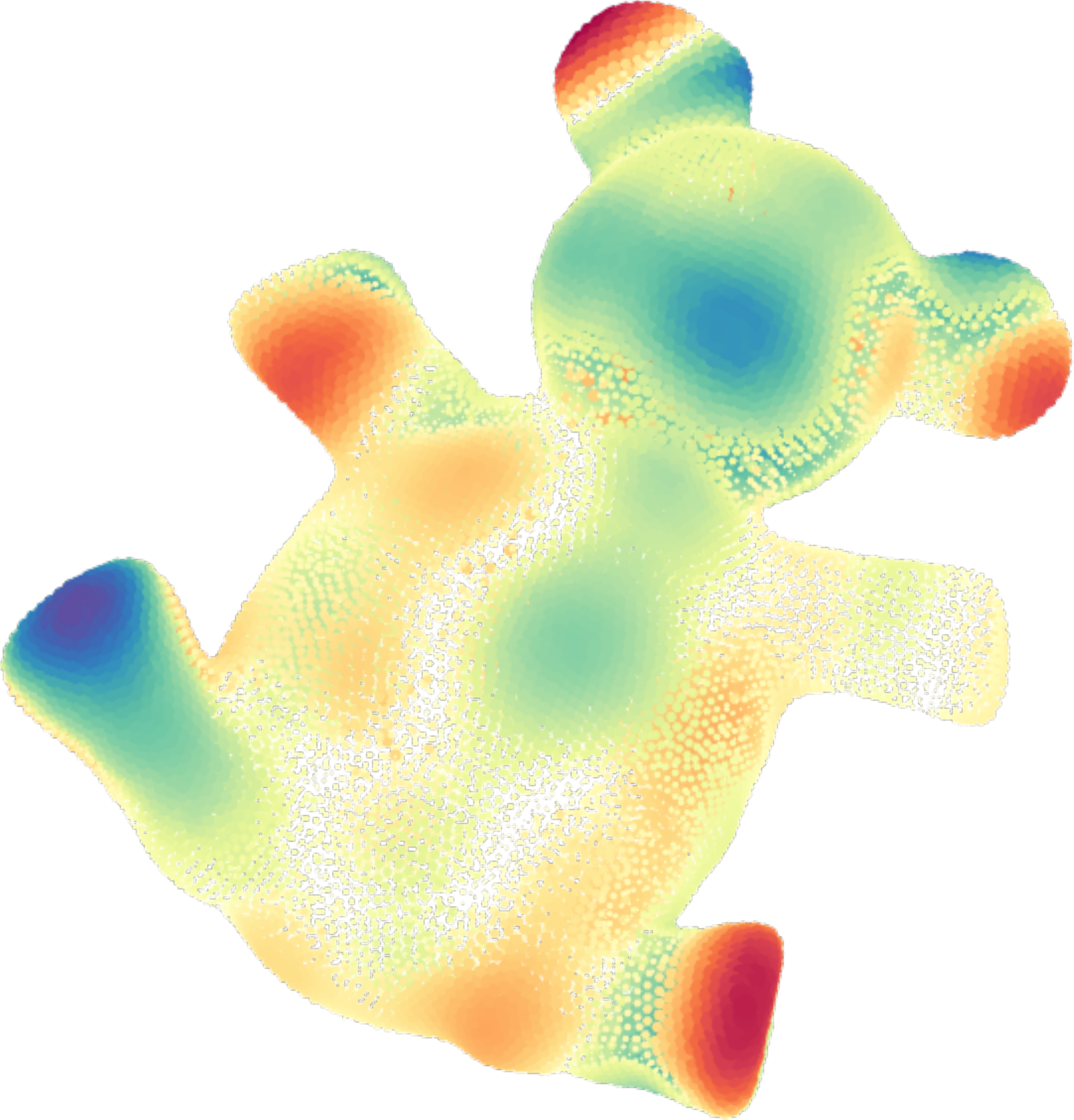} &
\includegraphics[width=\imagewidth]{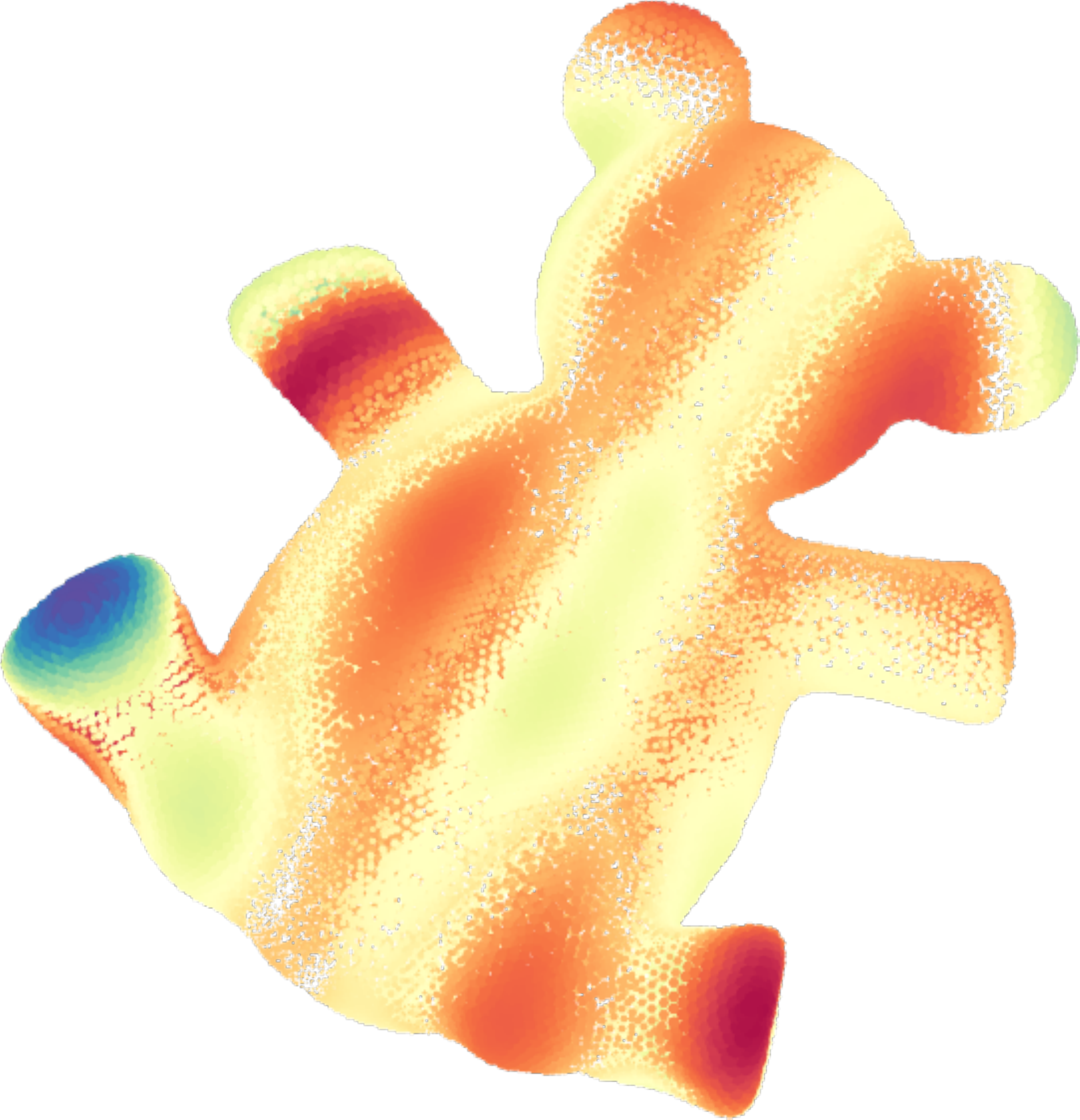} &
\includegraphics[width=\imagewidth]{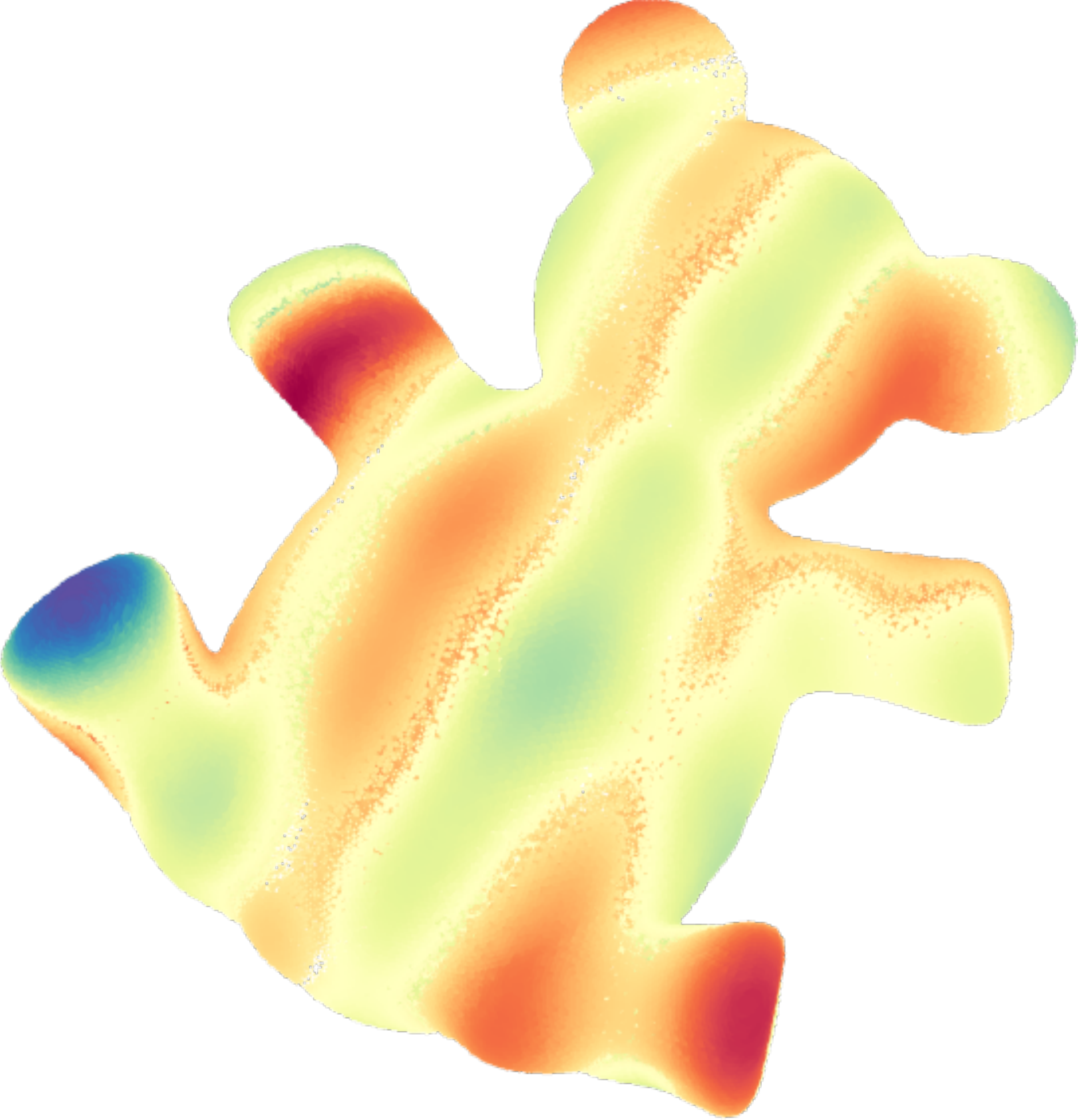} \\
\includegraphics[width=\imagewidth]{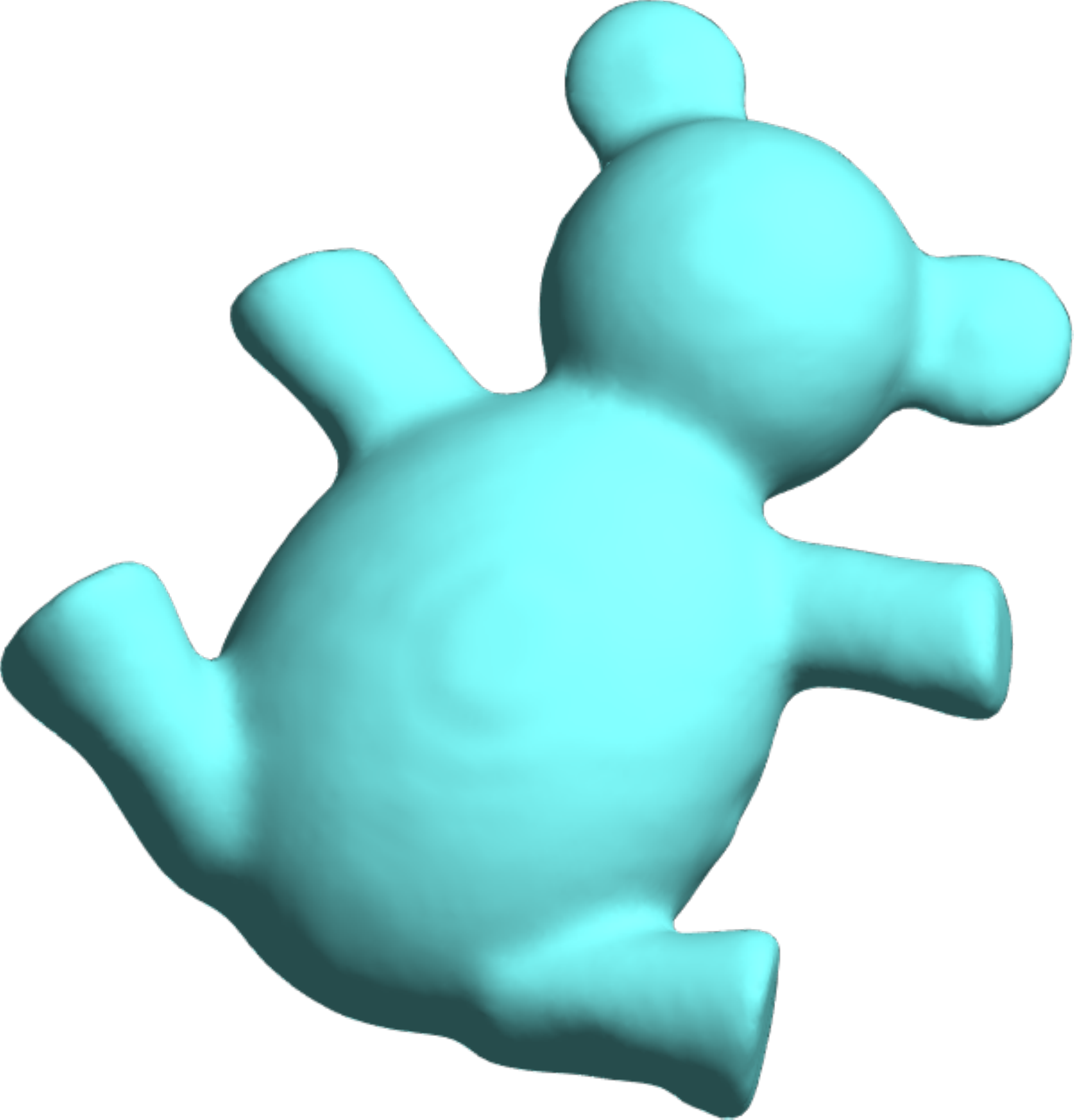} &
\includegraphics[width=\imagewidth]{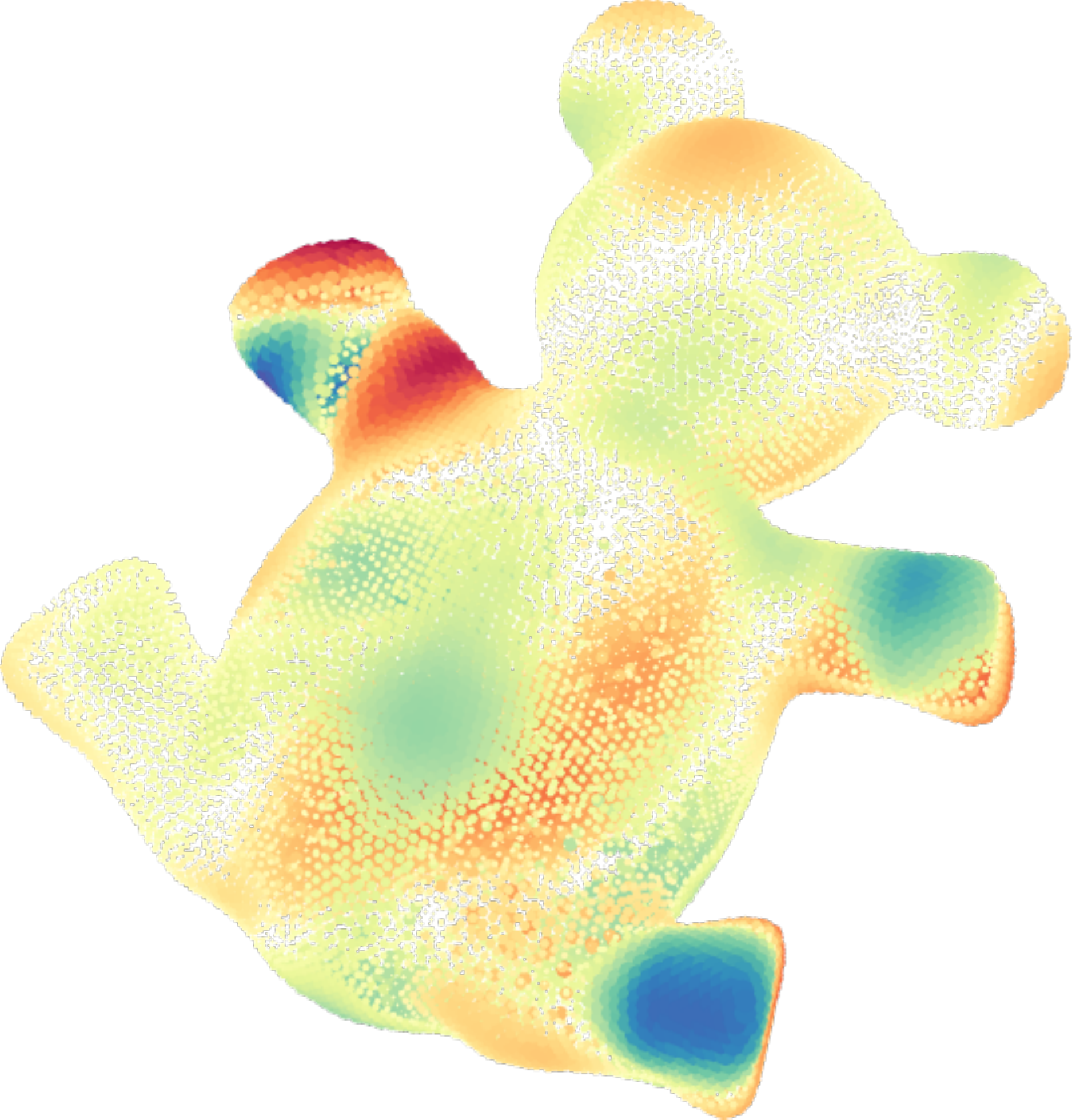} &
\includegraphics[width=\imagewidth]{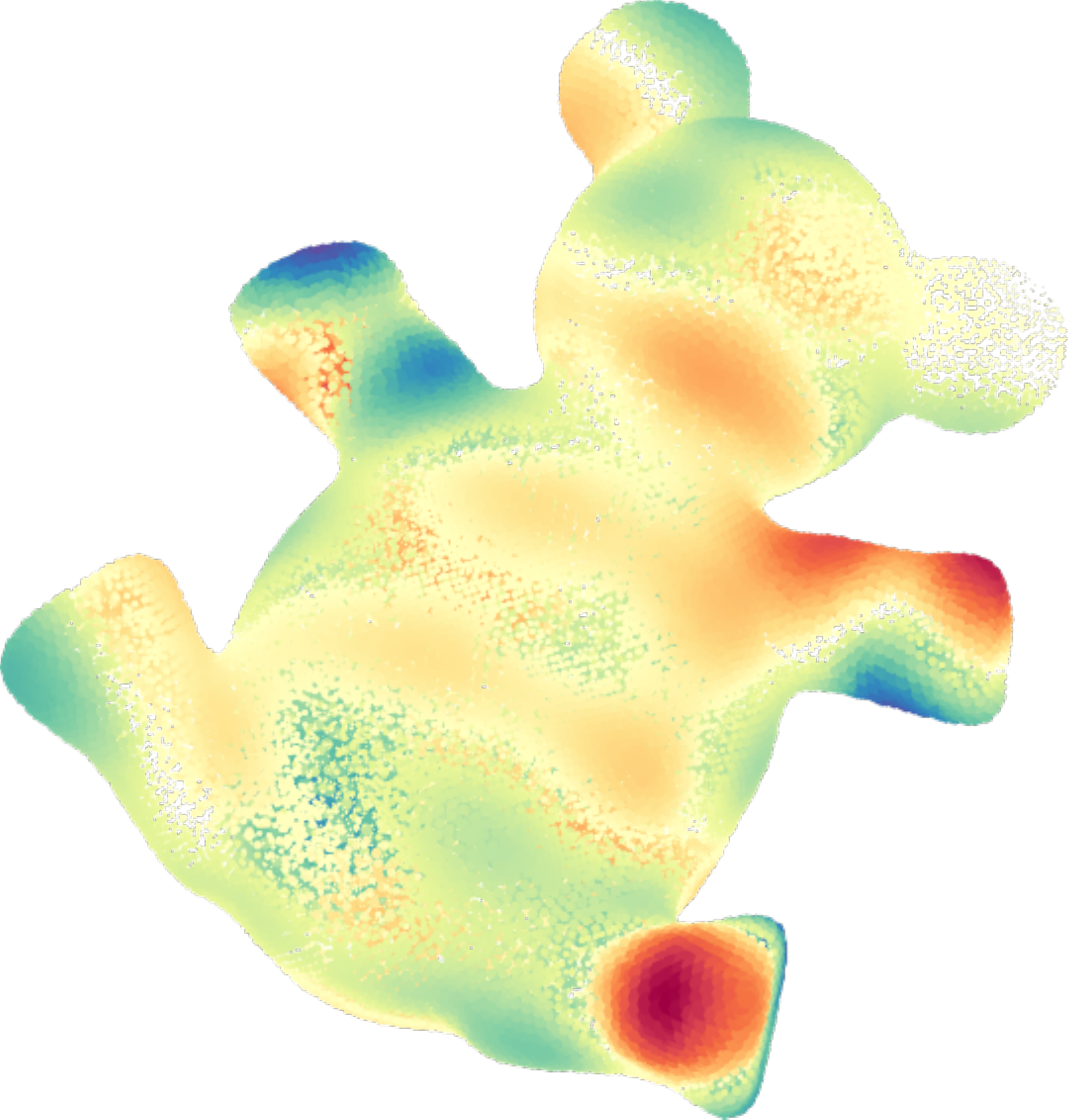} &
\includegraphics[width=\imagewidth]{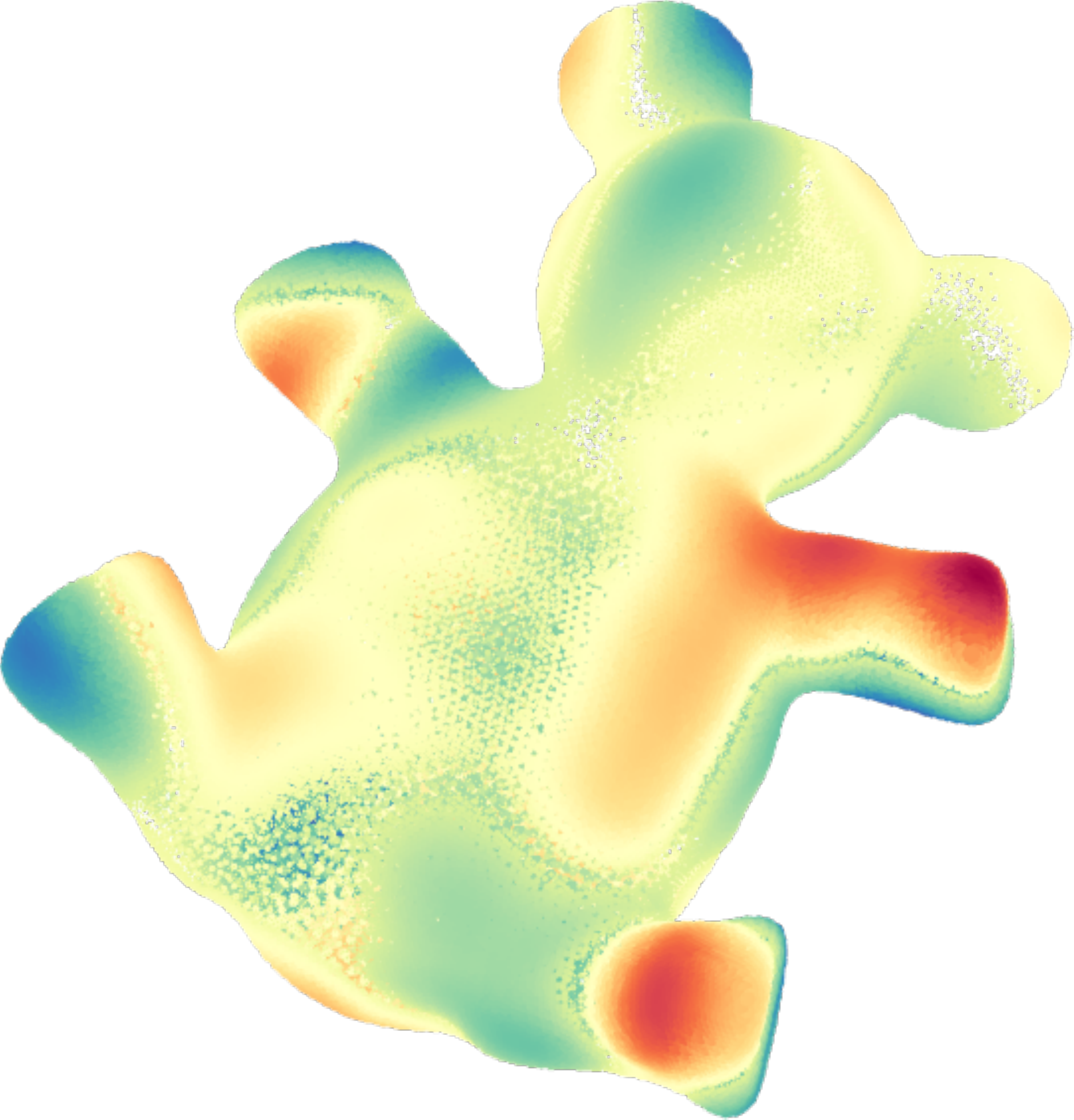} \\
& {\footnotesize $L_{\text{mean}} = \num{0.0373}$} & {\footnotesize $L_{\text{mean}} = \num{0.0287}$} & {\footnotesize $L_{\text{mean}} = \num{0.0233}$}
\end{tabular}
\caption{Comparison of frame field operator eigenfunctions corresponding to $\lambda_{16}$ (top), $\lambda_{32}$ (middle), and $\lambda_{48}$ (bottom) across various mesh resolutions. The frame field, its singular curves, and the domain boundary are shown at left. Note how the lower-frequency eigenfunctions appear to stabilize at higher resolutions. $L_{\text{mean}}$ indicates mean mesh edge length.}
\label{fig:ef-3d}
\end{figure}

\section{Discretization} \label{sec.discretization}
As the functional $\mathcal{E}_{T,\epsilon}$ is quadratic in the second derivatives of $u$, we pursue a mixed finite element discretization that follows the discretization of the Hessian energy by Stein et al.\ \cite{Stein2018}. Unlike their work, however, we do not necessarily want the natural boundary conditions of our functional. For example, we might want to impose Neumann boundary conditions in distance computation applications, or for computing eigenfunctions that have ridgelines on the boundary. Rather than clamping function values on boundary triangles, which can be numerically unstable, we show how to impose boundary conditions in a weak sense, i.e., by clamping a Lagrange multiplier instead.

\subsection{Mixed FEM Lagrangian}
In the mixed finite element method (mixed FEM), the degree of the finite element basis is too low to represent even the derivatives that appear in the variational or weak formulations of a PDE. Instead, we replace higher derivatives with coupled lower-order PDEs enforced via Lagrange multipliers.
As the frame field functional and operator generalize the Hessian energy and Bilaplacian, respectively, we adopt the mixed FEM approach of \cite{Stein2018}, in which functions, Hessians, and Lagrange multipliers are represented in the linear FEM basis on a triangle mesh (or, in our case, a tetrahedral mesh).
We begin by reformulating the problem $\min_u \mathcal{E}_{T,\epsilon}(u)$ into the equivalent constrained optimization problem
\begin{equation}
\begin{alignedat}{2}
  & \text{minimize } & & \int_\Omega \frac{1}{2}V:T^{\epsilon}:V \; d\Omega \\
   & \text{subject to }& \quad & \nabla^2u = V.
\end{alignedat}
\end{equation}
Enforcing the constraint $\nabla^2 u = V$ via the Lagrange multiplier $\Lambda$, a second-order symmetric tensor field, we obtain the Lagrangian
\begin{equation} \mathcal{L}_{T,\epsilon}(u,V) = \int_\Omega \left[\frac{1}{2}V:T^{\epsilon}:V + \Lambda:(V - \nabla^2u)\right] \; d\Omega.
\end{equation}
Now integrating by parts, we see that $\mathcal{L}_{T,\epsilon}$ can be rewritten
\begin{equation} \begin{aligned}
\mathcal{L}_{T,\epsilon}(u,V) &= \frac{1}{2} \int_\Omega \left[ V:T^{\epsilon}:V + (\nabla \cdot \Lambda) \cdot \nabla u + \Lambda : V \right]  \; d\Omega\\ &\quad + \int_{\partial\Omega}n^\top \Lambda \nabla u \; dA, \end{aligned} \label{eq.mixed-ibp} \end{equation}
where $\nabla \cdot V$ denotes the symmetric tensor divergence of $V$. Observe that $\mathcal{L}_{T,\epsilon}$ now includes only \emph{first derivatives} of $u$ and $\Lambda$, which can be represented faithfully in the linear FEM basis.

\subsection{Weak Boundary Conditions}
Boundary conditions on $V$ can now be imposed weakly by constraining on the Lagrange multiplier $\Lambda$. In particular, setting $\Lambda$ such that the normal $n(x)$ is an eigenvector of $\Lambda(x)$ for $x \in \partial\Omega$ will transform the boundary term in \eqref{eq.mixed-ibp} into the form
\begin{equation}
\int_{\partial\Omega}n^\top \Lambda \nabla u \; dA = 
\int_{\partial\Omega} \phi n^\top \nabla u \; dA \end{equation}
for an arbitrary function $\phi : \partial \Omega \to \R$, which has the form of a homogeneous Neumann boundary term. An equivalent way to write the constraint on $\Lambda$ is
\begin{equation}
(I - n(x) n(x)^\top) \Lambda(x) n(x) = 0 \quad x \in \partial \Omega.
\end{equation}
This equation is linear and homogeneous---in particular, it can be expressed in the form $B(x) \operatorname{vec}\Lambda(x) = 0$, where $B(x)$ is a matrix-valued field on $\partial \Omega$, and $\operatorname{vec} \Lambda(x)$ denotes the coefficients of $\Lambda(x)$ arranged in a vector.

To obtain \emph{natural boundary conditions}, we instead set $\Lambda$ to zero on the boundary, thus eliminating the boundary term from \eqref{eq.mixed-ibp} entirely.

\begin{figure}
\centering
\newcommand{\imagewidth}{0.19\columnwidth}
\setlength\tabcolsep{1.5pt}
\begin{tabular}{@{}c|cccc@{}}
\includegraphics[width=\imagewidth]{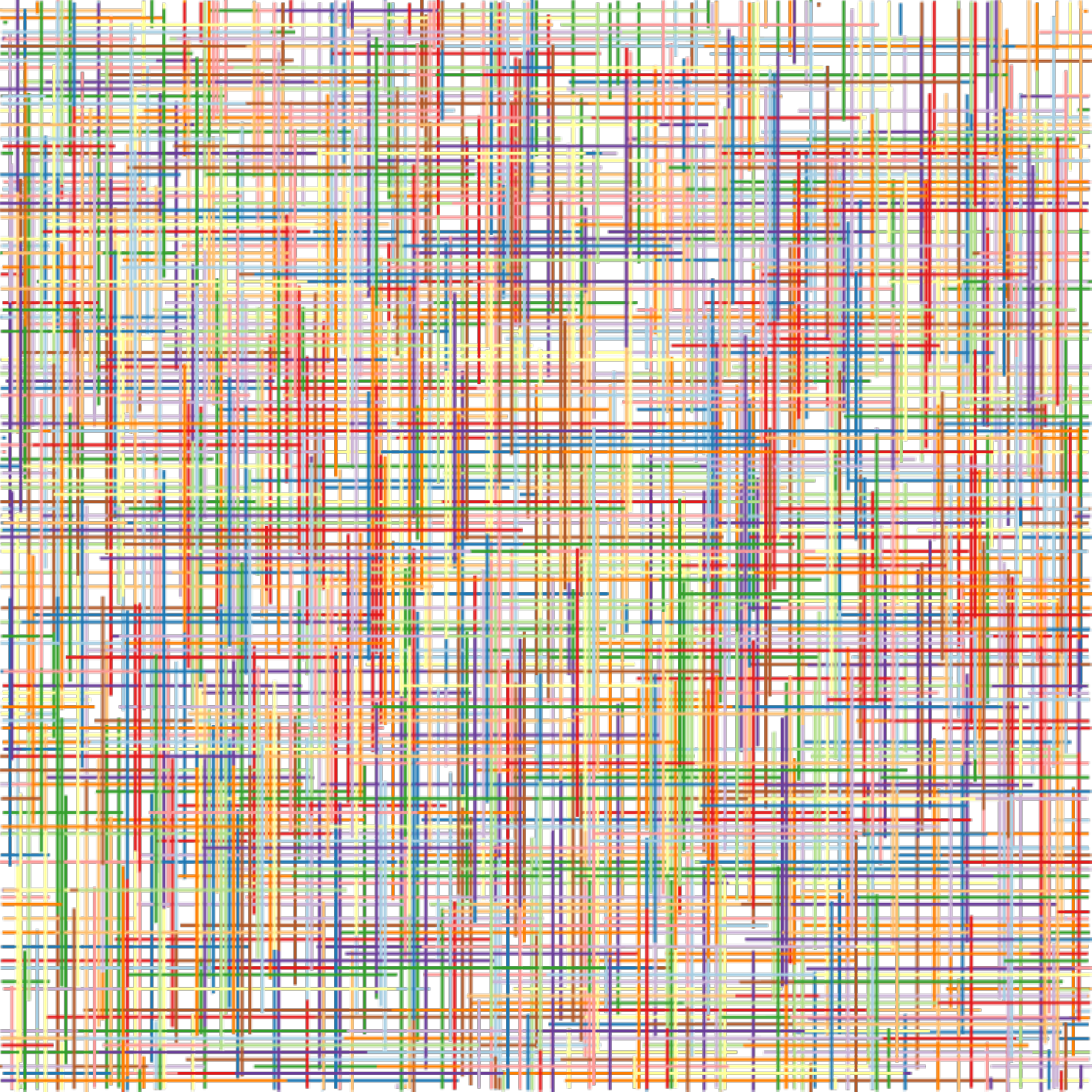} &
\includegraphics[width=\imagewidth]{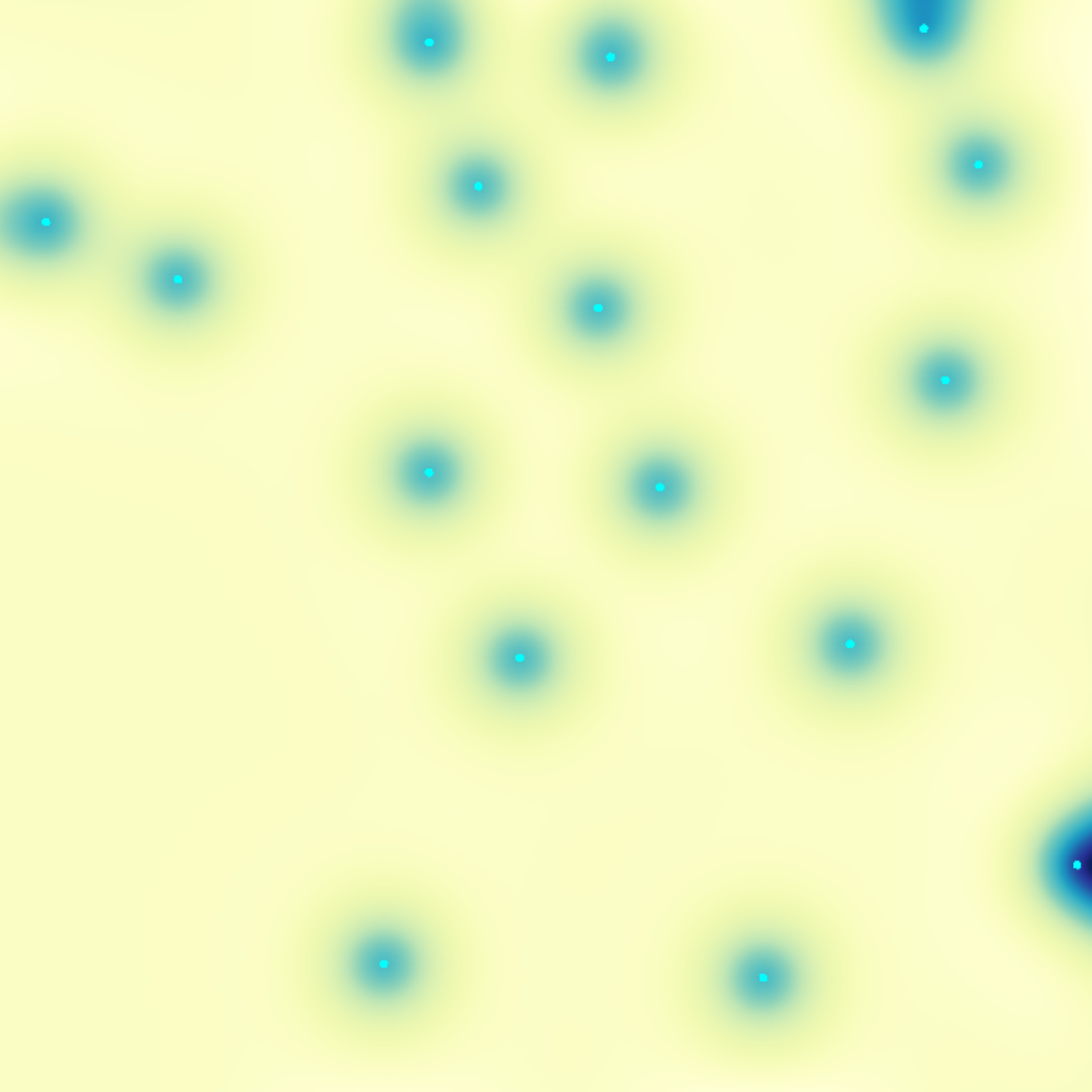} &
\includegraphics[width=\imagewidth]{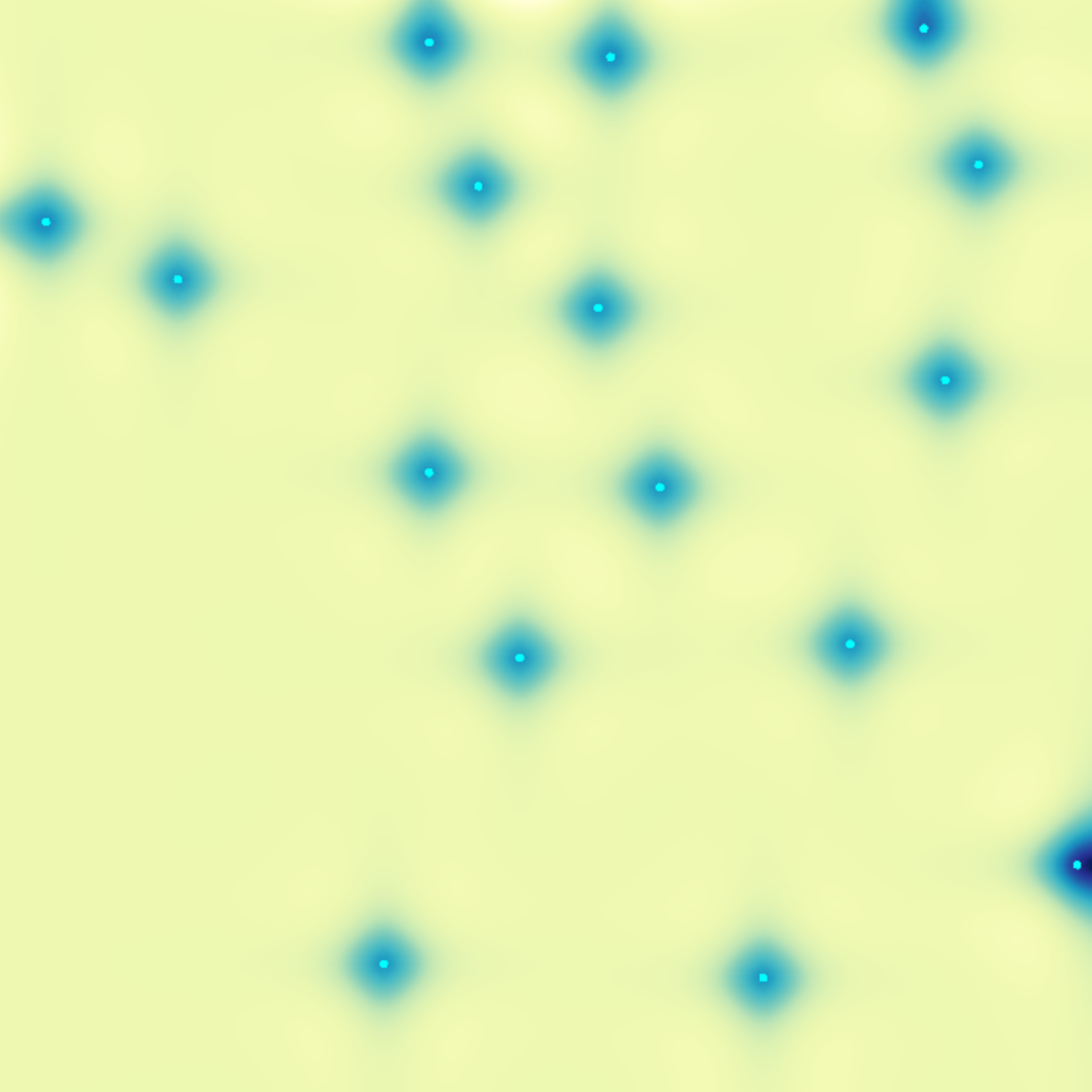} &
\includegraphics[width=\imagewidth]{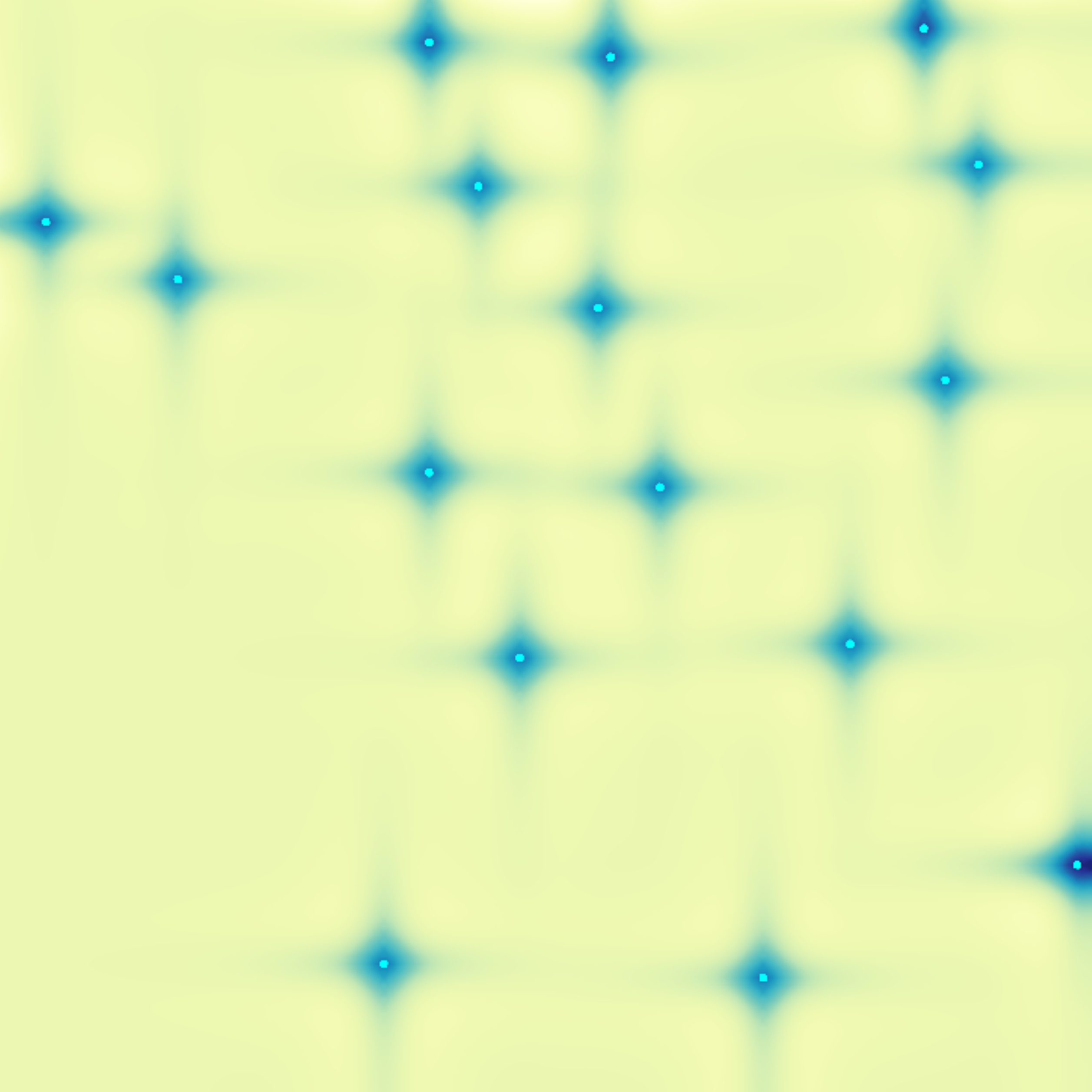} &
\includegraphics[width=\imagewidth]{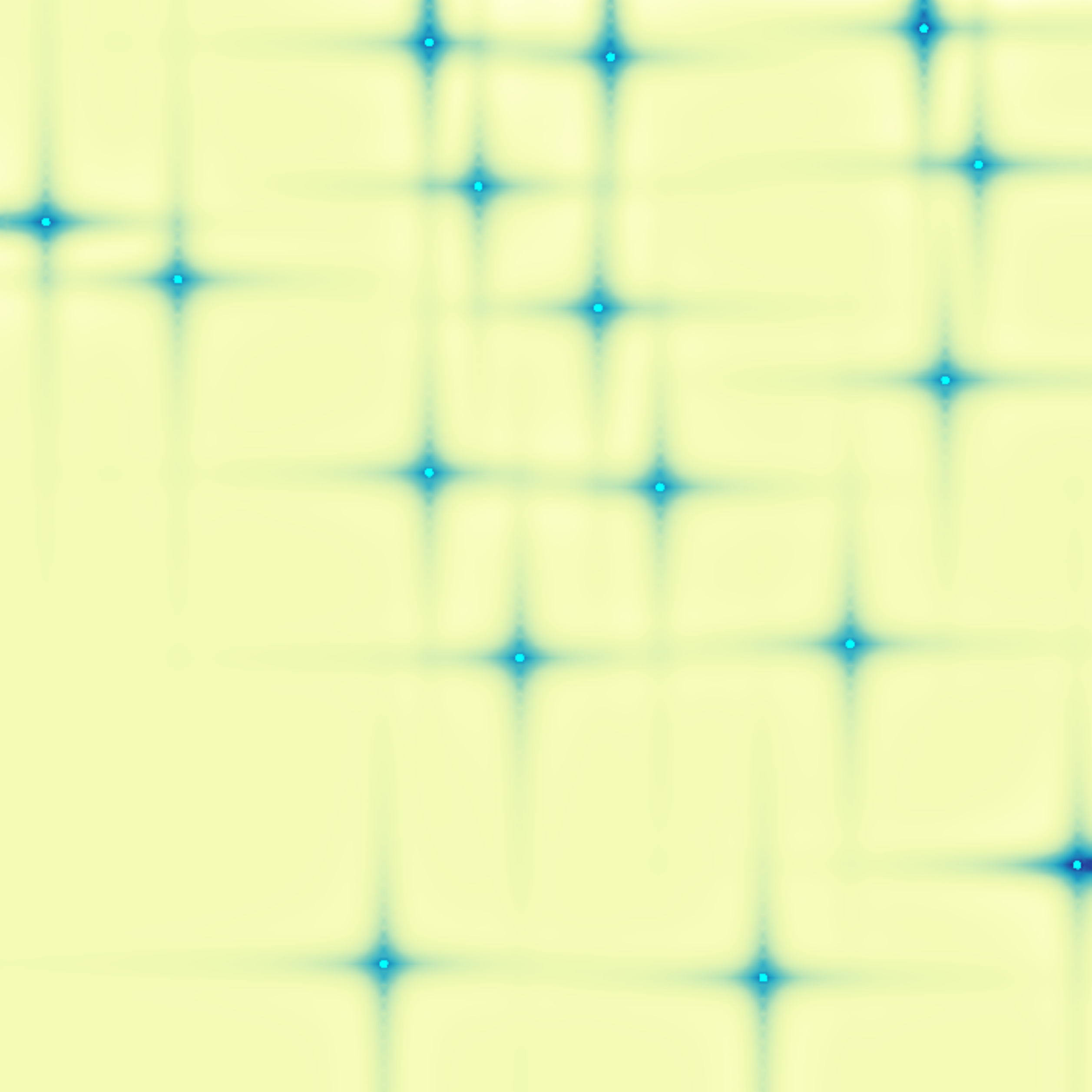} \\
\includegraphics[width=\imagewidth]{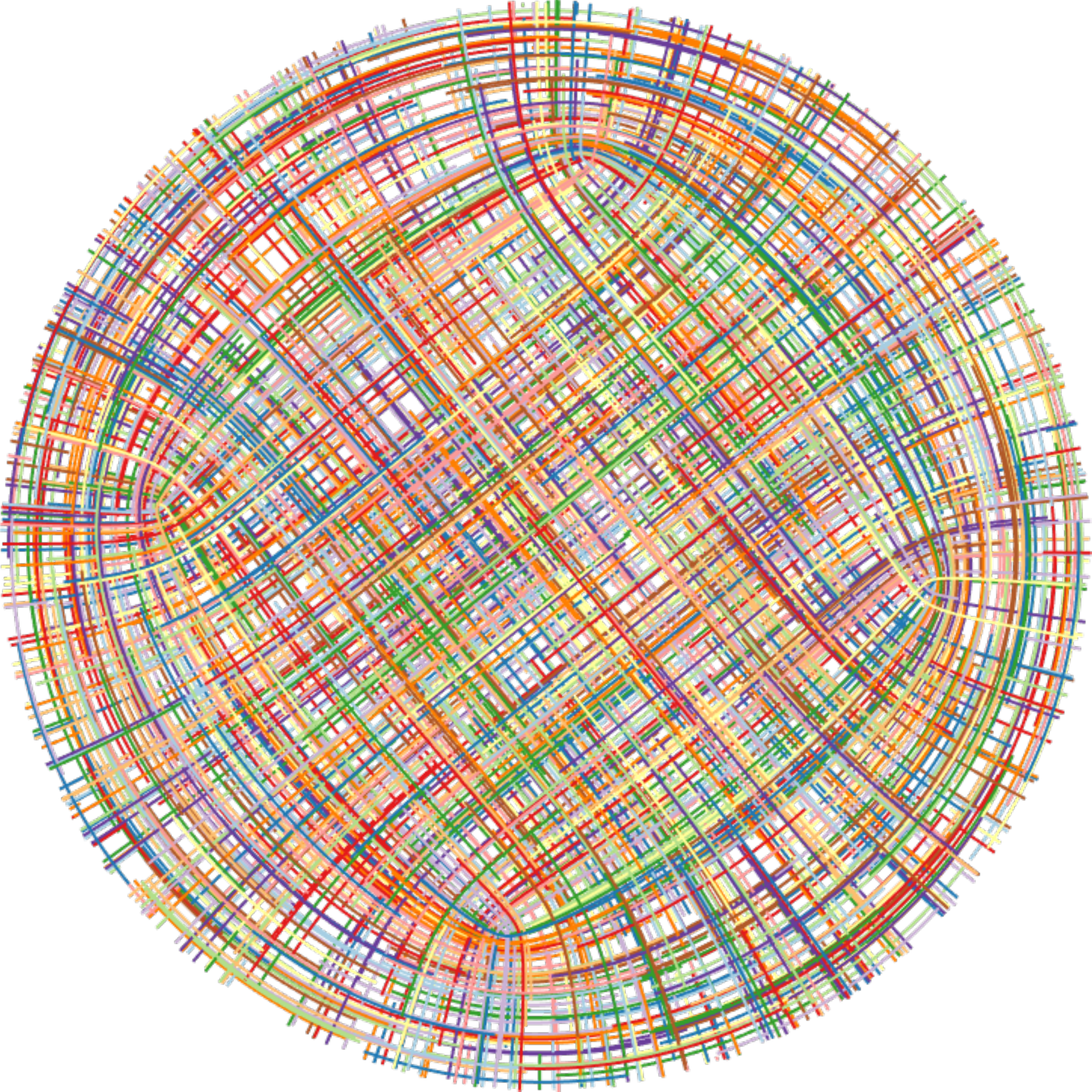} &
\includegraphics[width=\imagewidth]{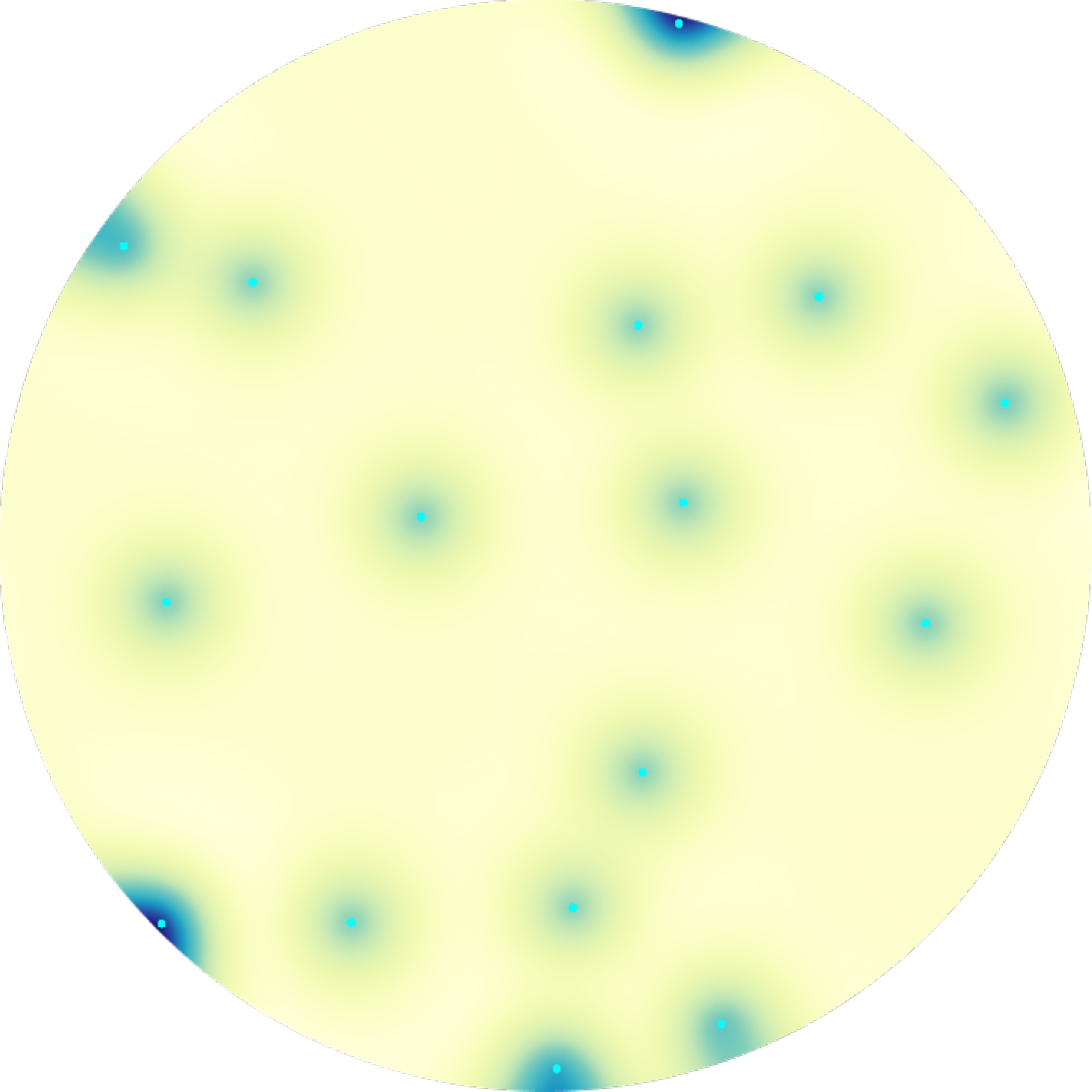} &
\includegraphics[width=\imagewidth]{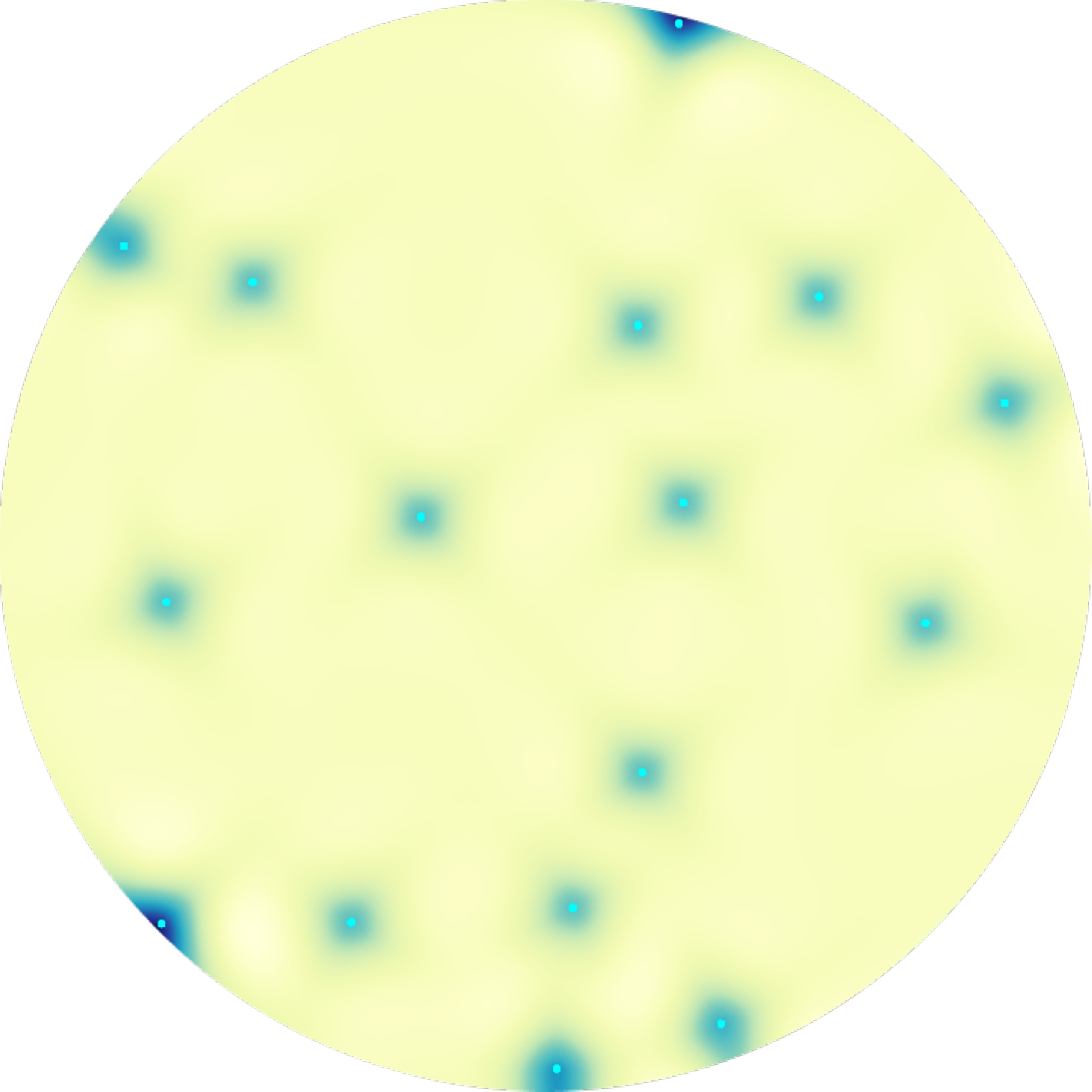} &
\includegraphics[width=\imagewidth]{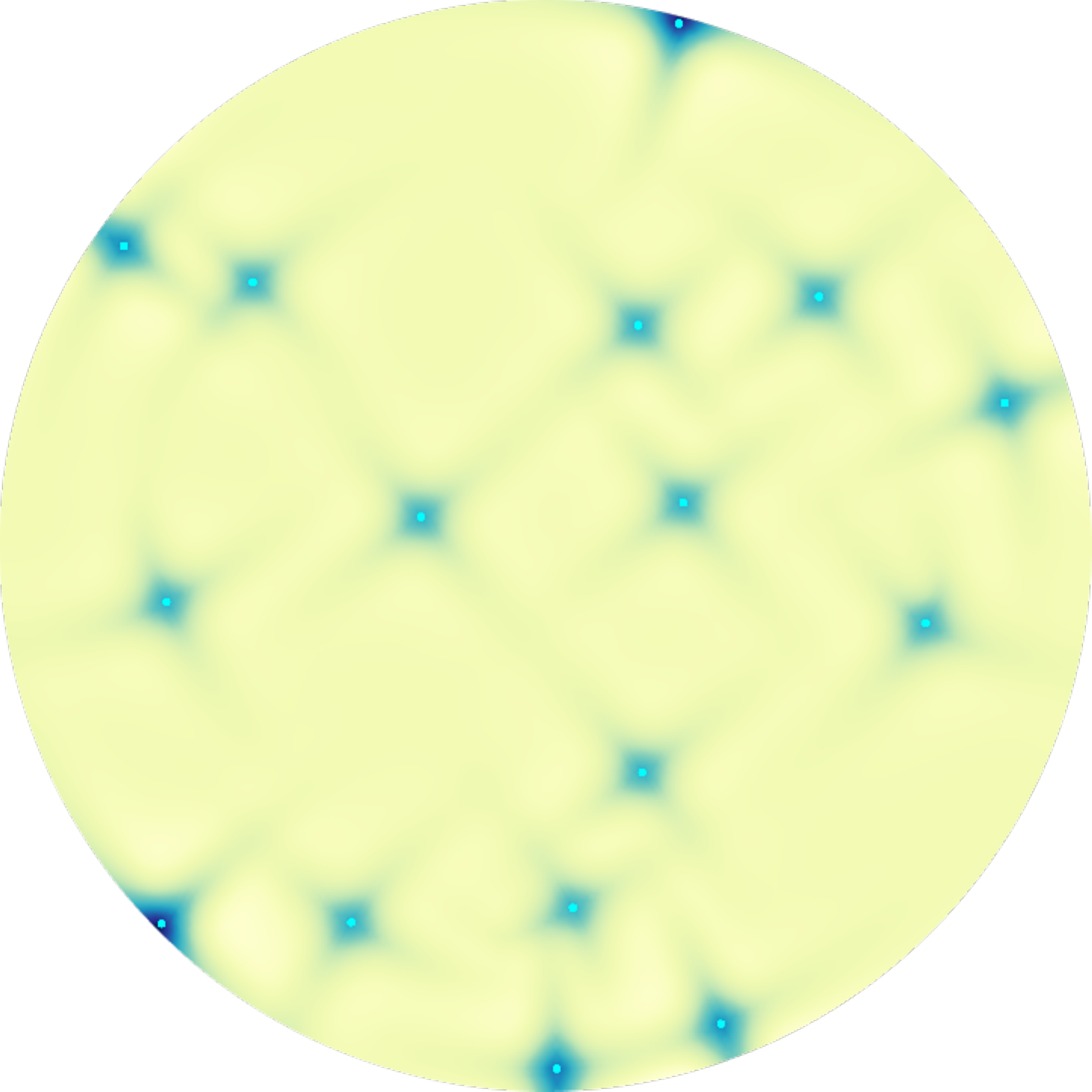} &
\includegraphics[width=\imagewidth]{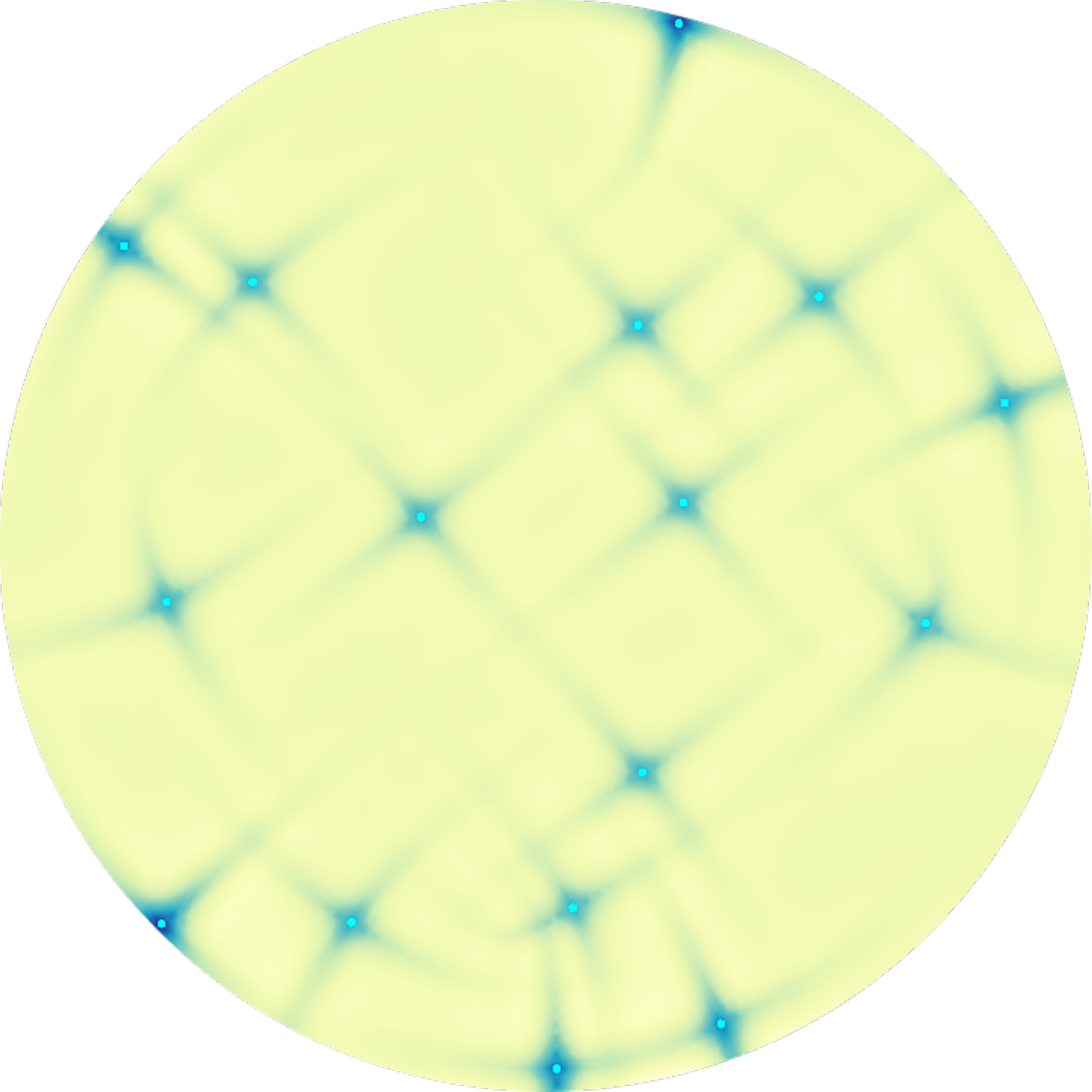} \\
\includegraphics[width=\imagewidth]{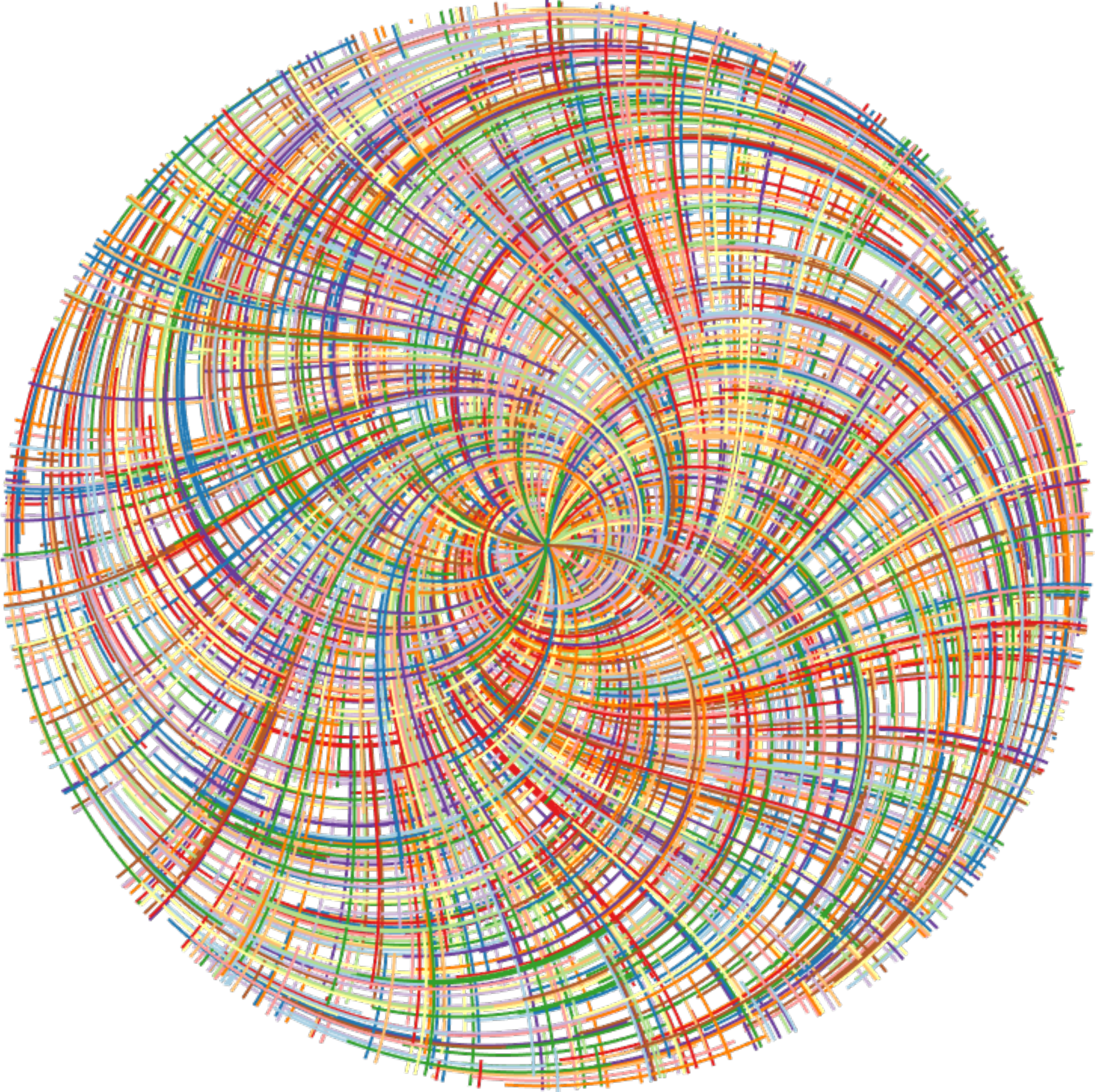} &
\includegraphics[width=\imagewidth]{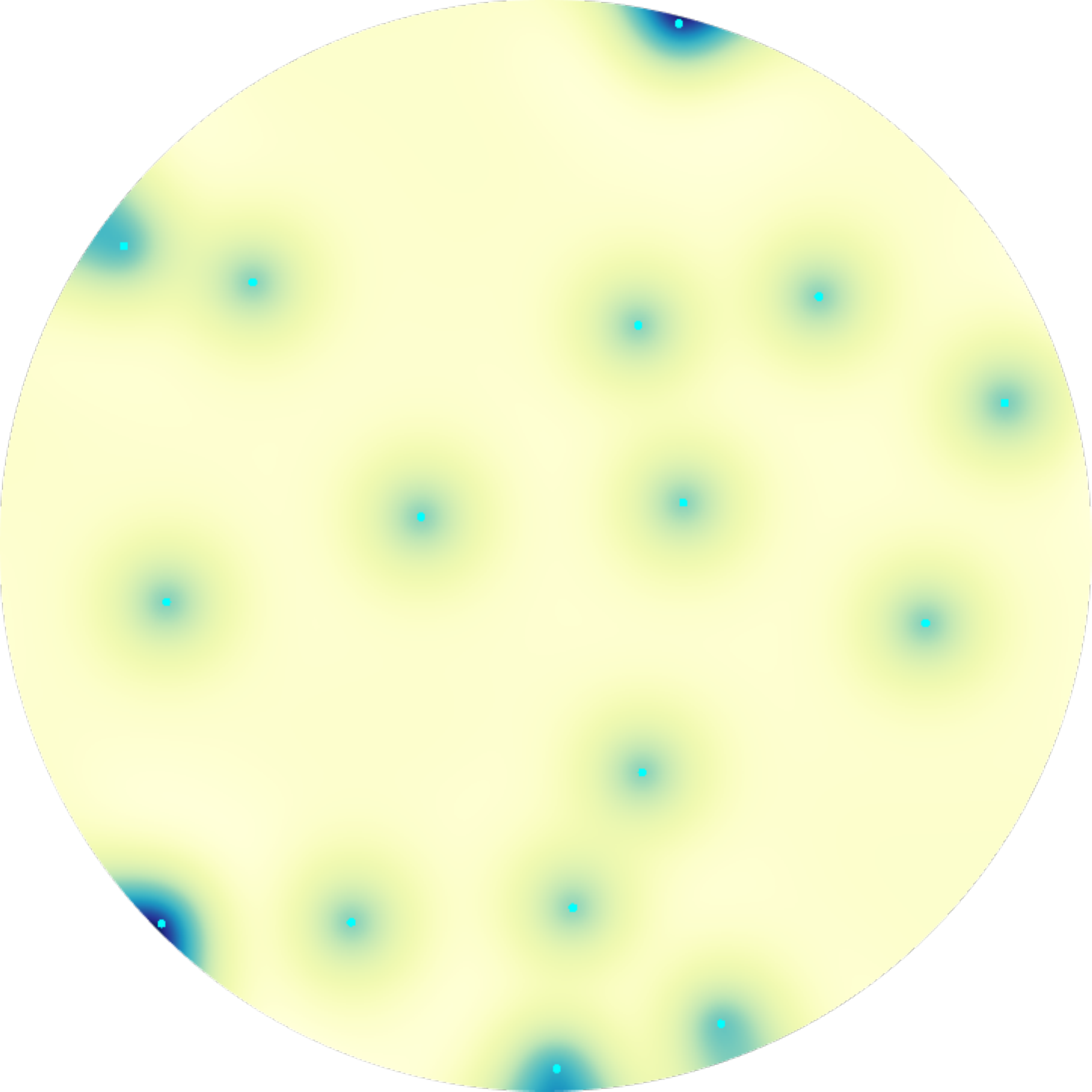} &
\includegraphics[width=\imagewidth]{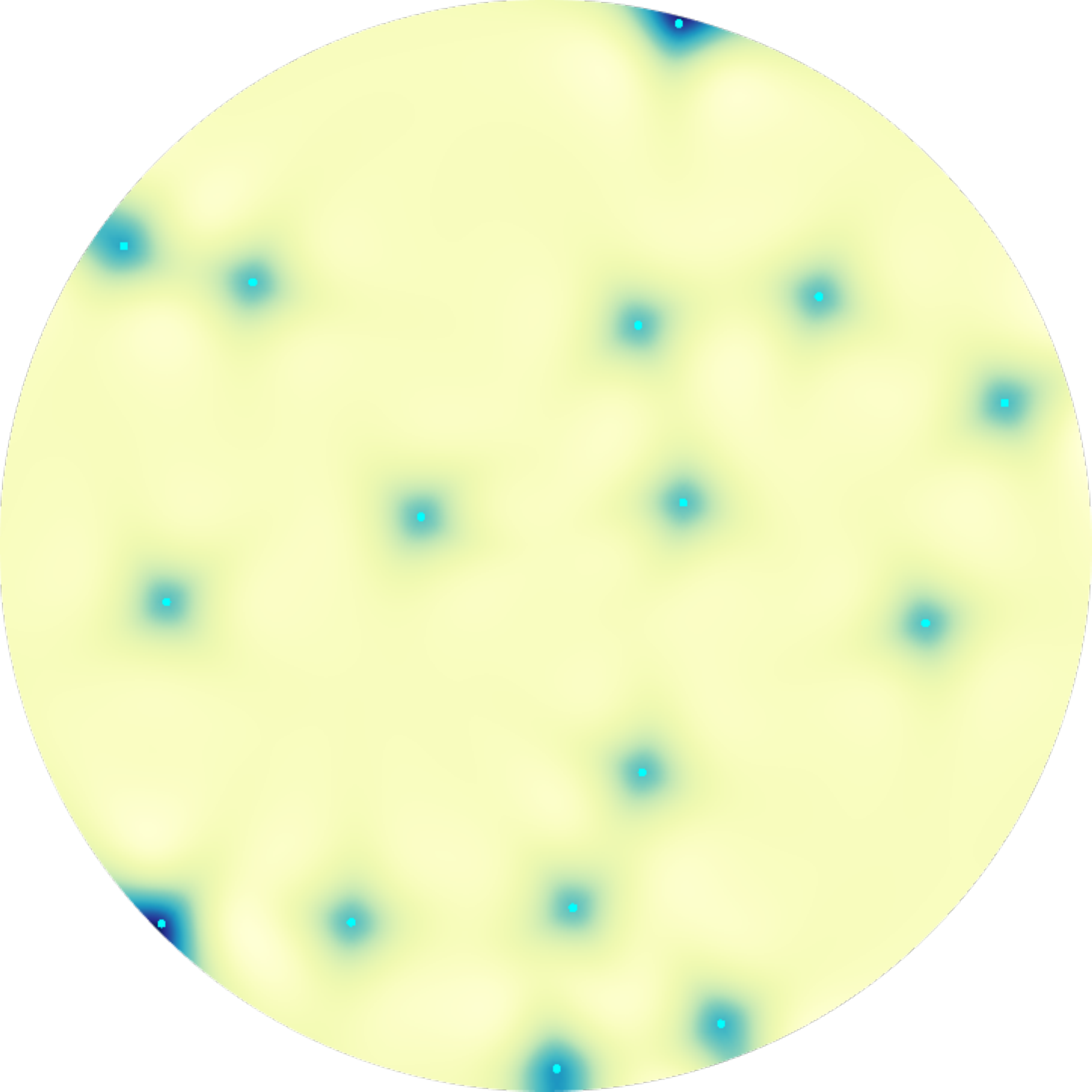} &
\includegraphics[width=\imagewidth]{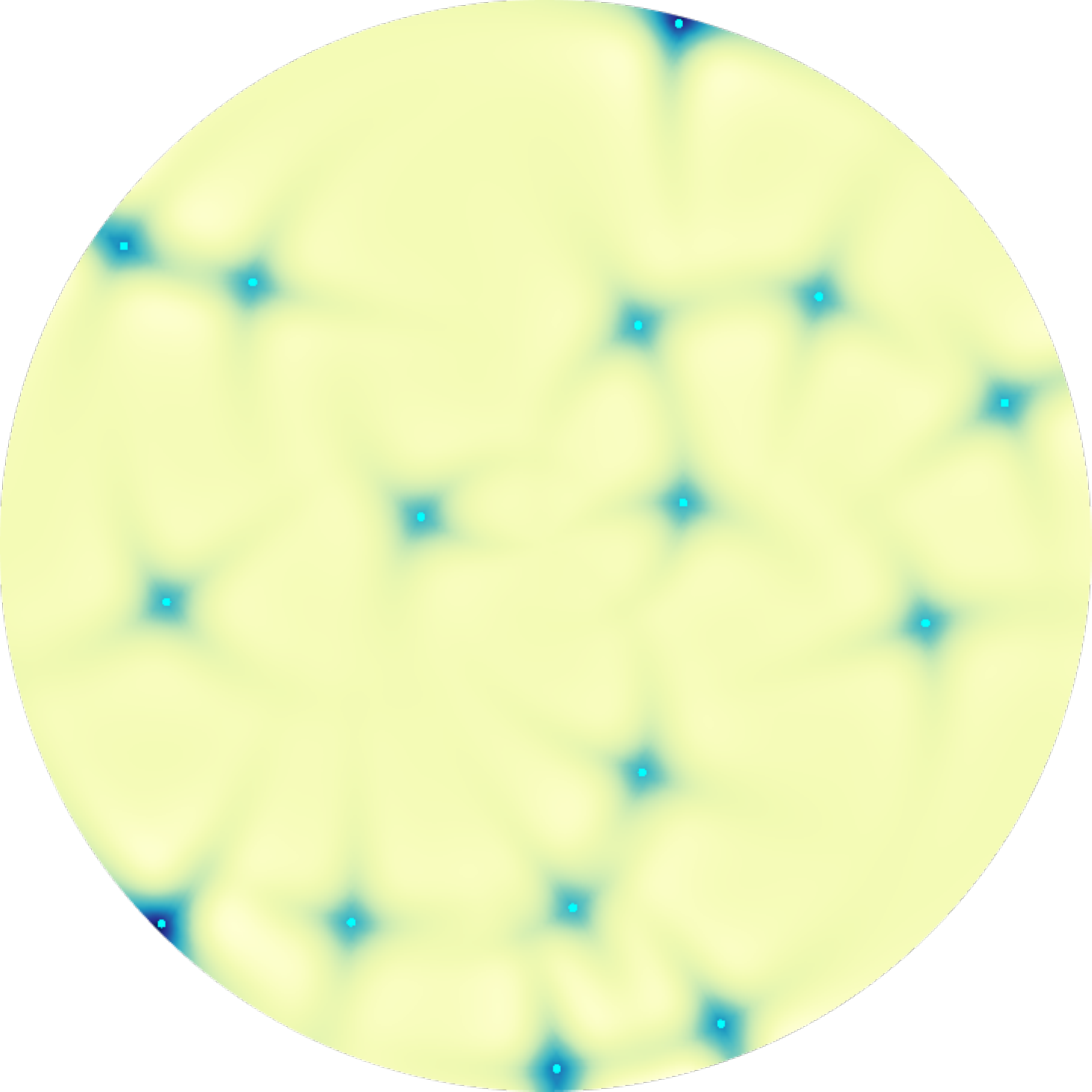} &
\includegraphics[width=\imagewidth]{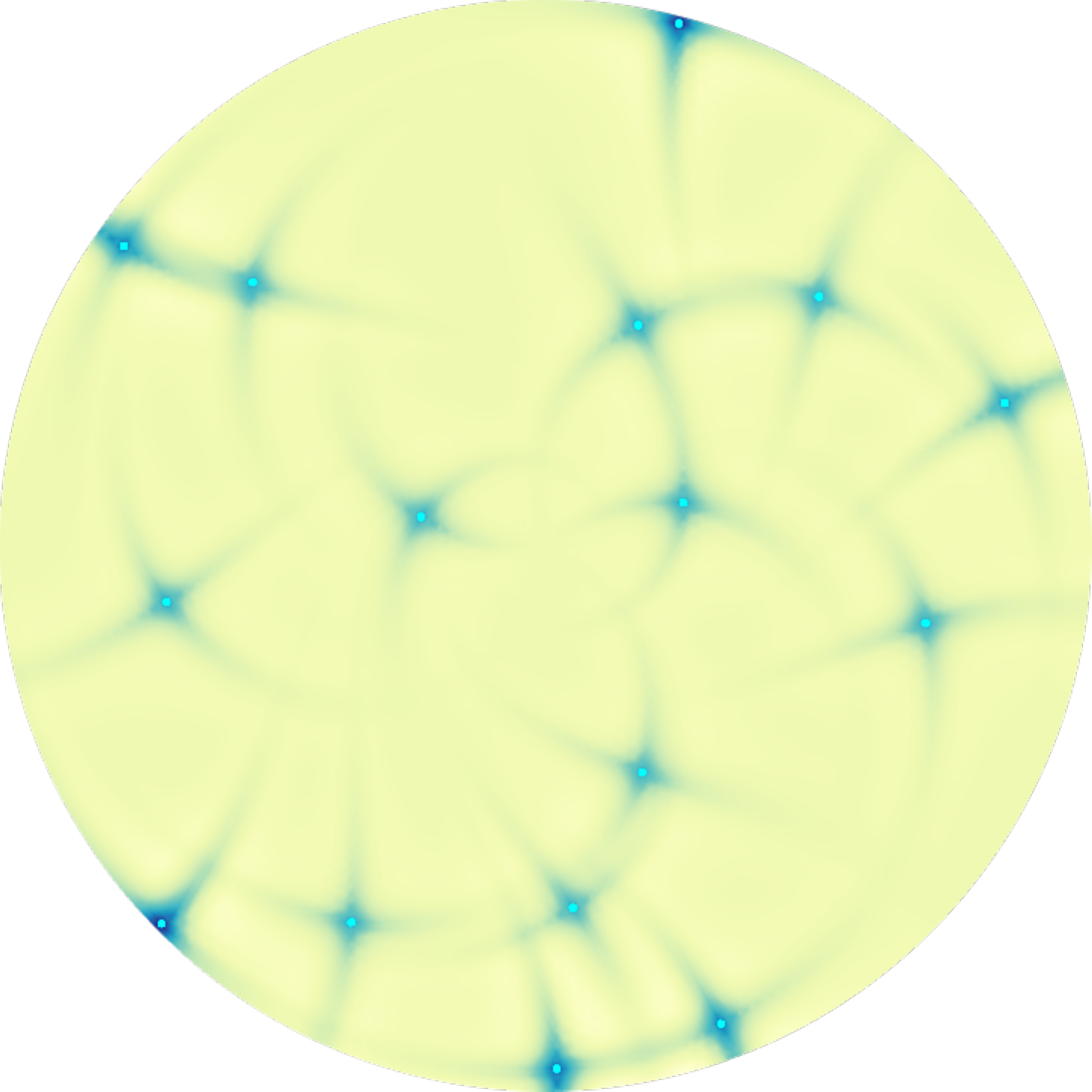} \\
\includegraphics[width=\imagewidth]{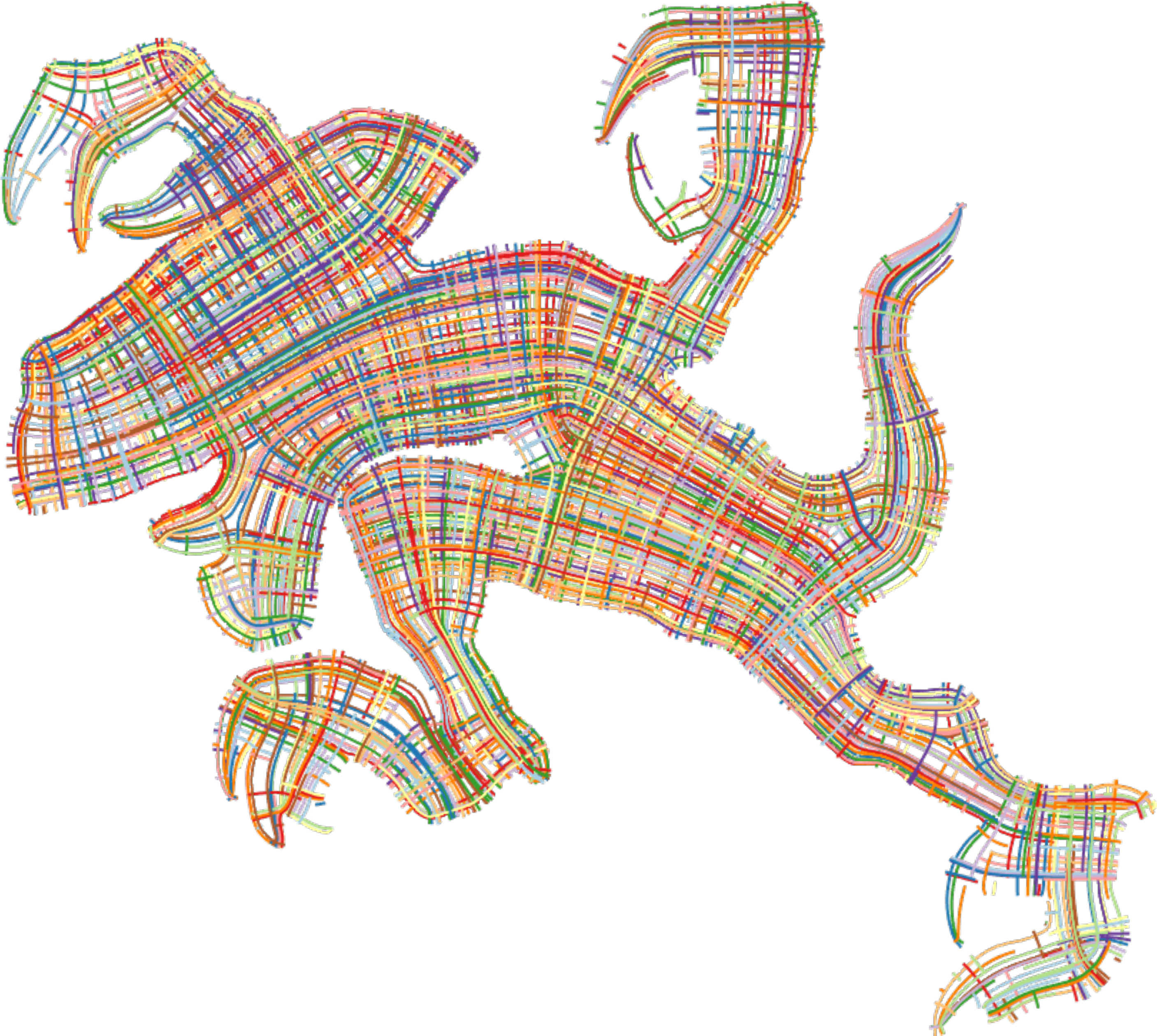} &
\includegraphics[width=\imagewidth]{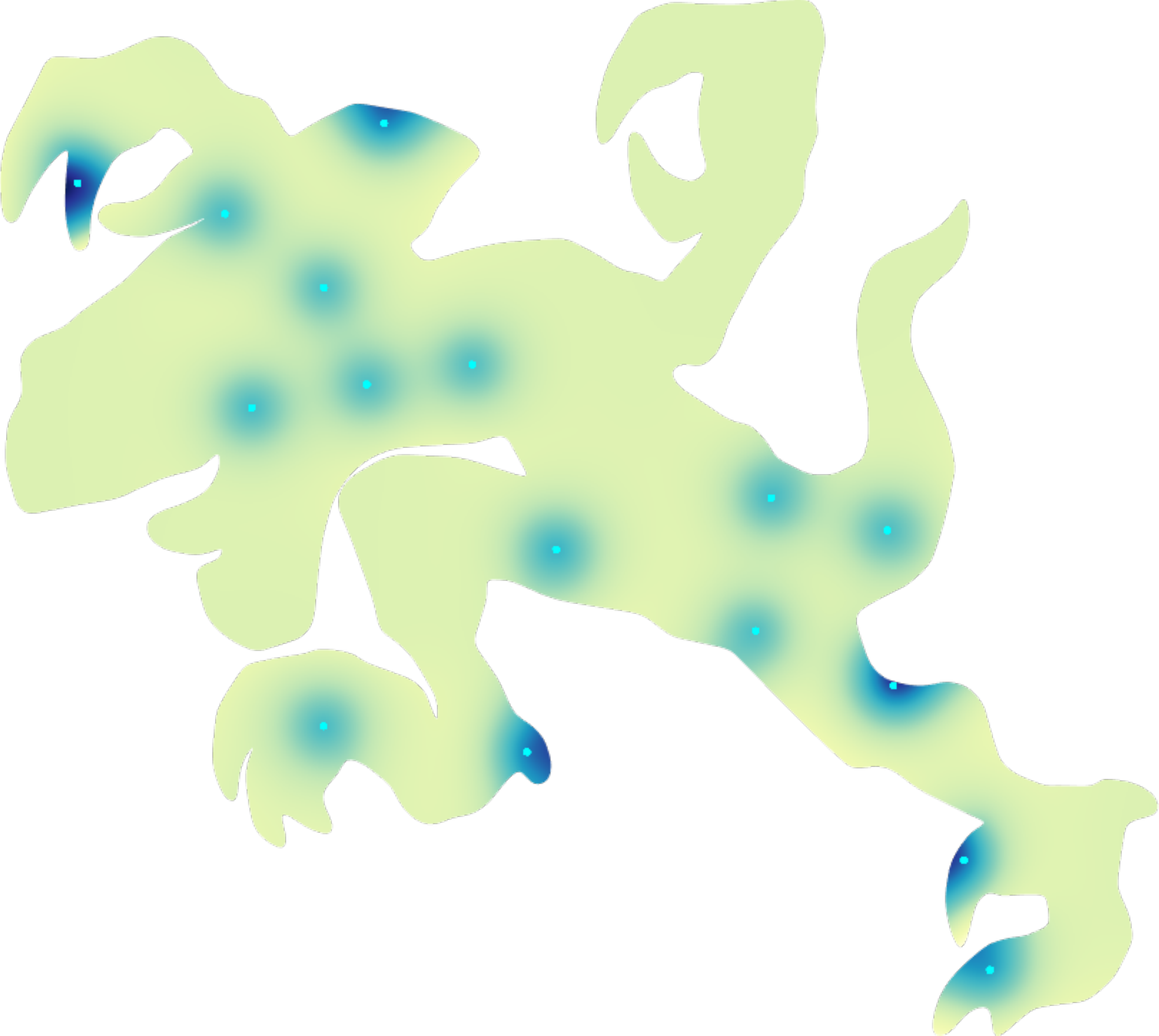} &
\includegraphics[width=\imagewidth]{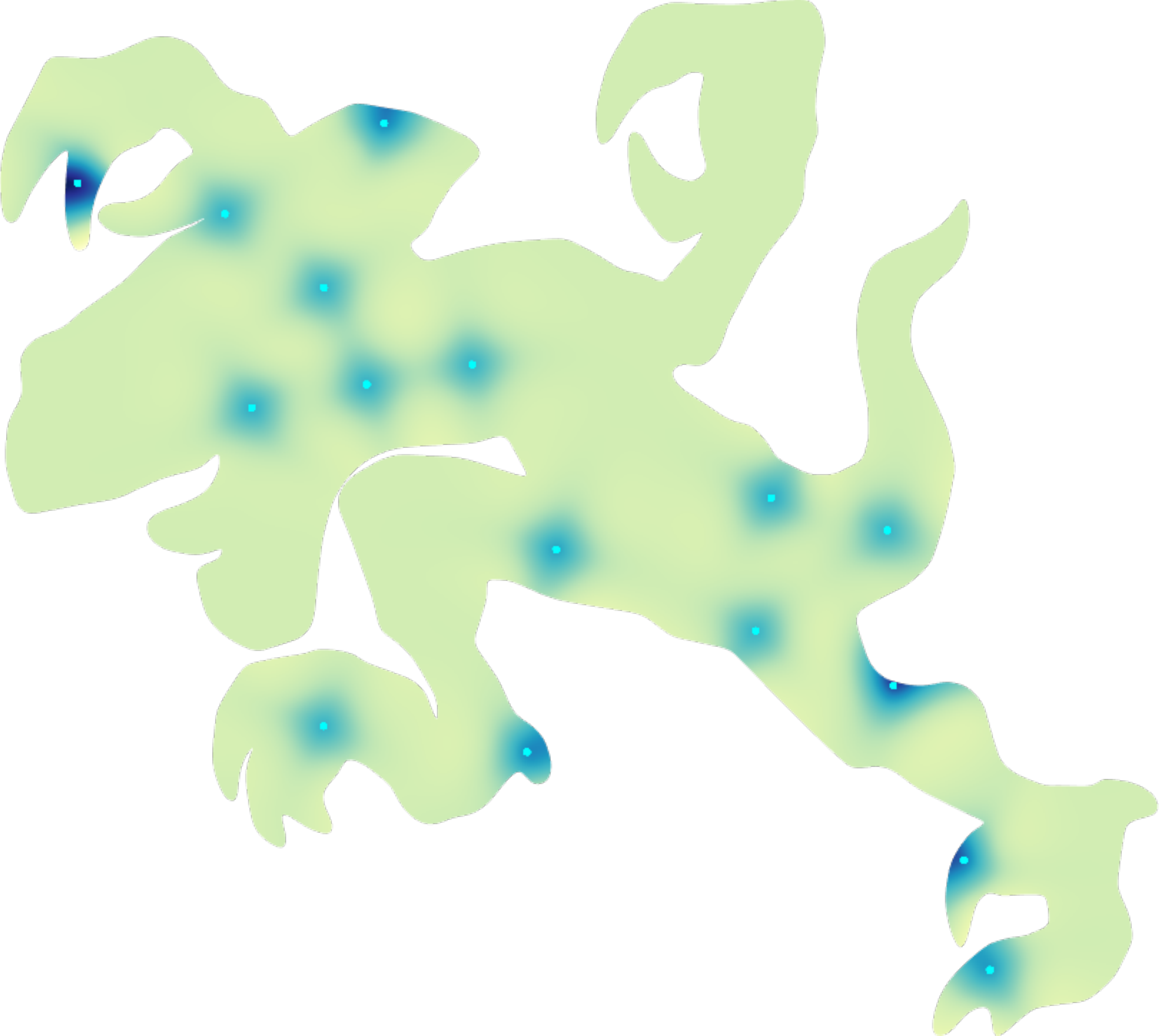} &
\includegraphics[width=\imagewidth]{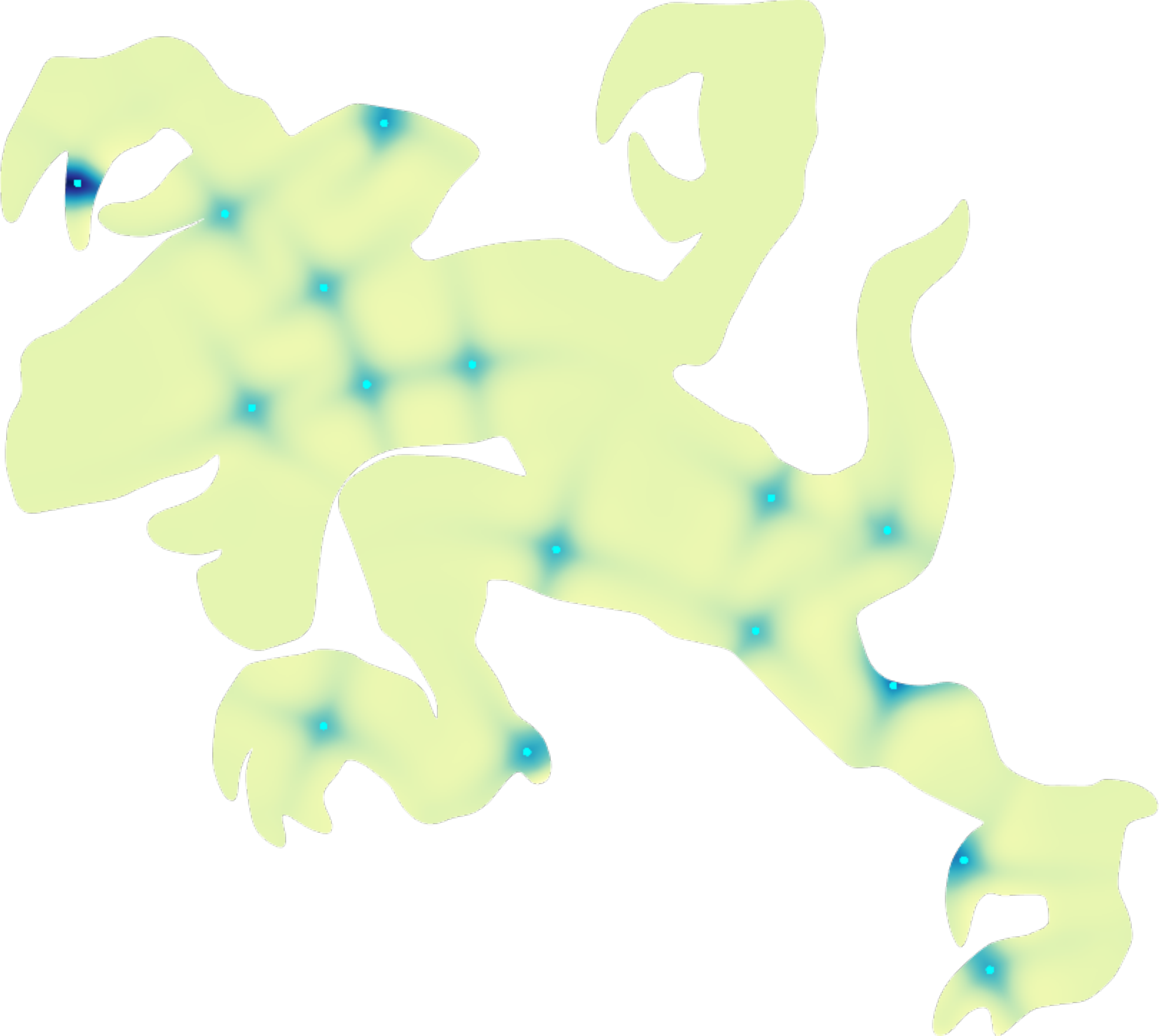} &
\includegraphics[width=\imagewidth]{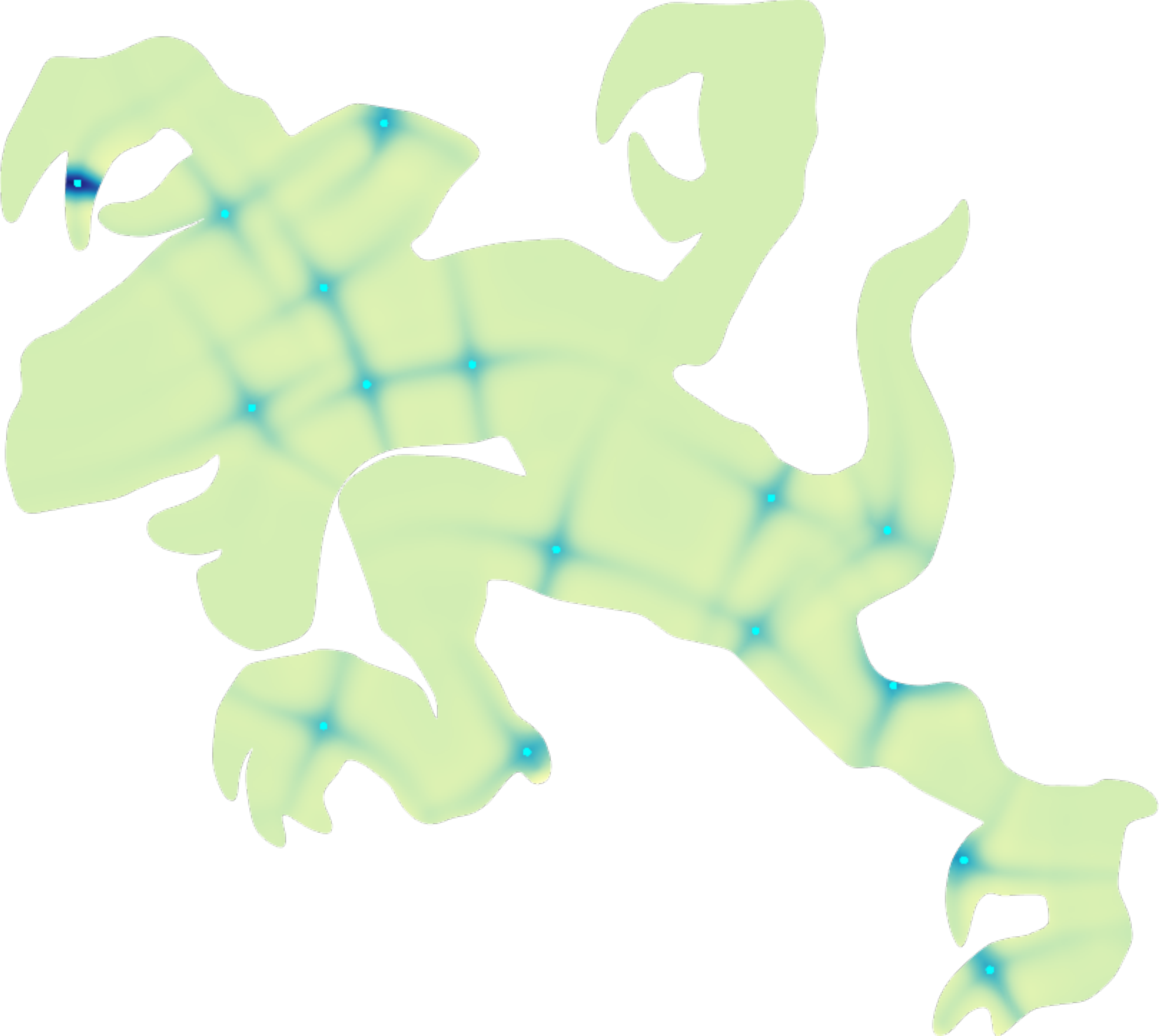} \\
\includegraphics[width=\imagewidth]{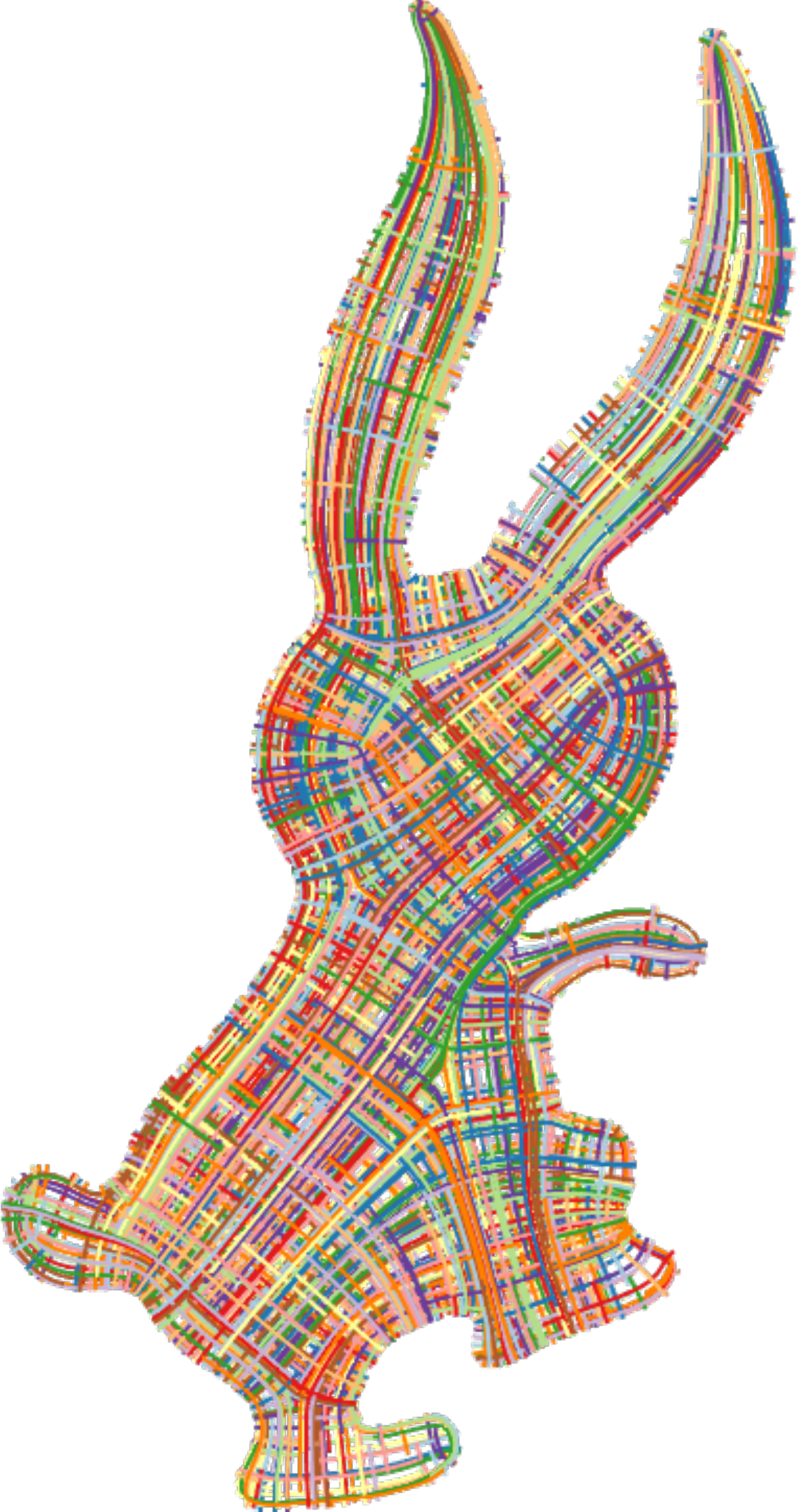} &
\includegraphics[width=\imagewidth]{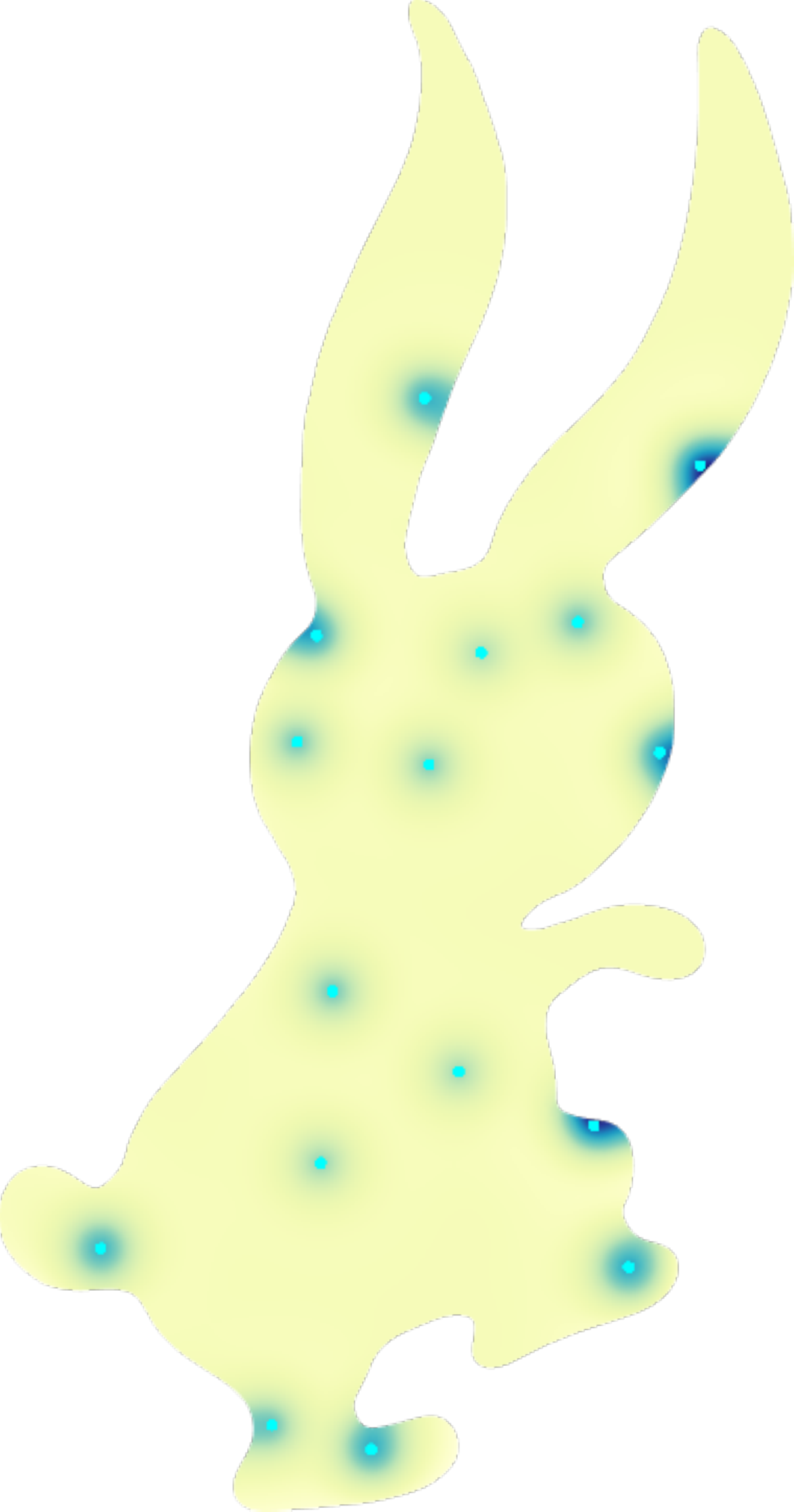} &
\includegraphics[width=\imagewidth]{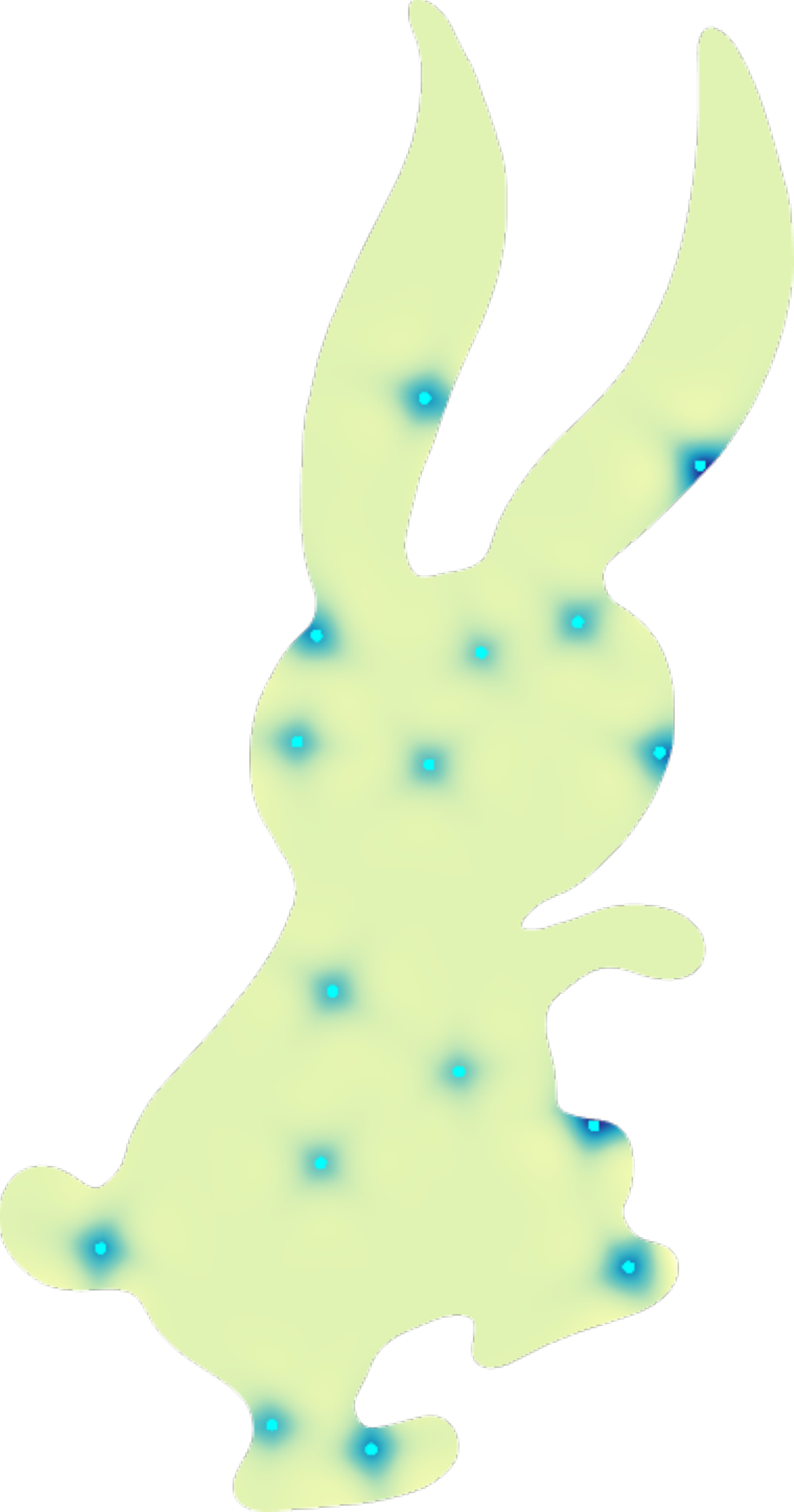} &
\includegraphics[width=\imagewidth]{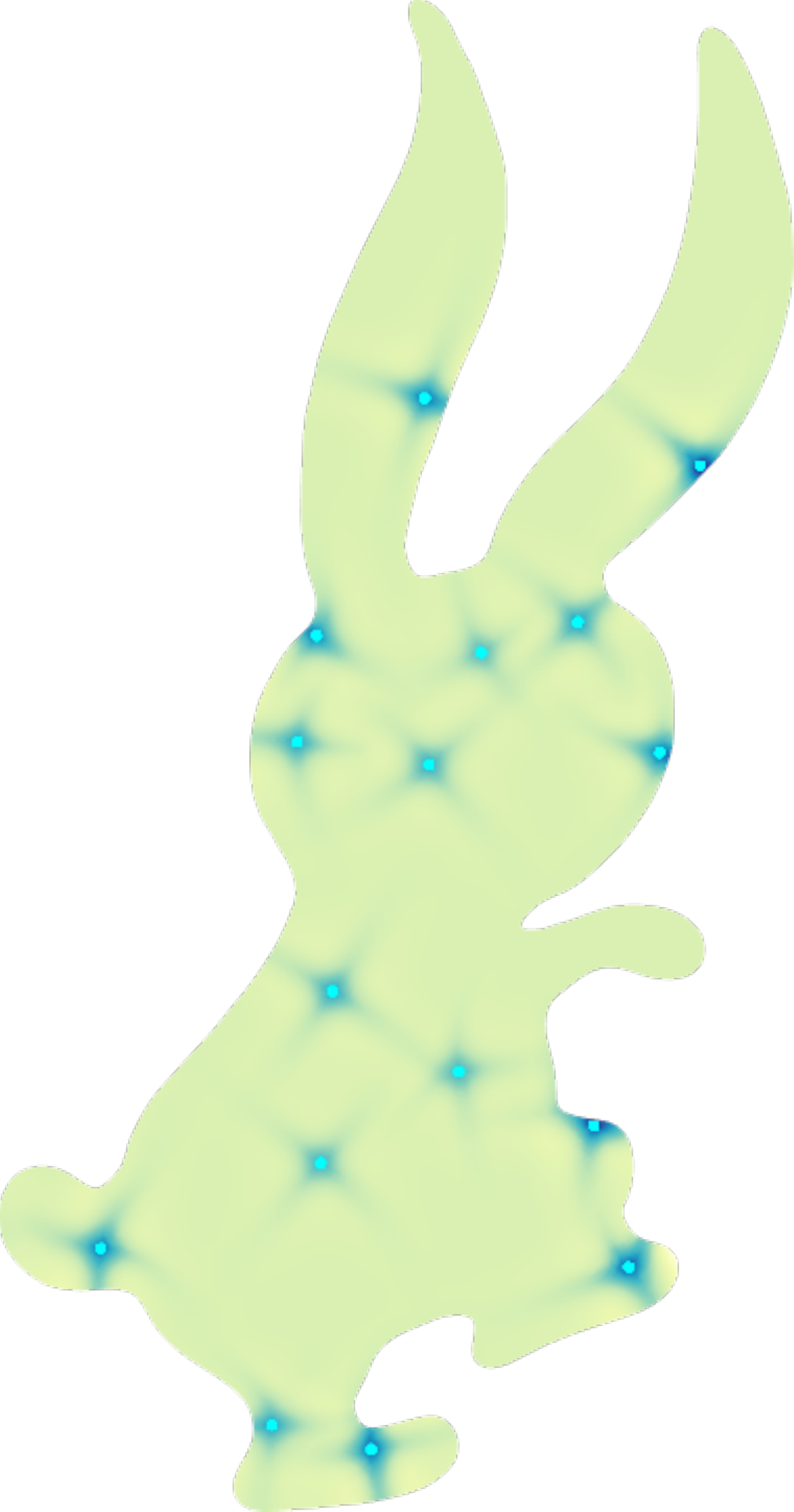} &
\includegraphics[width=\imagewidth]{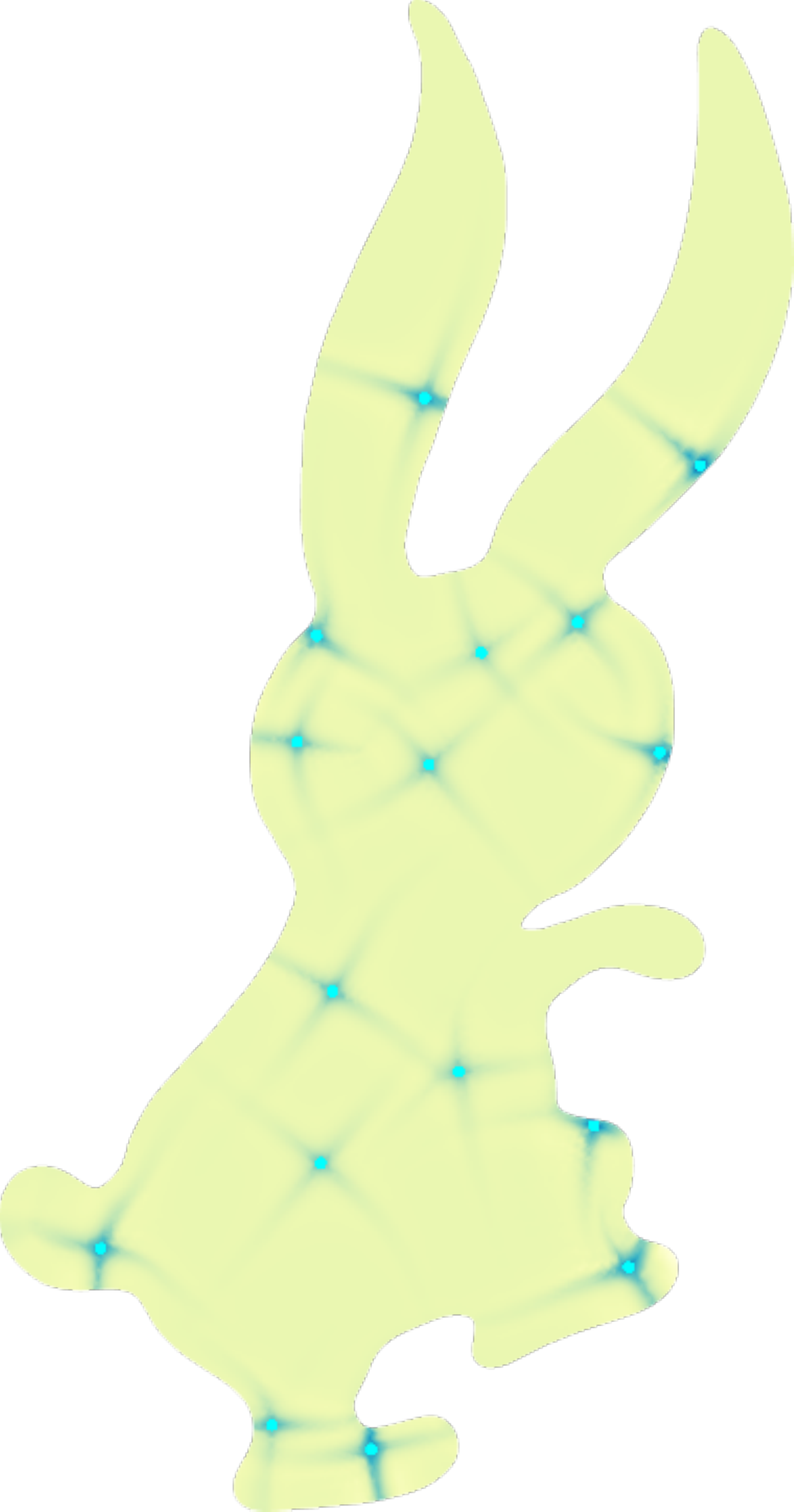} \\
& $\epsilon = 1$ &
$\epsilon = \num{2e-1}$ &
$\epsilon = \num{4e-2}$ &
$\epsilon = \num{8e-3}$
\end{tabular}
\caption{As the ellipticity parameter $\epsilon$ decreases, the operator becomes more anisotropic, as shown by the short-time solution to the ``diffusion'' equation $\partial_t u = \mathcal{A}_{T,\epsilon} u$ with initial condition set to a sum of Dirac deltas. Also note the differing impulse responses for two different frame fields on the disk. The diffusion time is set to $10^{-5}$ for the square and disk, $\num{2e-7}$ for the raptor, and $10^{-4}$ for the bunny.}
\label{fig:impulseresponse}
\end{figure}

\begin{figure}
\centering
\newcommand{\imagewidth}{0.3\columnwidth}
\setlength\tabcolsep{1.5pt}
\begin{tabular}{@{}c|cc@{}}
\includegraphics[width=\imagewidth]{figures/cylinder-ff1-int} &
\includegraphics[width=\imagewidth]{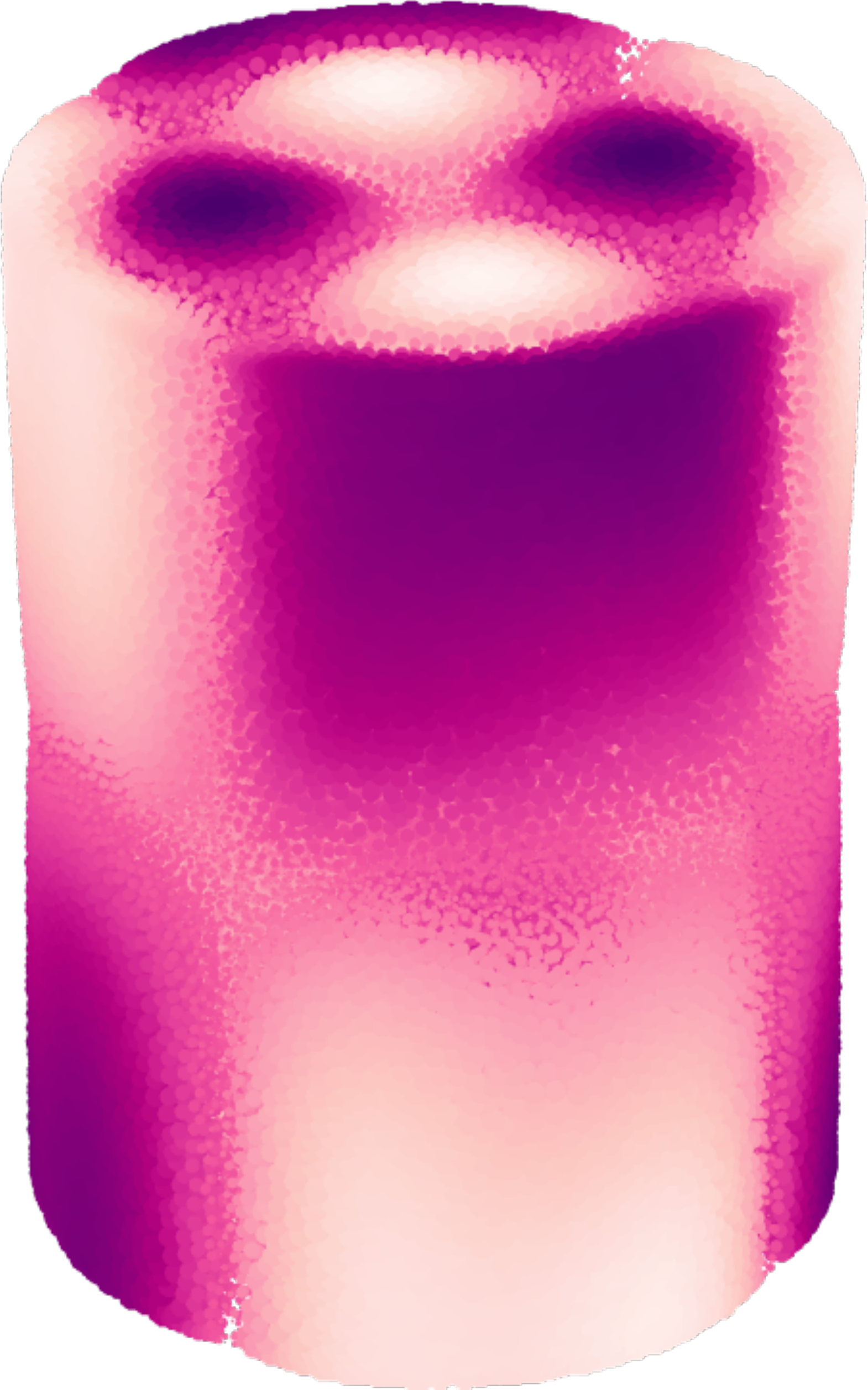} &
\includegraphics[width=\imagewidth]{figures/cylinder-ff1-ef1.082e-4-alt} \\
& $\lambda = \num{9.59e-5}$
& $\lambda = \num{1.082e-4}$ \\
\includegraphics[width=\imagewidth]{figures/cylinder-ff2-int} &
\includegraphics[width=\imagewidth]{figures/cylinder-ff2-ef9.16e-5-alt} &
\includegraphics[width=\imagewidth]{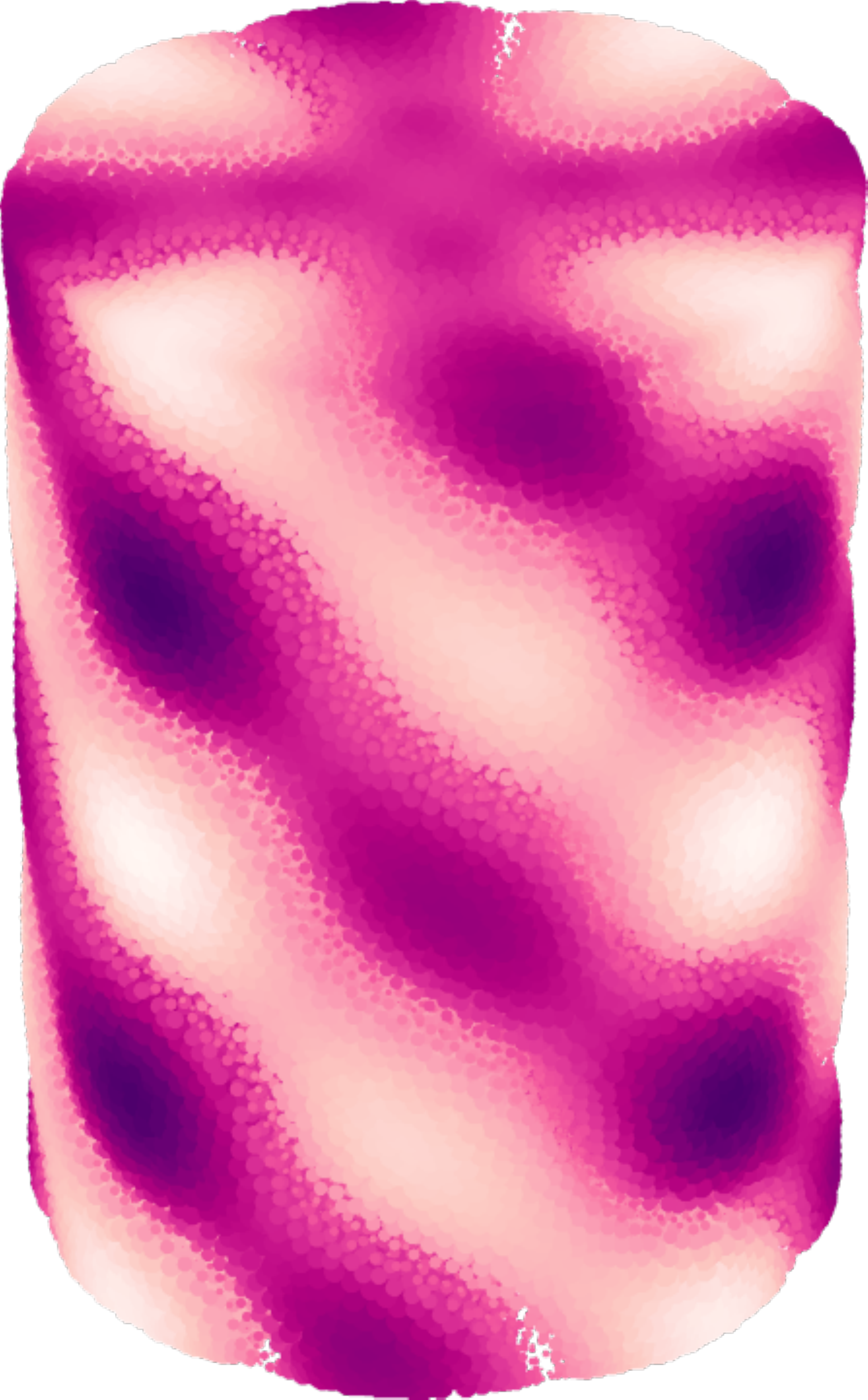} \\
& $\lambda = \num{9.16e-5}$
& $\lambda = \num{9.25e-5}$
\end{tabular}
\caption{Eigenfunctions of frame field operators for two different frame fields on the volumetric \textbf{cylinder}. Note how the oscillations follow the field lines.}
\label{fig:cylinder}
\end{figure}

\begin{figure*}
\newcommand{\imagewidth}{0.45\textwidth}
\newcommand{\imageheight}{0.1\textwidth}
\centering
\begin{tabular}{c|c}
\includegraphics[height=\imageheight]{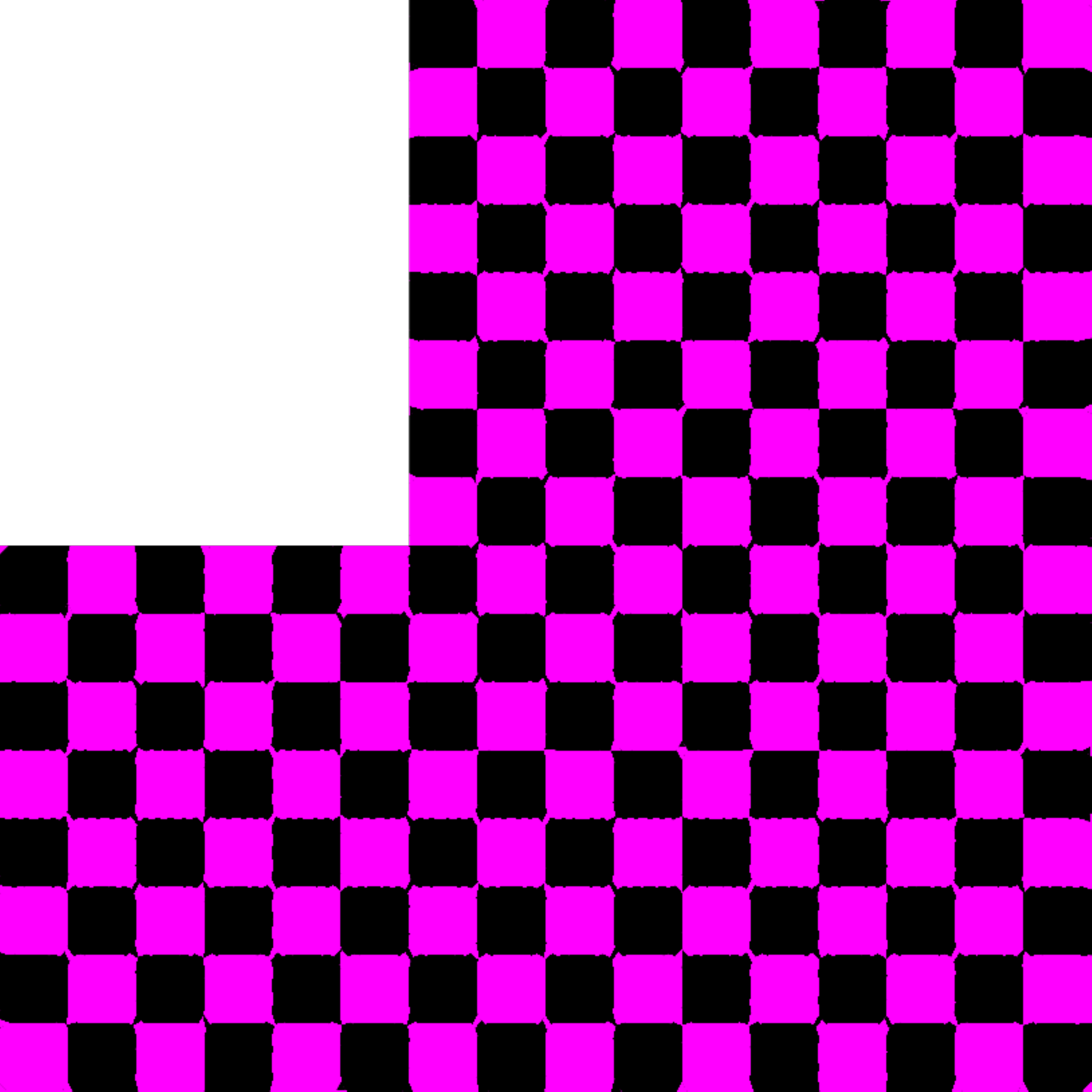} &
\includegraphics[height=\imageheight]{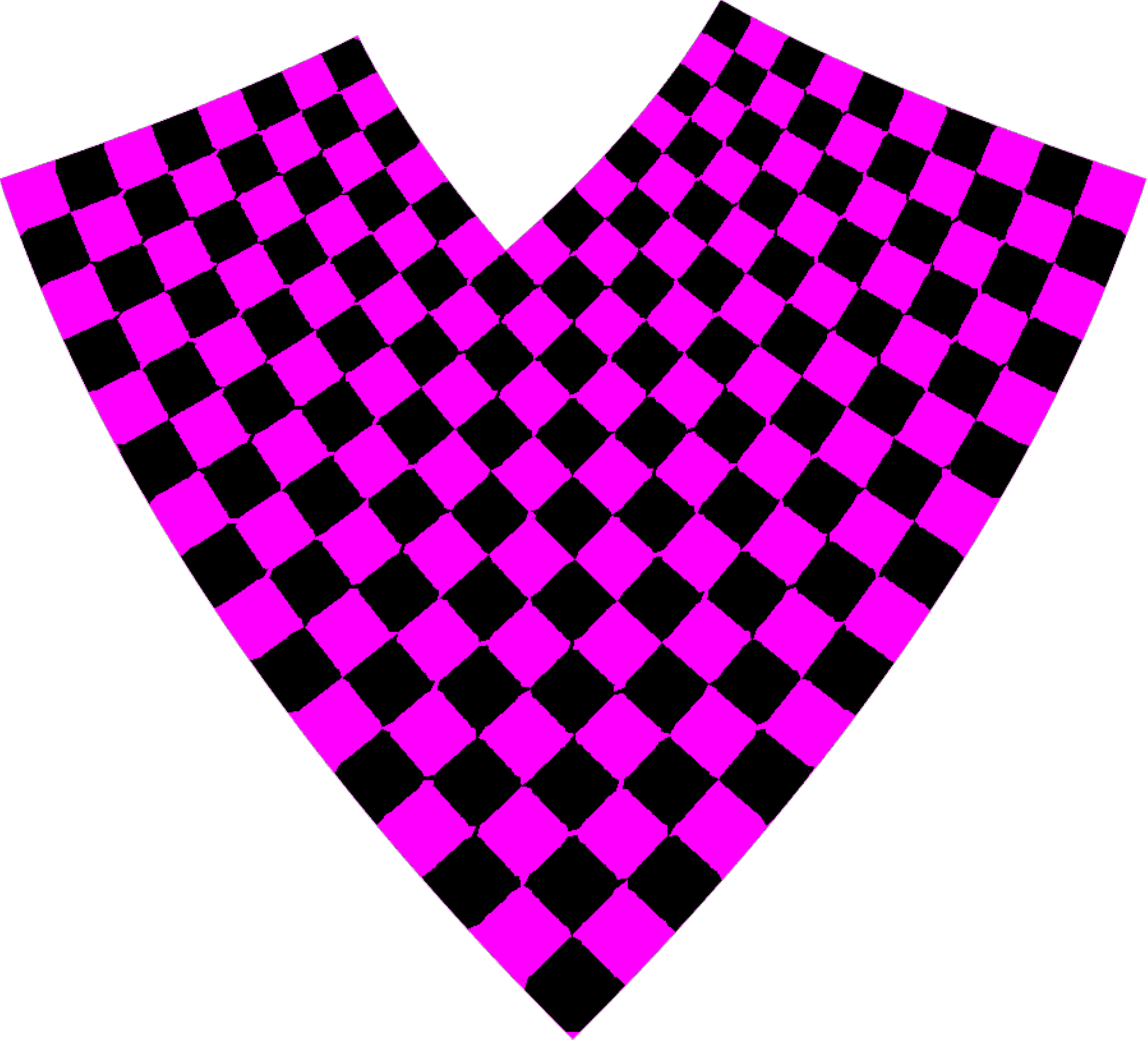}
\\
\includegraphics[width=\imagewidth]{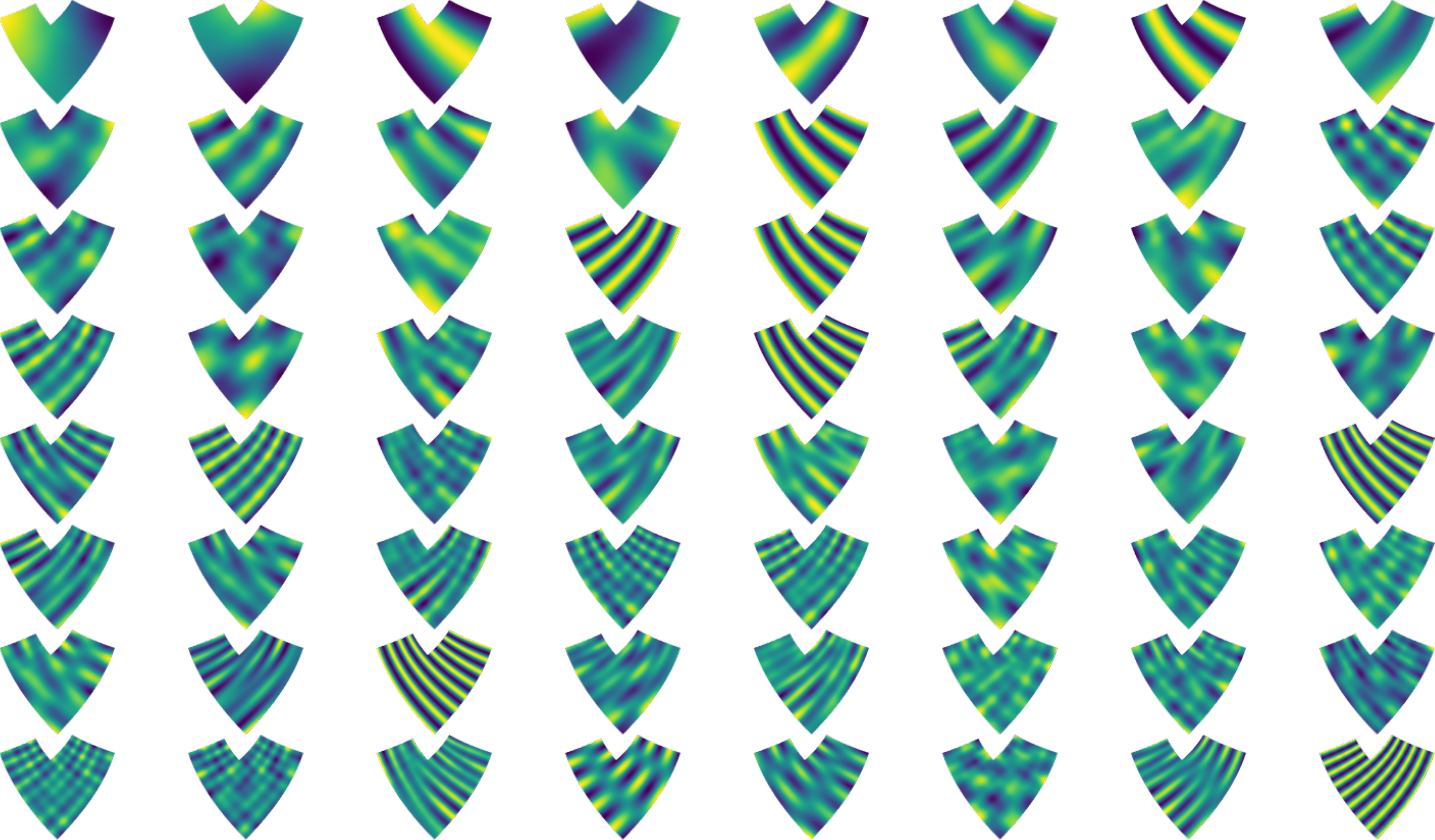} &
\includegraphics[width=\imagewidth]{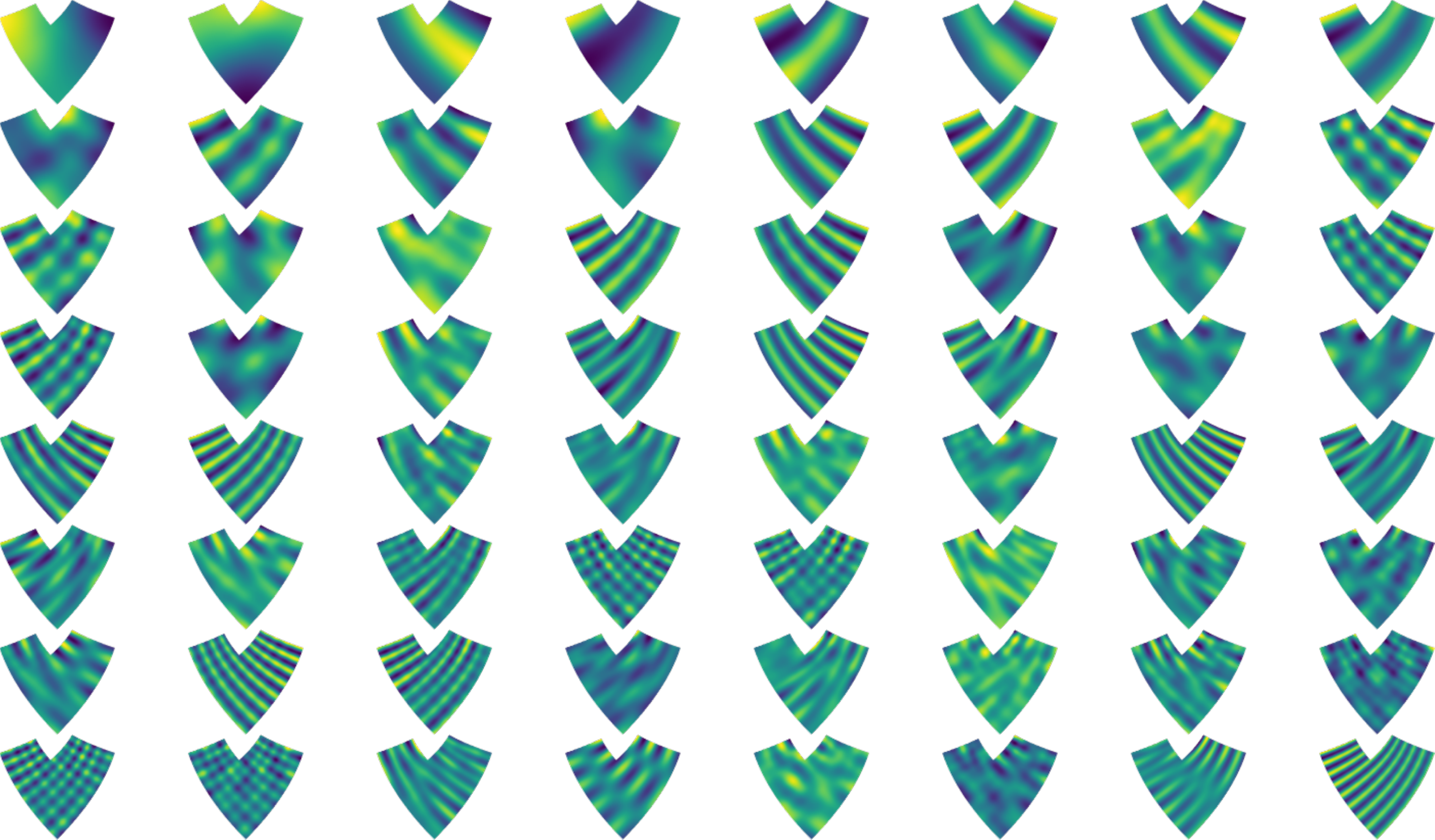}
\end{tabular}
\caption{Eigenfunctions of the operator associated to the constant axis-aligned frame field on the base domain (left) display similar qualitative behavior to those computed on a conformally warped domain with the conformal map coframe field operator (right), when both are displayed on the warped domain.}
\label{fig:warp}
\end{figure*}

\subsection{Matrix Representation}
Discretizing $u$, $V$, and $\Lambda$ in the piecewise linear hat basis, we obtain the matrix Lagrangian
\begin{equation} \footnotesize
\mathcal{L}_{T,\epsilon}(\bm{u}, \bm{V}) = \frac{1}{2} \bm{V}^\top \bm{M}_{T^{\epsilon}} \bm{V}^\top + \bm{\Lambda}^\top (\bm{D}^\top \bm{A} \bm{G} \bm{u} + \bm{M} \bm{V} + \bm{B}^\top \bm{\mu}), \end{equation}
where $\bm{M}_{T^{\epsilon}}$ is a block-diagonal matrix encoding the tensor field $T^{\epsilon}$ as a field of bilinear forms acting on symmetric second-order tensors scaled by the dual cell volumes, $\bm{G}$ and $\bm{D}$ are the piecewise-linear gradient and tensor divergence operators, respectively, $\bm{A}$ is a diagonal matrix of simplex volumes, and $\bm{M}$ is a diagonal matrix of dual cell volumes.
Note we have introduced a new term, $\bm{\mu}^\top \bm{B} \bm{\Lambda}$, which enforces the boundary constraint $\bm{B}\bm{\Lambda} = 0$ via yet another (discrete) Lagrange multiplier $\bm{\mu}$. $\bm{B}$ encodes the constraint $B(x) \operatorname{vec}\Lambda(x) = 0$ at each boundary vertex $x$.

The first order optimality conditions for the Lagrangian $\mathcal{L}_{T,\epsilon}$ are the following matrix equations:
%\begin{align} 0 &= \bm{D}^\top \bm{A}\bm{G}\bm{u} + \bm{M}\bm{V} + \bm{B}^\top \bm{\mu} \\
%0 &= \bm{M}_{T^{\epsilon}} \bm{V} - \bm{M} \bm{\Lambda} \\
%0 &= \bm{G}^\top \bm{A}\bm{D}\bm{\Lambda} \\
%0 &= \bm{B}\bm{\Lambda}, \end{align}
\begin{equation}\begin{pmatrix}
\bm{M}_{T^{\epsilon}} & \bm{M} & \bm{0} & \bm{0} \\
\bm{M} & \bm{0} & \bm{B}^\top & \bm{D}^\top \bm{A}\bm{G} \\
\bm{0} & \bm{B} & \bm{0} & \bm{0} \\
\bm{0} & \bm{G}^\top \bm{A}\bm{D} & \bm{0} & \bm{0}
\end{pmatrix} \begin{pmatrix}
\bm{V} \\ \bm{\Lambda} \\ \bm{\mu} \\ \bm{u}
\end{pmatrix} = \bm{0}, \end{equation}
which reduce to the single equation
\begin{equation}\footnotesize \bm{G}^\top \bm{A}\bm{D} \left(\overline{\bm{M}}_{T^{\epsilon}} - \overline{\bm{M}}_{T^{\epsilon}} \bm{B}^\top (\bm{B}\overline{\bm{M}}_{T^{\epsilon}} \bm{B}^\top)^{-1} \bm{B} \overline{\bm{M}}_{T^{\epsilon}}\right)\bm{D}^\top \bm{A}\bm{G}\bm{u} = \bm{0}, \end{equation}
where $\overline{\bm{M}}_{T^{\epsilon}} = \bm{M}^{-1} \bm{M}_{T^{\epsilon}} \bm{M}^{-1}$. We thus define the \textbf{discrete frame field operator} as
\begin{equation} \footnotesize
    \mathcal{A}_{T, \epsilon} \coloneqq \bm{G}^\top \bm{A}\bm{D} \left(\overline{\bm{M}}_{T^{\epsilon}} - \overline{\bm{M}}_{T^{\epsilon}} \bm{B}^\top (\bm{B}\overline{\bm{M}}_{T^{\epsilon}} \bm{B}^\top)^{-1} \bm{B} \overline{\bm{M}}_{T^{\epsilon}}\right)\bm{D}^\top \bm{A}\bm{G}.
\end{equation}
In case we want natural boundary conditions, we set $\bm{\Lambda} = 0$ on the boundary, so $\bm{B}$ becomes a matrix that selects out boundary vertex coordinates from $\bm{\Lambda}$. The matrix expression for $\mathcal{A}_{T,\epsilon}$ with natural boundary conditions thus reduces to
\begin{equation}
    \mathcal{A}_{T,\epsilon} = \bm{G}^\top \bm{A}\bm{D}^\circ \overline{\bm{M}}_{T^{\epsilon}}^\circ (\bm{D}^\circ)^\top \bm{A}\bm{G},
\end{equation}
where superscript $\circ$ denotes that boundary columns (and rows in the case of $\overline{\bm{M}}_{T^{\epsilon}}$) have been deleted.
This is similar to the expression for the Bilaplacian with natural boundary conditions in \cite{Stein2018}. In fact it reproduces the Bilaplacian exactly when $\epsilon = 1$ as $\overline{\bm{M}}_{T^{1}} = \bm{M}^{-1}$.

By adding a unit norm constraint on $\bm{u}$ to the Lagrangian, we can also obtain the discrete frame field eigenproblem
\begin{equation}\mathcal{A}_{T,\epsilon} \bm{u} = \lambda \bm{M}\bm{u}. \end{equation}

\section{Validation}\label{sec:validation}
In this section, we check that the discrete operator constructed in the previous section has the desired behavior---convergence under mesh refinement, controllable anisotropy, and behavior under pullback.
\paragraph*{Dirichlet Problem}
We first examine convergence of solutions to the frame field operator \emph{Dirichlet problem}
\begin{equation}
\begin{aligned}
    \mathcal{A}_{T,\epsilon} u &= 0 \\
    \nabla_n u \mid_{\partial\Omega} &= 0 \\
    u\mid_{\partial\Omega} &= u_0
\end{aligned}
\end{equation}
as we refine the underlying computational mesh. This is a standard test of convergence for finite element methods. We should expect to see the mixed FEM solution converge to the true solution.
The fact that the frame field operator has non-constant coefficients adds an extra complication. To test convergence of the PDE solutions, we first need to ensure that the underlying frame fields converge. To address this, we set up a hierarchy of frame field operators as follows: we first compute a boundary-aligned frame field at the finest resolution via MBO \cite{viertel2019}, then resample it to the coarser meshes, and finally renormalize so that frame fields at all levels are octahedral. A frame field operator is then constructed from the frame field at each level, using the same value of $\epsilon$.

\Cref{fig:cvg-2d} displays Dirichlet solutions over six levels of Loop subdivision on the \textbf{elephant} and \textbf{troll} meshes. Each successive subdivision halves edge lengths. The boundary values $u_0$ are set to square waves, which have components over many frequencies. At the coarsest levels, the high-frequency boundary data is highly aliased, and the solution appears muddy in the interior. After subdividing a few times, the solution quickly becomes smooth and displays the clear influence of the underlying frame field, as the sharp edges in the boundary data propagate along frame directions.

\paragraph*{Spectral Convergence}

Given a hierarchy of frame field operators at successive refinement levels, we can also test convergence of the spectra of the operators. For this experiment, we construct a hierarchy of operators with Neumann boundary conditions over six refinement levels on the \textbf{horse} domain, and we compute the first $64$ eigenvalues and eigenfunctions at each level. In \Cref{fig:eigs-2d}, we plot the error at each level $1$--$5$ against the eigenvalue at the finest level $6$, which we use as a proxy for the true spectrum. We drop the smallest eigenvalue because it is zero for Neumann boundary conditions. The error grows with the eigenvalue itself, but drops consistently at finer levels. \Cref{fig:eigf-2d} shows the same data in a different way, showing how the eigenvalue error drops with average edge length across a variety of eigenvalues. We also display the eigenfunction corresponding to the $64$th eigenvalue at each level. Note how the overall structure of oscillations remains consistent over many levels of refinement.

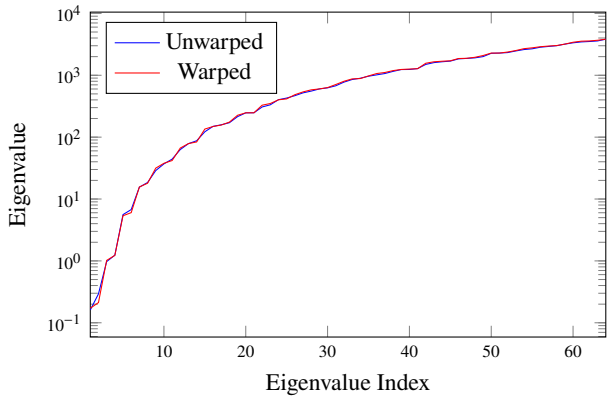
\begin{figure}
\pgfplotstableread[col sep=comma, skip first n=1]{figures/warp-evs.csv}{\warpevs}
\begin{tikzpicture}
    \begin{axis}[
        width = \columnwidth,
        height = 0.7\columnwidth,
        enlarge x limits = 0,
        ymode = log,
        grid = none,
        xlabel = {Eigenvalue Index},
        ylabel = {Eigenvalue},
        ylabel near ticks,
        every tick label/.append style = {font=\tiny},
        legend pos = north west]
    \addplot+ [mark = none] table [x expr=\coordindex+1, y index=0] {\warpevs};
    \addlegendentry{Unwarped}
    \addplot+ [mark = none] table [x expr=\coordindex+1, y index=1] {\warpevs};
    \addlegendentry{Warped}
    \end{axis}
\end{tikzpicture}
\caption{The spectra of the constant frame field operator (``unwarped'') and the map coframe field operator (``warped'') show broad agreement.}
\label{fig:warp-evs}
\end{figure}

\paragraph*{Analytic Ground Truth} In one special case, we can compare eigenvalues and eigenfunctions of the discrete frame field operator to their analytic counterparts. Consider the constant axis-aligned frame field on the square $[-1, 1]^2$, given by
\begin{equation} T = \sum_{\alpha = 1}^2 (e^\alpha)^{\otimes 4}, \end{equation}
where $e^\alpha$ is the standard basis in $\R^2$. The corresponding operator is
\begin{equation} \begin{aligned} \mathcal{A}_{T,\epsilon} u &= \partial^4 u_{ijij} - (1 - \epsilon)u_{iiii} \\
&= 2 u_{xxyy} + \epsilon (u_{xxxx} + u_{yyyy}). \end{aligned} \end{equation}
A Fourier basis component
\begin{equation} \phi_\omega = e^{i(\omega_x x + \omega_y y)} \end{equation}
is an eigenfunction of $\mathcal{A}_{T, \epsilon}$, since
\begin{equation} \begin{aligned} \mathcal{A}_{T,\epsilon} \phi_\omega &= [\omega_x^2\omega_y^2 + \epsilon (\omega_x^4 + \omega_y^4)] e^{i(\omega_x x + \omega_y y)} \\
&= \sigma_P(\mathcal{A}_{T,\epsilon})(\omega)e^{i(\omega_x x + \omega_y y)}. \end{aligned} \end{equation}
Thus, we can compute the analytic spectrum of the constant frame field operator on the square by evaluating the principal symbol $\sigma_P(\mathcal{A}_{T,\epsilon})$ on the Fourier lattice and then sorting the resulting eigenvalues.

In \Cref{fig:square}, we compare analytic eigenvalues computed this way to eigenvalues of discrete frame field operators generated from constant axis-aligned frame fields on meshes of the square at multiple levels of refinement. Observe that the error drops consistently with the mean edge length.

\begin{figure*}
\centering
\newcommand{\imagewidth}{0.15\textwidth}
\setlength\tabcolsep{5pt}
\begin{tabular}{@{}cc|cccc@{}}
\includegraphics[width=\imagewidth]{figures/rockerarm-ff-int} &
\includegraphics[width=\imagewidth]{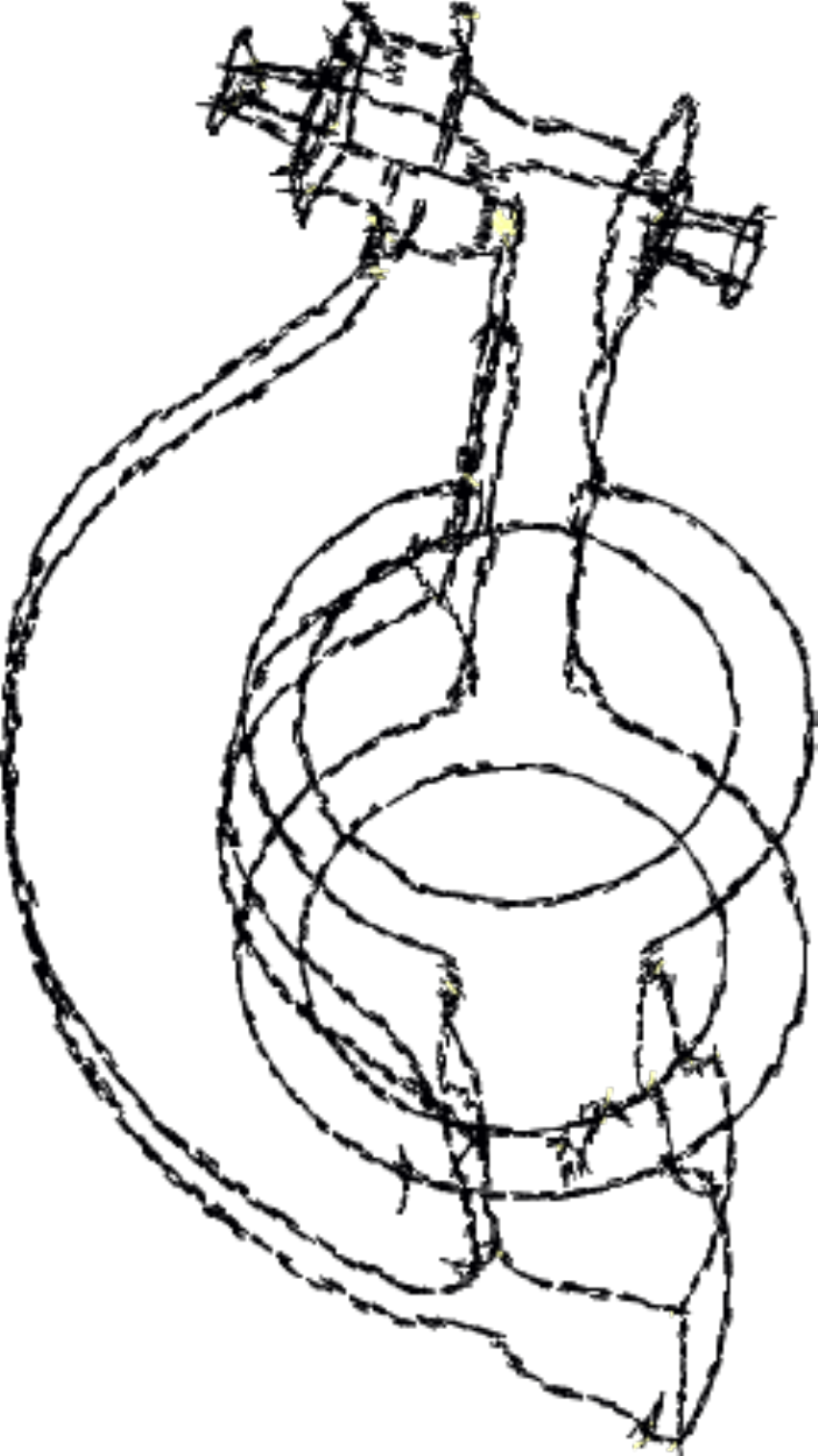} &
\includegraphics[width=\imagewidth]{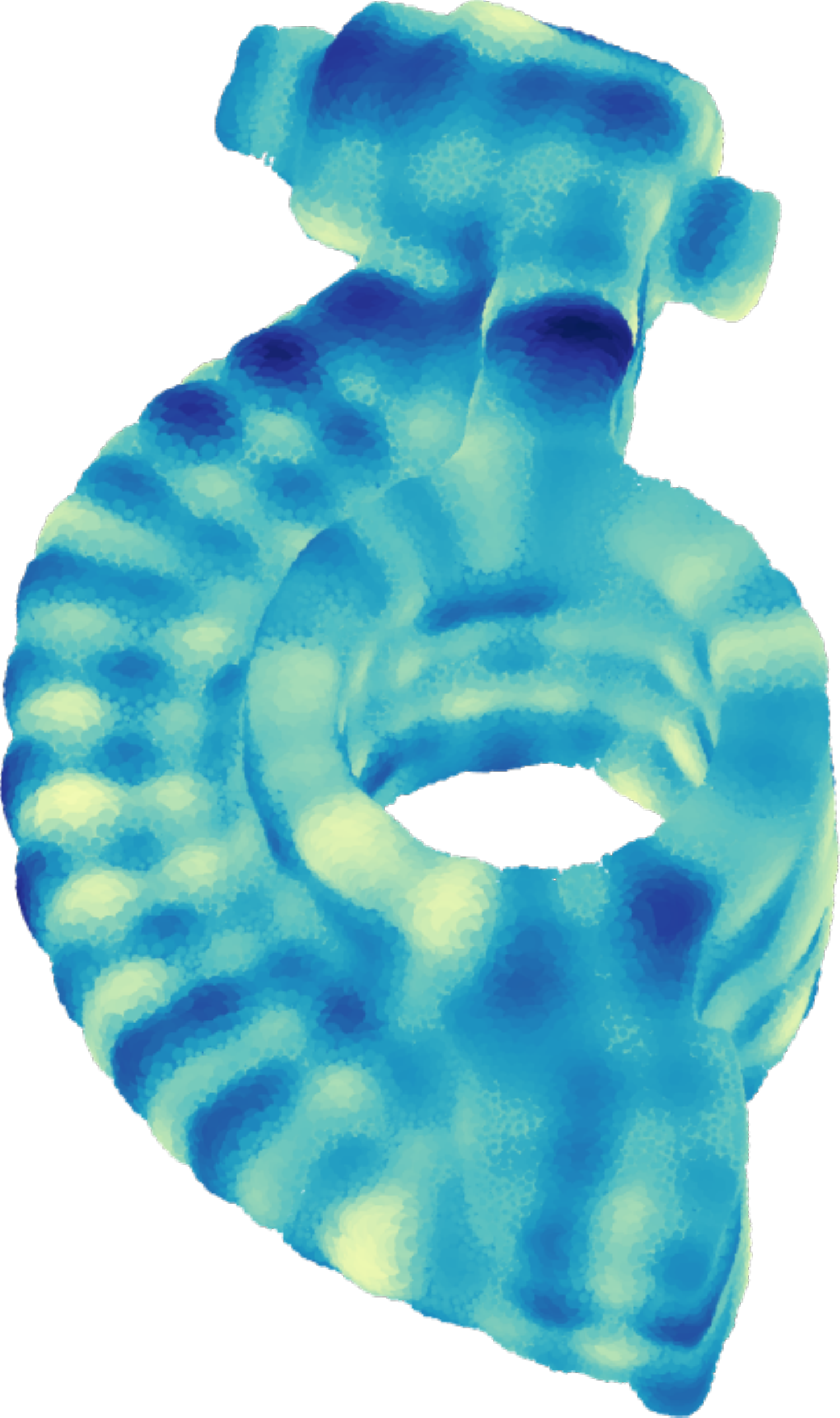} &
\includegraphics[width=\imagewidth]{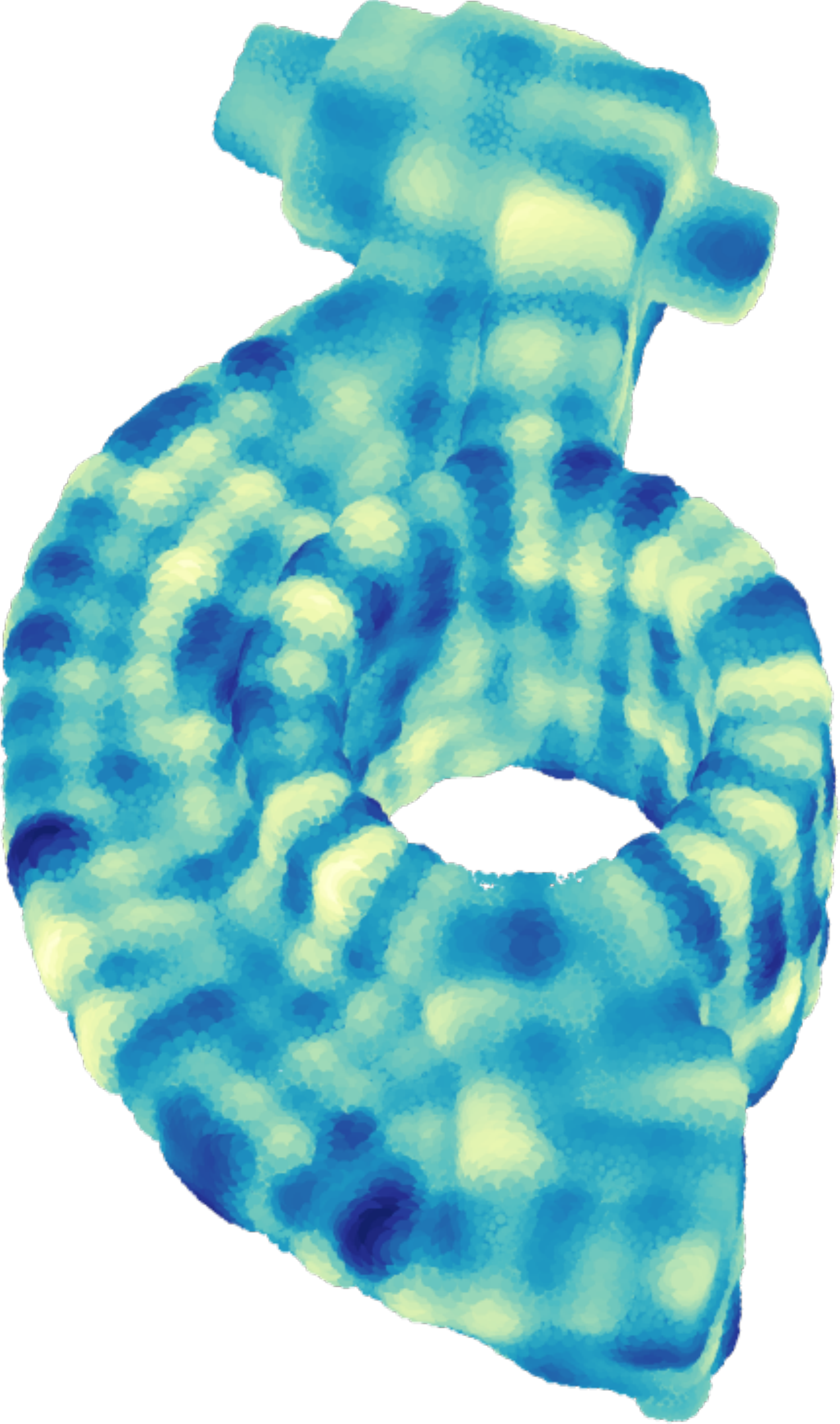} &
\includegraphics[width=\imagewidth]{figures/rockerarm-ef5-scatter} &
\includegraphics[width=\imagewidth]{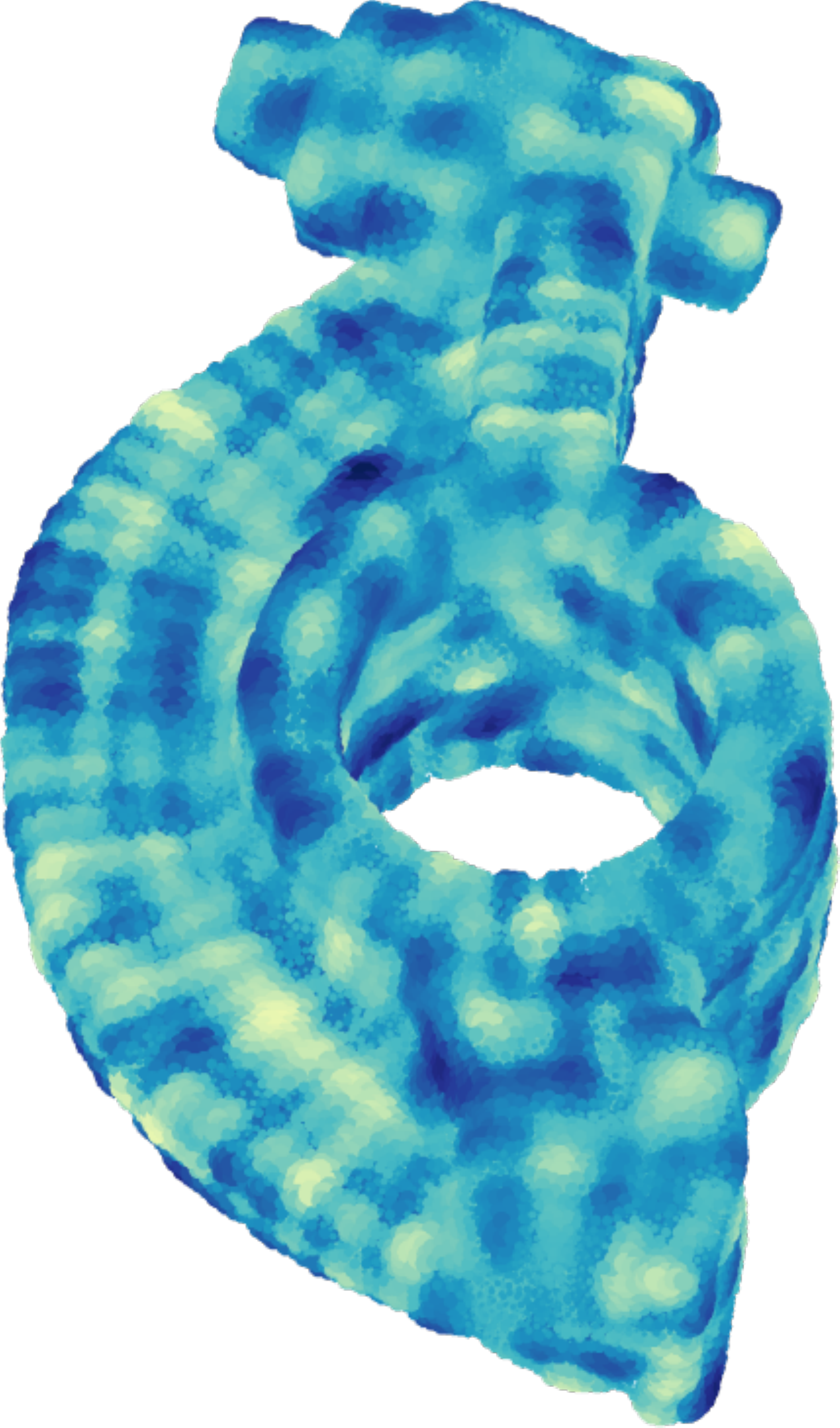} \\
 &  & $\lambda = 1.0048$ & $\lambda = 3.0105$ & $\lambda = 4.9904$ & $\lambda = 9.9805$
\end{tabular}
\caption{At several different frequencies, oscillations of frame field operator eigenfunctions on the volumetric \textbf{rockerarm} model align to the field directions. Integral and singular curves of the frame field are shown at left.}
\label{fig:eigf-3d}    
\end{figure*}

\paragraph*{Volumetric Spectral Convergence}
To test convergence of our operator on a volumetric domain, we perform a similar experiment to the one detailed above. However, we lack an equivalent to Loop subdivision for tetrahedral meshes that preserves tetrahedral angles and overall quality. Thus, to construct a hierarchy of frame fields, we use a sequence of (separately generated) tetrahedral meshes of different target edge lengths. We optimize a frame field on the finest mesh using volumetric frame field MBO \cite{palmer2020}. Then the field coefficients are linearly interpolated to the vertices of each coarser mesh, reprojected into the octahedral variety, and further optimized to ensure they are smooth at each level. This procedure should ensure that the overall structure of the frame field is consistent across levels.

\Cref{fig:cvg-3d} plots eigenvalue error for a frame field operator across various levels of refinement of the unit ball domain. The octahedral frame field and its singular structure are depicted in the inset. Error is measured against the eigenvalues at the finest level. There is a consistent decrease in error with decreasing mean edge length.

\Cref{fig:ef-3d} compares frame field operator eigenfunctions computed at various mesh resolutions on the \textbf{teddy}. They appear to stabilize as the mesh resolution increases, more so at lower frequencies.

\paragraph*{Controllable Anisotropy}
The frame field operator $\mathcal{A}_{T,\epsilon}$ has two parameters: a frame field $T$, which encodes the orientation of anisotropy, and a scalar $\epsilon$, which controls the degree of anisotropy and the uniform ellipticity bound. In \Cref{fig:impulseresponse}, we examine the effect of different settings of $T$ and $\epsilon$. The impulse response to a sum of delta functions $u_0$ is computed by solving a short-time diffusion problem $\partial_t u = \mathcal{A}_{T,\epsilon} u$ with natural boundary conditions via one step of implicit Euler integration---so the discrete equation is
\begin{equation}
    (\bm{M} + \tau \mathcal{A}_{T,\epsilon}) \bm{u} = \bm{M} \bm{u}_0,
\end{equation}
where $\tau$ is the diffusion time.
When $\epsilon = 1$, the frame field operator reduces to the Bilaplacian, and diffusion occurs isotropically. When $\epsilon < 1$, observe that diffusion occurs mostly along integral curves of the underlying frame fields. This is due to propagation along characteristic directions of the operator. The effect is accentuated as $\epsilon \to 0$. Also note how the impulse response differs for two different frame fields on the disk---displaying fine-grained control of anisotropy through the frame field.

\Cref{fig:cylinder} displays control of anisotropy in a volumetric setting. When the frame field is aligned to the axis of the cylinder, the eigenfunctions oscillate radially and along this axis. With a helical frame field, the eigenfunctions show a similar helical pattern.

\paragraph*{Map Frame Field} \Cref{fig:warp} tests the results of \Cref{sec:paramconnection}. We start with a constant axis aligned frame field on a base domain comprising a union of rectangles. The domain is then warped via a conformal map computed in closed form. The derivatives of the map are also computed and used to build the map coframe field on the warped domain, a conformal octahedral field. Eigenfunctions of the map frame field operator on the warped domain are compared to eigenfunctions of the constant frame field operator on the base domain, after the latter are remapped onto the warped domain. Note the overall qualitative similarity of the eigenfunctions, showing broad agreement even at relatively low frequencies. \Cref{fig:warp-evs} shows that the spectra of the two operators also agree.

\paragraph*{More Volumetric Examples}
\Cref{fig:eigf-3d} shows eigenfunctions on the \textbf{rockerarm} at various eigenvalues. Eigenfunctions of the frame field operator at several relatively high frequencies display unmistakable alignment to the frame field.

\section{Additional Experiments}\label{sec:applications}

In this section, we provide some additional experiments involving our new operator and its discretization.  In particular, we demonstrate how it can be substituted into two operator-based methods in geometry processing as a substitute for its isotropic counterparts, yielding output from these methods that is aware of the structure of the input frame field.

% \subsection{Meshing}
% \textcolor{red}{ODED: I find this section dangerous. We do not have meshing examples as figures, so why mention it here?}
% 

\begin{figure*}
\centering
\newcommand{\imageheight}{0.11\textwidth}
\setlength\tabcolsep{1.5pt}
\begin{tabular}{@{}c|ccccc@{}}
\includegraphics[height=\imageheight]{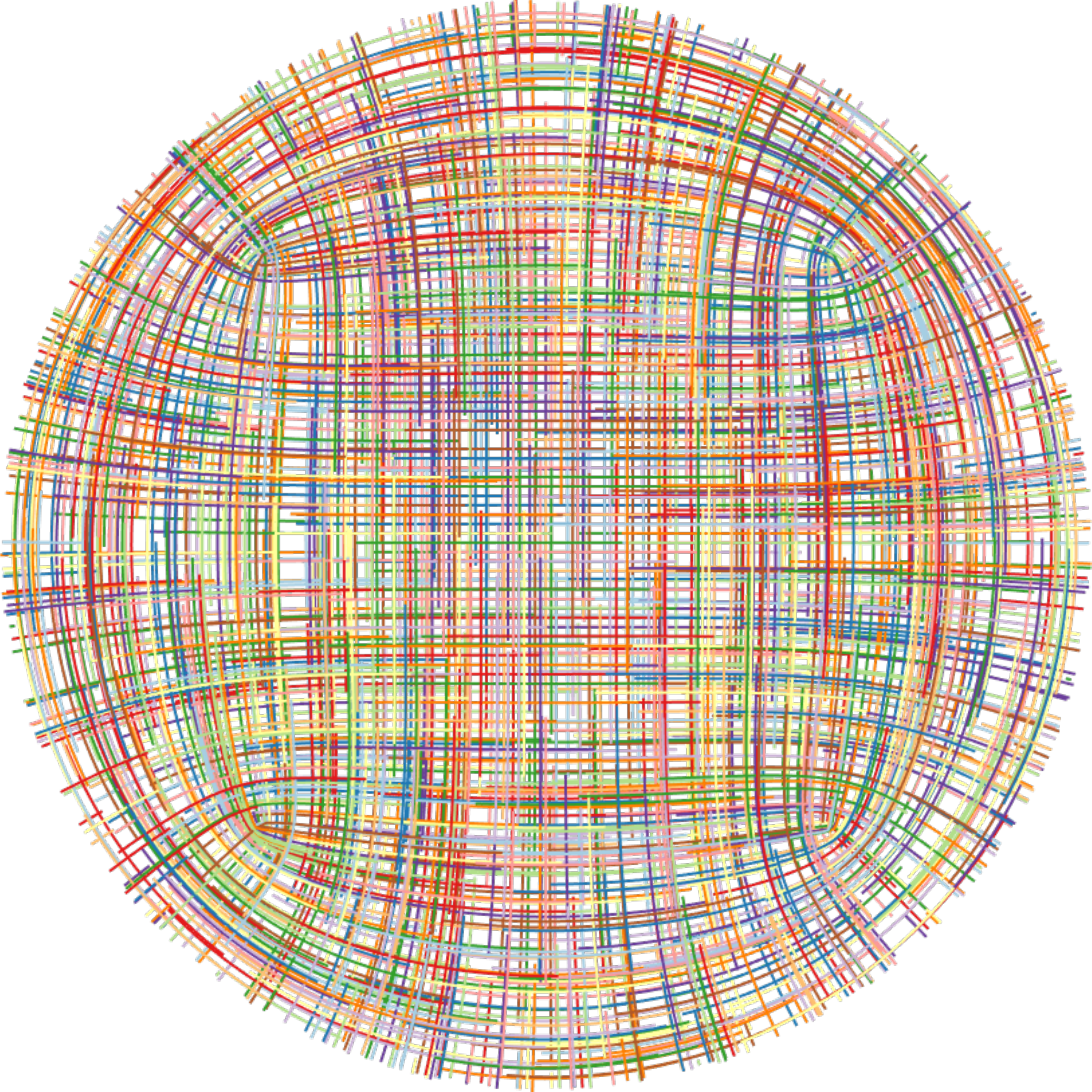} &
\includegraphics[height=\imageheight]{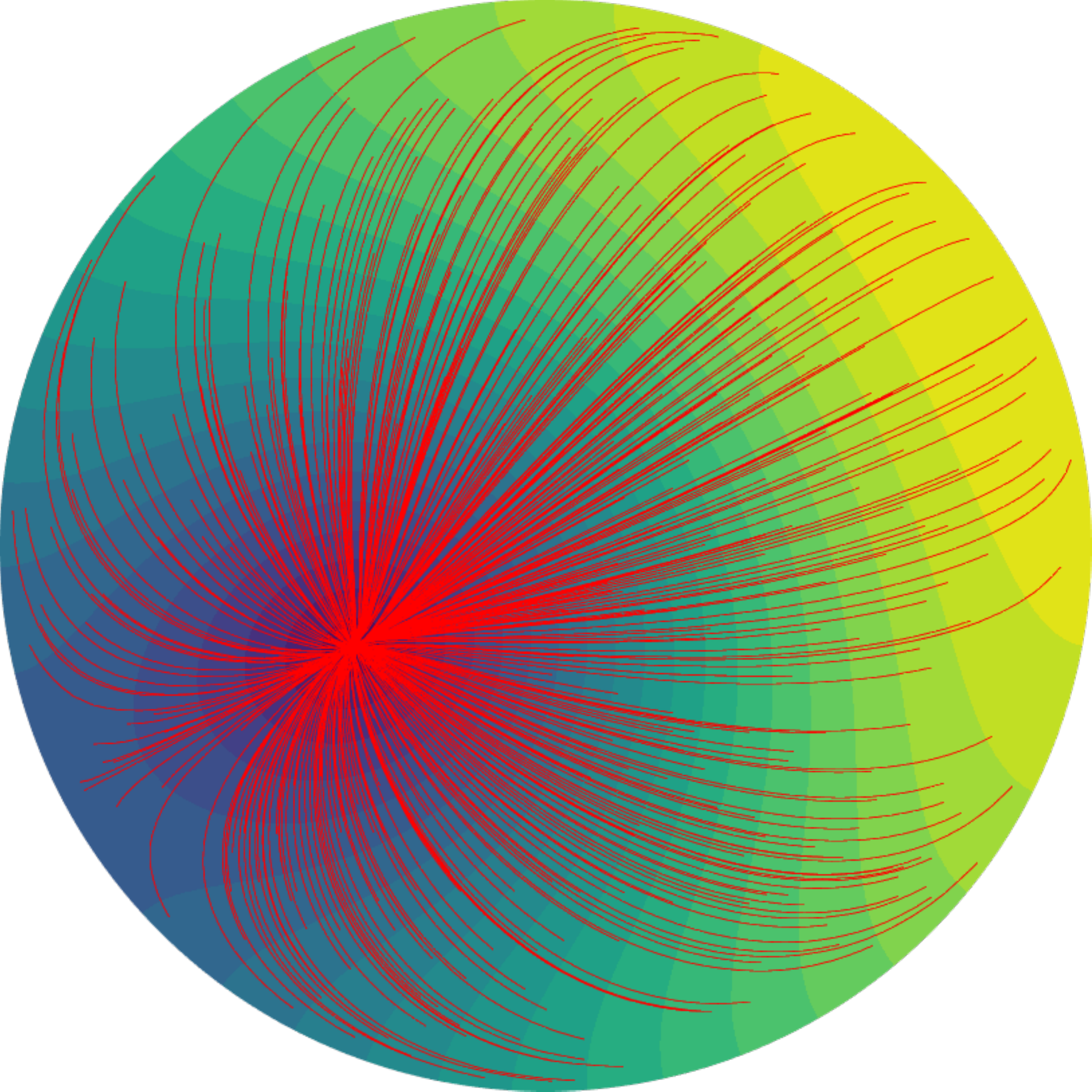} &
\includegraphics[height=\imageheight]{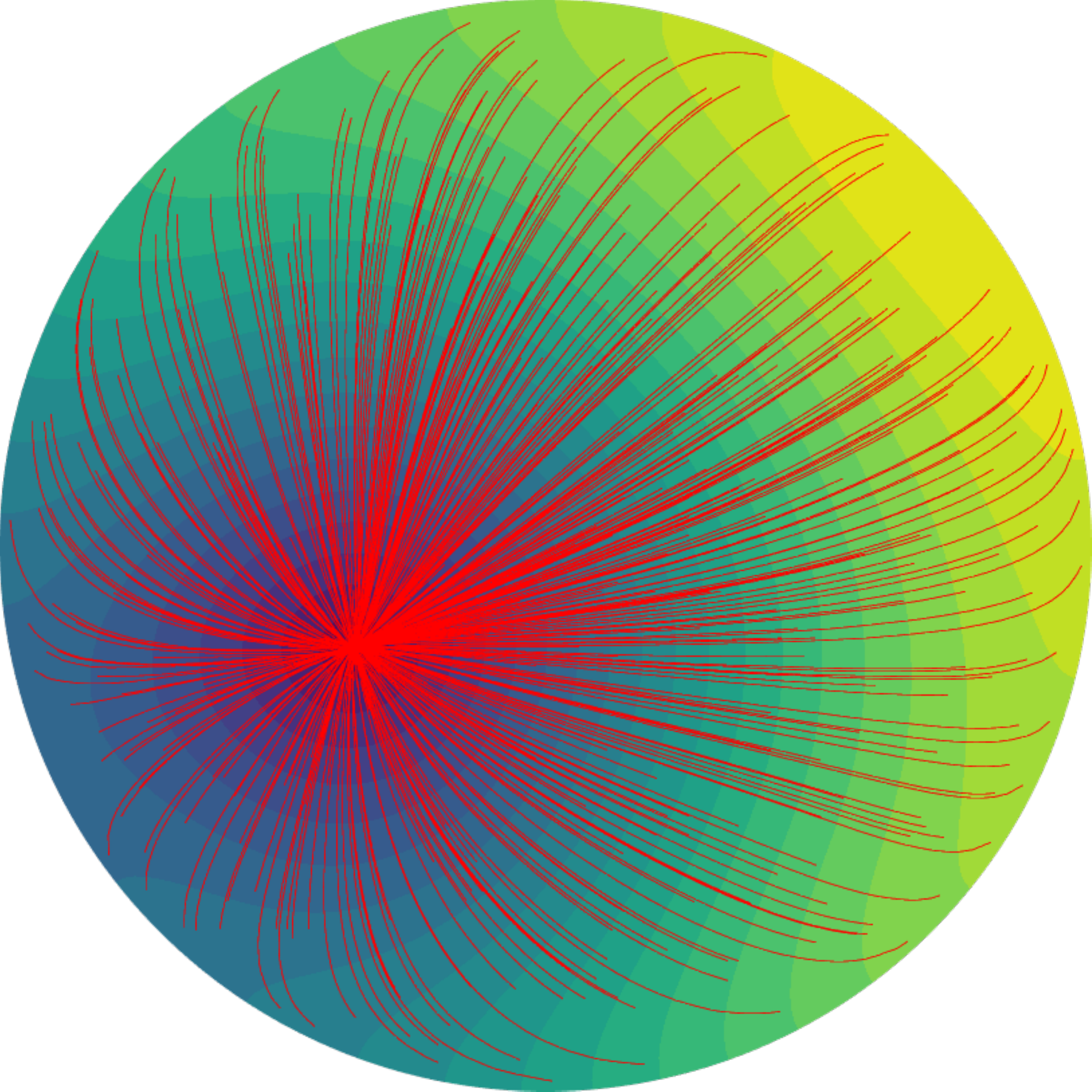} &
\includegraphics[height=\imageheight]{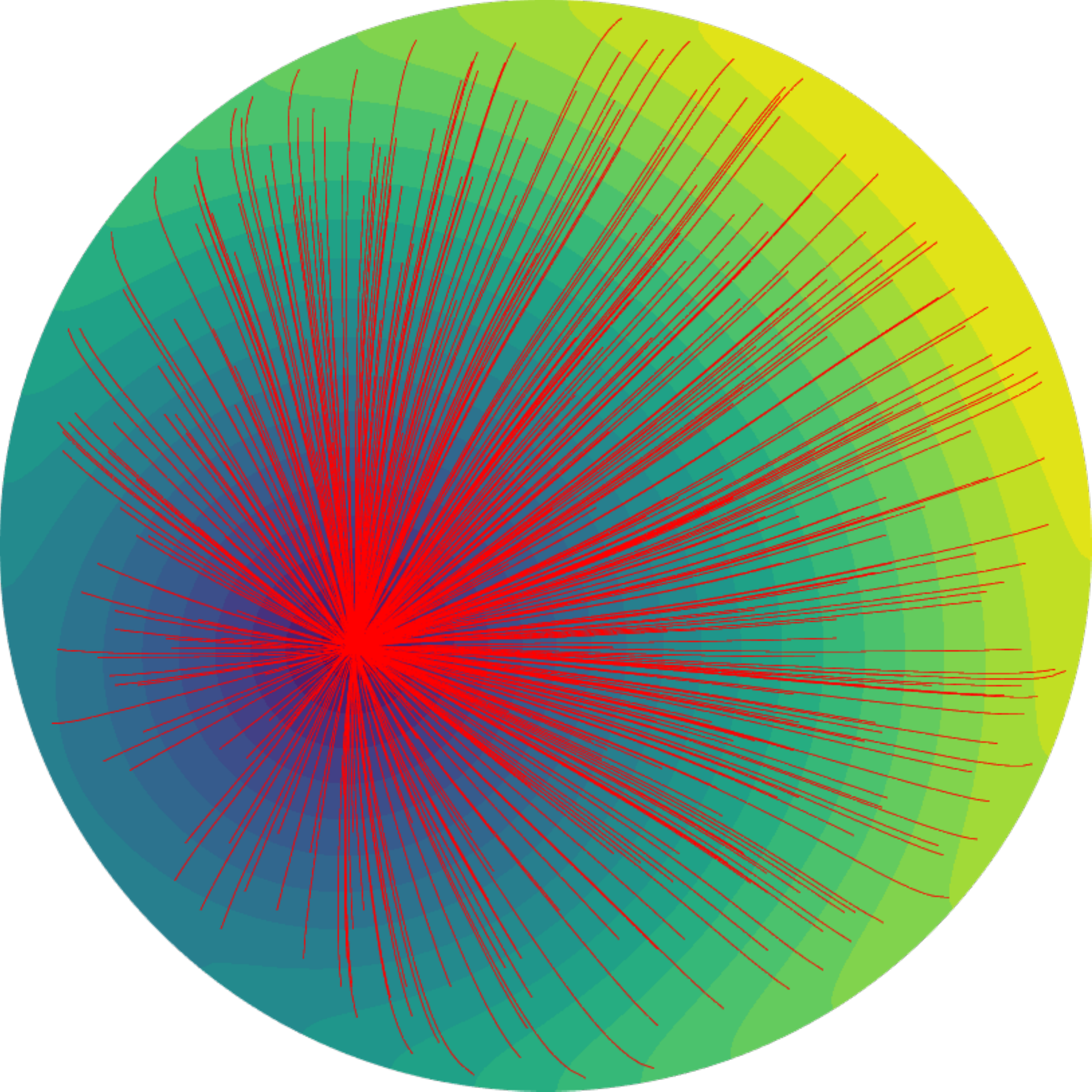} &
\includegraphics[height=\imageheight]{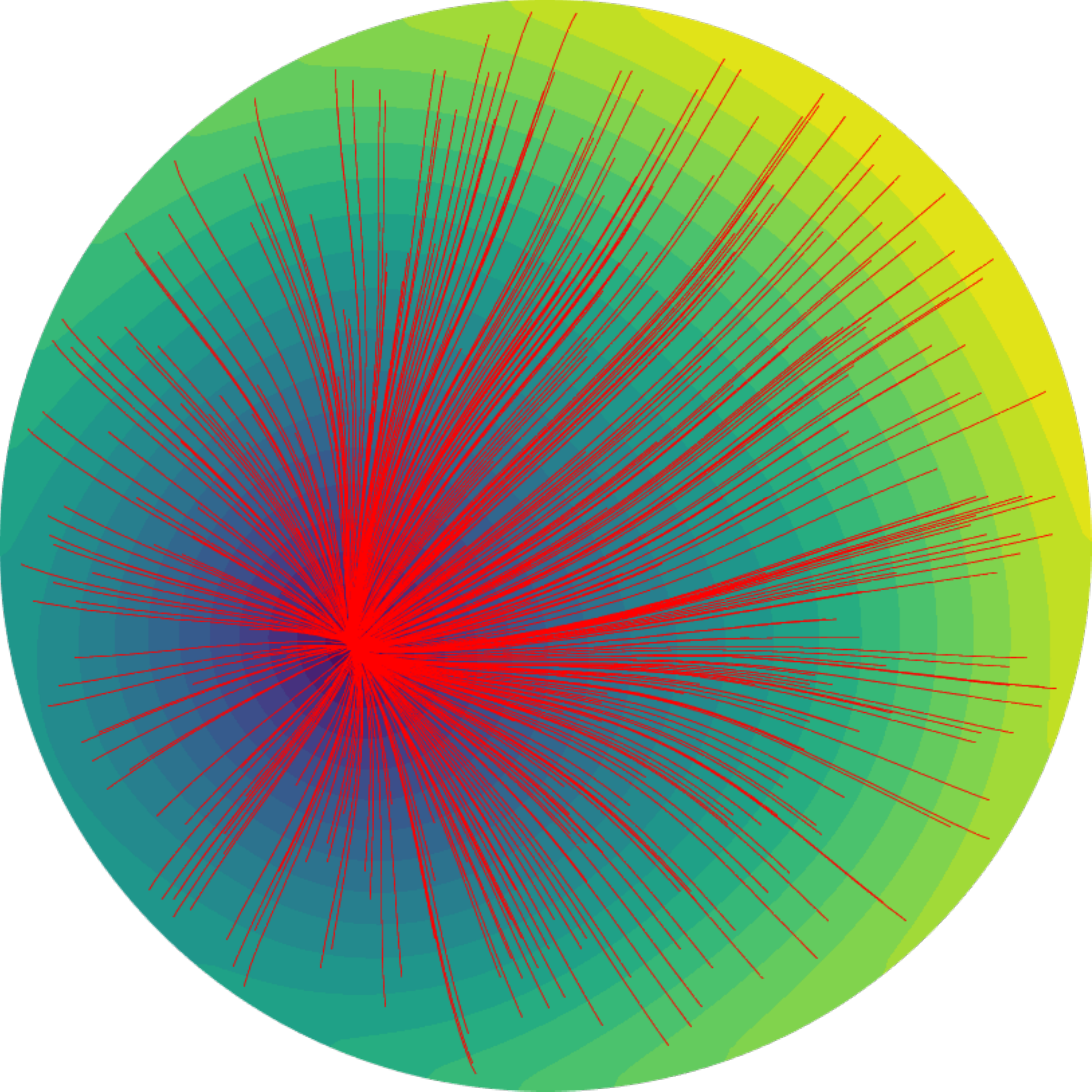} &
\includegraphics[height=\imageheight]{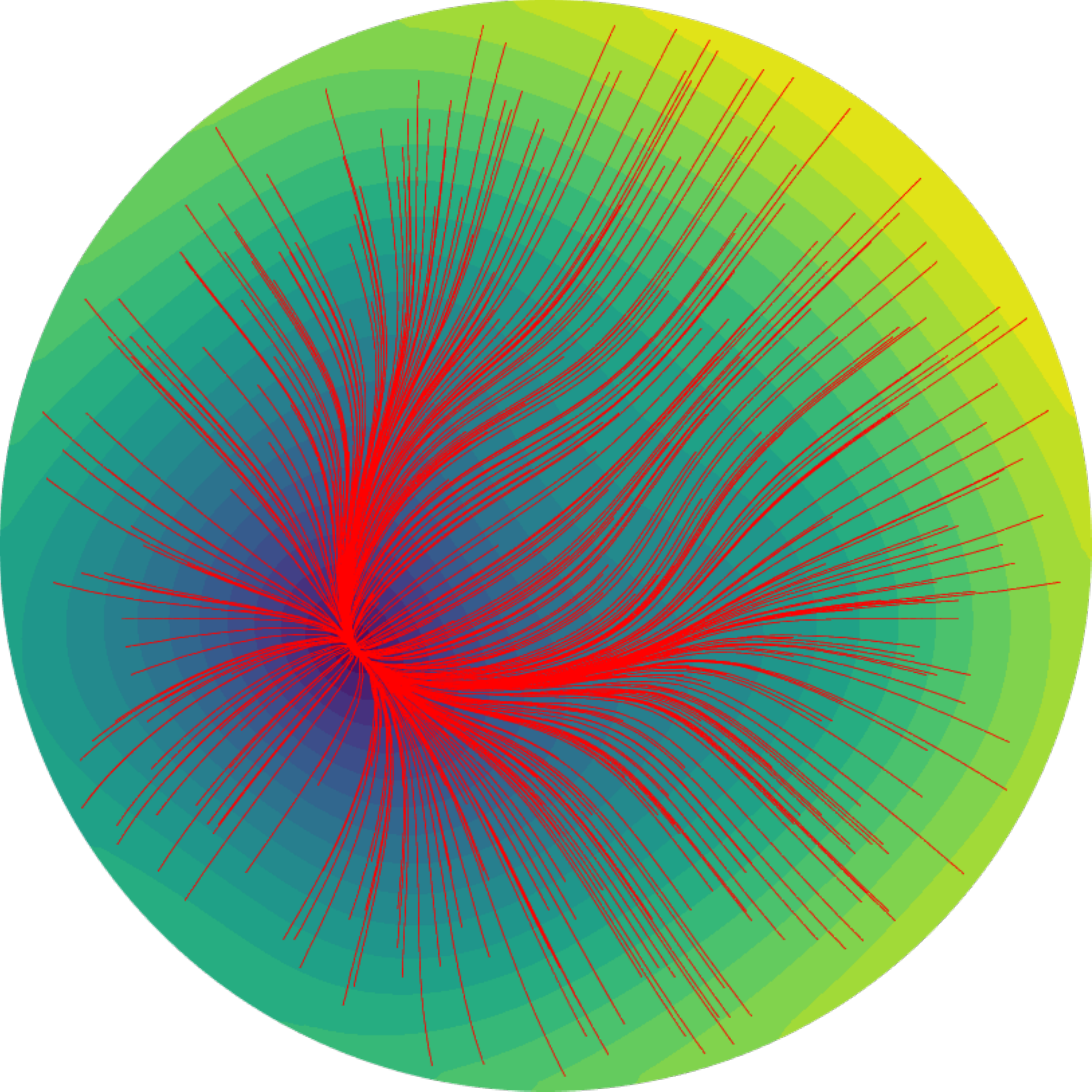} \\
\includegraphics[height=\imageheight]{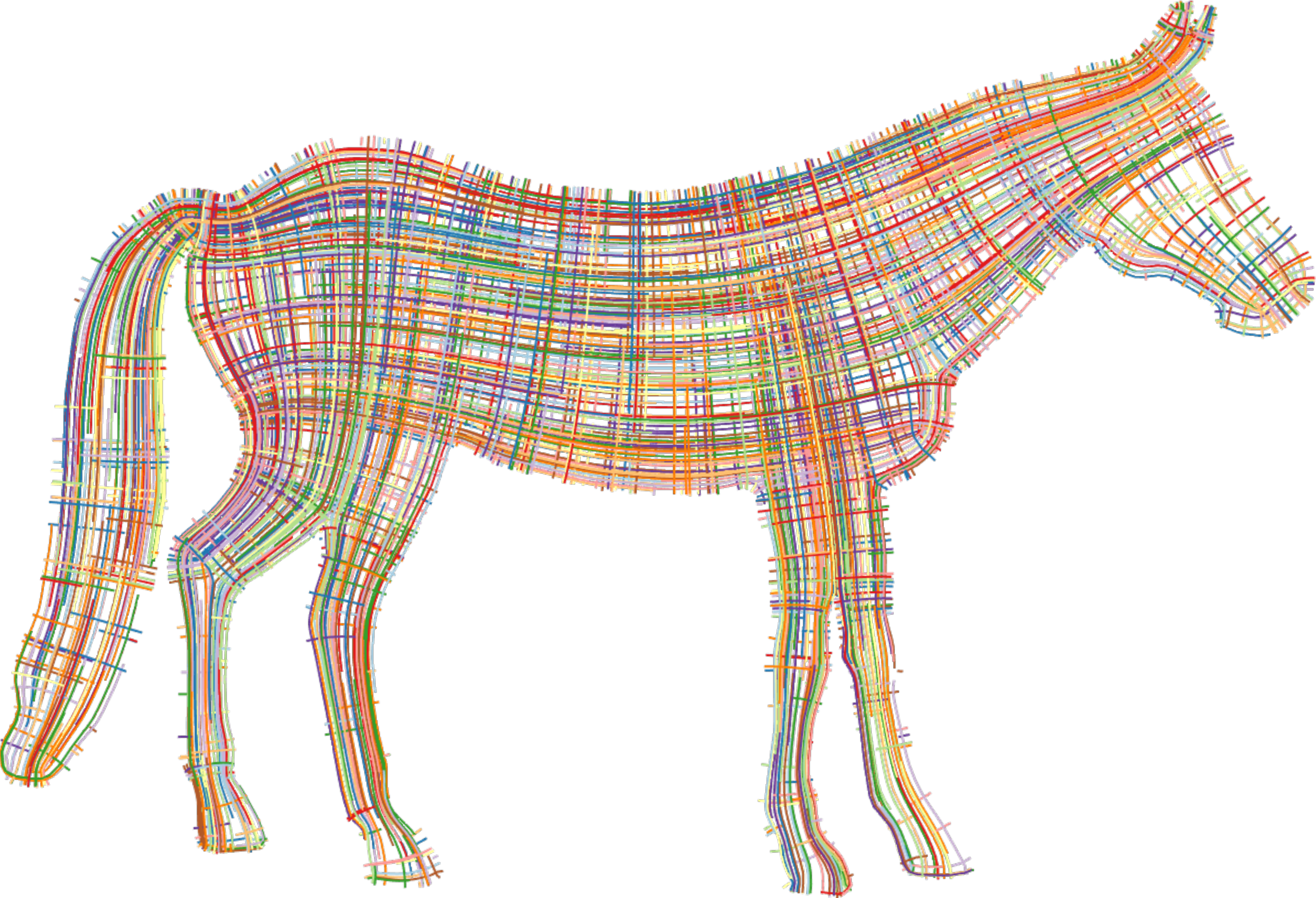} &
\includegraphics[height=\imageheight]{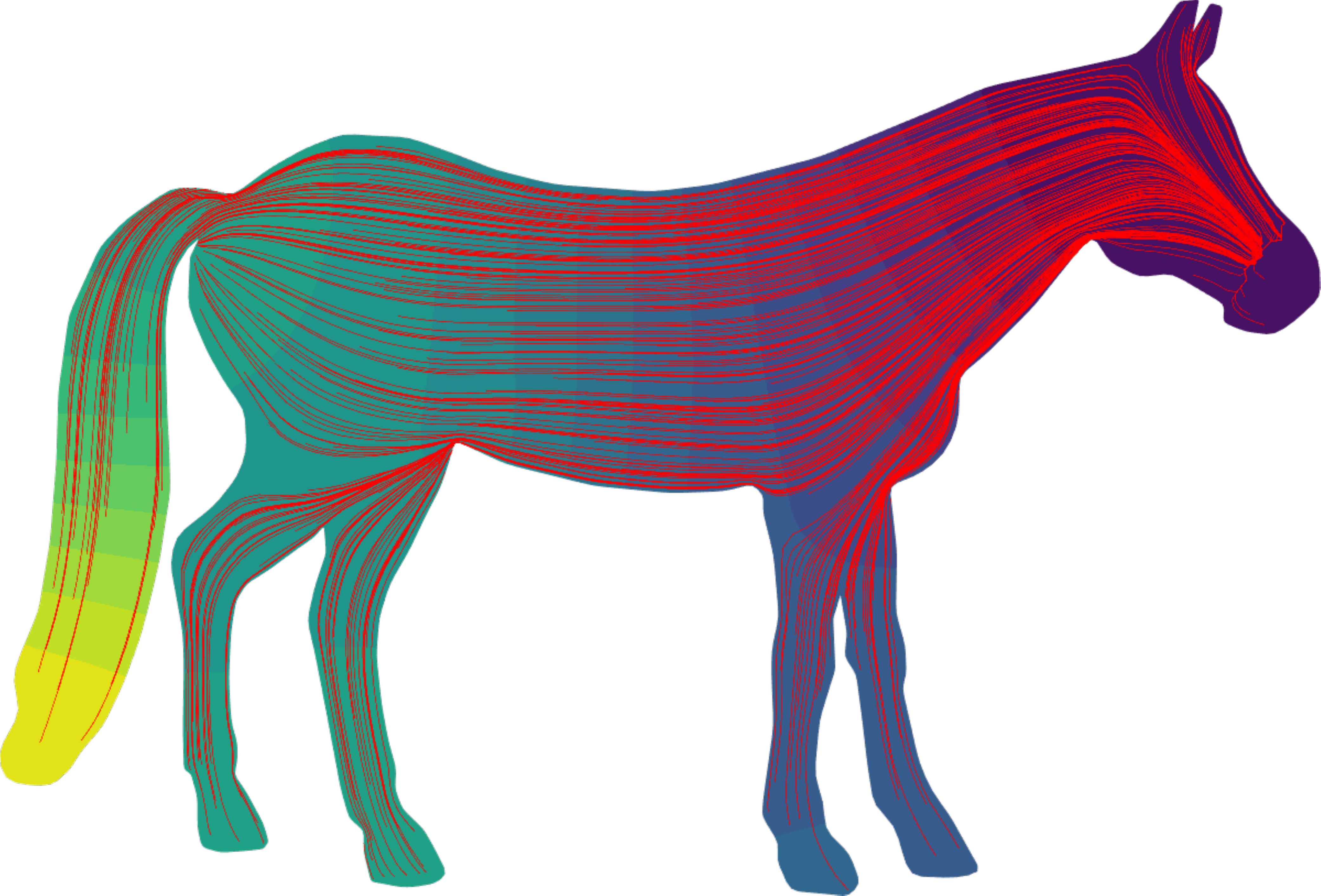} &
\includegraphics[height=\imageheight]{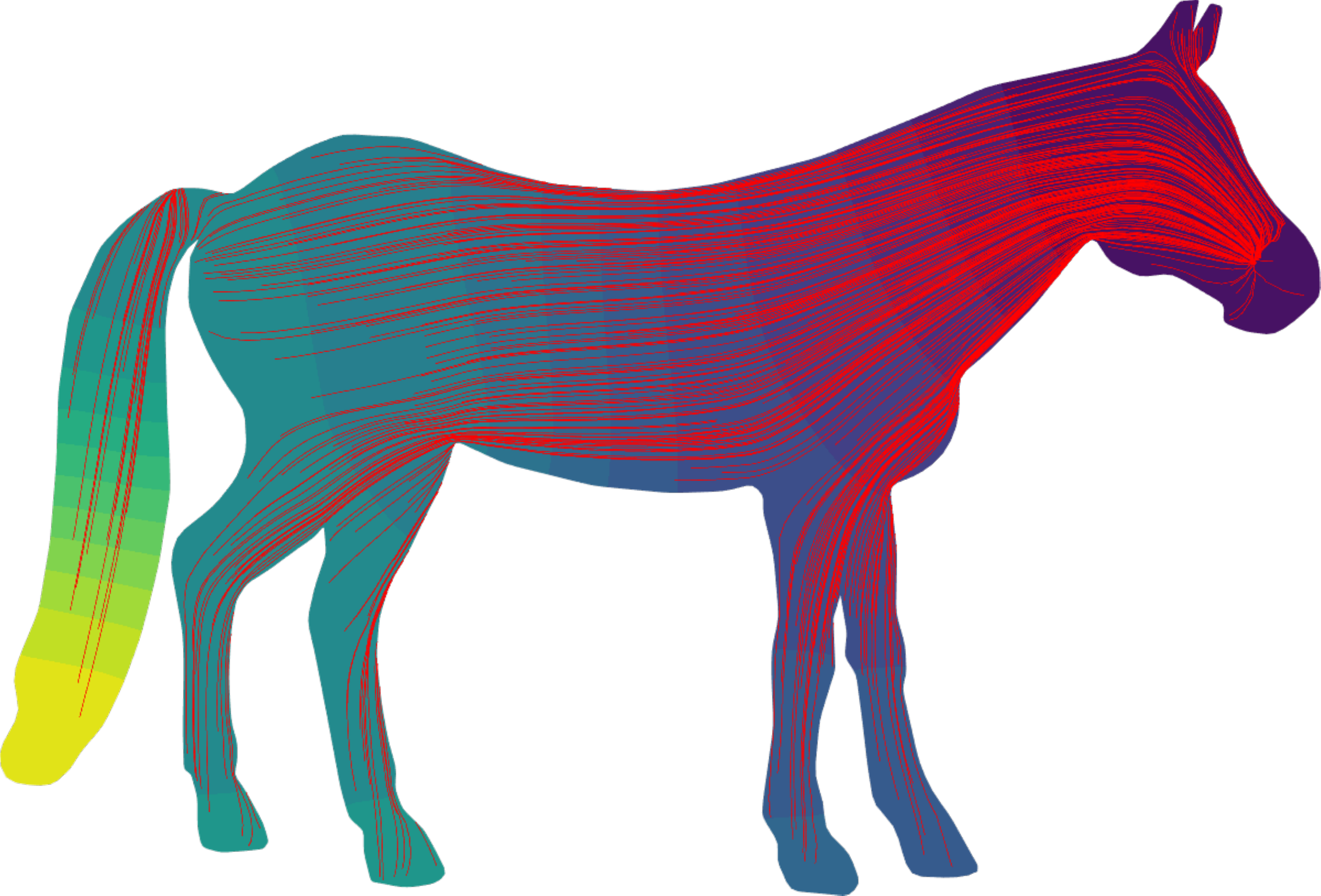} &
\includegraphics[height=\imageheight]{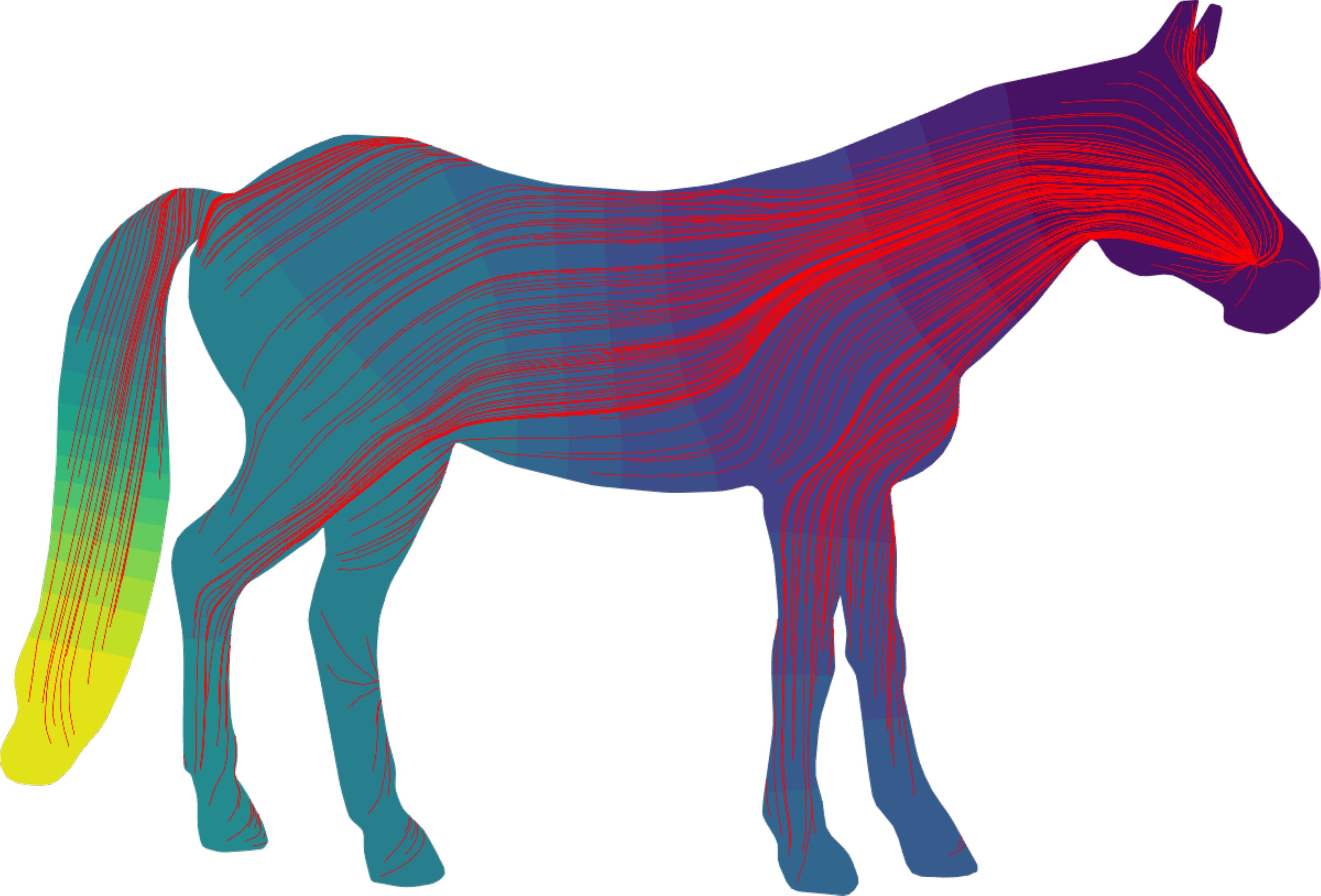} &
\includegraphics[height=\imageheight]{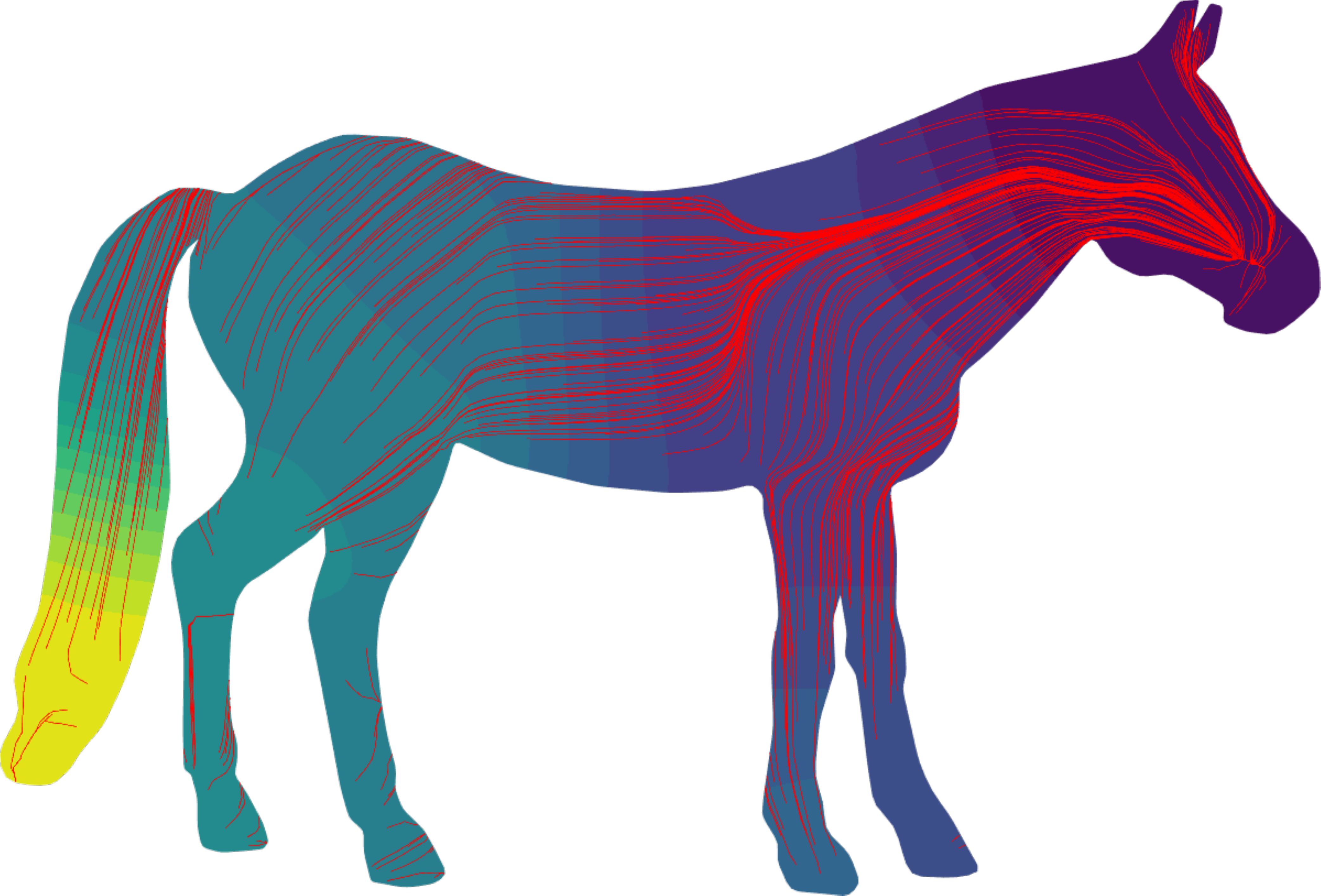} &
\includegraphics[height=\imageheight]{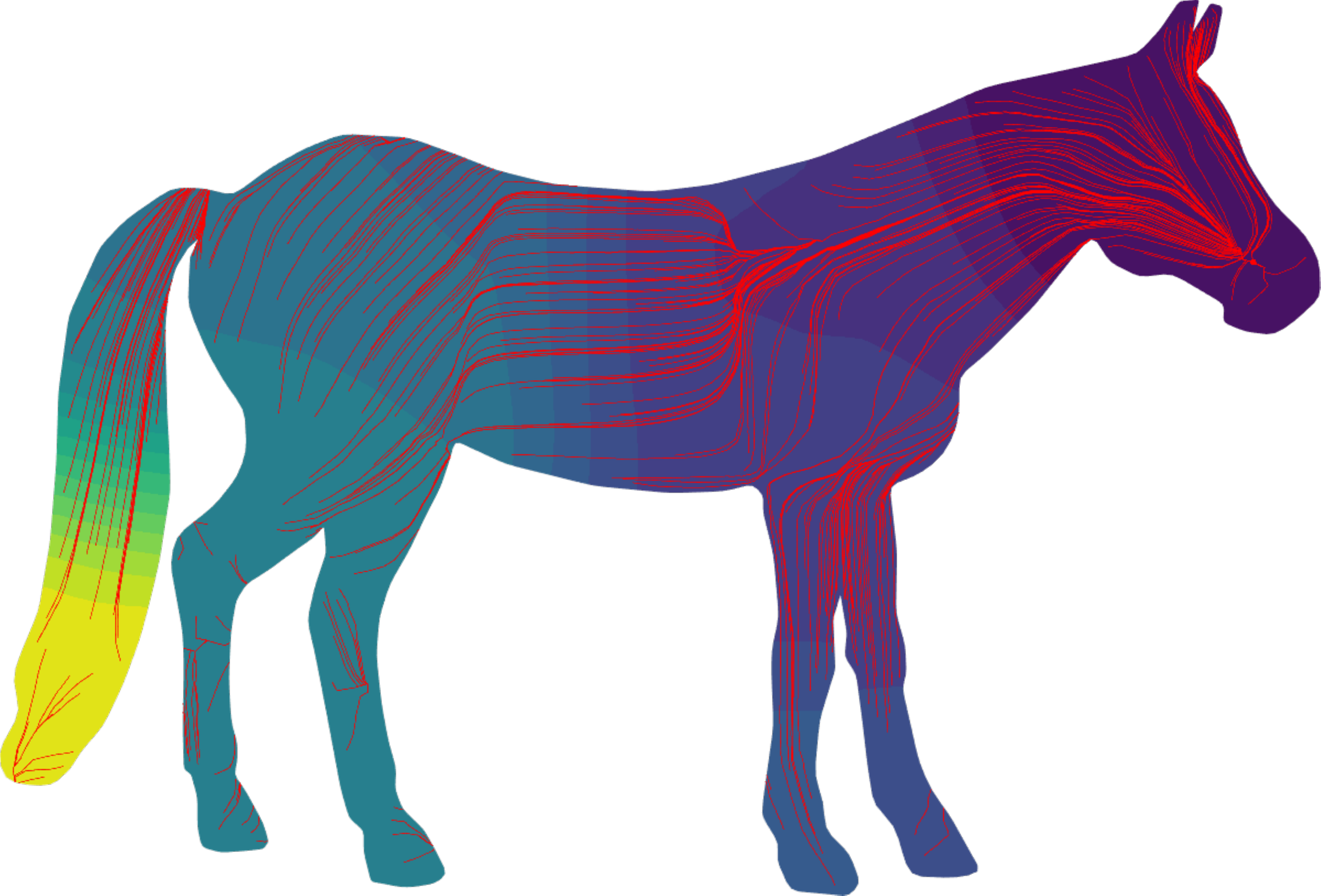} \\
& $\epsilon = 1$ &
$\epsilon = 10^{-1}$ &
$\epsilon = 10^{-2}$ &
$\epsilon = 10^{-3}$ &
$\epsilon = 10^{-4}$
\end{tabular}
\caption{Analogously to biharmonic distances, we can compute a smooth distance function from distances in the spectral embedding given by our operator with Neumann boundary conditions. When $\epsilon = 1$ we get biharmonic distances. As $\epsilon$ decreases, the distance functions become more anisotropic, and the shortest paths computed by gradient descent on distance become more aligned to the frame fields.}
\label{fig:dist}    
\end{figure*}

\subsection{Anisotropic Biharmonic Distance}

By analogy to the biharmonic distance \cite{Lipman2010}, we can design smooth anisotropic distance functions that exhibit ``Manhattan-like" behavior along a prescribed frame field. These distance functions might be used for example in navigation, where we want the robot to trace out a path along a grid that varies smoothly and aligns to domain boundaries. 

Inspired by the biharmonic distance, our frame field operator distances are computed as follows: first, the first $k = 1, \dots, N$ nonzero eigenvalues $\lambda_k$ and corresponding eigenfunctions $\phi_k$ of the frame field operator $\mathcal{A}_{T,\epsilon}$ are computed, discarding those where $\lambda_k = 0$. Then the frame field operator distance between points $p$ and $q$ is defined by:
\begin{equation}
d_{T, \epsilon}(p, q)^2 \coloneqq \sum_{k=1}^N \frac{|\phi_k(p) - \phi_k(q)|^2}{\lambda_k^2}.
\end{equation}
This is essentially computing distances in the spectral embedding corresponding to the inverse of the frame field operator.

%\justin{add a paragraph here with the formula for the distance between two points in terms of the eigenvalues/eigenfunctions}

\Cref{fig:dist} illustrates isolines of our frame field-aware distance along a disk and a horse model in the plane; we also show shortest paths in the domain from a set of randomly-chosen source points to a single source point, computed using gradient descent on the distance function.  When $\epsilon$ is fairly large, our distances behave similarly to the biharmonic distance. As $\epsilon\rightarrow0$, however, the level sets of the distance are roughly $45^\circ$ rotated from the field, as might be expected from computing $L^2$ distances between functions like the impulse responses illustrated in \Cref{fig:impulseresponse}.

%shows isolines of frame field distance from various points for a frame field on the disk.

\subsection{Coloring with Frame Fields}

Diffusion curves \cite{Orzan2008} define a way to propagate color information from a sparse set of user-defined curves to the remainder of an image; similar approaches exist with higher-order operators \cite{Finch2011}.
One can achieve similar results by prescribing color values at the boundary of a meshed domain
and then minimizing a smoothing energy to smoothly color the domain.

\begin{figure*}
\centering
\newcommand{\imagewidth}{0.23\columnwidth}
\setlength\tabcolsep{3pt}
\renewcommand{\arraystretch}{1.5}
\begin{tabular}{@{}cc|cc|cc|cc@{}}
\includegraphics[width=\imagewidth]{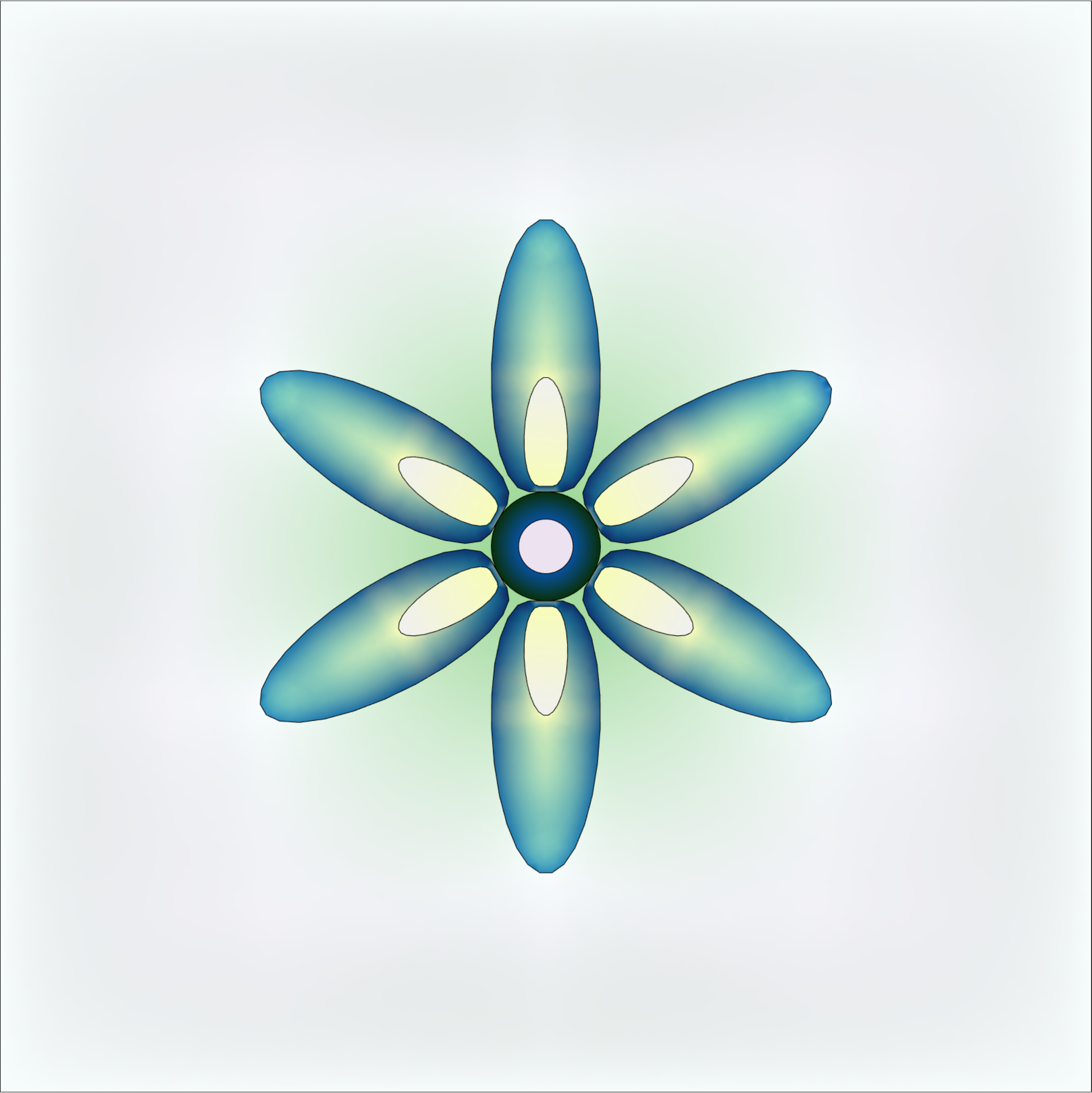} &
\includegraphics[width=\imagewidth]{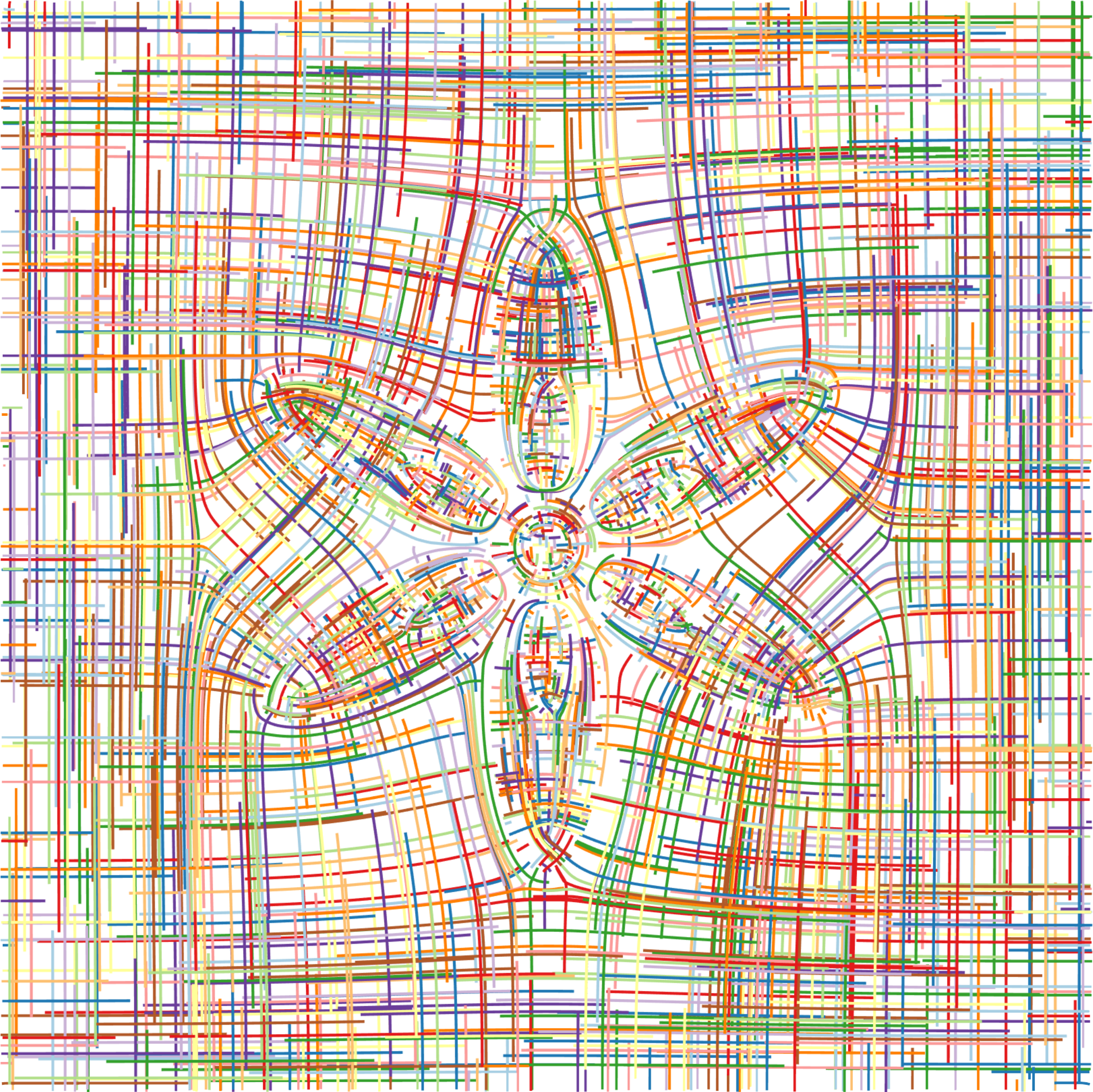} \;&\;
\includegraphics[width=\imagewidth]{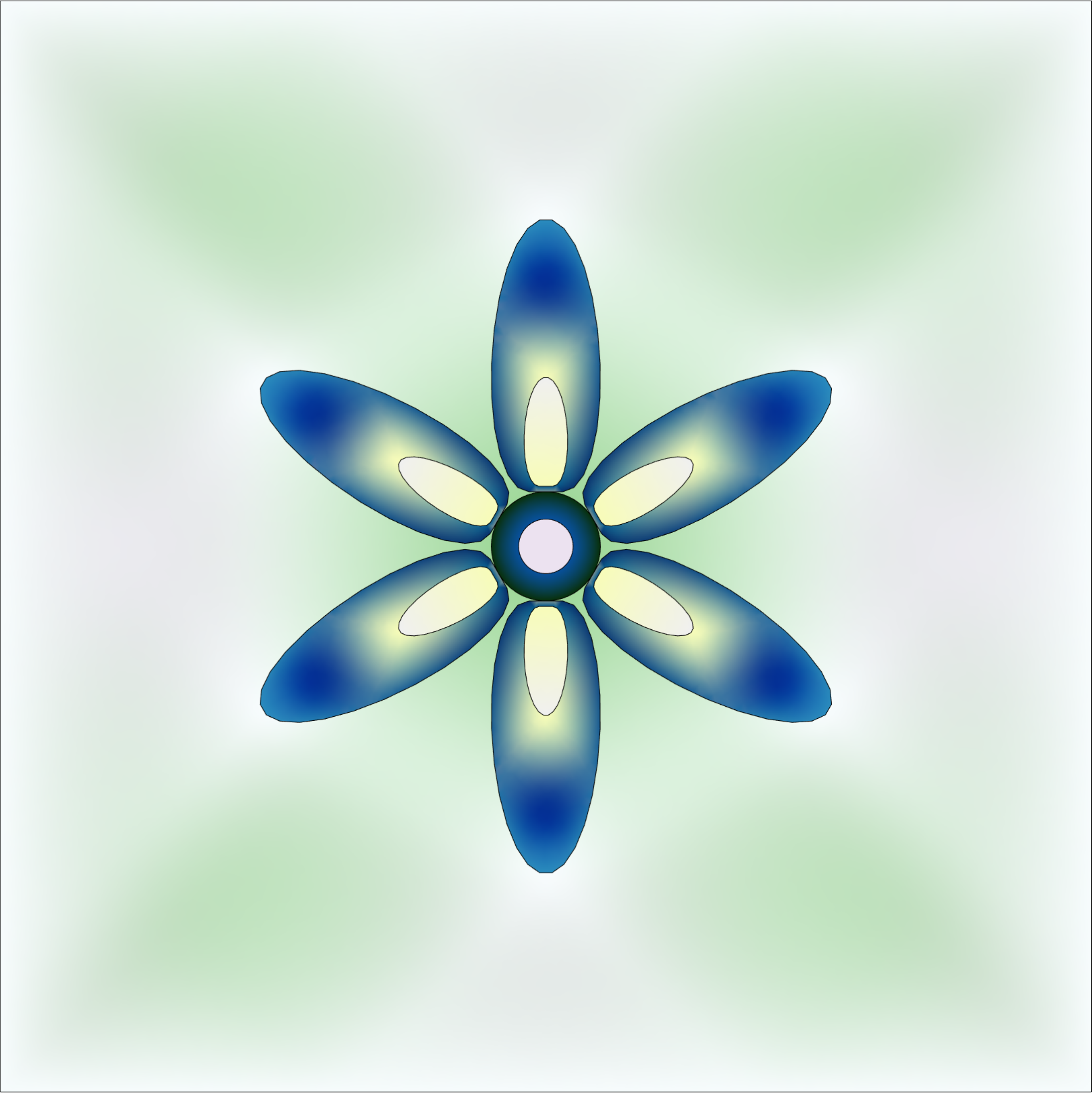} &
\includegraphics[width=\imagewidth]{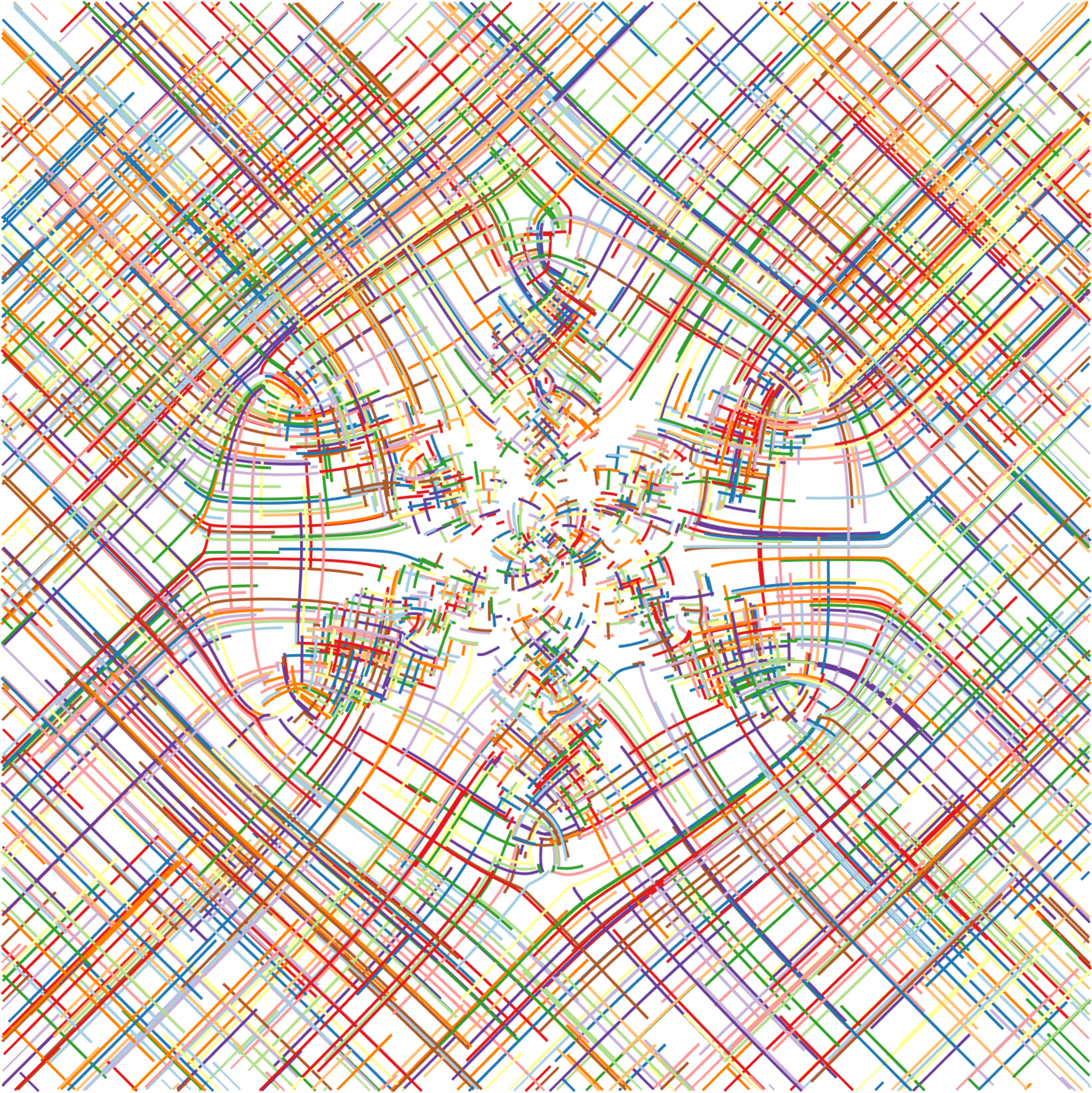} \;&\;
\includegraphics[width=\imagewidth]{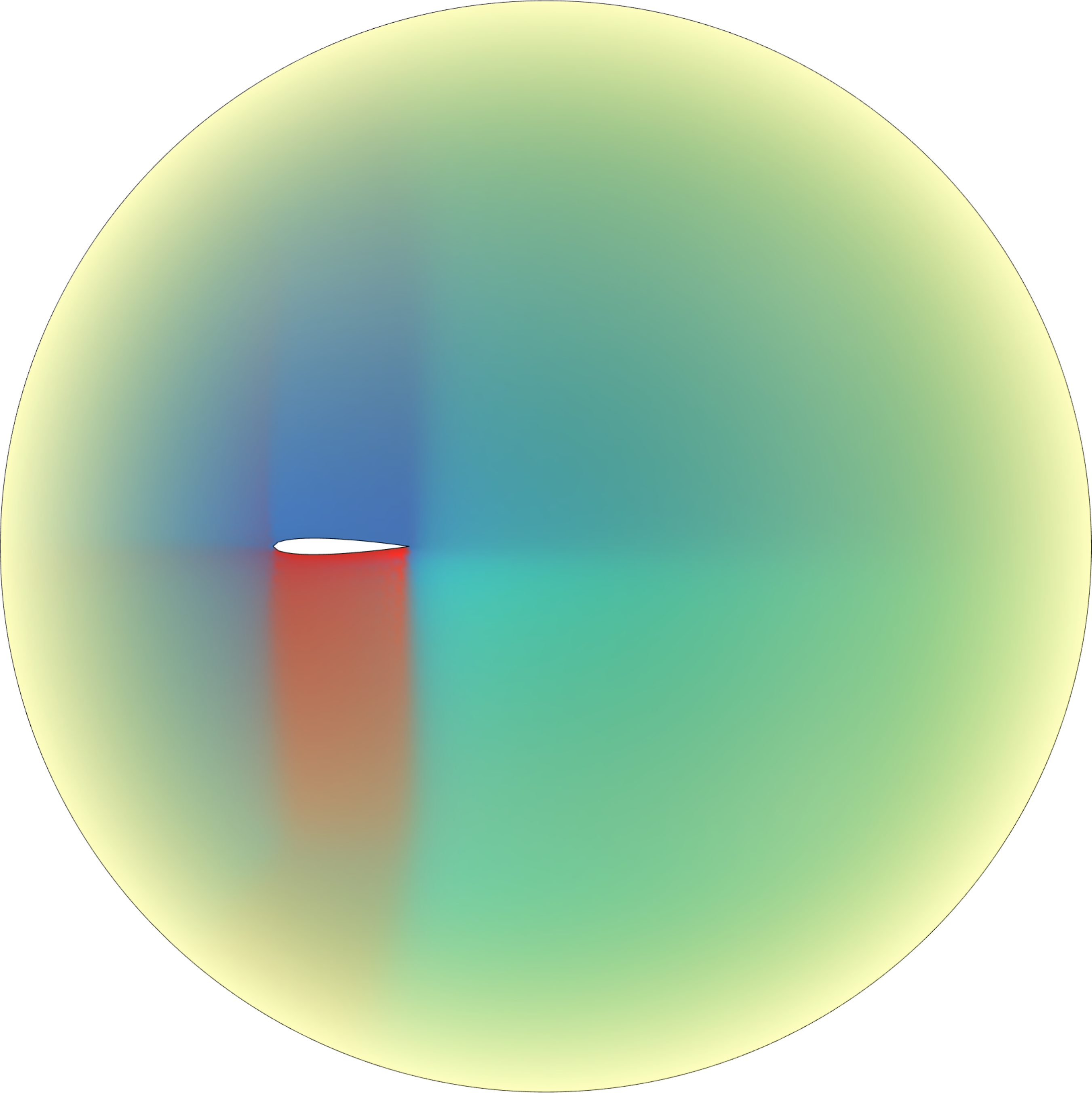} &
\includegraphics[width=\imagewidth]{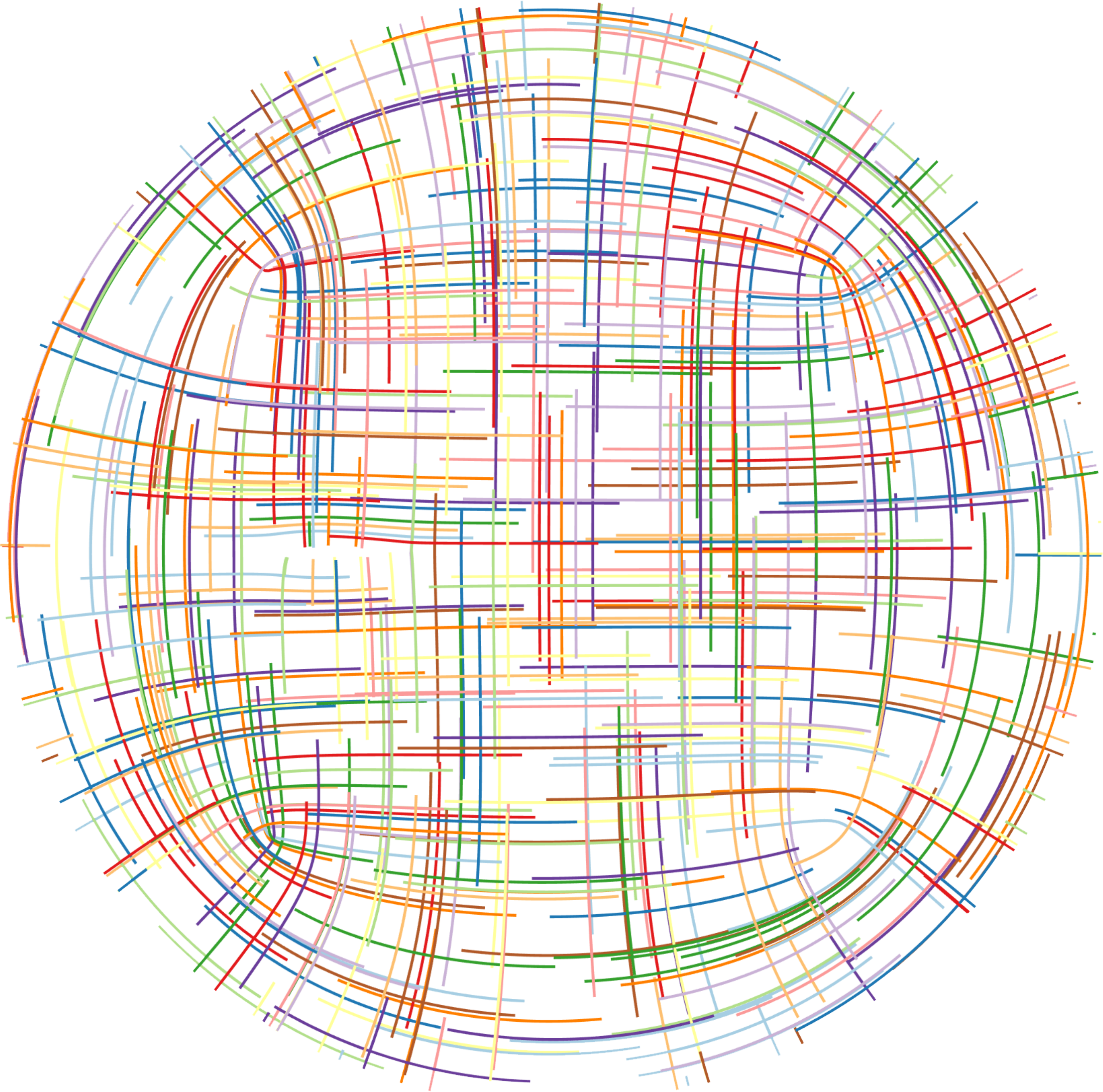} \;&\;
\includegraphics[width=\imagewidth]{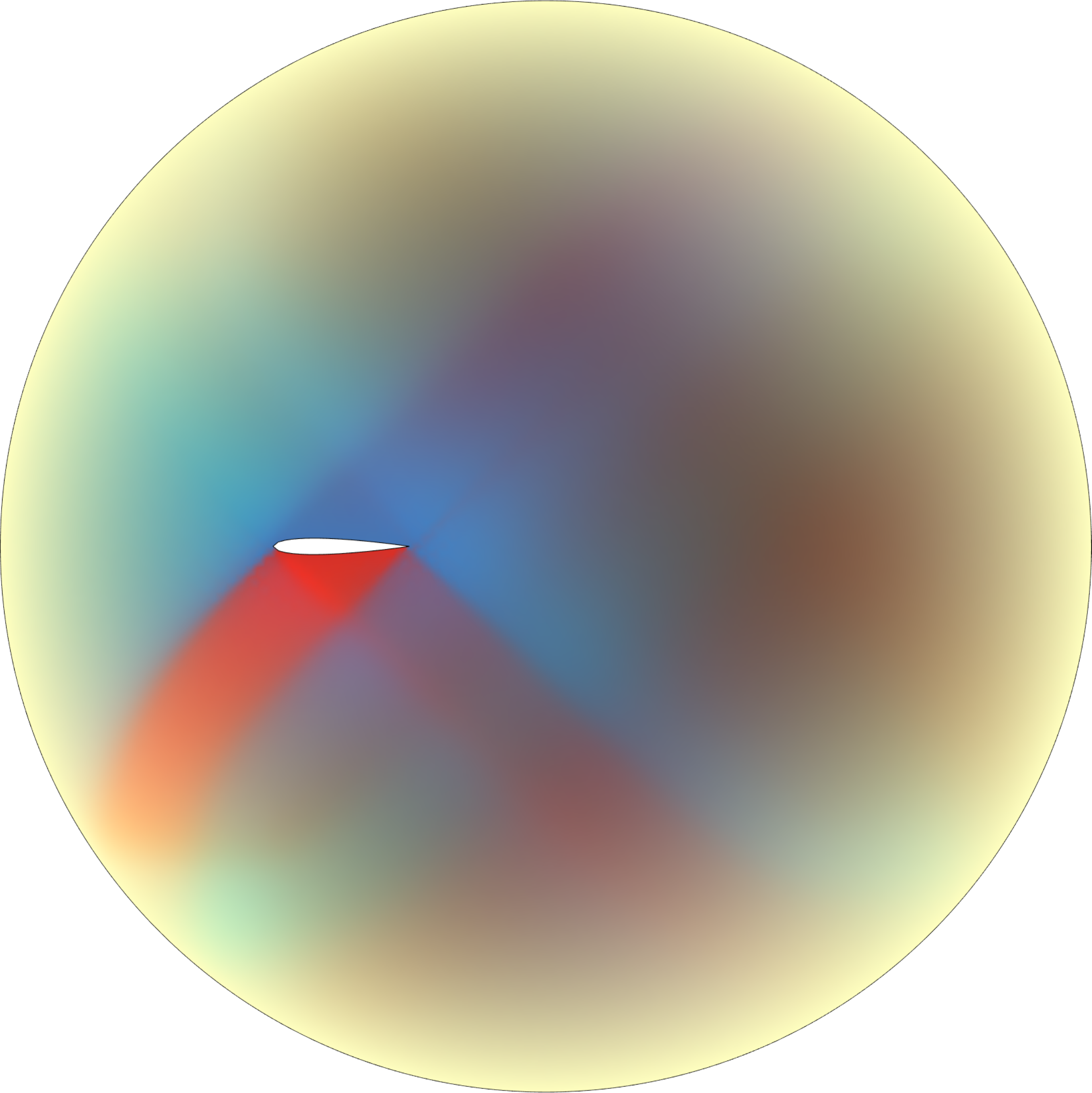} &
\includegraphics[width=\imagewidth]{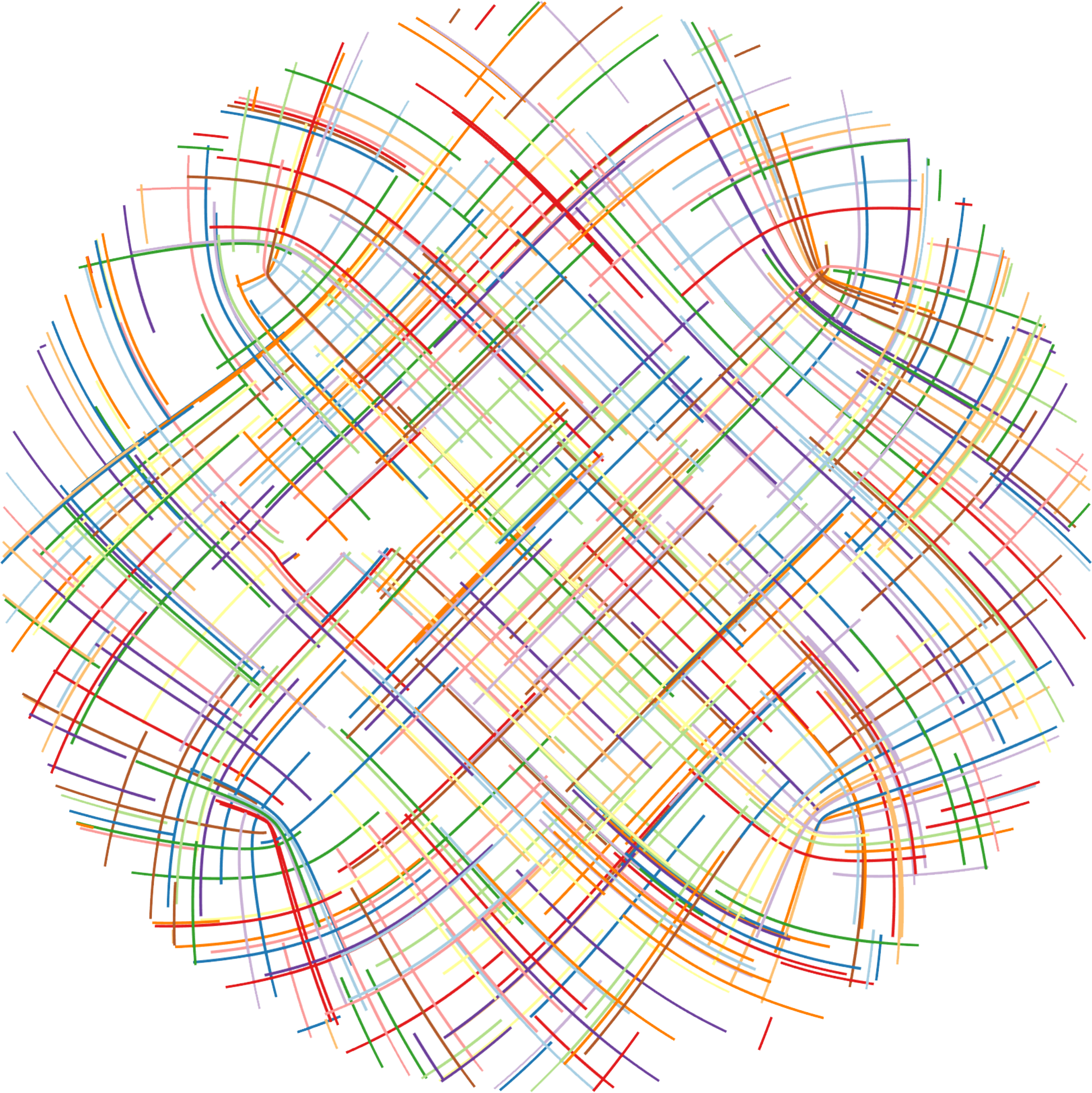}
\end{tabular}
\caption{This figure shows two different domains on which we solve the boundary value coloring problem \eqref{eq:diffusioncoloring} twice with the same boundary data, but with different frame fields.
The resulting coloring changes based on the frame field, as colors diffuse along the field directions.}
\label{fig:boundarycoloring}    
\end{figure*}

As an illustration of this technique, in \Cref{fig:boundarycoloring} we solve a quadratic programming problem in each RGB color channel to obtain the color value \(\bm{c}\):
\begin{equation}\label{eq:diffusioncoloring}
    \bm{c} = \argmin_{\bm{c}} \; \frac12 \bm{c}^\transp \mathcal{A}_{T,0.01} \bm{c},
    \quad
    \bm{l} \leq \bm{c} \leq \bm{u}
    \;\textrm{,}
\end{equation}
with the operator's natural boundary conditions using the primal version
of the operator and the inequality bounds set so that the
minimum and maximum values in each color channels occur at the boundary.

The choice of field heavily influences the result; the direction of color diffusion
follows the selected field.  Hence, we can view \eqref{eq:diffusioncoloring} using the frame field operator as a means of giving greater control to diffusion-based painting methods by linking this toolbox to frame field design.

% \begin{figure}
% \centering
% \newcommand{\imagewidth}{0.22\columnwidth}
% \setlength\tabcolsep{5pt}
% \renewcommand{\arraystretch}{1.5}
% \begin{tabular}{@{}cc|cc@{}}
% \includegraphics[width=\imagewidth]{figures/diffusioncurves-flower-odeco.png} &
% \includegraphics[width=\imagewidth]{figures/diffusioncurves-field-flower-odeco.png} &
% \includegraphics[width=\imagewidth]{figures/diffusioncurves-flower-odeco-rotpi.png} &
% \includegraphics[width=\imagewidth]{figures/diffusioncurves-field-flower-odeco-rotpi.png} \\
% \hline
% \rule{0pt}{62pt}
% \includegraphics[width=\imagewidth]{figures/diffusioncurves-airfoil-odeco.png} &
% \includegraphics[width=\imagewidth]{figures/diffusioncurves-field-airfoil-odeco.png} &
% \includegraphics[width=\imagewidth]{figures/diffusioncurves-airfoil-odeco-rotpi.png} &
% \includegraphics[width=\imagewidth]{figures/diffusioncurves-field-airfoil-odeco-rotpi.png}
% \end{tabular}
% \caption{This figure shows two different domains on which we solve the boundary value coloring problem \eqref{eq:diffusioncoloring} twice with the same boundary data, but with different frame fields.
% The resulting coloring changes based on the frame field, as colors diffuse along the field direction.}
% \label{fig:boundarycoloring}    
% \end{figure} 

\subsection{Conformal Octahedral Fields and Singularities}
\begin{figure*}
\centering
\newcommand{\imagewidth}{0.45\textwidth}
\begin{tabular}{@{}c|c@{}}
\includegraphics[width=0.08\textwidth]{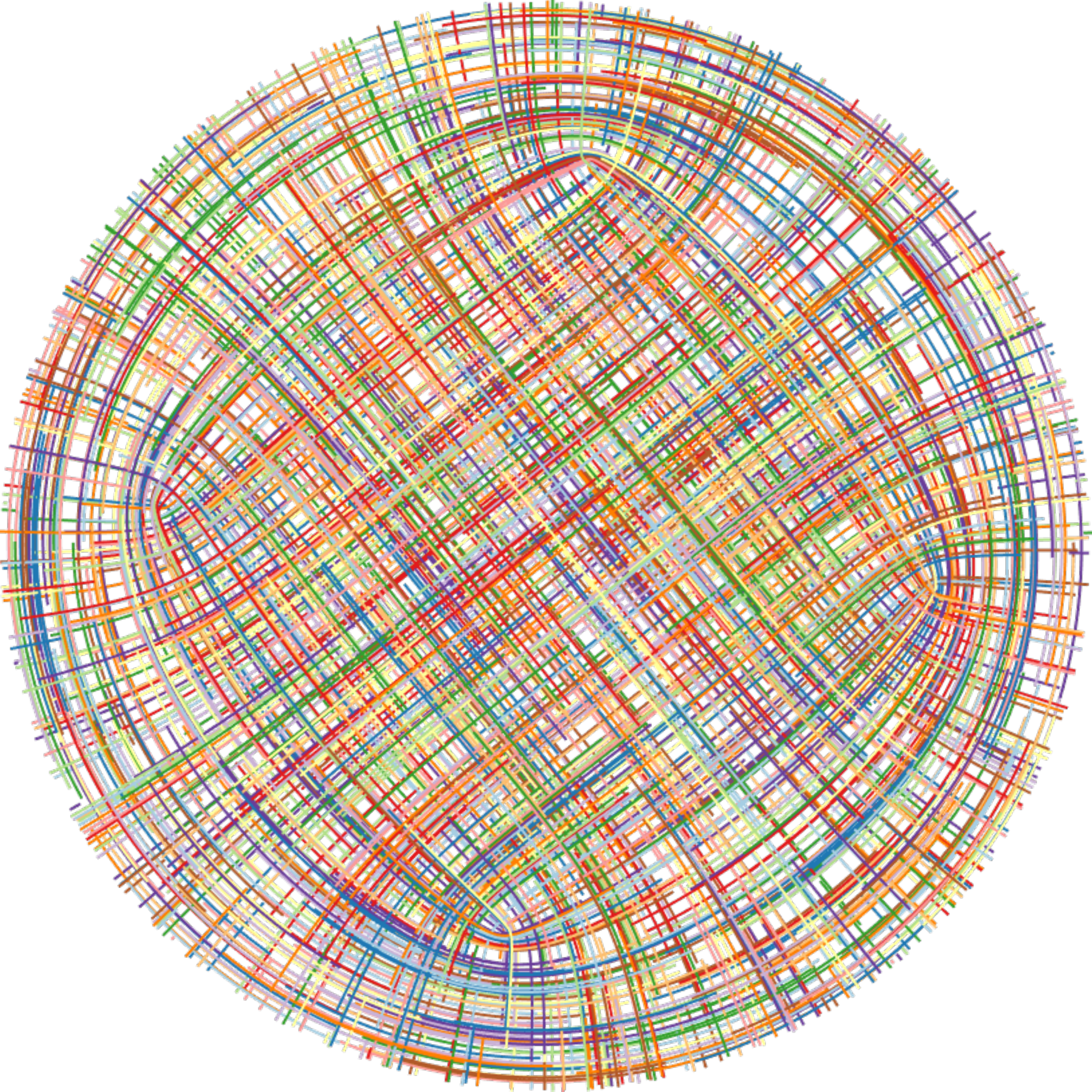} & \includegraphics[width=0.08\textwidth]{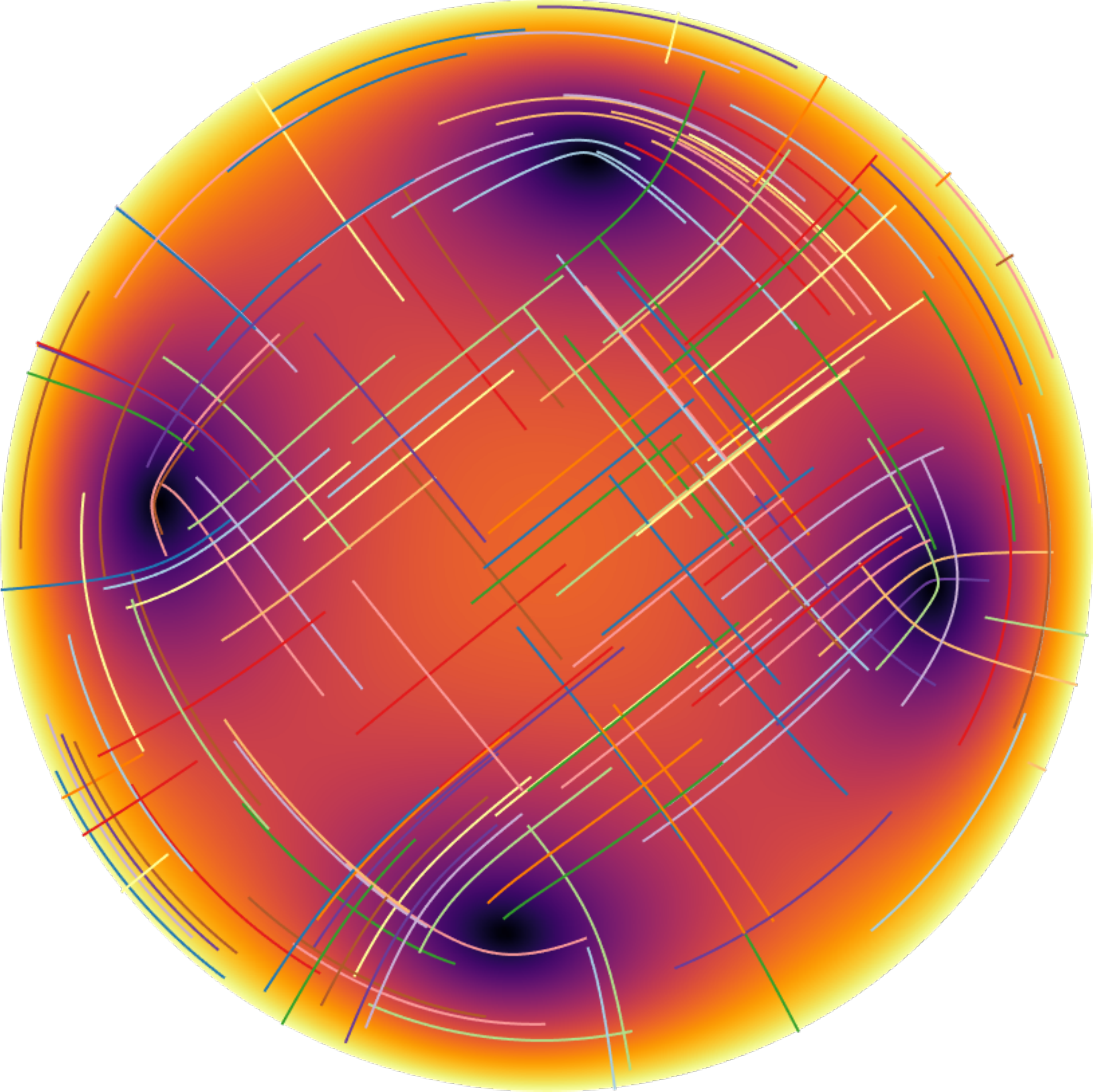} \\
\includegraphics[width=\imagewidth]{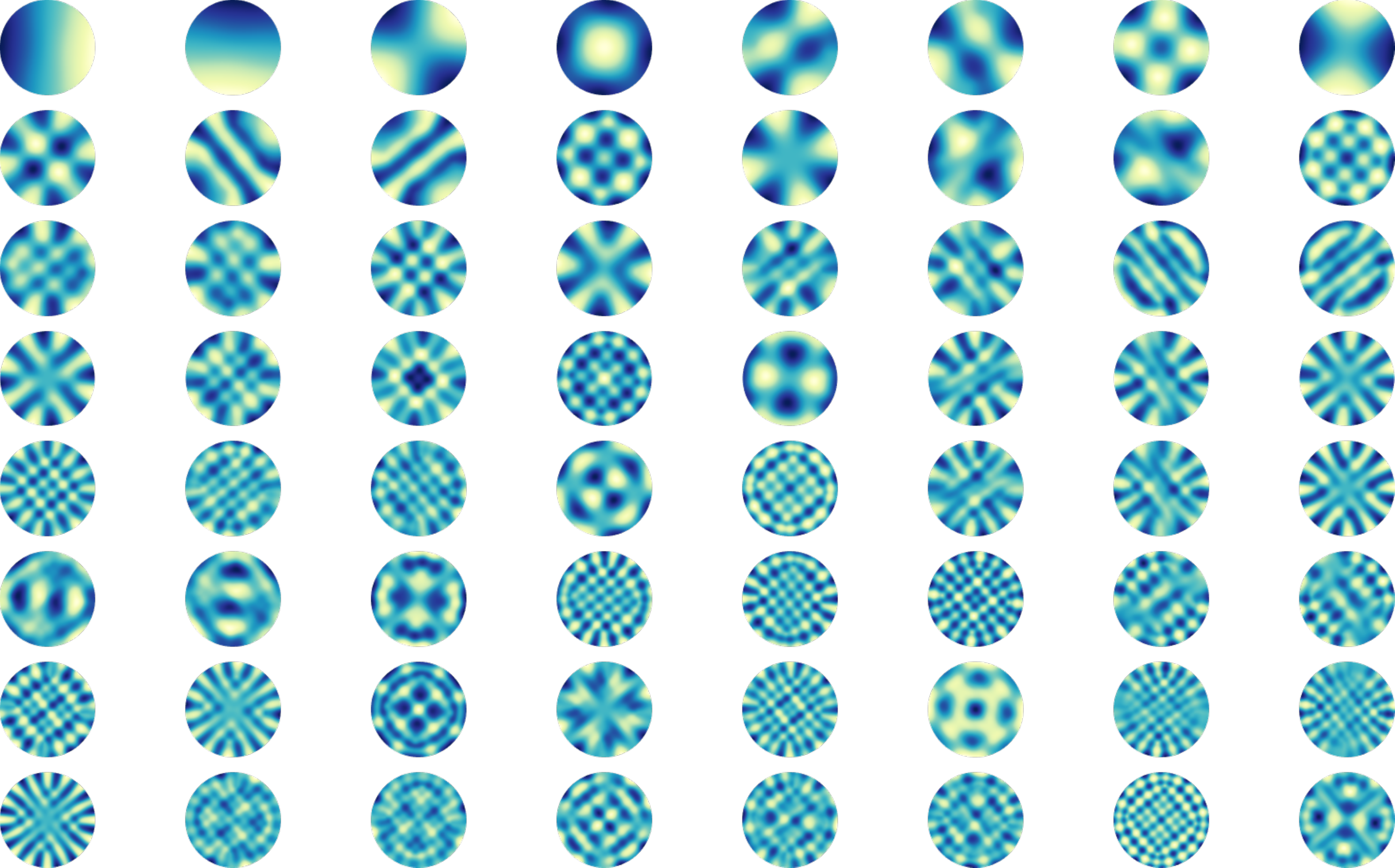} &
\includegraphics[width=\imagewidth]{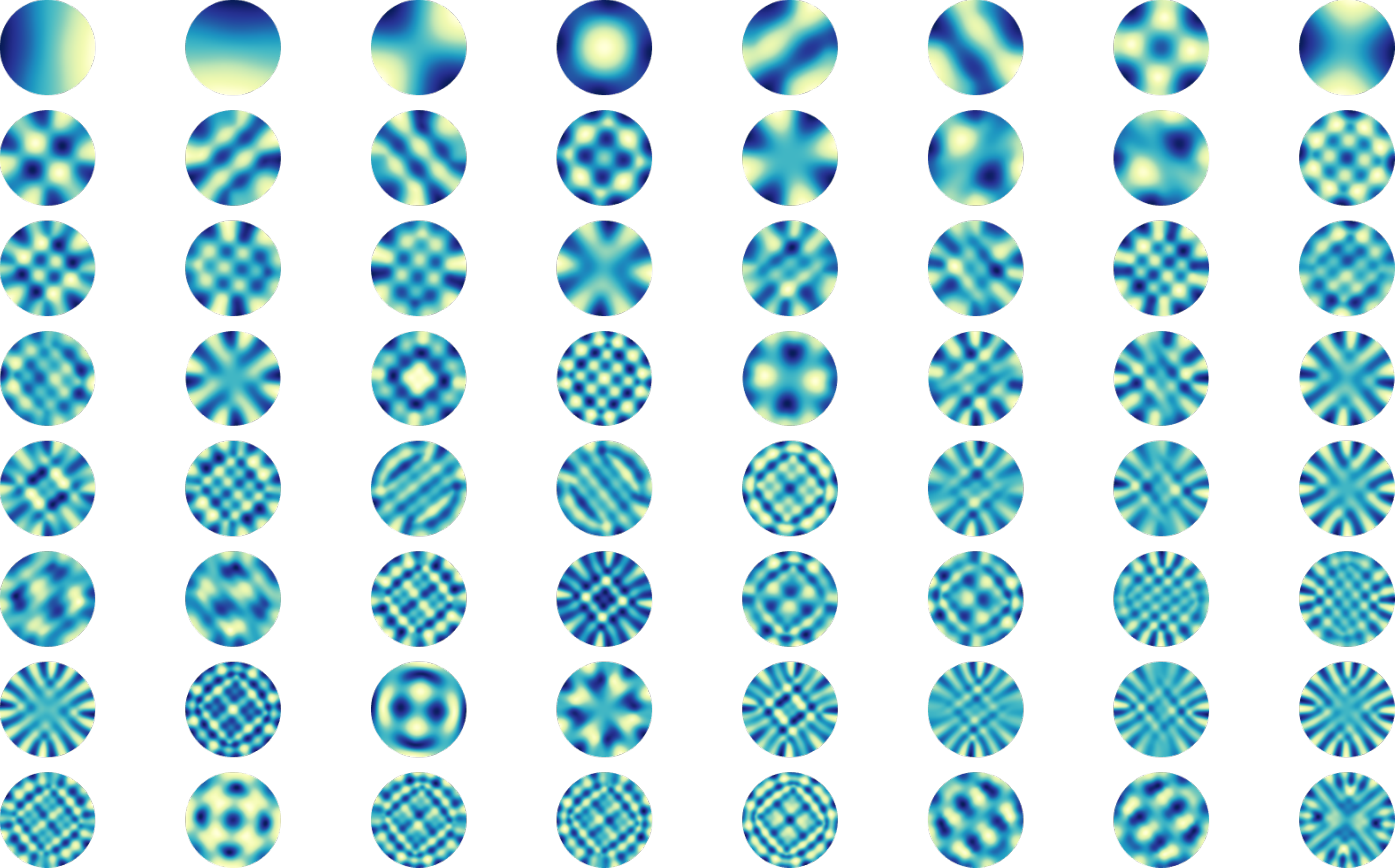}
\end{tabular}
\caption{This figure compares the first $64$ nonconstant eigenfunctions of operators associated to two frame fields that share the same structure, except that one is octahedral (left) and one is conformal octahedral (right), scaling to zero near its singularities. The eigenfunctions look very similar, but they appear in a different order.}
\label{fig:normed-comparison}
\end{figure*}

Many of the octahedral fields we have depicted in this paper include singularities, places where the field is ill-defined because octahedral fields have unit norm everywhere. Our theory does not explicitly deal with these singularities; instead, they are considered to be ``cut out'' of the domain $\Omega$. Conformal octahedral fields can explicitly represent singularities as zeroes (i.e., points $x$ where $\|T(x)\| = 0$). In \Cref{fig:normed-comparison}, we compare the frame field operators arising from a pair of fields, one of which is octahedral and the other conformal octahedral. The fields have identical structure because the octahedral field is simply given by normalizing the conformal octahedral field. Their eigenfunctions look similar, but they appear in a different order. A deeper investigation of the behavior of a frame field operator around singularities of its underlying frame field would be an intriguing topic for future work.

\section{Conclusion and Future Work}

Our work provides an initial link between two key areas of study in geometry processing:  spectral geometry and frame field design.  By moving from second-order to fourth-order, we are able to design a differential operator that captures the complex structure of frame fields in both planar regions and volumes.  

From a technical perspective, our work advances applications of mixed finite elements to a broader class of operators than have previously been considered in geometry processing, including a variety of boundary conditions.  While the experiments in \Cref{sec:validation} show that our discretization reaches the empirical standard of convergence needed for applications, theoretical proof of convergence in the limit of mesh refinement will be a challenging avenue for future research in numerical analysis, broadening the scope of isotropic results like \cite{scholz1978mixed,stein2019mixed}.   Our constructions also can be easily generalized to non-orthogonal frame fields, although design of such fields is largely an open problem in geometry processing.

Perhaps the most obvious next step of our research, however, will involve incorporating our operator into methods like those described in \Cref{sec:meshingrelatedwork} for quad and hex meshing. 
%
%In particular, our operator likely can be used to compute quad- and hex-dominant meshes by Morse-based methods. 
As our high-frequency eigenfunctions exhibit grid-like oscillatory behavior (see \Cref{sec:paramconnection}), we can introduce eigenproblems involving our operator into Morse-based meshing pipelines.  While engineering such a meshing pipeline may require substantial changes to heuristic steps of existing Morse-based algorithms, which depend heavily on the structure of the Laplacian operator specifically, the promise of linking Morse-based and field-based meshing is an enticing next-step beyond the simpler applications suggested in \Cref{sec:applications}.

It would also be interesting to explore frame field operators acting on vector or tensor fields. Replacing the Hessian in our variational problem with the differential of a vector field would be one simple way to do this, which would result in a second-order vectorial operator. We expect that the eigenfields of such operators would also show frame-aligned oscillations.

%Alternatively, we can solve the least squares problem $\min_u \|\mathcal{A}_{T,\epsilon} u - \lambda u\|_2^2$ subject to more general constraints, including second-order inhomogeneous boundary constraints such as that proposed in \citeme{}. Either approach results in a very simple and fast meshing strategy.

More broadly, our work suggests a new way to think about frame fields, direction fields, and other generalized vector fields studied in geometry processing. Associating operators to fields exposes a rich representation admitting a wide variety of tools for research and application---spectral methods and semidefinite programming come to mind. It may be possible to pose frame field design problems by optimizing spectral properties over the space of frame field operators. We hope that this new representation will enable new end-to-end methods in meshing and other domains.

\section*{Acknowledgments}

The authors would like to thank Mirela Ben-Chen and David Bommes for their thoughtful insights and feedback, as well as Xianzhong Fang and Jin Huang for helping with some speculative experiments.

David Palmer acknowledges the generous support of the Hertz Graduate Fellowship and the MathWorks Fellowship. This work is supported in part by the Swiss National Science Foundation's Early Postdoc.Mobility fellowship. The MIT Geometric Data Processing group acknowledges the generous support of Army Research Office grant W911NF2010168, of Air Force Office of Scientific Research award FA9550-19-1-031, of National Science Foundation grant IIS-1838071, from the CSAIL Systems that Learn program, from the MIT–IBM Watson AI Laboratory, from the Toyota–CSAIL Joint Research Center, from a gift from Adobe Systems, from an MIT.nano Immersion Lab/NCSOFT Gaming Program seed grant, and from the Skoltech–MIT Next Generation Program.

%-------------------------------------------------------------------------
% bibtex
% \bibliographystyle{eg-alpha-doi} 
% \bibliography{egbibsample}       

% biblatex with biber
\printbibliography
%-------------------------------------------------------------------------

\end{document}